\documentclass{aa}

\usepackage{natbib}
\bibpunct{(}{)}{;}{a}{}{,} %
\usepackage{amsmath}
\usepackage{wasysym}
\usepackage{marvosym}
\usepackage{txfonts}
\usepackage{mathtools}

\usepackage[pdftex]{graphicx}   
\usepackage{epstopdf}
\usepackage{graphicx}
\pdfoutput=1
\usepackage[pdftex]{color,xcolor}
\usepackage{hyperref}             
\hypersetup{pdfauthor=Christoph Mordasini}
\hypersetup{breaklinks=true,colorlinks=true,urlcolor=blue, linkcolor=blue,  citecolor=blue, bookmarksopen=true}
\usepackage{etoolbox}

\makeatletter

\patchcmd{\NAT@citex}
  {\@citea\NAT@hyper@{%
     \NAT@nmfmt{\NAT@nm}%
     \hyper@natlinkbreak{\NAT@aysep\NAT@spacechar}{\@citeb\@extra@b@citeb}%
     \NAT@date}}
  {\@citea\NAT@nmfmt{\NAT@nm}%
   \NAT@aysep\NAT@spacechar\NAT@hyper@{\NAT@date}}{}{}

\patchcmd{\NAT@citex}
  {\@citea\NAT@hyper@{%
     \NAT@nmfmt{\NAT@nm}%
     \hyper@natlinkbreak{\NAT@spacechar\NAT@@open\if*#1*\else#1\NAT@spacechar\fi}%
       {\@citeb\@extra@b@citeb}%
     \NAT@date}}
  {\@citea\NAT@nmfmt{\NAT@nm}%
   \NAT@spacechar\NAT@@open\if*#1*\else#1\NAT@spacechar\fi\NAT@hyper@{\NAT@date}}
  {}{}

\makeatother

\def\mearth{M_\oplus}

\def\msun{M_\odot}

\def\mcore{M_{\rm core}}

\def\f1{f_{\rm I}}
\def\mstar{M_*}

\def\rcore{R_{\rm core}}

\def\menv{M_{\rm env}}

\def\mpla{m_{\rm pla}}

\def\beq{\begin{equation}}
\def\eeq{\end{equation}}

\def\t2{\tau_{\rm II}}

\def\sigmas0{\Sigma_{\rm s,0}}
\def\mj{M_{\textrm{\tiny \jupiter }}}
\newcommand{\lj}{L_{\textrm{\tiny \jupiter}}}
\newcommand{\rj}{R_{\textrm{\tiny \jupiter}}}

\def\spf{s_{\rm pf}}
\def\lpf{L_{\rm pf}}

\def\lshock{L_{\rm shock}}
\def\lint{L_{\rm int}}

\def\kB{k_{\rm B}}

\newcommand{\mxy}{M_{\rm XY}}

\newcommand{\lsun}{L_{\odot}}

\newcommand{\mdotz}{\dot{M}_{\rm Z}}
\newcommand{\mdotxy}{\dot{M}_{\rm XY}}

\newcommand{\Z}{{\mathfrak{Z}}}
\newcommand{\rxy}{\rho_{\rm XY}}
\newcommand{\rz}{\rho_{\rm Z}}
\newcommand{\Epot}{E_{\textnormal{pot}}}
\newcommand{\mzenv}{M_{\rm Z,env}}
\newcommand{\rdotz}{\dot{\rho}_{\rm Z}}
\newcommand{\rdotxy}{\dot{\rho}_{\rm XY}}

\def\tauMisch{\tau_{\rm mix,~env}}
\def\tauMischlok{\tau_{\rm mix,~loc}}
\def\tauEin{\tau_{\rm impact,~glob}}
\def\tauEinlok{\tau_{\rm impact,~loc}}

\def\tauKH{\tau_{\rm KH}}

\def\mpla{m_{\rm pmal}}
\def\dpla{d_{\rm pmal}}
\def\rhopla{\rho_\bullet}
\def\HP{H_P}

\defcitealias{marleyfortney2007}{M07}
\defcitealias{spiegelburrows2012}{SB12}

\def\apjl{ApJ}                %
\def\apjs{ApJS}               %
\def\ao{Appl.Optics}          %
\def\aap{A\&A}                %

\def\({\left(}
\def\){\right)}
\def\<{\left<}
\def\>{\right>}

\definecolor{refco}{HTML}{547687}

\defcitealias{mordasinialibert2012b}{Paper~I}
\defcitealias{mordasinialibert2012c}{Paper~II}

\begin{document}

\title{Characterization of exoplanets from their formation III:\\ The statistics of planetary luminosities }

\author{C. Mordasini\inst{1}  \and G.-D. Marleau \inst{1} \and P. Molli\`ere\inst{2} }%

\institute{Physikalisches Institut, Universit\"at Bern, {Gesellschaftsstrasse 6}, CH-3012 Bern, Switzerland  \and Max-Planck-Institut f\"ur Astronomie, K\"onigstuhl 17, D-69117 Heidelberg, Germany }

\offprints{Christoph Mordasini, \email{christoph.mordasini@space.unibe.ch}.}
\date{Received 17.11.2016 / Accepted 01.08.2017}

\abstract
{This paper continues a series in which we predict the main observable characteristics of exoplanets based on their formation. In Paper I we described our global planet formation and evolution model that is based on the core accretion paradigm. In Paper II we studied the planetary mass-radius relationship with population syntheses.}
{In this paper we present an extensive study of the statistics of planetary luminosities during both formation and evolution. Our results can be compared with individual directly imaged extrasolar (proto)planets and with statistical results from surveys.}
{We calculated three populations of synthetic planets assuming different efficiencies of the accretional heating by gas and planetesimals during formation. We describe the temporal evolution of the planetary mass-luminosity relation. We investigate the relative importance of the shock and internal luminosity during formation, and predict a statistical version of the post-formation mass vs. entropy ``tuning fork'' diagram. Because the calculations now include deuterium burning we also update the planetary mass-radius relationship in time.}
{We find significant overlap between the high post-formation luminosities of planets forming with hot and cold gas accretion because of the core-mass effect. Variations in the individual formation histories of planets can still lead to a factor 5 to 20 spread in the post-formation luminosity at a given mass. However, if the gas accretional heating and planetesimal accretion rate during the runaway phase is unknown, the post-formation luminosity may exhibit a spread of as much as 2--3 orders of magnitude at a fixed mass. As a key result we predict a flat log-luminosity distribution for giant planets, and a steep increase towards lower luminosities due to the higher occurrence rate of low-mass ($M\lesssim10$--$40~M_\oplus$) planets. Future surveys may detect this upturn.}
{Our results indicate that during formation an estimation of the planetary mass may be possible for cold gas accretion if the planetary gas accretion rate can be estimated. If it is unknown whether the planet still accretes gas, the spread in total luminosity (internal+accretional) at a given mass may be as large as two orders of magnitude, therefore inhibiting the mass estimation. Due to the core-mass effect even planets which underwent cold accretion can have large post-formation entropies and luminosities, such that alternative formation scenarios such as gravitational instabilities do not need to be invoked. Once the number of self-luminous exoplanets with known ages and luminosities increases, the resulting luminosity distributions may be compared with our predictions.}


\titlerunning{Characterization of exoplanets from their formation. III.}
\authorrunning{C.\ Mordasini et al.}

\maketitle

\section{Introduction}

A relative newcomer to the toolbox of exoplanet discovery and characterization techniques,
direct imaging occupies an important niche with a potential for growth in the near future.
Indeed, direct detections are particularly sensitive to planets at large separations ($\approx$10-100~AU) from their host star, thus probing the outer architecture of planetary systems and complementing
radial-velocity, transit, and microlensing observations.
This can inform models of the migration and build-up of planets also in the inner regions ($\lesssim$10~AU),
with orbital and compositional consequences for instance for hot Jupiters or close-in super-Earths.
Moreover, the detection of photons originating in the (non-irradiated) atmosphere of these objects
affords precious information about their composition, chemistry \citep[e.g.,][]{laviemendoca2016}, and
various microphysical processes such
as dust growth and sedimentation, cloud formation, or possibly even lightning
(e.g., \citealp{buenzlimarley2015,buenzlisaumon2015,bonnefoyzurlo2016,baileyhelling2014}).
Spectroscopy can also serve to measure the temporal variability and,
at high resolution, the rotation period of these young ($\lesssim100$~Myr) objects \citep[e.g.,][]{snellenbrandl2014,zhouapai2016,allersgallimore2016,schwarzginski2016},
which are connected to their atmospheric dynamics.
A yet unexplored facet of exoplanet formation, the spin may also reflect the angular-momentum accretion history.

Several past and ongoing surveys have revealed a scarce but interesting population
of young gas giants on wide orbits \citep[e.g.,][]{chauvinlagrange2010,tamura2016,rameauchauvin2013,galichermarois2016}.
These surveys, which 
have various observing strategies and target different stellar masses\footnote{
On a related note, binary stars are usually excluded,
but the SPOTS survey \citep{thalmanndesidera2014}
targets them explicitly.
See also \citet{rodigasweinberger2015} and \citet{thomasbelikov2015}.
},
and other searches as well (e.g.,\ \citealp{durkhanjanson2016,bryanbowler2016}),
consistently find
that low-mass companions are rare.
In a recent meta-analysis,
\citet{bowler2016} derived a companion fraction
of  less than roughly one percent around all stars for masses from 5 to 13~$\mj$ and semi-major axes of 10 to 100 or 1000~AU \citep[see also][]{lannierdelorme2016}.
Ongoing surveys, principally the SPHERE GTO program, GPIES, LEECH, and Project~1640
\citep{beuzitfeldt2008,macintoshgraham2014,skemerhinz2014,hinkleyoppenheimer2011},
as well as the upcoming SCExAO system \citep{jovanovicmartinache2015},
should be able to increase the sample size and statistical significance of the results.

Direct imaging discoveries include
the well-studied $\beta$~Pic~b \citep{lagrangebonnefoy2010}
and HR~8799~bcde objects \citep{maroismacintosh2008,maroiszuckerman2010},
but also a handful of exciting recent additions:
HD~95086~b and
51~Eri~b with small lower mass limits of a few Jupiter masses, and HIP~65426~b \citep{rameauchauvin2013b,rameauchauvin2013c,derosarameau2016,macintoshgraham2015,samlandmolliere2017,chauvindesidera2017}.
As advances both in observational   %
and data-reduction techniques   %
make it possible to probe
closer in to the host stars, one could expect an increasing number of discoveries $N$
if the radial-velocity result that $dN/dP\approx-0.7<0$ \citep{cummingbutler2008},
derived for masses $M=0.3$--10~$\mj$ and periods $P=2$--2000~days,
holds also at separations relevant for direct imaging.  %

Direct observations measure an object's flux but the fundamental quantity relevant
for understanding its formation is its mass.
Therefore, interpreting these observations intrinsically relies on models.
In principle, one can derive mass and radius from the spectroscopy or photometry
by comparing to theoretical spectra.
In practice, however, there are several objects whose spectral appearance
currently cannot be explained, with non-equilibrium chemistry (linked to vertical mixing)
and patchy or variably thick cloud decks of ill-known composition (leading to a dusty photosphere)
thought to play a role.
Also, constraining the surface gravity of these objects has proven difficult
since atmospheric properties are relatively insensitive to it.
A more robust approach consists of
using model atmospheres to derive a bolometric luminosity
or, as in \citet{morzinskimales2015} for $\beta$~Pic~b,
deriving it most empirically by combining photometry of a sufficiently large part of its spectrum.
However, the issue is that to derive a mass from the bolometric luminosity,
the post-formation luminosity needs to be known;
theoretical studies of the last decade have made clear that this is still an open issue,
and in fact the single most important one in the field.

Predicting the mass--post-formation luminosity relationship of gas giants
requires modeling the energetics of gas accretion.
In particular, as first pointed out by \citet{marleyfortney2007},
one of the key aspects is the nature of the planetary gas accretion shock sitting on the surface of the planet during giant planet formation.
The details are reviewed in Sect.~\ref{sec:acc thermo}, and the extreme outcomes are
``cold starts'' and ``hot starts''
(``start'' referring to the beginning of the cooling, i.e.,\ the end of the accretion).
The issue is that the difference between the two, which in the first several Myr
can be as large as a factor $\sim1000$ in the luminosity \citep{marleyfortney2007}
or  {up to 8}~mag in the $K$ band {(\citealt{fortneymarley2008}}; \citealt{spiegelburrows2012}),
persists for the initial Kelvin--Helmholtz (cooling) timescale;
for cold starts, this is often larger than the age of the observed systems.
This is an important issue because, as \citet{marleaucumming2014} explore in detail,
a given brightness can be either explained by a low mass with a hot start,
a high mass with a cold start, or an intermediate combination.
Without further information on the mass,
this degeneracy between the initial luminosity
and the mass derived from a measurement of the current luminosity
cannot be lifted. %
Depending on the system, this leads to an uncertainty of a few to several $\mj$.
Therefore, any prediction of the post-formation luminosity of planets
would represent an important step.

In only the last few years,
cutting-edge observations
have caught
a small number of
putative low-mass companions
in their mass-assembly phase
(around LkCa~15, HD~142527, HD~100546, and HD~169142; \citealp{krausireland2012,sallumfollette2015,closefollette2014,quanzamara2013,reggianiquanz2014}).
These fascinating observations of planet formation as it happens provide an additional motivation to study
the luminosity of planets already during the formation phase,
whether or not their radiation can leave relatively unimpeded their natal circumstellar and -planetary disk(s).
Indeed, a part of the accretion luminosity could be directly visible in H\,$\alpha$ (e.g.,\ \citealp{sallumfollette2015}), or in the infrared  \citep{vanboekelhenning2016}
and the planet's radiative feedback is expected to change the local thermal and density structure of the disk \citep{montesinoscuadra2015,klahrkley2006} and thus its chemistry \citep{cleevesbergin2015}.
In the near future, this should be accessible to the unprecedented resolution and exquisite sensitivity
of the Atacama Large Millimeter
Array (ALMA; \citealp{wolf2008,cleevesbergin2015}). Future instruments like METIS at the ELT \citep{brandlfeldt2014} may be able to detect the vast population of low-mass planets during formation as found by comparing the detection capability of METIS with the luminosity of forming low-mass planets predicted here \citep{vanboekelhenning2016}.

Previous theoretical studies have looked at the evolution of a planet's luminosity,
from its beginnings as a rocky core to its cooling on Gyr scales, for different masses
\citep{marleyfortney2007,mordasinialibert2012b,bodenheimerdangelo2013,mordasini2013}.
However, this was only for often artificial simplifications such as a prescribed
gas accretion rate and a fixed semi-major axis, and for a limited set of conditions
(e.g., a given stellar mass, nebula temperature, and planetesimals surface density),
while \citet{mordasini2013} showed that the post-formation luminosity depends
critically on these assumptions, in particular on the planetesimal surface density.
Therefore,
to make realistic predictions
and enable
a statistically meaningful comparison to observations,
one needs to sample the whole range of input parameter values
and to analyze the outcome statistically.
This is precisely what population synthesis makes possible
and the subject of this work. The resulting fundamental statistical predictions like the planetary luminosity distribution  in time may be compared in future with the results of direct imaging surveys.

The contents of this paper are as follows. In Sect.~\ref{sect:model} we shortly present our global planet formation and evolution model. We also list the parameters that were used to generate the synthetic populations. Sect.~\ref{sect:synthplanpopus} describes the three populations studied in this work, the cold-nominal, the hot, and the cold-classical population. In Sect. \ref{sect:formevo5MJ} the simulation of the formation and evolution of a 5 $\mj$ planet taken from the cold-nominal synthetic population is discussed. Many of the findings for this individual planet are useful to understand the statistical population-wide findings in Sect. \ref{sect:statresults}, which is the main part of this paper. In this part we discuss three fundamental statistical properties, which is first the planetary mass--luminosity relation both during formation and evolution (Sect. \ref{sect:MLrelation}), the mass--entropy diagram at the moment when the protoplanetary disk disappears (Sect. \ref{sect:spflpf}),  and finally the luminosity distribution as a function of time (Sect. \ref{sect:Ldistro}). We furthermore revisit the mass--radius relation that was extensively discussed in \citetalias{mordasinialibert2012c} in Sect. \ref{sect:MRrelation}, now including the effect of deuterium burning. In Sect. \ref{sect:conclusions} we summarize our findings and give the conclusions. {The Appendices \ref{sect:fitS0} and }\ref{app:fit10MyrLSD} contain fits to the post-formation properties of the synthetic planets which are of interest as initial condition for evolution models. {In Appendix \ref{app:coremasseffect} we study the energy deposition into a protoplanet by planetesimal impacts which is important for the post-formation entropy of the planets.}

\section{Formation and evolution model}\label{sect:model}
Our global planet formation and evolution model was described in detail in several earlier publications (in particular \citealt{alibertmordasini2005}; \citealt[hereafter \citetalias{mordasinialibert2012b}]{mordasinialibert2012b};  \citealt[][hereafter \citetalias{mordasinialibert2012c}]{mordasinialibert2012c}). Therefore we  only give here a  short summary. Based on the core accretion paradigm, our planet formation model combines three standard elements:
\begin{enumerate}
\item A classical giant planet formation model very similar to the one described in \citet{pollackhubickyj1996} and \citet{bodenheimerhubickyj2000}. It calculates the growth of the solid core by the accretion of planetesimals based on the Safronov equation as well as the gas accretion rate and structure of the gaseous envelope by solving the 1D internal structure equations. The envelope structure is calculated both in the attached (or nebular) phase for planets with masses less than $10-100\mearth$ and in the detached (or transition) phase. The internal structure calculations in particular also yield the luminosity. In the attached phase the planet's gaseous envelope smoothly transitions into the background nebula. The planet's outer radius in this phase is on the order of the planet's Hill sphere or Bondi radius, whichever is smaller.  At the beginning of the detached phase, the planet's outer radius detaches from the disk as the disk can no more deliver enough gas given the contraction of the envelope to keep the envelope and disk in contact. This contraction becomes increasingly rapid as the core grows (runaway accretion). The outer radius then rapidly decreases to a value that is much smaller than the Hill sphere \citepalias[\citealt{bodenheimerhubickyj2000}, ][]{mordasinialibert2012b}, and the planet now has an actual surface. In the attached phase, the Kelvin--Helmholtz contraction of the envelope sets the gas accretion rate, while in the detached phase, the accretion rate is controlled by the protoplanetary disk rather than the planet itself  (disk-limited gas accretion phase, see \citetalias{mordasinialibert2012b}). After a short transition phase with Bondi-limited accretion, the gas accretion rate in this phase is set equal to 0.9 the non-equilibrium gas accretion rate in the local protoplanetary disk \citepalias{mordasinialibert2012b}.   
\item  A standard 1+1D $\alpha$ model for the protoplanetary gas disk \citep{lynden-bellpringle1974,papaloizouterquem1999} including stellar irradiation \citep{fouchetalibert2012} and internal and external photoevaporation  \citepalias{mordasinialibert2012c}. 
\item Orbital migration due to angular momentum exchange with the protoplanetary gas disk \citep{goldreichtremaine1980,linpapaloizou1986}, including both non-isothermal type I and type II migration \citep{dittkristmordasini2014}. 
\end{enumerate}
After the protoplanetary disk has disappeared, the planet evolution model which was described in detail in \citetalias{mordasinialibert2012b} and \citetalias{mordasinialibert2012c}  calculates the long-term thermodynamical evolution (cooling and contraction)  up to an age of 10 Gyr. It is self-consistently linked to the formation model by directly using the internal structure of the planets at the end of the formation phase. This direct link  yields in particular the post-formation entropy $\spf$ in the deep convective zone of a giant planet. In the evolutionary model  simple gray outer boundary conditions are employed which yield however cooling tracks that are generally in good agreement  with non-gray atmospheric models \citepalias[\citealt{bodenheimerhubickyj2000},][]{mordasinialibert2012b}. In the convective regions standard zero-entropy gradient convection is assumed. Deuterium burning is included \citep{mollieremordasini2012} while atmospheric evaporation is neglected as we are interested in massive planets which are mostly unaffected by this process \citep{jinmordasini2014}. The most important settings and parameters of the formation and evolution model can be found in Table \ref{tab:popsynth}.

\subsection{Thermodynamics of the accretion process}
 \label{sec:acc thermo}
The thermodynamics of the accretion process of gas and planetesimals set the post-formation entropy and therefore luminosity, radius, {and magnitudes} of a planet  (\citealt{bodenheimerhubickyj2000,fortneymarley2005,marleyfortney2007}; {\citealt{fortneymarley2008}}; \citealt{spiegelburrows2012,mordasinialibert2012c,mordasinialibert2012b,mordasini2013,bodenheimerdangelo2013,owenmenou2016,berardocumming2016,szulagimordasini2016,marleauklahr2016}). 

Regarding  the accretion of gas, during the disk-limited gas accretion phase ($M_{\rm planet}>10-100\mearth$) where giant planets gain most of their mass, gas falls, in the 1D spherically symmetric picture, in near-free fall from the Hill sphere onto the planet's surface where it shocks. The recent 3D global radiation-hydrodynamic simulations of  \citet{szulagimordasini2016} show that (polar) accretion shocks on the circumplanetary disk and protoplanet are crucial for a planet's thermodynamic evolution also in a more realistic accretion geometry.

In the current absence of detailed radiation-hydrodynamic simulations of the planetary gas accretion shock with a realistic EOS \citep[see][for 1D results with an ideal EOS]{marleauklahr2016},  we consider the limiting cases of either completely cold accretion, where the entire gas accretion shock luminosity is radiated away in a supercritical shock, or of  completely hot accretion where no radiative losses occur (subcritical shock), as described in \citetalias{mordasinialibert2012b}. As for stars \citep{hartmanncassen1997,baraffechabrier2009}, such cold (hot) accretion lead to the accretion of low (high) entropy gas which in turn leads to giant planets with low (high) post-formation luminosities and small (large) radii. It is interesting to note here that for stars, radiation-hydrodynamic simulations show that the accretion shock onto the first Larson core ($M\sim200\mj$, $R\sim10$ AU, $\dot{M}\sim10^{-5}$ $M_{\odot}$/yr) is supercritical \citep{commerconaudit2011b,vaytetaudit2012}, while the accretion shock of the second Larson core is subcritical and almost all the energy from the infalling material is absorbed into it \citep{tomidatomisaka2013,vaytetchabrier2013}. The second Larson core initially shares some properties with an accreting giant planet ($M\sim 1\mj$, $R\sim6\rj$, compare with Fig. \ref{fig:MRhot}), but with $\dot{M}\sim10^{-1}$ $M_{\odot}$/yr its accretion rate is about seven orders of magnitude higher. The recent study of the planetary gas accretion shock in the 1D spherically symmetric approximation by \citet{marleauklahr2016} has shown that for an ideal EOS, the accretion shock is supercritical, but that it may still lead to a significant advection of thermal energy into the planet due to energy recycling by the infalling gas. Therefore, both limiting cases of completely hot and cold accretion are currently of interest.

Regarding the accretion of planetesimals,  we assume in all except one population (see below) that  planets continue to normally accrete planetesimals in the detached phase as found from the Safronov equation, and that the planetesimals always sink to the solid core {(sinking approximation, \citealt{pollackhubickyj1996})}. As demonstrated  in \citet{mordasini2013} and \citet{bodenheimerdangelo2013}, a high   planetesimal accretion luminosity associated with the formation of a massive core leads to a high post-formation entropy (and luminosity) even for a completely cold accretion of the gas due to the self-amplifying mechanism found in \citet{mordasini2013}. For high core masses ($\sim 100 \mearth$), the post-formation luminosities even approach those in classical hot start simulations.  The very low post-formation luminosities of \citet{marleyfortney2007} were in contrast obtained under the assumption that the accretion of planetesimals  artificially stops when the planet detaches, resulting in low-mass cores. 

As an additional source of luminosity, deuterium burning in massive planets \citep{baraffechabrier2008a} is also included as described in \citet{mollieremordasini2012}. The (small) radiogenic core luminosity is included as well \citepalias{mordasinialibert2012b}, but the thermal cooling of the core is currently neglected. During the formation phase, the luminosity resulting from the accretion of planetesimals is anyway usually much bigger than the core's cooling. But it leads to an underestimation of the luminosity of low-mass, core-dominated planets with  masses $\lesssim30\mearth$ during the evolutionary  phase at constant mass \citep{lopezfortney2014}. We therefore restrict this work mostly to giant planets where the luminosity is strongly dominated by the cooling of the H/He envelope also during evolution. %

\begin{table}
\caption{Parameters and settings for the model.}\label{tab:popsynth}
\begin{center}
\begin{tabular}{l c}
\hline \hline
Quantity & Value  \\ \hline
Stellar mass & 1 $\msun$\\
Disk viscosity parameter $\alpha$ & $7\times 10^{-3}$\\
Disk photoevaporation  &  int. \& external (\citetalias{mordasinialibert2012c}) \\
Irradiation for disk temperature profile & included \\
Number of embryos per disk & 1 \\
Embryo starting mass & 0.6 $\mearth$ \\
Core accretion rate & \citet{alibertmordasini2005} \\
Planetesimal size & 100 km\\
Fate of dissolved planetesimals & sink to core interface \\
Core density & variable  (\citetalias{mordasinialibert2012c}) \\
Envelope type & primordial H$_{2}$/He\\
Envelope evaporation & not included  \\
Atmosphere type during evolution& Eddington gray (\citetalias{mordasinialibert2012b})\\ 
ISM grain opacity reduction factor & 0.003 \\
Radiogenic luminosity & included\\ 
Deuterium burning &  included  \\ 
Accretion shock luminosity treatment & fully cold/hot accretion \\
$\partial l/\partial r$ in the envelope & zero \\
Simulation duration & 10 Gyr \\ \hline
\end{tabular}
\end{center}
\end{table}%

In this study we work in a strictly 1D spherically symmetric approximation, meaning in particular that we do not consider the presence of a circumplanetary disk. Multi-dimensional hydrodynamic simulations have found more complex accretion geometries involving preferential accretion at high latitudes \citep[e.g.,][]{AyliffeBate2012,TanigawaOhtsuki2012,GresselNelson2013,SzulagyiMorbidelli2014,szulagimordasini2016}. While the total accretional luminosity should remain similar in such more complex geometries for a fixed planetary mass, radius, and gas accretion rate, the characteristic surface over which it is radiated may change, meaning that the SED would be different than when assuming that the accretional luminosity originates homogeneously from the planet's entire surface. This could have important observational consequences \citep[e.g.,][]{ZhouHerczeg2014,eisner2015,zhu2015,szulagimordasini2016} and must therefore be critically kept in mind.

\subsection{{Simplifications and limitations of the internal structure model}}\label{sect:limintstruct}
{In this work we have modeled the interiors of the planets and their evolution under the standard assumptions of  fully convective adiabatic interiors without net rearrangements of matter. In the solar system, the evolution to the present-day luminosity of 2 of the 4 giant planets (Jupiter and Neptune) can be well reproduced with such simple models \citep{fortneyikoma2011}. For the other two, namely Saturn which is brighter relative to these models by about 60\% \citep{fortneyikoma2011} and Uranus which is fainter by at least one order of magnitude \citep{guillotgautier2014}, additional physical effects must play a role that we do not consider here. These effects could be: a demixing of different chemical species followed by gravitational settling causing a net rearrangement of matter like a helium rain, which could explain Saturn \citep{stevensonsalpeter1977}, compositional gradients \citep[e.g.,][]{vazanhelled2016} that can lead to semiconvection and non-adiabatic interiors, which could explain both Saturn \citep{lecontechabrier2013} and Uranus \citep{nettelmannhelled2013}, and finally core erosion where a part of the planet's luminosity is used to dredge up core material \citep[e.g.,][]{guillotstevenson2004}. }

\subsubsection{Simplifications for the EOS and statistical imprints}
{Furthermore, we have assumed that the envelope can be describe by the H/He equation of state of \citet{saumonchabrier1995}, i.e., we have neglected the enrichment by heavy elements. For low-mass planets, this enrichment mostly by water (for a formation outside of the iceline) can be very high \citep{fortneymordasini2013}. For giant planets that are at the focus of this paper because of their better detectability, the planet metallicity $Z$ should usually be rather low at least during the evolution phase with $Z\sim0.1$ \citep{ThorngrenFortney2016}. }

{These processes, if  occurring frequently in a statistical sense,  imprint into the  statistical quantities like the luminosity distribution or mass-luminosity relation that we study here. Our theoretical results can then serve as a baseline comparison sample the difference to which could be used as a diagnostics of the aforementioned processes. A challenge could be that in contrast to the solar system, the data on extrasolar planets is often incomplete. For example, to understand whether a luminosity differs from the prediction of simple models, a sufficiently exact  knowledge of the planetary age or mass is necessary. }

\subsubsection{{Uncertainties related to the core-mass effect}}
{Another simplification in the model is the sinking approximation that is used for all solids that are accreted into the planets.}

{During the formation of a giant planet, the accretion of planetesimals can go through a (second) maximum at the moment when the planet starts to accrete gas in a runaway fashion. At this moment, envelope growth leads via the increasing Hill sphere to an extension of the feeding zone of planetesimals \citep[][]{zhoulin2007,shiraishiida2008,mordasini2013}. This process can significantly increase the total mass of solids that is accreted into the planet \citep{helledlunine2014}. It also leads, at least in the sinking approximation used here, to a significant energy input into the planet during the runaway and early disk-limited gas accretion phase. This has, as mentioned, important effects on the  post-formation entropy via the core-mass effect: as found independently  by \citet{mordasini2013} and \citet{bodenheimerdangelo2013}, the stronger pressure support in a planet with higher planetesimal accretion luminosity $L_{\rm pla}$ leads during the early disk-limited gas accretion phase to a larger radius of the planet and therefore to a lower gas accretion shock luminosity $L_{\rm shock}$. A lower $L_{\rm shock}$ means for cold gas accretion that gas of higher entropy is incorporated into the planet which in turn leads to a slower decrease of the radius, such that the mechanism is self-amplifying (see also \citealt{berardocumming2016} for the importance of the entropy value at detachment). }

{This core-mass effect means that high luminosities are found for large core masses even for fully cold gas accretion where the entire gas accretion shock luminosity is radiated away. The high luminosities are almost comparable to those found for hot gas accretion (see Sect. \ref{sect:spflpf}). This could mean that the expected outcome for core accretion are quite bright planets, and not the very low  luminosities found in \citet{marleyfortney2007} where only low core masses $\lesssim17\mearth$ were considered. In the nominal population synthesis discussed below, we find core masses that are usually significantly higher (see Fig. \ref{fig:MLPostMcoreCD777}). These high heavy element contents agree at least in an approximate way with the observationally inferred values \citep{ThorngrenFortney2016} as discussed in detail in Sect. \ref{sect:impactcoremass}. So from this point of view, the core-mass effect could indeed be at work in reality.}

{However, there is an important caveat to this: High luminosities because of the core-mass effect require not only high heavy element contents nowadays, but also that the solids are accreted rapidly during the late attached and early detached phase, and that these solids sink quickly deep into the potential well to provide a strong accretional heating of the planet's interior \citep{mordasini2013}. }

{This has not yet been studied with giant planet formation models tracking the thermodynamical and compositional evolution of the interior (planetesimal accretion, dissolution, sinking, mixing), representing an important open question. First works have recently started to study the compositional aspect \citep{venturinialibert2016,lozovskyhelled2017}, but they do not yet address the consequences for the luminosity. Because of this, it is currently also not clear if, for the core-mass effect to work efficiently, a high total heavy element content today (in the envelope and/or core) is sufficient as implicitly assumed in the argument above. This is because it is currently unclear whether a centrally concentrated distribution of heavy elements (in a solid core or a strongly enriched inner region) at the beginning of detachment could shift to a more homogeneous distribution over time as heavy elements are soluble in H/He under conditions typical for giant planet interiors \citep[e.g.,][]{soubiranmilitzer2015}, or alternatively, that heavy elements sink over long timescale to the center. The results of \citet{vazanhelled2016} indicate that no full mixing occurs if the initial compositional gradient is strong enough. If the heavy element mass fraction as a function of enclosed mass $Z(M)$ in a planet today (see the Juno spacecraft data, \citealt{wahlhubbard2017}) therefore reflects at least partially the structure during build-up, this would open an interesting avenue to constrain the ratio of the solid accretion to the gas accretion rate as a function of planet mass, $\dot{M}_{\rm Z}/\dot{M}_{\rm XY}(M)$. This would obviously be of high interest in the context of the core-mass effect.}

{To investigate the efficiency of the core-mass effect if the solids do not sink to the core, we present in Appendix~\ref{app:coremasseffect} a preliminary study based on analytical considerations using polytropic models and numerical results. We find that a homogeneous mixing of the solids into the envelope instead of sinking still provides a significant energy source, at least for the cases we studied. The reduction factors of the heating relative to the sinking case (2-3 for homogeneous mixing, 3-8 for no sinking) are not very large. This could imply that the highest luminosities caused by the core-mass effect are reduced, but that the luminosities still do not become very low as found for no impact heating at all as in \citet{marleyfortney2007}. This is the observationally most relevant question, as hot and ``hotter'' starts converge rapidly, whereas cold starts differ for a long time. However, it is clear that future work must address this important question more thoroughly. In that sense, the high luminosities because of the core-mass effect found here must also be confirmed by future work.}

\subsection{Nomenclature for the luminosities}
In the following sections, several different sources of luminosity are addressed, for which we use the following nomenclature:  $L_{\rm HHe}$, the usual luminosity resulting from the cooling and contraction of the gaseous H/He envelope (material that is already part of the planet); $L_{\rm radio}$, the radiogenic luminosity in the solid core;  $L_{\rm D}$, the deuterium burning luminosity; $L_{\rm pla}$, the luminosity due to the accretion of planetesimals during the formation phase; and $L_{\rm shock}$, the  luminosity due to the gas accretion shock during the formation phase. With internal luminosity $L_{\rm int}=L_{\rm HHe}+L_{\rm radio} + L_{\rm D}+L_{\rm pla}$ we denote the luminosity that is generated in the planet's interior. The total luminosity $L$ which we address most frequently additionally includes the accretion shock luminosity that is generated at the planet's surface, so that for cold accretion $L=L_{\rm int}+L_{\rm shock}$. It should be noted that for hot accretion $\lshock=0$, as no shock luminosity escapes in this case. The accretional energy is rather deposited into the planet and radiated later on. In the evolutionary phase, i.e., after the dissipation of the protoplanetary disk, when $L_{\rm shock}=0$ and $L_{\rm pla}$=0, $L=L_{\rm int}$.

\section{Synthetic planetary populations}\label{sect:synthplanpopus}
For the statistical study of the planetary luminosities, and in particular the global impact of cold and hot gas accretion and of massive cores, we conduct planetary population syntheses. For them, four fundamental initial conditions are varied in a Monte Carlo way: the disk metallicity, the initial disk gas mass, the external photoevaporation rate which controls together with viscous dissipation the disk lifetime, and the initial starting position of the planetary embryo. The \citet{shakurasunyaev1973} disk viscosity parameter $\alpha$ is fixed to $7\times10^{-3}$. The population synthesis method and the probability distributions from which the four Monte Carlo variables are drawn were described in \citet{mordasinialibert2009a}. We  consider the following three synthetic planetary populations:

\begin{enumerate}
\item \textbf{The cold-nominal population CD753.} It assumes completely cold gas accretion, i.e., supercritical shocks radiating all accretional luminosity, and includes the accretion of planetesimals in the detached phase, assuming that planetesimals always sink rapidly to the  core. This population is identical to the nominal one presented in \citetalias{mordasinialibert2012c} with the difference that deuterium burning is now included.
\item \textbf{The hot population CD752.} It is identical to the cold-nominal population with the only difference that completely hot gas accretion is assumed (subcritical shocks, all accretional energy is added to the planets' interior). In \citetalias{mordasinialibert2012b} it was found that under this assumption, giant planet formation by core accretion leads to high post-formation entropies that are traditionally often associated only with a formation via the direct collapse (gravitational instability) mechanism \citep[e.g.,][]{galvagnihayfield2012}.  This shows that for the post-formation entropy the structure of the accretion shock is as important as the fundamental formation mechanism. In our model the treatment of the accretion shock energetics does not influence the gas accretion rate, therefore some giant planets in the cold and hot populations can be cross-matched (see Fig. \ref{fig:mlevocoldhot}). Note however that the total number of planets differs in the two populations. The accretion rate of planetesimal is in contrast slightly influenced by the thermodynamics: for hot accretion, the planet's capture radius for planetesimals remains larger during the detached phase, allowing the planets to accrete more planetesimals. The impact is however very small, with differences in the final core masses of 5\% or less. 
\item \textbf{The cold-classical population CD777.} As for the cold-nominal population, it assumes that the gas accretion is completely cold. But in contrast to the cold-nominal population, it is here assumed that planetesimal accretion stops artificially once a giant planet enters the disk-limited gas accretion (detached) phase. This causes the maximal core masses to be lower than in the cold-nominal population. Therefore the effect that massive cores make hot planets \citep{mordasini2013} is of reduced importance in this population. Furthermore, orbital migration is not included in this population so that planets form in situ. Both these assumptions are the same as in the classical work of \citet{marleyfortney2007} where planets with a very low entropy were found to form. This population was already studied in \citet{mordasiniklahr2014}, where a further description can be found.
\end{enumerate}

The populations CD752 and CD753 are publicly available on DACE, the Data Analysis Centre for Exoplanets of the NCCR PlanetS reachable at \url{https://dace.unige.ch}. Additional populations will be added in future. DACE yields both interactive snapshots of the entire population at a given moment in time like the population-wide $a-M$ or $M-R$ diagrams and formation tracks of individual planet (e.g., $M(t)$, $R(t)$, etc.) for all synthetic planets.

An important idealization of the syntheses presented here is that they assume that only one embryo per disk forms (see \citealt{alibertcarron2013} for the impact of the concurrent formation of many embryos). While this should not usually affect the thermodynamics of the accretion process itself, it can change the formation tracks in particular in systems where several giant planets form. Future work will study the impact of multiplicity on the planetary luminosities and address in particular the predictions for the number and luminosities of giant planet scattered to large orbital distances.

\section{Formation and evolution of a 5 $\mj$ planet}
  \label{sect:formevo5MJ}
We now turn to the results obtained with the framework introduced in the previous two sections. But before we address the statistical population-wide results, we study the formation and evolution of an individual planet taken from the cold-nominal population. This detailed analysis is useful to understand the statistical results presented later in Sect. \ref{sect:statresults}. 

\begin{figure*}
\begin{center}
\begin{minipage}{0.5\textwidth}
	      \centering
        \includegraphics[width=0.95\textwidth]{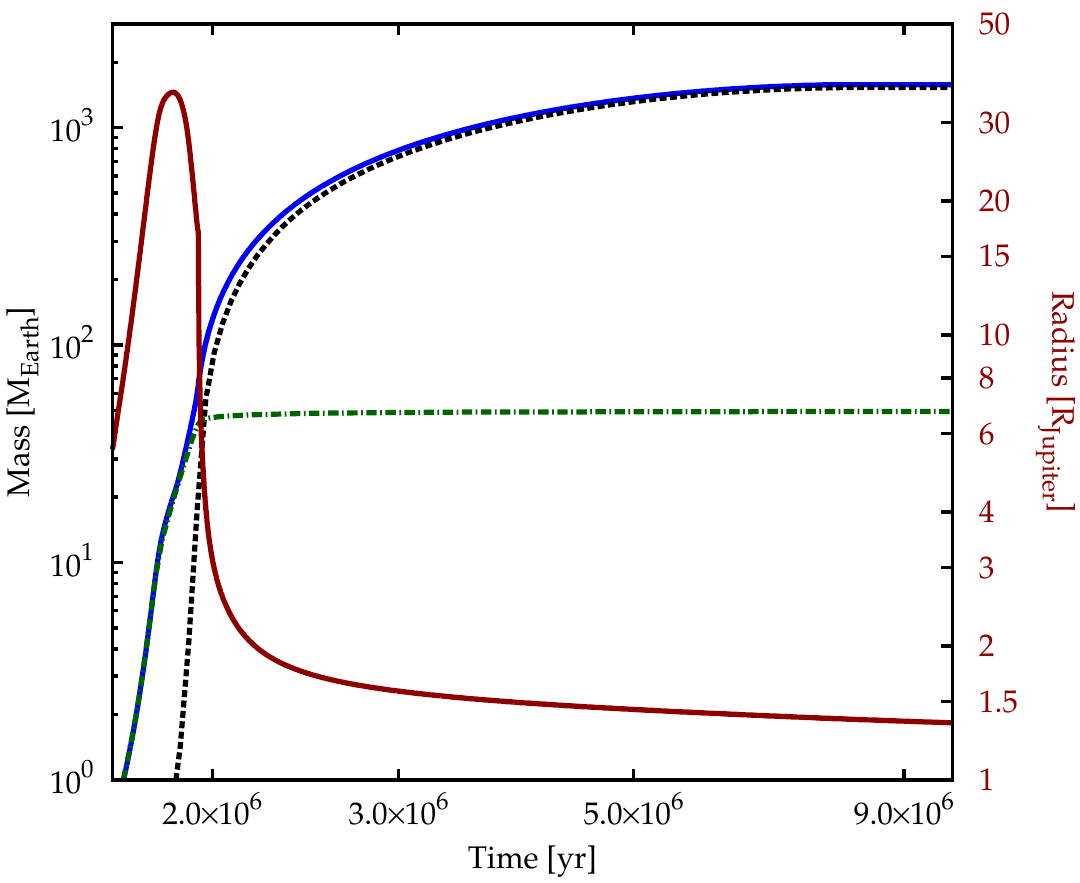}
     \end{minipage}\hfill
     \begin{minipage}{0.5\textwidth}
      \centering
       \includegraphics[width=0.9\textwidth]{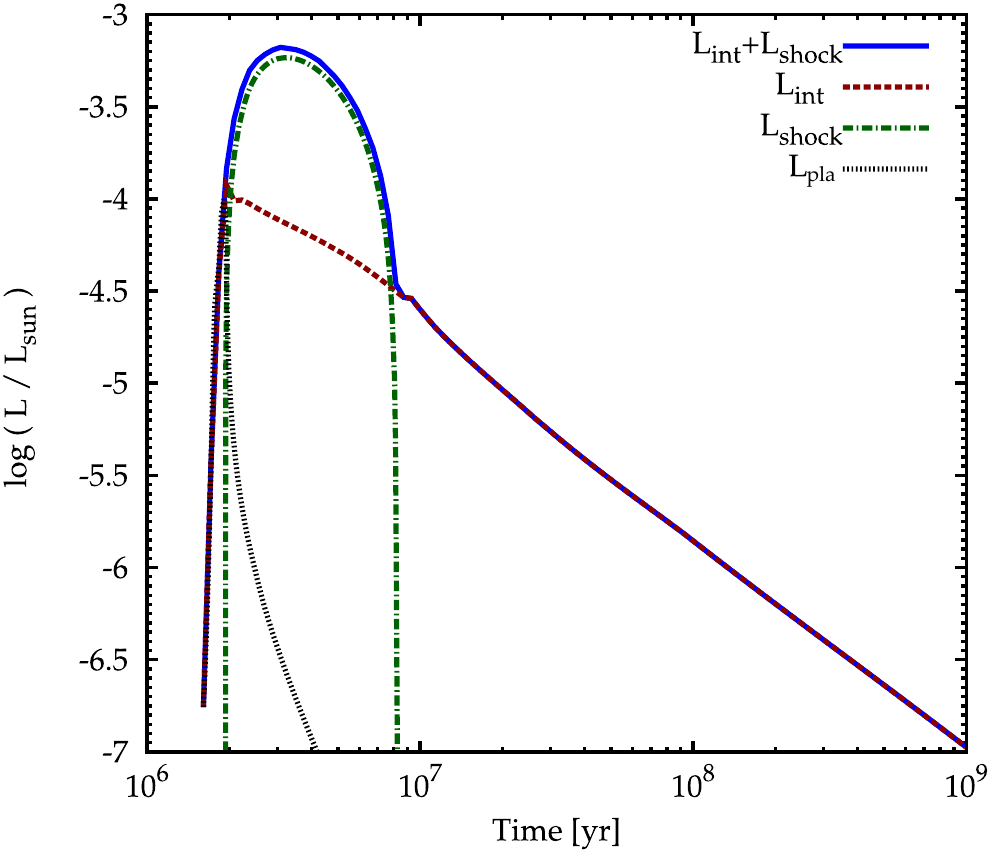}
     \end{minipage}
\caption{Illustrative example of the formation and evolution of a 5 $\mj$ planet. \textit{Left panel}: Total (blue solid), core (green dash-dotted), and envelope mass (black dashed line) as a function of time during the formation phase. These lines belong to the left y-axis. The solid brown line shows the planetary radius (right y-axis). \textit{Right panel}: Luminosity as a function of time during the formation and subsequent evolutionary phase. The total luminosity (blue solid) and the individual contributions from the internal luminosity (brown dashed), the accretion shock (green dash-dotted), and the planetesimal accretion (black dashed line) are shown. }\label{fig:tML5MJ}
\end{center}
\end{figure*}
\subsection{Temporal evolution of the mass and radius}
The left panel of Figure \ref{fig:tML5MJ} shows the temporal evolution of the core, envelope, and total mass of a giant planet starting to form at 4.6 AU. As typical for the core accretion mechanism, in a first step, the core is formed. Thanks to migration, no lengthy ``phase II'' occurs as no isolation of the core happens \citep{alibertmordasini2004}. Instead, due the large feeding zone and the high surface density in this disk (about 12 times the MMSN) the core continues to grow and reaches a mass of 45 $\mearth$ at the crossover point which occurs at 1.96 Myr. At the crossover point, core and envelope mass are equal. In contrast to the simulations of \citet{pollackhubickyj1996}, because of the large core mass, the crossover point occurs slightly after the moment when the planet enters the disk-limited gas accretion phase at 1.94 Myr. The gas accretion rate reaches a maximum of about 0.0012 $\mearth$/yr at this point, about an order of magnitude smaller than the often assumed 0.01 $\mearth$/yr in the disk limited phase  \citep[e.g.,][]{hubickyjbodenheimer2005}.  The total mass of the planet is then 66 $\mearth$.  Afterwards, the gas accretion rate decreases roughly as $1/t^{2}$ as the disk dissipates because of accretion onto the star and photoevaporation. These rather low gas accretion rates have important implications as discussed below. As in earlier simulations, we do not reduce the planetary gas accretion rate due to gap formation because of the eccentric instability found by  \citet{kleydirksen2006}, but assume that the planetary accretion rate is simply 0.9 times the local gas accretion rate in the protoplanetary disk \citepalias{mordasinialibert2012b}.  The disk disappears at about 8.6 Myr so that the accretion of gas and solids stops. This relatively long disk lifetime \citep[][but see also \citealt{pfalznersteinhausen2014}]{haischlada2001} reflects the correlation of long disk lifetimes and a high probability of the formation of a massive giant planet \citep{mordasinialibert2012a}. The final total mass of the planet is about 1578 $\mearth$ (4.97 $\mj$), while the final core mass is 49.4 $\mearth$. 

The temporal evolution of the planetary radius is also shown in the left panel. During the attached phase, the radius is approximately the smaller of the Bondi and of one third of the Hill sphere radius \citep{lissauerhubickyj2009} that is given for a planet of mass $M$ and semimajor axis $a$ around a star mass of mass $M_{\star}$ as 
\beq
R_{\rm H}=\left(\frac{M}{ 3 M_{\star}}\right)^{1/3}a.
\eeq
 It therefore initially increases as the planet grows in mass. But after about 1.8 Myr, it decreases as the reduction of the Hill sphere by  the planet's inward migration becomes dominant. At 1.94 Myr the planet detaches from the disk as it enters the disk limited gas accretion phase, which is visible as a change in slope in the curve. The radius then decreases rapidly \citep{bodenheimerhubickyj2000} to reach a value of about 1.4 $\rj$ at the end of the formation phase, which agrees very well with the value given by \cite{spiegelburrows2012} for a planet of this mass and entropy. This radius is intermediate between the radii predicted by \citet{marleyfortney2007} which are about 1.25 $\rj$ for a classical (very) cold start, and about 2.1 $\rj$ for a hot start. 

\subsection{Temporal evolution of the luminosity}
The right panel of Fig. \ref{fig:tML5MJ} shows the luminosity as a function of time during both the formation and subsequent evolution phase at constant mass. The internal, shock and planetesimal accretion luminosity are shown separately. For this planet both the radiogenic and deuterium burning luminosity are of negligible importance. At the beginning, the planet's luminosity is dominated by the accretion of the planetesimals $L_{\rm pla}$. Since we assume that planetesimals sink to the core  
\beq
L_{\rm pla}\approx\frac{G M_{\rm core} \dot{M}_{\rm core}}{R_{\rm core}}
\eeq
where $G$ is the gravitational constant, $M_{\rm core}$ the core mass, $\dot{M}_{\rm core}$ the protoplanet's planetesimal accretion rate, and $R_{\rm core}$ the radius of the solid core. {As discussed in Appendix \ref{app:coremasseffect} this expression is strictly speaking only applicable when the envelope mass is negligible compared to the core mass.} The planetesimal accretion luminosity reaches a maximal value of about $10^{-4} \lsun$, about a factor 10 higher than in \citet{pollackhubickyj1996} which is due to the higher planetesimal surface density and therefore higher $\dot{M}_{\rm core}$, and the higher $M_{\rm core}$ itself. After about 2 Myr, $L_{\rm pla}$ decreases again because of two reasons: first, it is a consequence of the decrease of the planetesimal capture radius in the detached phase \citepalias{mordasinialibert2012b} which enters quadratically in  $\dot{M}_{\rm core}$. Second, at about this moment, the planet passes into slower type II migration which also reduces the core accretion rate because the planet now migrates slower into parts of the disk with a high planetesimal surface density. 

{The} high energy input by planetesimal accretion during the runaway and early disk-limited gas accretion phase has {via the core-mass effect (see discussion in Sect. \ref{sect:limintstruct}) the important} consequence that despite the cold accretion of gas, the planet has a post-formation entropy and therefore luminosity that is significantly higher than in the classical (very) cold start simulation of \citet{marleyfortney2007}. In this latter paper, only low core masses of 17-19 $\mearth$ were considered. In the specific case, the post-formation luminosity is approximately $\log(L/\lsun)=-4.5$, whereas \citet{marleyfortney2007} had for 5 $\mj$ about $\log(L/\lsun)=-5.6$ or about a factor 13 less, an observationally  relevant difference.

Figure \ref{fig:tML5MJ} also shows that during the disk-limited gas accretion phase where the planets gains almost all of its mass, the gas accretion shock luminosity dominates over the internal luminosity by almost one order of magnitude (factor 8.5). The accretion shock luminosity of a planet with mass $M$ and radius $R$ is calculated as
\beq
L_{\rm shock}=G M \dot{M}_{\rm XY}\left(\frac{1}{R}-\frac{1}{R_{H}}\right)
\eeq
where $\dot{M}_{\rm XY}$ is the planet's gas accretion rate. For cold accretion, this dominance of $L_{\rm shock}$ is  typical  for giant planets in their main growth phase as we will see in the statistical analysis below (Sect. \ref{sect:lshocklint}). 

The maximum gas accretion rate in this simulation is, as mentioned, only 0.0012 $\mearth$/yr at the beginning of the disk-limited phase at 1.94 Myrs, followed by even lower $\dot{M}_{\rm XY}$ during the subsequent 7 Myrs during which the planet grows from 66 $\mearth$ to its final mass. This means that the time-averaged gas accretion rate is only about $2\times10^{-4}$ $\mearth$/yr. Such a gas accretion rate is clearly lower than the value of $10^{-2}$ $\mearth$/yr assumed in \citet{marleyfortney2007}. In the syntheses, it is found that such lower disk-limited gas accretion rates (a few $10^{-4}$ to $10^{-3}$ $\mearth$/yr) are actually the typical case (see Fig. \ref{fig:MLLacc3Myr}). This has important implications for the observability of the planets in this phase: instead of short spike of very high luminosity as in \citet{marleyfortney2007}, the planets are typically less bright in our simulations, but retain this level during several million years. For a 5 $\mj$ planet for example, \citet{marleyfortney2007} find a luminosity of up to $\log(L/\lsun)=-1.5$ using an arbitrary gas accretion rate of  $10^{-2}$ $\mearth$/yr, but the high luminosity phase ($\log(L/\lsun)>-4$) only lasts about $3\times10^{5}$ years. In our simulation where the gas accretion rate is given self-consistently by the disk model, the planet only reaches a peak luminosity of about $\log(L/\lsun)=-3.15$, but retains a $\log(L/\lsun)>-4$ during more than 5 Myrs. The specific form of $L_{\rm shock}$ as a function of time (Fig. \ref{fig:tML5MJ}, right panel) is given by the interplay of the gas accretion rate and the planet mass as $L_{\rm shock}\propto M \dot{M}_{\rm XY}$. Initially $L_{\rm shock}$ is lower because the planet's mass is still low. As its mass grows, also $L_{\rm shock}$ increases. But at the same time $\dot{M}_{\rm XY}$ decreases in time as the disk gradually dissipates which becomes the dominant effect after some time, such that $L_{\rm shock}$ again diminishes after the maximum at about 3.1 Myr is reached. Finally, when the gas disk disappears, which marks the end of the formation phase, $L_{\rm shock}$ vanishes on a relatively short timescale, and the evolutionary phase starts. Also this temporal behavior of the luminosity during the formation phase is typical, as found again in the syntheses.

As an additional effect not seen in this simulation, for planets more massive than about 13 $\mj$ \citep{baraffechabrier2008a,spiegelburrows2011},  deuterium burning delays the luminosity decrease (for hot accretion) respectively re-increases it (for cold accretion) after the planets have stopped accreting during a timescale of $10^{7}$ to a few $10^{8}$ years, depending on the mass of the planet \citep{mollieremordasini2012,bodenheimerdangelo2013}.

\subsection{Post-formation entropy $\spf$}
A practical measure to quantify the thermodynamical state of a giant planet is the specific entropy in its deep convective zone \citep[e.g.,][]{marleaucumming2014}. In this simulation we find a post-formation entropy $\spf$ of about 9.1 $k_{\rm B}$/baryon\footnote{\label{Fussnote:DeltaS}All entropy values reported in this work use the version of the \citet{saumonchabrier1995} equation of state which, for a given planet mass and luminosity, leads to an entropy lower by $(1-Y)\ln 2=0.52$ $k_{\rm B}$/baryon than the published tables, as in \citet{burrowsmarley1997}, \citet{marleyfortney2007}, \citet{spiegelburrows2012}, and \citet{mollieremordasini2012}. This makes no physical difference but needs to be stated. The MESA code \citep{paxton11,paxton13,paxton15}, \citet{marleaucumming2014}, and \citet{berardocumming2016} use the published version of the tables. See Appendix~B and Fig.~1 of \citet{marleaucumming2014}.}.  In the classical (very) cold starts studied by \citet{marleyfortney2007}, a value of 8.5 $k_{\rm B}$/baryon was found, which leads as seen above to a difference in the luminosity of more than one order of magnitude. The difference is due to the mentioned core-mass effect. But the value of 9.1 $k_{\rm B}$/baryon is also clearly lower than the 10.2 $k_{\rm B}$/baryon found in \cite{mordasini2013} for a planet with $M=5$ $\mj$ and $M_{\rm core}=49$ $\mearth$, i.e., for a very similar core mass as in the simulation here. However, in \cite{mordasini2013}, the gas accretion rate in the disk limited phase was (arbitrarily) $10^{-2}$ $\mearth$/yr, while here it is more than one order of magnitude less. As shown by \citet{spiegelburrows2012}, lower gas accretion rates lead to lower $\spf$ because the planet has more time to cool while it is still accreting. This means that the gas accretion rate not only directly affects $L_{\rm shock}$ as seen before, but indirectly, and even more importantly, also the post-formation luminosity. \citet{spiegelburrows2012} show that a reduction of $\dot{M}_{\rm XY}$ by a factor 30 leads to decrease of  $\spf$ by about 1 $k_{\rm B}$/baryon. In our simulations, we also see this effect: if the simulation of \cite{mordasini2013} for $M=5$ $\mj$ and $M_{\rm core}=49$ $\mearth$ is repeated with a disk limited gas accretion rate of $10^{-3}$ instead of $10^{-2}$ $\mearth$/yr, then a $\spf$=8.9 $k_{\rm B}$/baryon results. This is relatively close to the value found in the simulation here. An exactly identical result cannot be expected, because in the in situ simulations in \cite{mordasini2013}, the overall temporal evolution of both $\dot{M}_{\rm core}$ and $\dot{M}_{\rm XY}$ differ significantly from the case simulated here.  We thus see  that the post-formation entropy can only be predicted reliably if all aspects of a planet's formation tracks (temporal evolution of the gas and solid accretion rate, semimajor axis, opacity, etc.) are taken into account. 

This  dependency of $\spf$ and thus the post-formation luminosity on the individual formation track of a planet suggests that we should expect a significant spread in post-formation properties for giant planets, even for a fixed radiative efficiency of the gas accretion shock. With the population syntheses presented below, we confirm this, and can partially quantify it. In Figure \ref{fig:MSmarleycomp} we for example present the population wide mass-$\spf$ relation, finding that even for completely cold gas accretion, there is a wide spread of cold, warm, and even relatively hot starts. Since the radiative efficiency of the accretion shock will likely also vary depending on the planet's properties (for example via its Mach number, \citealt{marleauklahr2016}), an even larger spread should exist in reality, and additional factors like the opacity will also play a role. This shows that  statistical constraints will be among the most useful (and necessary ones) to better understand giant planet formation from direct imaging observations. 

\section{Statistical results}\label{sect:statresults}
We now turn to main subject of this paper which is the statistical study of the luminosity of forming and young giant planets. We discuss the mass-luminosity relation, the statistics of the post-formation properties, the luminosity distribution, and the mass-radius relationship.

\subsection{The planetary mass - luminosity relation}\label{sect:MLrelation}
Figures \ref{fig:mlform} and \ref{fig:mlformhot}  show one of the most important results of this study. It is the  mass-luminosity relation of giant planets during the formation phase for cold and hot gas accretion, respectively. 

While the mass-luminosity relation during the evolutionary phase at constant mass after formation has been studied in many works like the classical models of \citet{burrowsmarley1997} or \citet{baraffechabrier2003}, the mass-luminosity relation during formation has in contrast not been studied in a systematic way. Recently, several discoveries of (candidate) low-mass companions (low-mass stars and protoplanets) embedded in the circumstellar disk of young stars were announced: around LkCa 15 \citep{krausireland2012}, HD 142527  \citep{billerlacour2012}, HD 100546 \citep{quanzamara2013}, and HD 169142 \citep{reggianiquanz2014,billermales2014}. While the exact nature of some of these sources needs to be further investigated, some of these objects  are probably indeed observed in the act of formation as indicated by the H$\alpha$ emission of LkCa 15 b \citet{sallumfollette2015} (see also HD 142527B, \citealt{closefollette2014}). This warrants an extension of the $M-L$ relation to the formation phase. 

\subsubsection{Formation phase: cold-nominal population}\label{sect:MLformcold}
\begin{figure*}
\begin{center}
\begin{minipage}{0.5\textwidth}
	      \centering
         \includegraphics[width=0.95\textwidth]{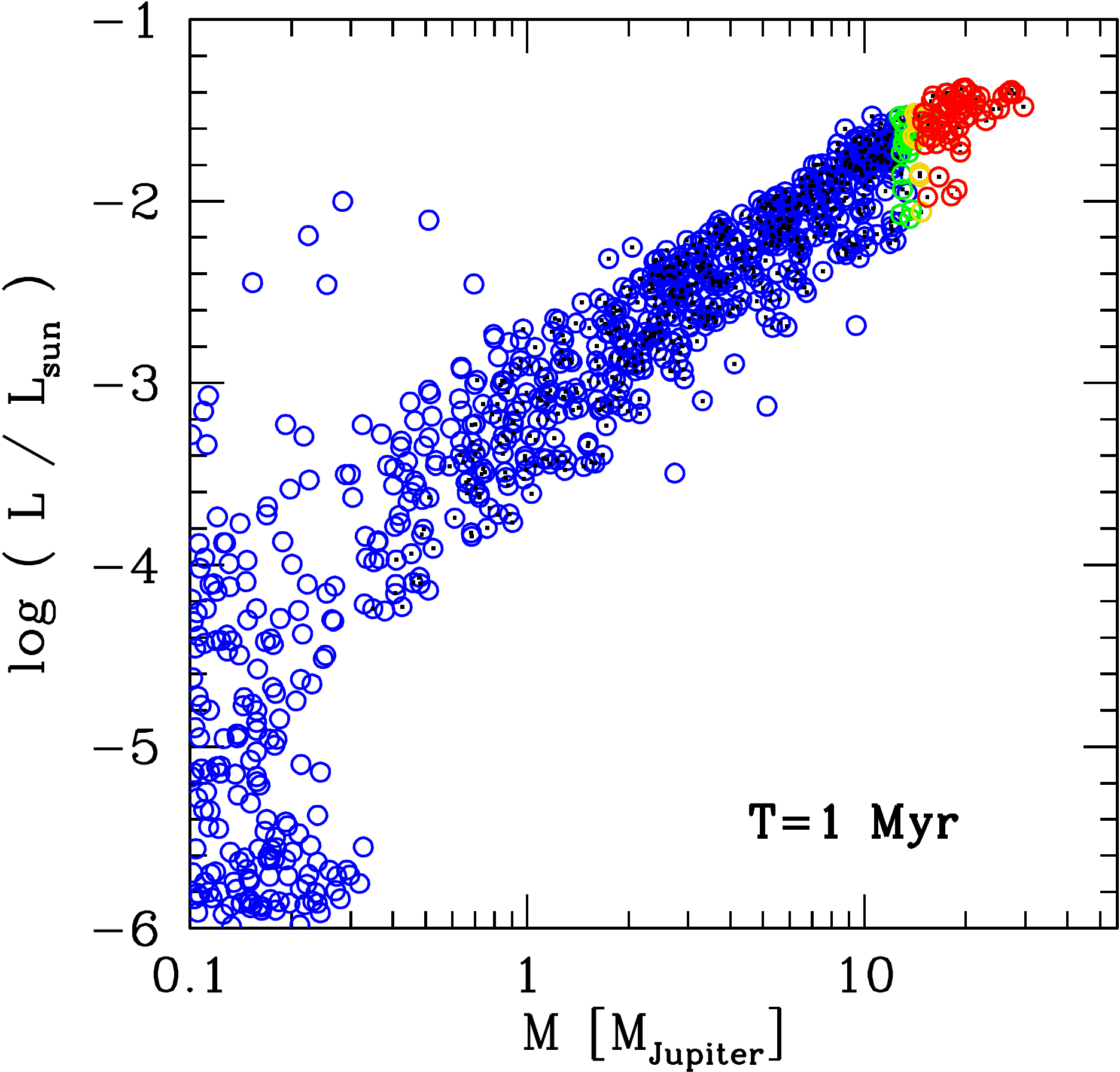}
        \includegraphics[width=0.95\textwidth]{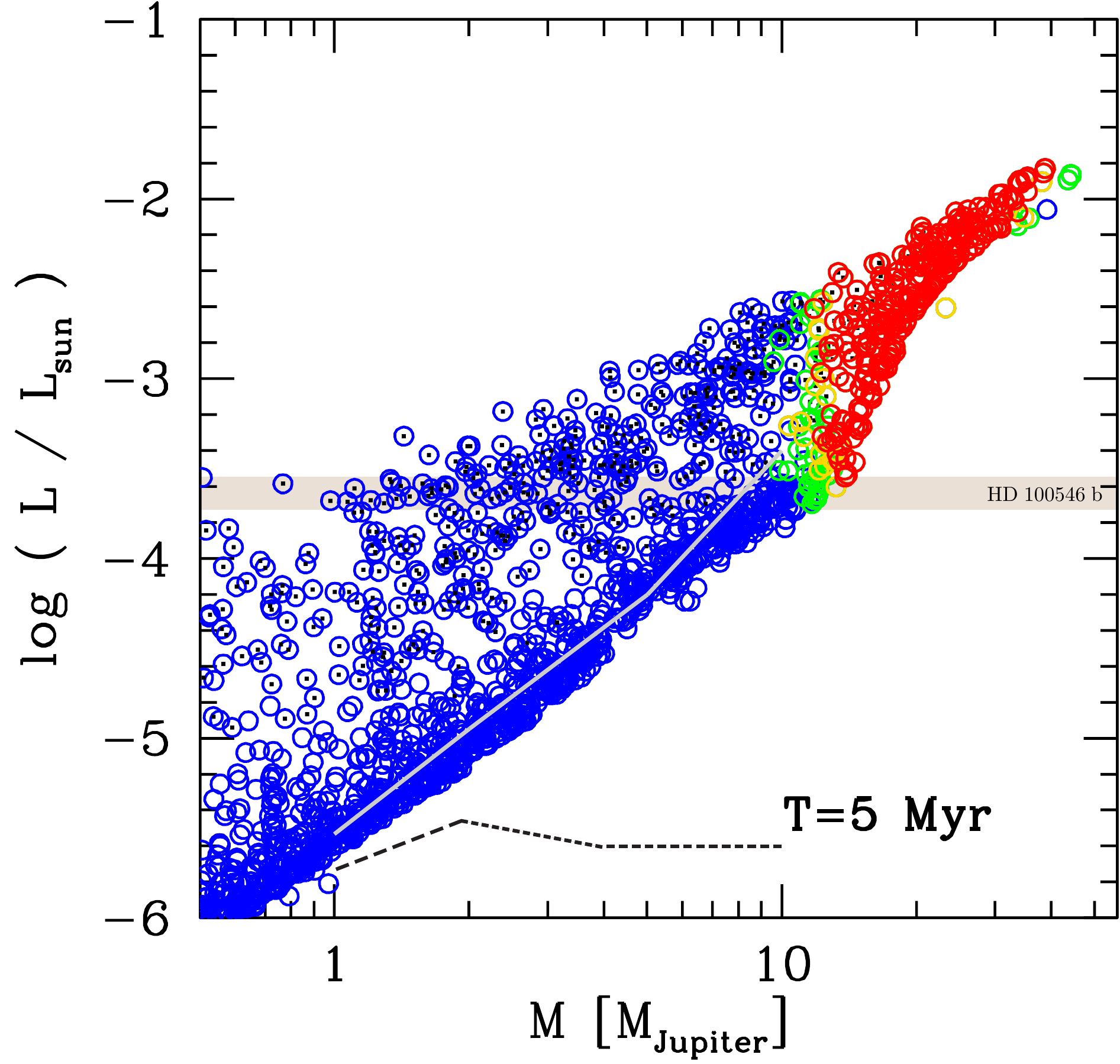}
     \end{minipage}\hfill
     \begin{minipage}{0.5\textwidth}
      \centering
       \includegraphics[width=0.95\textwidth]{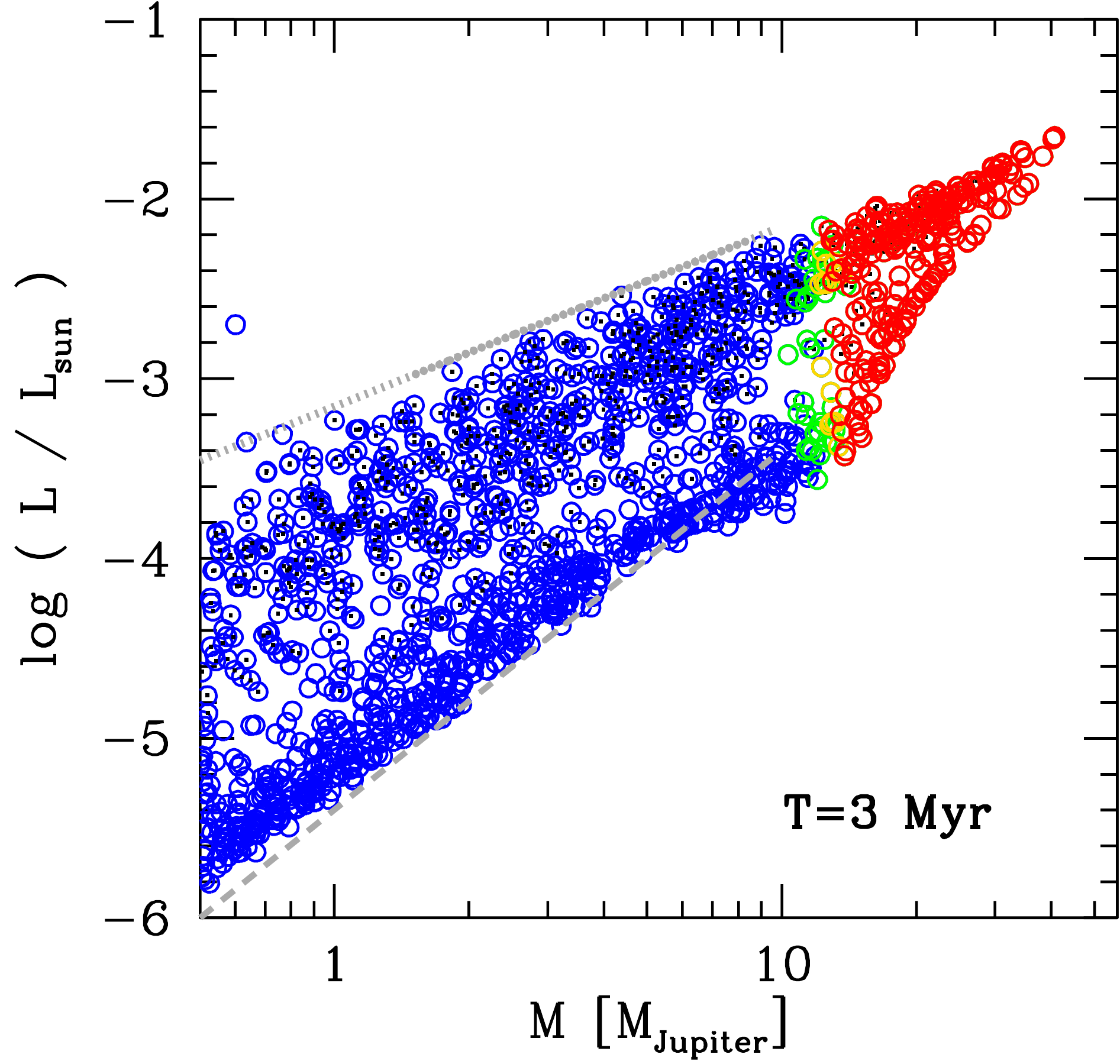}
      \includegraphics[width=0.95\textwidth]{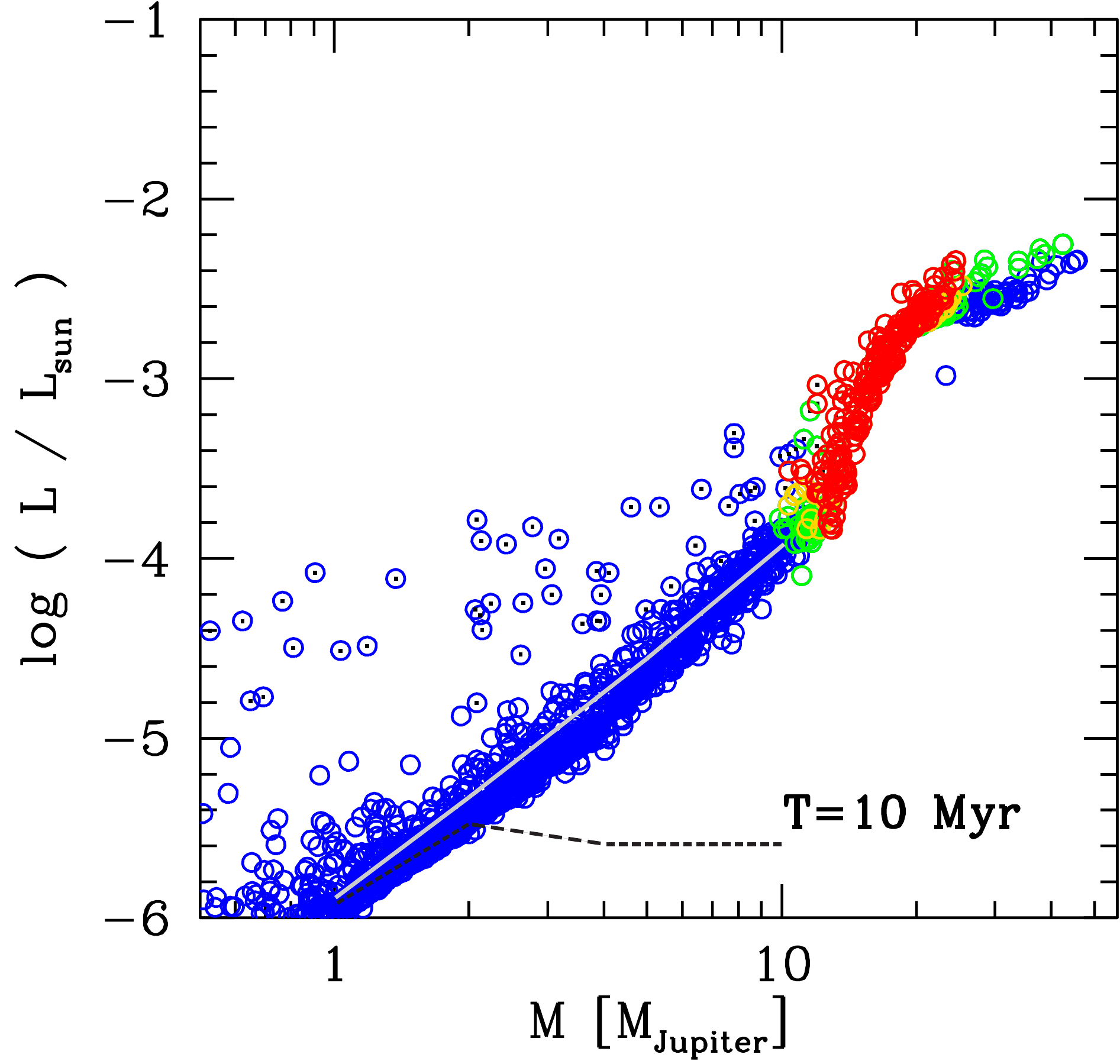}
     \end{minipage}
\caption{The planetary mass - luminosity  relation during the formation phase for cold gas accretion (cold-nominal population). The luminosity $L$ is the sum of the internal luminosity $L_{\rm int}$ and (for accreting planets) the accretion shock luminosity $L_{\rm shock}$. The planets with a small black dot in the center have $L_{\rm shock}/L_{\rm int}>1$, i.e., their luminosity is dominated by the accretion shock luminosity. Red, yellow, and green dots correspond to planets with a fractional contribution of the deuterium burning luminosity of  $L_{\rm D}/L_{\rm int}$ of at least 0.5, 0.1, and 0.05, respectively. In the panel at 3 Myrs the gray dotted (dashed) lines show the linear (quadratic) $M-L$ relations discussed in the text. In the two bottom panels, the gray solid line shows the mass-luminosity relation in the classical hot start tracks of \citet{burrowsmarley1997}, while the black dashed line shows the classical cold start simulations of \citet{marleyfortney2007}. Despite the cold gas accretion, the synthetic planets have luminosities almost as high as in the hot start models, a consequence of the core-mass effect. At 5 Myr, the luminosity of HD 100546 b \citep{quanzamara2015} is shown.} \label{fig:mlform}
\end{center}
\end{figure*}

Figure \ref{fig:mlform} shows the total  luminosity $L= \lint + \lshock $ as a function of mass for all giant synthetic planets in the cold-nominal population at 1, 3, 5 and 10 Myrs after the start of the simulation, which can approximately be associated with the stellar age. We note the problem of a potential time delay between stellar age and planetary age, as encountered when comparing observations with purely evolutionary models not considering formation \citep{fortneymarley2005a,krausireland2012}, does not exist here, because the time needed to form the planets is self-consistently included.
In the plot, planets for which the accretion shock luminosity is dominant over the internal luminosity ($L_{\rm shock} >  L_{\rm int}$) are marked with a black point in the middle. The separate contributions of $\lint$ and $\lshock$ are discussed in Sect. \ref{sect:lshocklint}. The colors indicate the importance of the deuterium burning luminosity. We now discuss each panel in turn.

\textbf{1 Myr}. In the first panel at 1 Myr, the luminosity of almost all giant planets is clearly dominated by the accretion shock luminosity as visible from the black points. Almost all giant planets are in the detached, disk-limited gas accretion phase. No protoplanetary disk has yet dissipated. At this relatively early time, the accretion rate in the disks is still high, therefore (see Sect. \ref{sect:lshocklint}) also the accretion rates of the planets is high (up to $5\times10^{-3}$ $\mearth$/yr), and thus also $\lshock$. High accretion rates in the disks are even more likely because to form a giant planet on a short timescale of 1 Myr, a high disk mass is required. Within the constant $\alpha$ viscosity disk model, such massive disks also have high accretion rates \citep{lynden-bellpringle1974}.  %

\begin{figure}%
\begin{center}
	      \centering
            \includegraphics[width=\columnwidth]{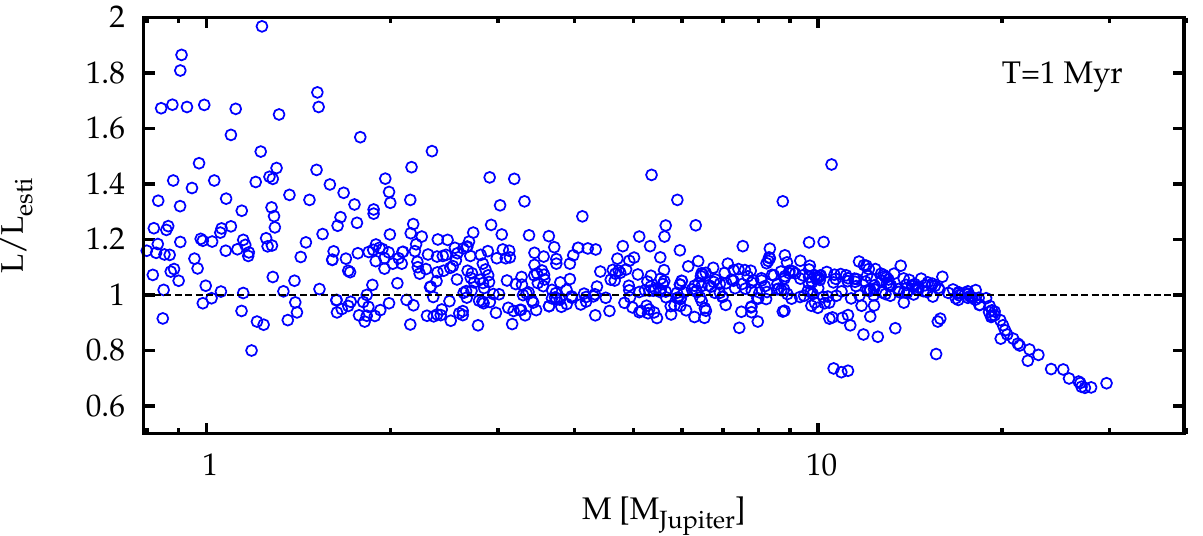}     
\caption{Ratio of the actual total luminosity at 1 Myr (shown in the top left panel of Fig. \ref{fig:mlform}) to the simple estimation given by equation \ref{eq:lest}, assuming that the giant planets all have a radius of 1.8 $\rj$. The gas accretion rate is taken from the simulation.}\label{fig:LLest}
\end{center}
\end{figure}

Because of the dominance of $\lshock$, and because $R\ll R_{\rm H}$ (except for a short time directly after detachment) the  luminosity can here be approximated simply as 
\beq\label{eq:lest}
L_{\rm est}=\frac{G M \dot{M}_{\rm XY}}{R}.
\eeq
For cold accretion,  the radius of giant planets at 1 Myr is almost independent of mass between about 1 and 10 $\mj$ (see Fig. 8 in \citetalias{mordasinialibert2012c} and Fig. \ref{fig:MRcold} below) and approximately equal to 1.8 $\rj$. This facilitates estimating $L$. Figure \ref{fig:LLest} compares the luminosity estimated with Eq. \ref{eq:lest} assuming for all planets a fixed radius of 1.8 $\rj$ (but the accretion rate from the simulation) with the luminosity found in the simulation. One sees that the agreement is relatively good for masses between about 3 to 17 $\mj$ with a typical difference of 20\% or less. For less massive planets, the actual luminosity is larger than $L_{\rm est}$, because the contribution from $\lint$ is not negligible (see Sect. \ref{sect:lshocklint}). For more massive planets, the actual luminosity is in contrast smaller then $L_{\rm est}$. This is due to the increase of the radius $R$ for these massive planets that burn deuterium in their interior as can be seen in Fig. \ref{fig:MRcold}. This reduces $\lshock\propto1/R$.

When calculating $L_{\rm est}$ in Fig. \ref{fig:LLest}, we have taken the $\dot{M}_{\rm XY}$ of the planet from the simulation. In reality, the gas accretion rate of a planet is usually unknown. However, the accretion rate of the host star is more often known \citep[e.g.,][]{rigliaconatta2012}. If the planetary and stellar accretion rates are of the same order of magnitude as suggested by  hydrodynamic simulations \citep{lubowdangelo2006}, then the independency of $R$ on $M$ for cold accretion can be used to estimate, via $L_{\rm est}$, the planet's mass provided that $\lshock$ can be observationally obtained. This approach (see also \citealt{krausireland2012,closefollette2014}) then yields a constraint on the planet's mass besides other indicators such as the gap morphology \citep[e.g.,][]{kanagawamuto2015} or the local disk temperature \citep{montesinoscuadra2015} and scale height \citep{klahrkley2006}.  For hot accretion it is more complex to estimate the planet's mass in this way, as the radius becomes then also a function of mass (Fig. \ref{fig:MRhot}). 

The dominance of $\lshock$ in this phase, the independency of $R$ on $M$, and our assumption that the planet's gas accretion rate is proportional to the disk accretion rate, i.e., a priori independent of the planet mass (no reduction due to gap formation) mean that $L\approx\lshock$ should roughly increase linearly with planet mass for strongly accreting planets. The slope of the envelope of points in Fig. \ref{fig:mlform} indeed increases approximately as $M^{1.2}$. The slightly larger exponent is due to a weak positive correlation of the gas accretion rate and planet mass. It is an indirect consequence of the fact that more massive planets form in more massive disks that have a higher gas accretion rate. One furthermore notes a spread of about one order of magnitude in $L$ at a given mass.

Regarding deuterium burning, $L_{\rm D}/\lint$ is equal to 0.05, 0.1, 0.5, and 1  at 12.6, 13.2,  15.8, and 16.7 $\mj$. There is thus no single mass where deuterium burning starts  but a smooth transition to higher $L_{\rm D}$ contributions \citep{spiegelburrows2011}. Planets with masses above  16.7 $\mj$ have $L_{\rm D}/\lint>1$, meaning that part of the deuterium burning luminosity is spent on re-inflating their radius \citep{mollieremordasini2012}. 

Near the left border of the panel, one notes six planets of low mass (between about 0.15 and 0.7 $\mj$) that are quite luminous  with $\log ( L / \lsun)\gtrsim-2.5$. A similar case is also seen in the panel at 3 Myr. The absence of black dots shows that their high luminosity is not due to the gas accretion shock. Instead, these are planets which are just in the  phase where the envelope contracts rapidly \citepalias[but still quasi-hydrostatically,][]{mordasinialibert2012b} immediately after the detached phase starts. The high planetesimal accretion rate in this phase \citepalias[][ \citealt{dangelopodoak2015}]{mordasinialibert2012b} plus the fast contraction powers their luminosity. Since this phase typically lasts for only a few $10^{4}$ yr, only a  low number of planets are caught in this stage at any given moment in time. But it is  an interesting phase where even low-mass planets (typical masses of 50-200 $\mearth$) can reach high $L$ for a short period. 

\textbf{3 Myr}. In the second panel at 3 Myr, the maximal gas accretion rate has fallen to about $1\times 10^{-3}$ $\mearth/$yr, and, in contrast to 1 Myr, a significant number of protoplanetary disks have already dissipated. This has the following interesting consequence: there are now two classes of planets. First those that are still accreting in the detached disk-limited regime, and second those that have already stopped accreting because the disk has dissipated. They are already in the evolutionary phase at constant mass. The two groups can be identified in the figure by the presence/absence of  black dots. The planets that are still accreting (the ``accreting sequence'') have higher luminosities because of $\lshock$,  while the non-accreting planets (the ``evolving sequence'') are already on standard cooling tracks, and have lower $L$. Between the two sequences, there is a region which is less populated. This is due to the fact that because of photoevaporation, the final dissipation of the protoplanetary disk, and with it the decrease of the planetary $\dot{M}_{\rm XY}$ and thus $\lshock$ occurs on short timescale (``two-time-scale'' behavior, \citealt{clarkegendrin2001}). This means that at a specific  moment in time not many planets are in the final phase where $\lshock$ falls to zero. 

The two sequences are characterized by different $M-L$ scalings: as at 1 Myr, the accreting planets have roughly speaking $L\approx\lshock\propto M$, while the non-accreting planets follow roughly speaking $L=\lint\propto M^{2}$, as expected analytically \citep{burrowsliebert1993,marleaucumming2014}. To illustrate these scalings, we have added in the panel two gray lines which encompass the cloud of points: for planets without significant D-burning, the envelope of points is approximately bounded by an upper limit given by those planets of the accreting sequence that accrete at the highest rate in the most massive disks. A fit to the upper limit (by eye) gives
\beq
\frac{L}{L_{\odot}}\approx 7\times10^{-4}\left(\frac{M}{\mj}\right).
\eeq
This is shown by the gray dotted line in the figure. This is about a factor 5 smaller than the highest luminosities at 1 Myr because of the decrease of the maximal $\dot{M}_{\rm XY}$ by the same factor, a consequence of disk evolution. The lower limit (the least luminous planet in the evolving sequence) is given by the dashed line which  roughly follows
\beq
\frac{L}{L_{\odot}}\approx 4\times 10^{-6} \left(\frac{M}{\mj}\right)^{2}.
\eeq
The lines show that at a given mass, the spread of associated total luminosities is very large, about two orders of magnitude. This is a consequence of the fact that the planets not only have different entropies depending on their formation history (see Sect. \ref{sect:spflpf}) which causes a spread in $\lint$ (for the evolving sequence), but also that planets of a given mass can have different $\dot{M}_{\rm XY}$, which leads to a spread in $\lshock$. This obviously means that if only the sum of $\lint$ and $\lshock$ can be measured, it is very difficult to derive the mass from the observed luminosity.

Concerning D-burning, at 3 Myr the boundary for  a $L_{\rm D}/\lint$ equal to 0.05, 0.1, 0.5, and 1  is at 11.4, 11.9, 13.4, and 14.2 $\mj$, about 1-2 $\mj$ lower than at 1 Myr. This decrease is due to the cooling of the planets which has the well-known but counterintuitive effect \citep{kippenhahnweigert1994,mollieremordasini2012} that their central temperature increases (the interiors are not yet strongly degenerate). Planets more massive than about 20 $\mj$ are now in their D-burning main sequence, defined as the phase where $\lint=L_{\rm D}$ \citep{mollieremordasini2012}. We also see that for D-burning planets, a weaker decrease of $L$ occurs when accretion stops. There is narrower depleted region between the ``accreting'' and the evolving sequence than for $M\lesssim10\mj$. The reason for this is twofold: first, $\lshock$ approaches a constant value as a function of mass for D-burning planets at a given gas accretion rate as they have an increasing radius with mass (see Fig. \ref{fig:MRhot}), and second, $\lint$ is larger because of $L_{\rm D}$. 

\textbf{5 Myr}. The global structure at 5 Myr is quantitatively similar to the one at 3 Myr, with the difference that the accreting sequence is increasingly less populated, while the evolving sequence becomes more populated, as more and more protoplanetary disks dissipate. We also see that the luminosities have further decreased. For the accreting planets because of a decrease of the gas accretion rates (now less than $5\times10^{-4}$ $\mearth$/yr) , while for the evolving sequence it is because of cooling, at least for those planets that have already stopped accreting a certain time ago. Additionally, at late times, planets accrete at a lower gas accretion rate, which allows them to (partially) cool already while they accrete \citep{spiegelburrows2012}. The mass limits for a contribution of  $L_{\rm D}/\lint$  of 0.05, 0.1, 0.5, and 1 has again decreased slightly and is now at 10.8, 11.4, 12.7, and 13.5 $\mj$. As indicated by the blue and green color of the most massive planets ($\sim 40 \mj$), these planets which start intense D-burning already during formation (\citealt{mollieremordasini2012}) have already burnt most of their deuterium because of their high central temperatures and the extreme dependency of the D-burning rate approximately $\propto T^{11.8}$ \citep{stahler1988}. Thus, they have already started to  simply cool. The only planets for which the luminosity has not systematically decreased between 3 and 5 Myr are the those in the lower arc-like structure at about 20 $\mj$: they are still in the D-burning main sequence, where the thermostatic nature of D-burning \citep[see, e.g.,][]{mollieremordasini2012} keeps their luminosity approximately constant for some time. 

\textbf{10 Myr}.
At  10 Myr, almost all protoplanetary disk have disappeared, so that almost all planets have entered the non-accreting evolving sequence. Only a low number of  planets in long-lived disks are still accreting, giving them luminosities up to 2 orders of magnitude higher than their non-accreting counterparts.  
The minimum mass for D-burning has again very slightly decreased, while the planets more massive than about 20 $\mj$ have now already almost fully consumed their D-reservoir, so that they simply cool.

\subsubsection{Comparison with classical hot and cold start models: with sinking, core accretion makes hot planets}
In the panels at 3 at 5 Myr, the gray solid line shows the mass-luminosity relation in the classical hot start models of \citet{burrowsmarley1997} at these times. The \citet{baraffechabrier2003} would give very similar results. These purely evolutionary models start at $t=0$ with a fully formed planet of arbitrarily high specific entropy of the interior adiabat ($\gtrsim$10-12 k$_{\rm B}$/baryon). The black dotted line shows the mass-luminosity relation predicted by the formation and evolution models of \citet{marleyfortney2007}, also at 3 and 5 Myr. They  assumed as in our simulation here cold gas accretion and found low specific entropies of $\lesssim$ 8-9 k$_{\rm B}$/baryon. As it is well known,  the \citet{marleyfortney2007} cold start models predict luminosities that are much lower then the hot start ones,  by up to 2 orders of magnitudes at 10 $\mj$.

Comparing our non-accreting ``finished'' synthetic planets with these lines shows that they are, at first sight  surprisingly, almost as luminous as in the hot start simulations. {We thus find that despite the completely cold accretion of gas, the luminosity of the synthetic planets in the evolving sequence is very similar to the classical hot start models of the same age.} This already holds as early as at 3 Myrs. The planets are thus much more luminous than originally predicted by the classical core accretion models of \citet{marleyfortney2007}. As will be further illustrated below in Sect. \ref{sect:impactcoremass}, this is a consequence of the core-mass effect \citep{mordasini2013,bodenheimerdangelo2013}: especially the more massive planets (several $\mj$) in our simulations have core masses clearly larger ($\gtrsim 50 \mearth$, see \citealt{mordasiniklahr2014}) than assumed in \citet{marleyfortney2007} who only had core masses of 17-19 $\mearth$. This has a large effect on the luminosity because of the self-amplifying mechanism described in \citet{mordasini2013}. This has the following important consequence: if the core-mass effect is efficiently acting in reality {(see Sect. \ref{sect:limintstruct})}, then {core accretion actually predicts hot start planets even if the gas accretion shock is completely cold}. This has the important implication that based on a high luminosity alone, core accretion cannot be excluded as the formation mechanism. %

\subsubsection{Formation phase: hot gas accretion}\label{sect:formationphasehotgasaccretion}
\begin{figure*}
\begin{center}
\begin{minipage}{0.5\textwidth}
	      \centering
        \includegraphics[width=0.95\textwidth]{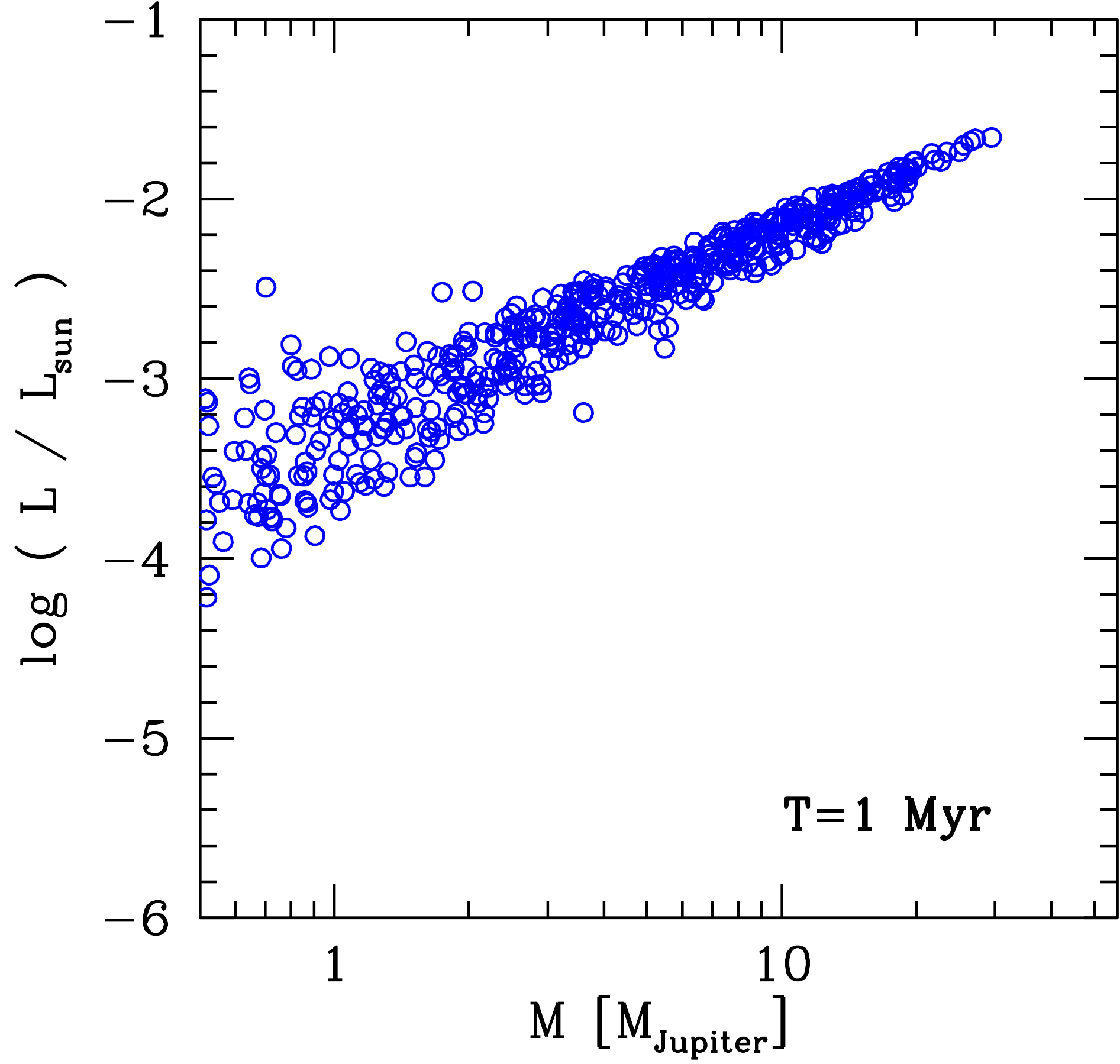}
        \includegraphics[width=0.95\textwidth]{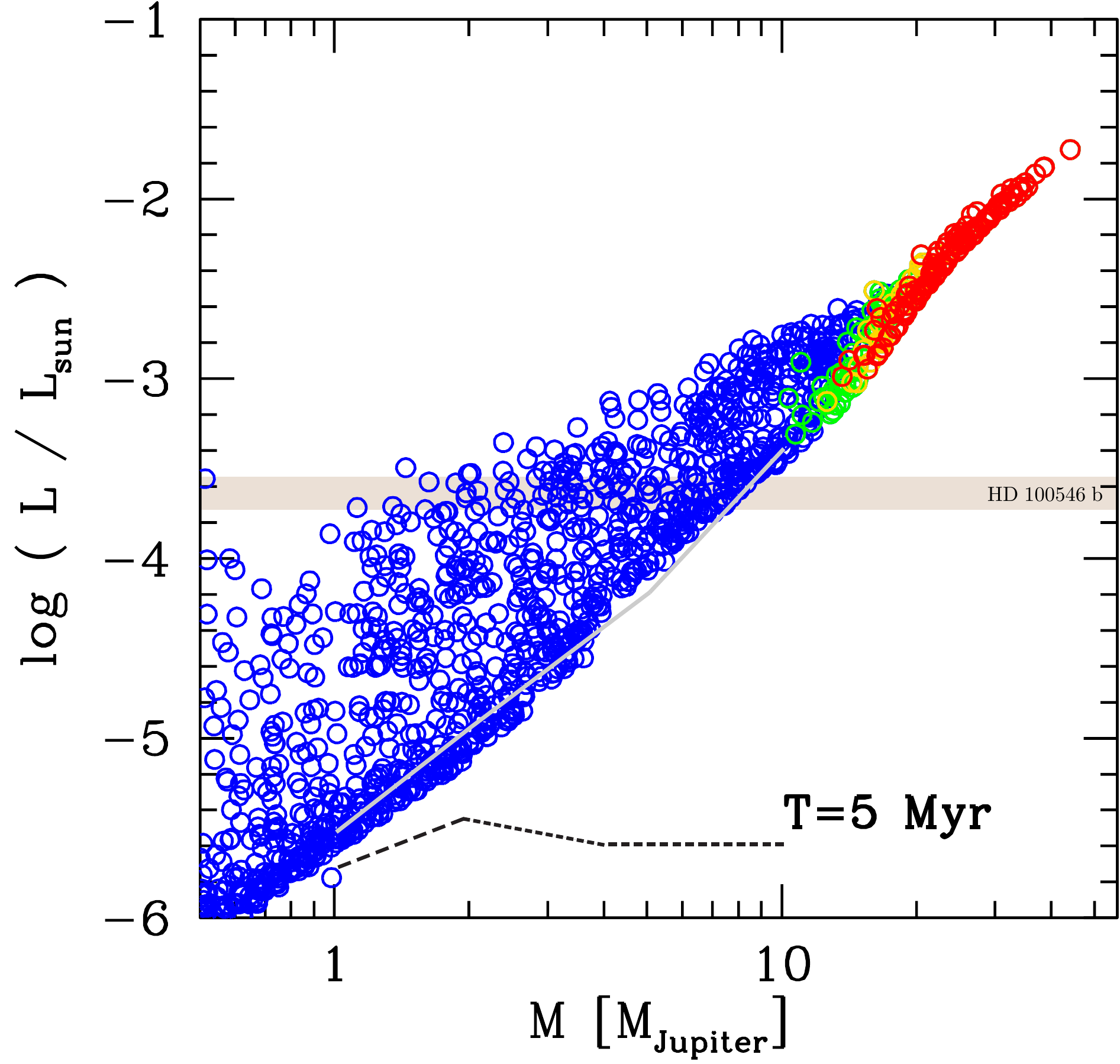}
     \end{minipage}\hfill
     \begin{minipage}{0.5\textwidth}
      \centering
       \includegraphics[width=0.95\textwidth]{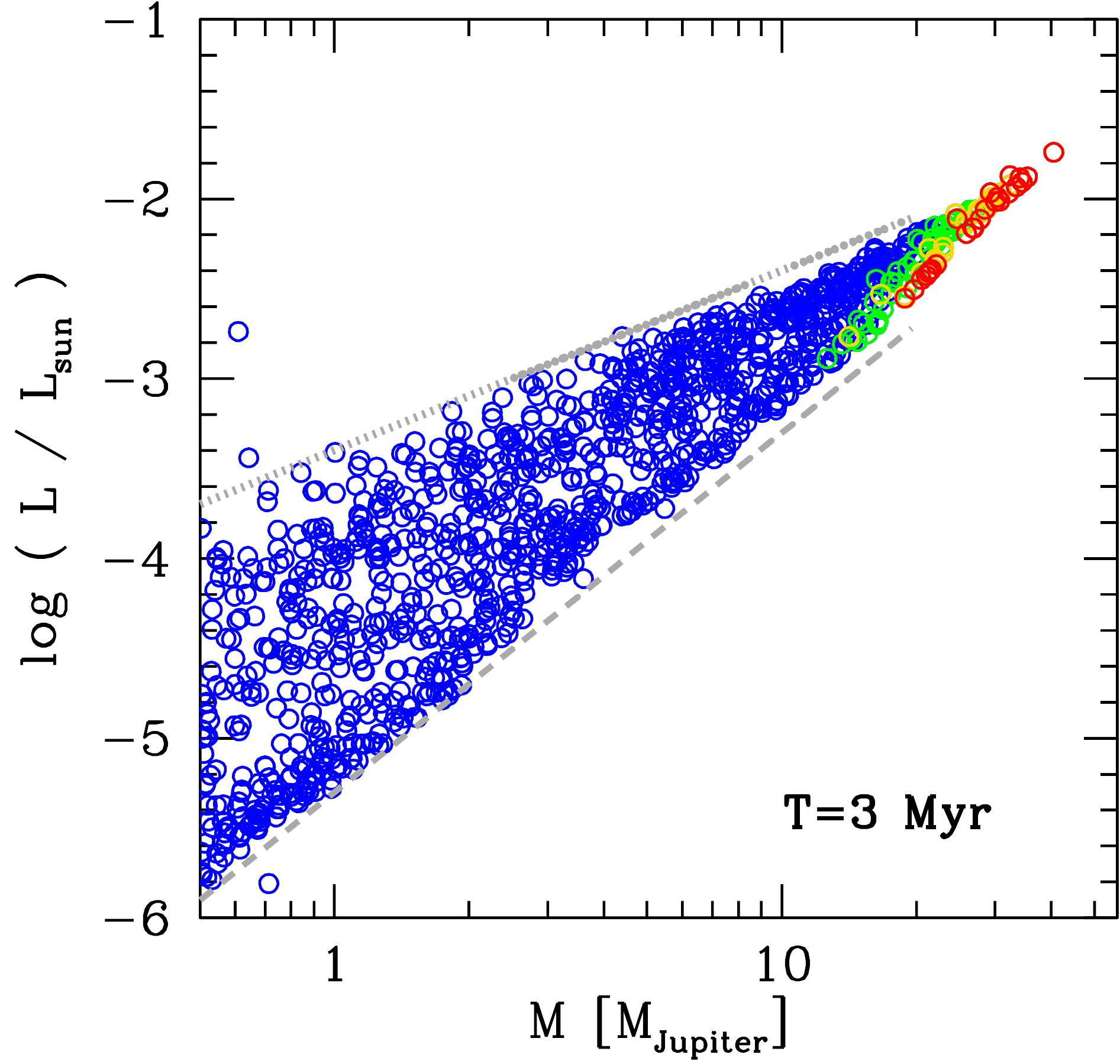}
      \includegraphics[width=0.95\textwidth]{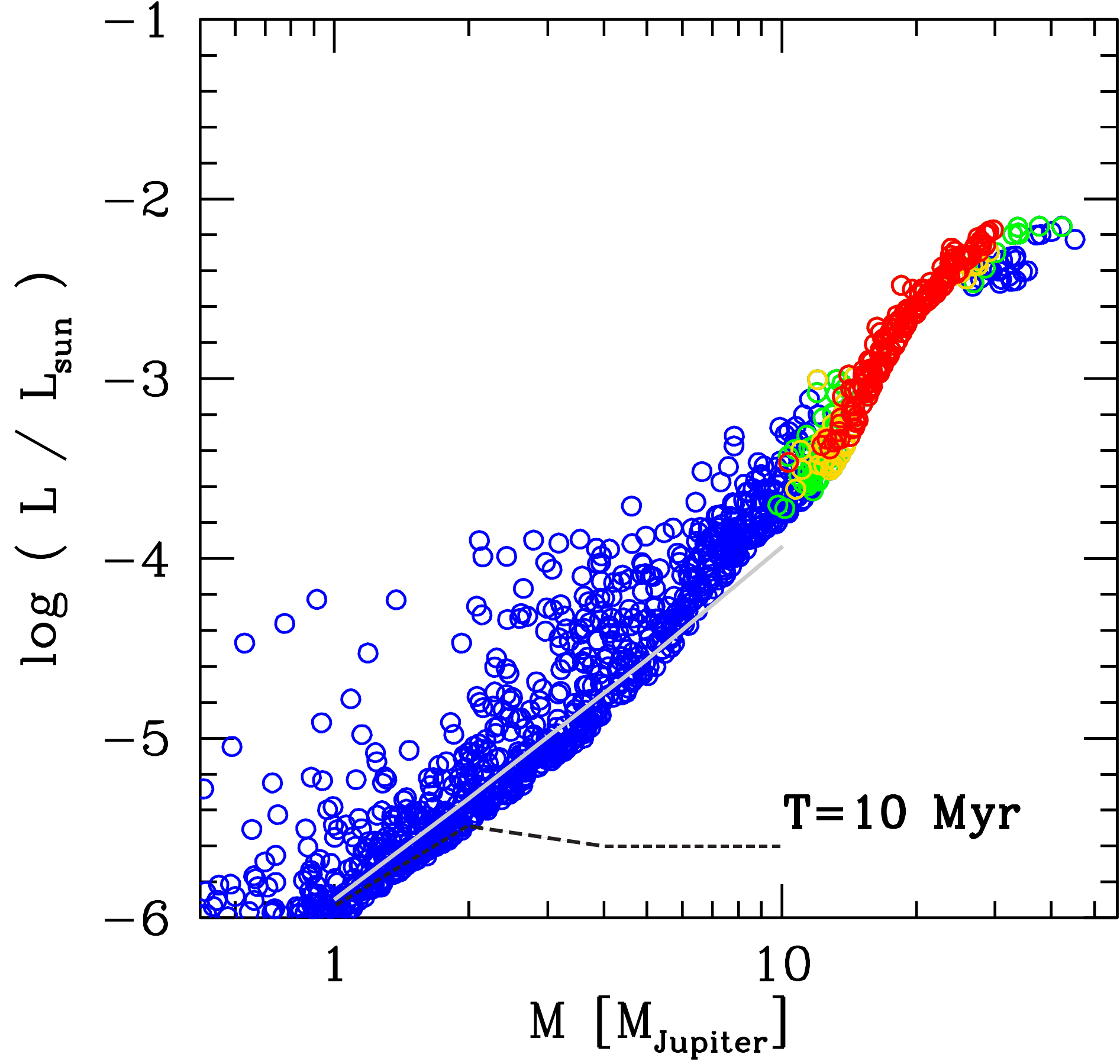}
     \end{minipage}
\caption{The planetary mass - total luminosity relation during the formation phase for hot gas accretion, analogous to Fig. \ref{fig:mlform}. The lines in the panel at 3 Myr  differ from the ones in Fig. \ref{fig:mlform}, while those at 5 and 10 Myr are identical. Compared to the cold-nominal population there are three differences that are particularly well visible in the panel at 3 Myr: (1) no split in an accreting and evolving sequence, (2) at a given mass, for accreting planets, the highest luminosities are lower, and (3) a temporally later beginning of D-burning. Furthermore, the  luminosities of planets that are no more accreting are still more luminous in the hot population than in the cold-nominal population (which are already high because of the the core-mass effect) especially for planets with masses between $\sim5\mj$ and the deuterium burning limit. }\label{fig:mlformhot}
\end{center}
\end{figure*}

Figure \ref{fig:mlformhot} shows the total luminosity $L$ as a function of mass for the hot population, analogous to Fig. \ref{fig:mlform} for the cold-nominal population. We recall that $L=\lint$ and $\lshock=0$ in this population, either because there is no shock, or because the shock is radiatively inefficient. Rather, it is assumed that newly accreted gas is directly added to the planet's gravothermal energy reservoir without radiative losses, assuming a radially constant luminosity as described in \citetalias{mordasinialibert2012b}. It is clear that the actual accretional process could be much more complicated, involve strong luminosity gradients in the accreting envelope, and depend in particular on the relation between the planet's initial internal entropy and the entropy of the material added by the accretion shock, as recently shown by \citet{berardocumming2016}. The results of this study, combined with those of works predicting the post-shock entropy (for shocks on the circumplanetary disk and then on the planet itself; see respectively \citealt{szulagimordasini2016} and \citealt{marleauklahr2016}) should allow us to obtain more realistic theoretical models for accreting young planets in the future. These can then in turn be used for future statistical studies as conducted here.

The panels show that the general trend of decreasing luminosities in time and of higher luminosities for more massive planets found for the cold-nominal population also holds for this population, as expected. However, one can also identify three important differences that are best visible in the panel at 3 Myr: 

\textbf{First}, there is no split into two distinct groups of planets in an ``accreting'' and ``evolving'' sequence as in the cold-nominal population. This is readily understood: in the cold-nominal population, at the end of the disk lifetime, the total luminosity ($L=\lint+\lshock$) decreases rapidly and significantly, as $\lshock$ disappears at this moment such that only $\lint$ is left, as visible in Fig. \ref{fig:tML5MJ} here  and Fig. 10 of \citetalias{mordasinialibert2012b}. In the hot population, no such sudden drop of $L$ occurs, first because there is no  $\lshock$ during accretion, and second because the $\lint$ in the hot population is higher. The $\lint$ in the cold-nominal population is also high when compared to the cold-classical population considered below because of the core-mass effect, but still less than in the hot population (see also Fig. \ref{fig:MLtdisk}). Rather, the envelope of points now covers without much substructure a triangular region in the $M-L$ space with strongly accreting planets in the upper part, and weakly or non-accreting planets close to the lower boundary of the triangle. 

\textbf{Second}, one sees that when comparing the highest total luminosities at a given mass (planets that are  accreting strongly), planets in the cold-nominal have a higher total luminosity ($L=\lint+\lshock$) than the brightest planets in the the hot population of the same mass, which only have $\lint$. This inversion (cold-accretion planets being brighter than hot-accretion planets during formation) relative to the situation during the evolution is also visible in Fig. 10 of  \citetalias[][]{mordasinialibert2012b}. This can be readily be understood from energy conservation arguments (see also \citealt{lissauerhubickyj2009}): During formation, the hot-accretion planets are heated by the kinetic energy influx originating from the accreting gas; this leads to larger radius and therefore to a higher gravothermal energy reservoir that can be radiated later during the  evolutionary phase. Consequently, they are less luminous during formation, as the material gets accreted at a larger radius (see Sect. \ref{sect:MRrelation}). For cold accretion, it is the opposite: these planets are not heated by gas accretion and accrete the gas therefore onto a smaller radius, resulting in a high $\lshock$ that is radiated away. This makes them  brighter during formation but dimmer during evolution because their gravothermal energy reservoir is smaller. For hot accretion, the same small radius is only reached at some later moment during evolution. This also means that the effective temperature of accreting planets is higher for cold accretion than for hot accretion at a given planetary mass.   Note finally that during formation, if the $\lint$ and $\lshock$ components of the total luminosity $L$ can be measured separately via a tracer of $\lshock$ such as H\,$\alpha$ emission, we should see  that the $\lint$ of cold-accretion objects tends to be lower than the $\lint$ of hot-accretion objects.

On the other hand, when comparing the planets that have already stopped accreting and have the lowest luminosity at a given mass (at the bottom of the triangular region in the hot population and in the ``evolving'' sequence in the cold-nominal population), we see the opposite, namely that the planets of the hot population have a higher $\lint$ than those of the cold-nominal population. The typical difference is however not very large,  as the core-mass effect heats the planet in the cold-nominal population almost as strongly as hot gas accretion, as discussed in the previous section. The difference is best visible when comparing the envelope of points with the gray lines at 5 Myr. The difference can be much larger for planets in the cold-nominal population as will be shown below. 

To illustrate these findings, we have added two empirical $M-L$ scalings in the panel at 3 Myr for the cold-nominal population that enclose the envelope of points and form the aforementioned triangular region. The upper dotted line is given by 
\beq
\frac{L}{L_{\odot}}\approx 4\times10^{-4}\left(\frac{M}{\mj}\right).
\eeq
while the lower boundary given by the dashed line approximately follows
\beq
\frac{L}{L_{\odot}}\approx 5\times 10^{-6} \left(\frac{M}{\mj}\right)^{2}.
\eeq
We see that the upper envelope thus corresponds to a lower luminosity than in the cold-nominal population, while it is the opposite for the lower envelope. The differences are, however,  not very large. But we also note that the simple power-law fits lead to a certain underestimation in the difference of  $\lint$ between the cold-nominal and the hot population for massive planets with masses between about 5-7 $\mj$ and the lower limit for D-burning. Here, the actual difference in $\lint$ at 3 Myr can actually be half an order of magnitude.

\textbf{Third}, there is a difference in the extent and timing of deuterium burning and also the shape of the  $M-L$ relation in the D-burning regime. Regarding the extent and timing  when D-burning occurs, we see that at 1 Myr in the cold-nominal population, a significant number of planets is already intensively burning deuterium as shown by the red points, while no significant D-burning occurs in the hot population. At 3 Myr, strong D-burning is ongoing in all planets more massive than about 13 $\mj$ in the cold-nominal population, whereas in the hot population only a handful more massive planets already burn significant deuterium. At even later times, strong D-burning exists in both populations, but in the hot population it is restricted to higher masses relative to the cold-nominal population. In summary we thus see that in the hot population, deuterium burning sets in later, as can also be seen in the simulations of \citet{mollieremordasini2012}. This is not related to a different mass of the planets as a function of time, as the accretion of the planets is unaffected from the hot/cold assumption in our model, as mentioned above. It is rather related to their different thermodynamic state and in particular central temperature, because of the following mechanism: for cold accretion, the planets have a smaller radius at a given mass.   Crudely estimating the (central) pressure $P$ and density $\rho$ of a planet  with mass $M$ and radius $R$ \citep[e.g.,][]{CoxGiuli1968}
\beq
\rho \sim \frac{M}{R^{3}} \, \,  \, \, \, \, \, \,  P \sim G\frac{M^{2}}{R^{4}}
\eeq
where $G$ is the gravitational constant, and because the planets are at this time not yet strongly degenerate so that the ideal gas law applies \citep{mollieremordasini2012}, one estimates a central temperature as a function of planetary radius ($\mathcal{R}$ is the gas constant and $\mu$ is the mean molecular weight),
\beq
T=\frac{\mu}{\mathcal{R}}\frac{P}{\rho}\sim\frac{G \mu}{\mathcal{R}}\frac{M}{R}\propto \frac{1}{R}.
\eeq
The smaller radius thus means that the central temperature of the cold accretion objects is higher in the first phase of D-burning when the thermostatic D-burning main sequence (where $L\approx L_{\rm D}$, see \citealt{mollieremordasini2012}) is not yet reached. As the deuterium  nuclear energy generation rate is approximately $\varepsilon \propto \rho T^{11.8}$ \citep{stahler1988} this means that in the cold start objects D-burning sets in earlier during formation, and uses up some of the deuterium which they then lack later on, during evolution. For hot accretion, sufficiently high central temperatures are, in contrast, only reached later on when the planetary radii have decreased, and at this moment also more deuterium is still available, so that deuterium burning occurs later and then also more intensively. The latter point is manifested by the fact that the radii of the planets during the D-burning main sequence is somewhat larger for hot accretion than for cold accretion despite the thermostatic effect of D-burning. This is found in the tracks of the planets in the syntheses and is visible in Fig. 7 of \citet{mollieremordasini2012}.

\textbf{Finally}, regarding the shape of $M-L$ of D-burning planets we see that for the cold-nominal population, the $M-L$ of such planets takes an arc-like form bending upwards with a higher $d L / d M$, whereas for the hot population, the $M-L$ relation  is more an approximately straight extension of the $M-L$ relation of lower mass planets not burning deuterium with a lower $dL / d M$. The reason for this is the following: in the cold-nominal population, the mass interval where D-burning occurs must connect non-deuterium burning planets with a lower $L$ than in the hot population (compare the right end of the gray line in the panel at 10 Myr) with D-burning planets that have a similar, higher $L$ in the two populations. This is because deuterium burning tends to remove the hot-cold difference because of its thermostatic nature, although it does not do so completely, as discussed above and in \citet{mollieremordasini2012}.

\subsubsection{Formation phase: cold-classical population}
 \label{sect:formationphasecoldclassicalgasaccretion}
\begin{figure*}
\begin{center}
\begin{minipage}{0.5\textwidth}
	      \centering
         \includegraphics[width=0.95\textwidth]{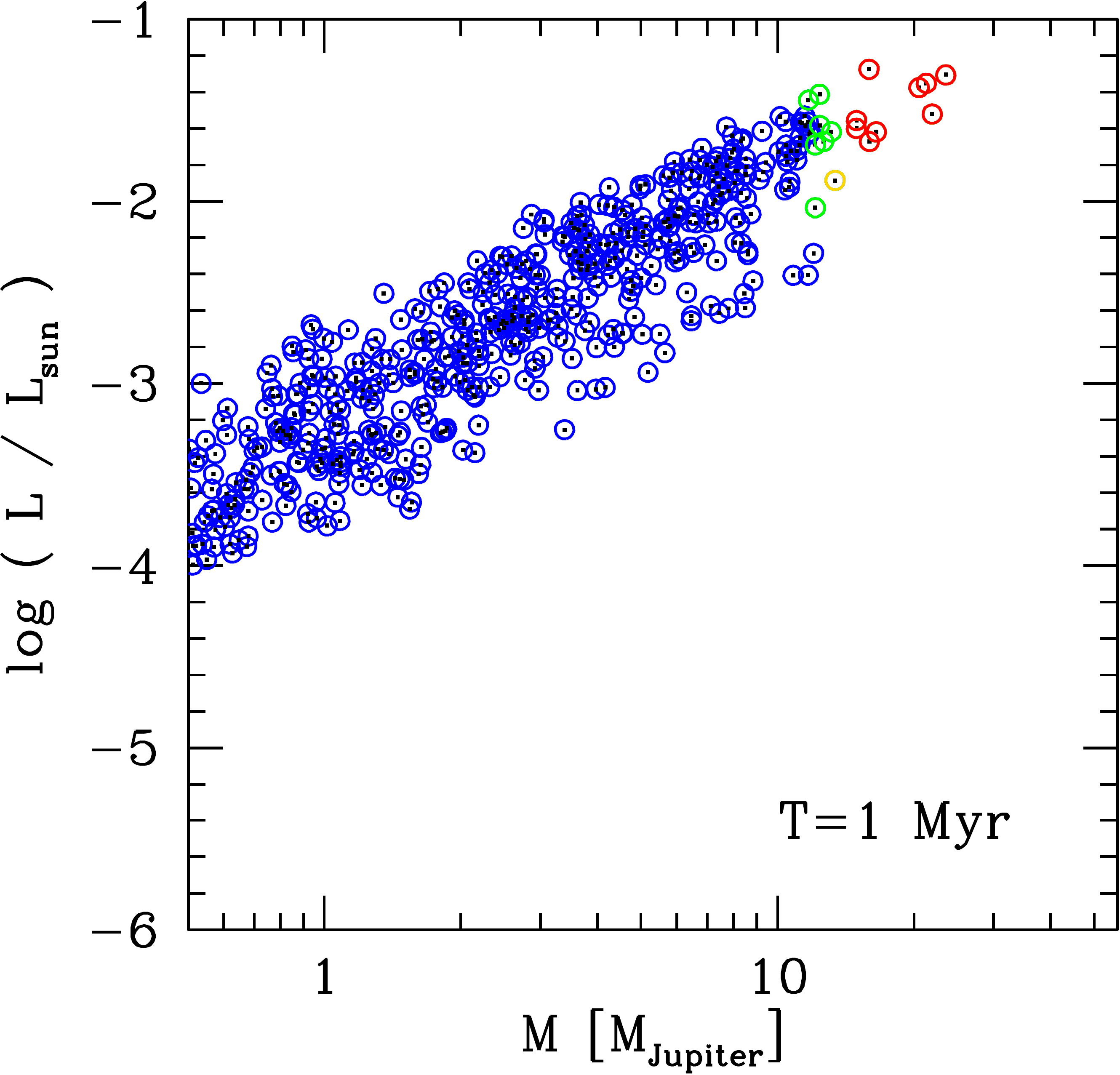}
        \includegraphics[width=0.95\textwidth]{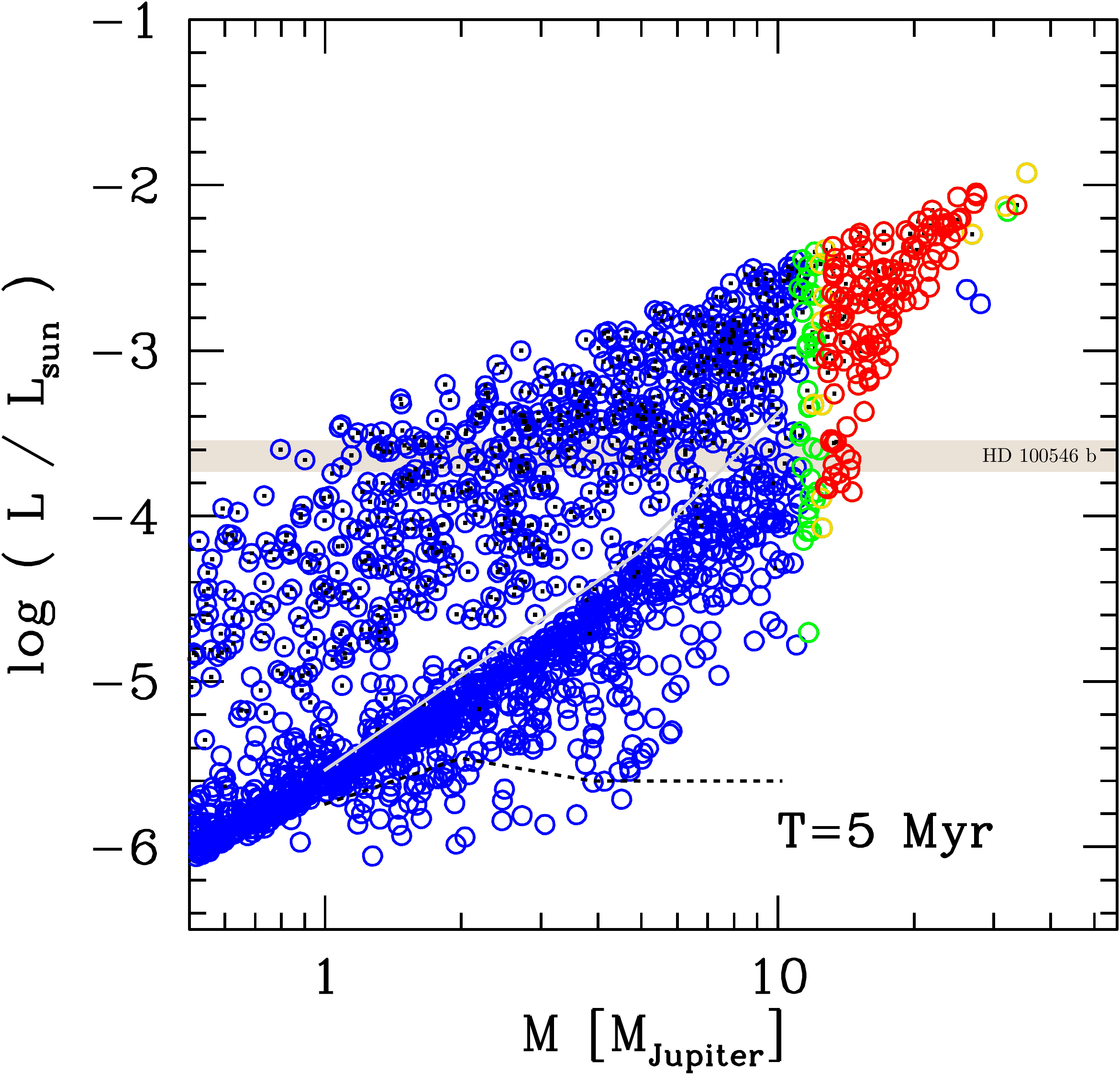}
     \end{minipage}\hfill
     \begin{minipage}{0.5\textwidth}
      \centering
       \includegraphics[width=0.95\textwidth]{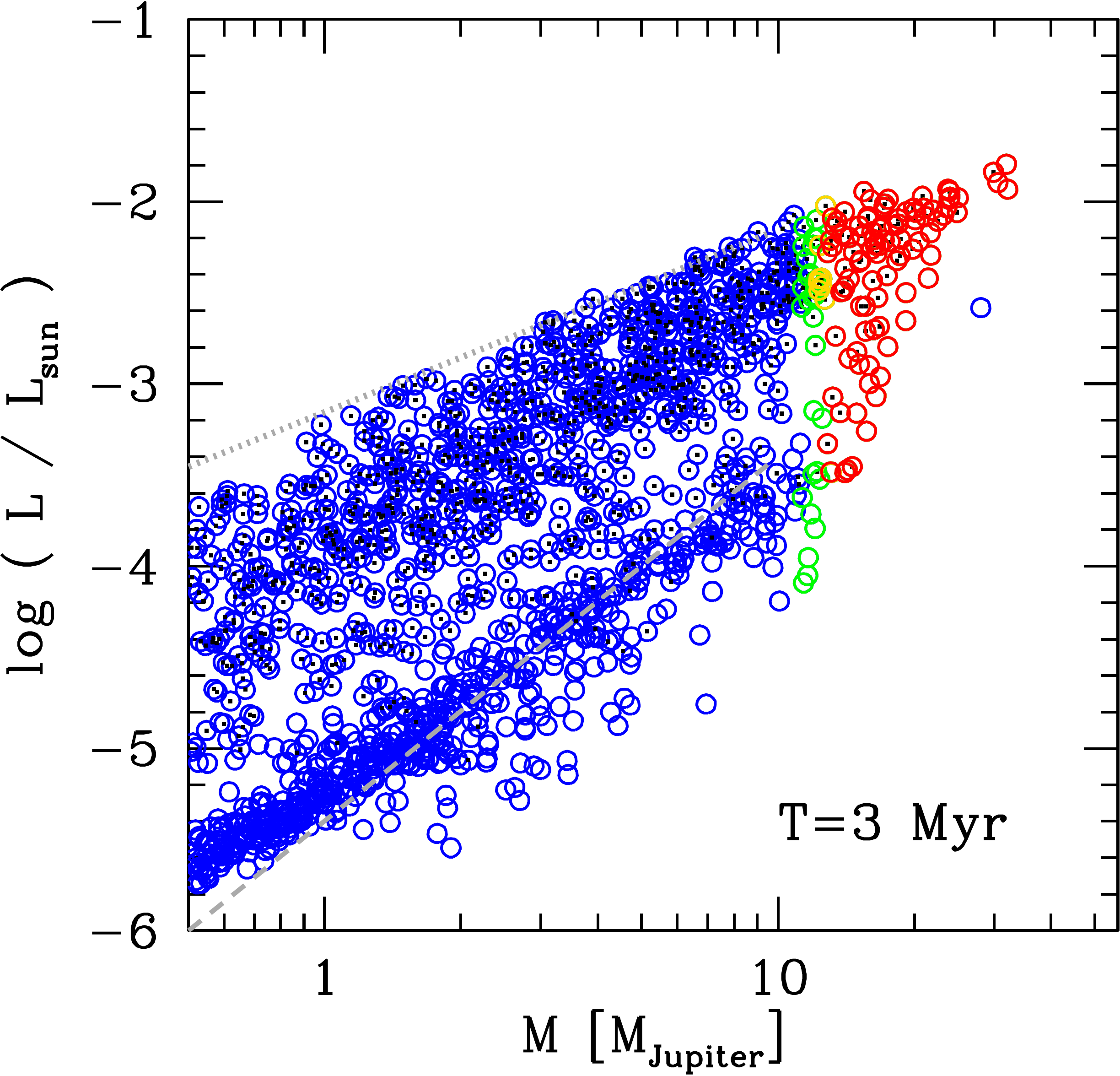}
      \includegraphics[width=0.95\textwidth]{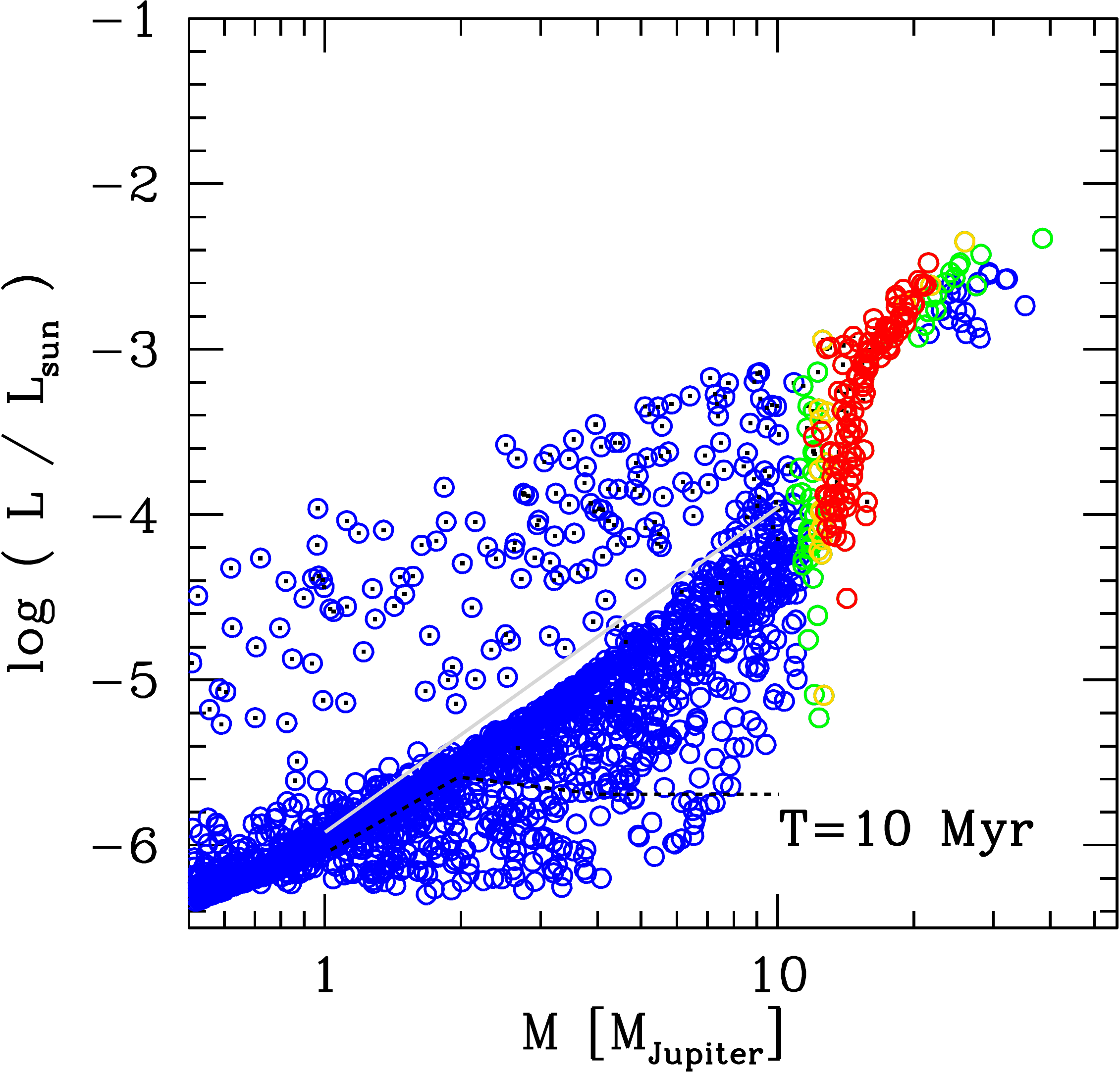}
     \end{minipage}
\caption{The  mass--luminosity relation during the formation phase for the cold-classical (low core mass) population, analogous to Fig. \ref{fig:mlform}. The gray and black lines in all four panels are identical as in Fig. \ref{fig:mlform}. Note the giant planets with very low post-formation luminosities, even lower than in the classical cold start simulations of \citet{marleyfortney2007} which are shown with the dotted black lines at 5 and 10 Myr. } \label{fig:mlformlimited}
\end{center}
\end{figure*}

Figure \ref{fig:mlformlimited} shows the mass-luminosity relation for the third population, the cold-classical population. As no accretion of planetesimals occurs in this population after detachment (as in \citealt{marleyfortney2007}), giant planets can have much lower core masses than in the cold-nominal population \citep[see Fig. \ref{fig:MLPostMcoreCD777} and ][for the core masses]{mordasiniklahr2014}. As the gas accretion is fully cold, this has, via the core-mass effect, the consequence that some planets are born with very low luminosities. The cold-classical population thus departs from the cold-nominal population in the opposite direction than the hot population does. This has the following consequences:

(1) First, generally speaking, this is the population with the lowest luminosities $\lint$ after formation. This is best seen when comparing the envelope of non-accreting planets (no black dots) with the gray solid line in the panel at 10 Myr. During the intense accretion phase, the total luminosity $\lint+\lshock$ is in contrast even (slightly) higher than in the cold-nominal population (and thus also in the hot population). The reason is the same as discussed in the previous section (total energy conservation). 

(2)  In contrast to the two other populations, there is a group of planets that have (essentially) stopped accreting and do not exhibit a positive correlation between planet mass and luminosity. In the other two populations, this positive correlation holds both during formation and evolution, despite a significant scatter around the mean $M-L$ relation. In the cold-classical population there is, in contrast, a significant group of planets  with masses between about 1 to 7 $\mj$ that have an $\lint$ that is concentrated around $\log ( L / \lsun) \sim -6$ independent of mass (see the panel at 5 and 10 Myr). This is very reminiscent of the \citet{marleyfortney2007} result who also found such a nearly mass-independent post-formation luminosity, which is not surprising as the cold-classical population is built on the same key assumptions. The $M-L$ relation of this work is shown in the 5 and 10 Myr panels with the black dotted line. We see that some of the synthetic planets have luminosities that are even lower than in \citet{marleyfortney2007}. Most planets have, however, not such extremely low luminosities, they rather fall into the warm start regime (see Fig. \ref{fig:Ldist10Myrgiants} below). Taken at face value, this population thus predicts that most giant planets have a warm start, while a smaller but significant group also has very low luminosities (cold start). 

(3) The low $\lint$ and high $\lint+\lshock$, as well as the very large spread in $\lint$ alone means that this is the population with the largest variation in the (total) luminosity at a given mass. The spread in $\lint$ is caused by the spread in the core mass of the planets. This is shown in Sect. \ref{sect:coldclassicalMLMcore}. A large spread could also result if the core-mass effect is not as efficient as assumed here (leading to lower $\lint$, {see Sect. \ref{sect:limintstruct}}), but if the efficiency of the accretion shock in radiating away the accretional energy is not 100\%. It will be interesting to repeat the statistical analysis once predictive models for the shock's efficiency \citep[e.g.,][]{marleauklahr2016} are coupled to planet formation and evolution models. 

(4) These points have the two following secondary consequences:  During the main formation phase (at 3 and 5 Myr) there is even a bigger separation between the accreting and evolving sequence compared to the cold-nominal population, as there is even a bigger decrease of the total luminosity at the moment when $\lshock$ vanishes.

(5) There is an even steeper $L-M$ relation in the deuterium burning mass domain at 10 Myr compared to the cold-nominal population, as in this mass domain an even lower $\lint$ (compared to the cold-nominal population) for the non-burning planets must be connected with the high $L\approx L_{\rm D}$ of the planets in the D-burning main sequence which is approximately equally high as in the cold nominal population because of the thermostatic effect.

\begin{figure}%
\begin{center}
	      \centering
            \includegraphics[width=\columnwidth]{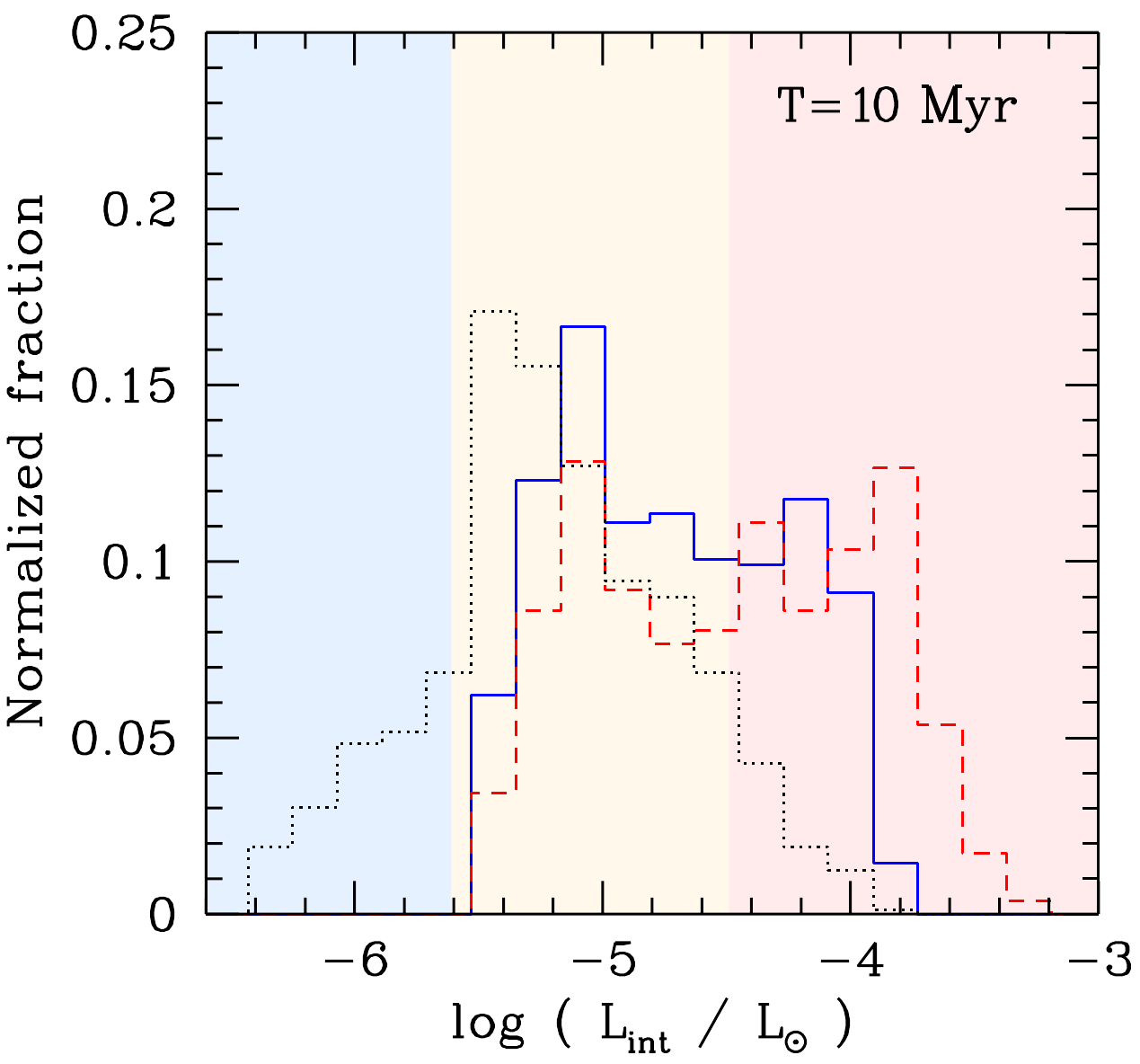}     
\caption{Distribution of the internal luminosity of giant planets with masses between 2 and 10 $\mj$ at 10 Myr in the cold-nominal (blue solid), hot (red dashed), and cold-classical population (black dotted lines). The colored regions  indicate representative cold, warm, and hot starts luminosities, even though the specific boundaries are in reality mass dependent. }\label{fig:Ldist10Myrgiants}
\end{center}
\end{figure}

To quantify the difference in the luminosities we show in in Fig. \ref{fig:Ldist10Myrgiants} the luminosity distribution of giant planets at 10 Myr. The plot shows the $\lint$ of giant planets with masses between 2 and 10 $\mj$ for the cold-nominal, hot, and cold-classical population. The plot also indicates roughly the luminosity domains that may be associated with cold, warm, and hot starts. It is clear that there is, in principle, not a single luminosity that corresponds to the cold/warm/hot transitions as the transition  at least for warm and hot starts depends on the mass \citep[e.g.,][]{spiegelburrows2012}. Therefore we have taken approximative limiting $\lint$ values for a planet of about 5 $\mj$. Given that the mass distribution is  similar in the three populations (see Fig. \ref{fig:histoMcomp}), the plot nevertheless makes clear that there is an obvious shift in the luminosity distribution between the three populations that comes from the different thermodynamics during accretion. One sees the shift from the cold-classical population that contains a significant number of cold cases, to the intermediate cold-nominal population, and finally to the hot population with a number of planets that have very high luminosities.

\subsubsection{Accretion shock and internal luminosities}\label{sect:lshocklint}
\begin{figure*}%
\begin{center}
	      \centering
            \includegraphics[width=\textwidth]{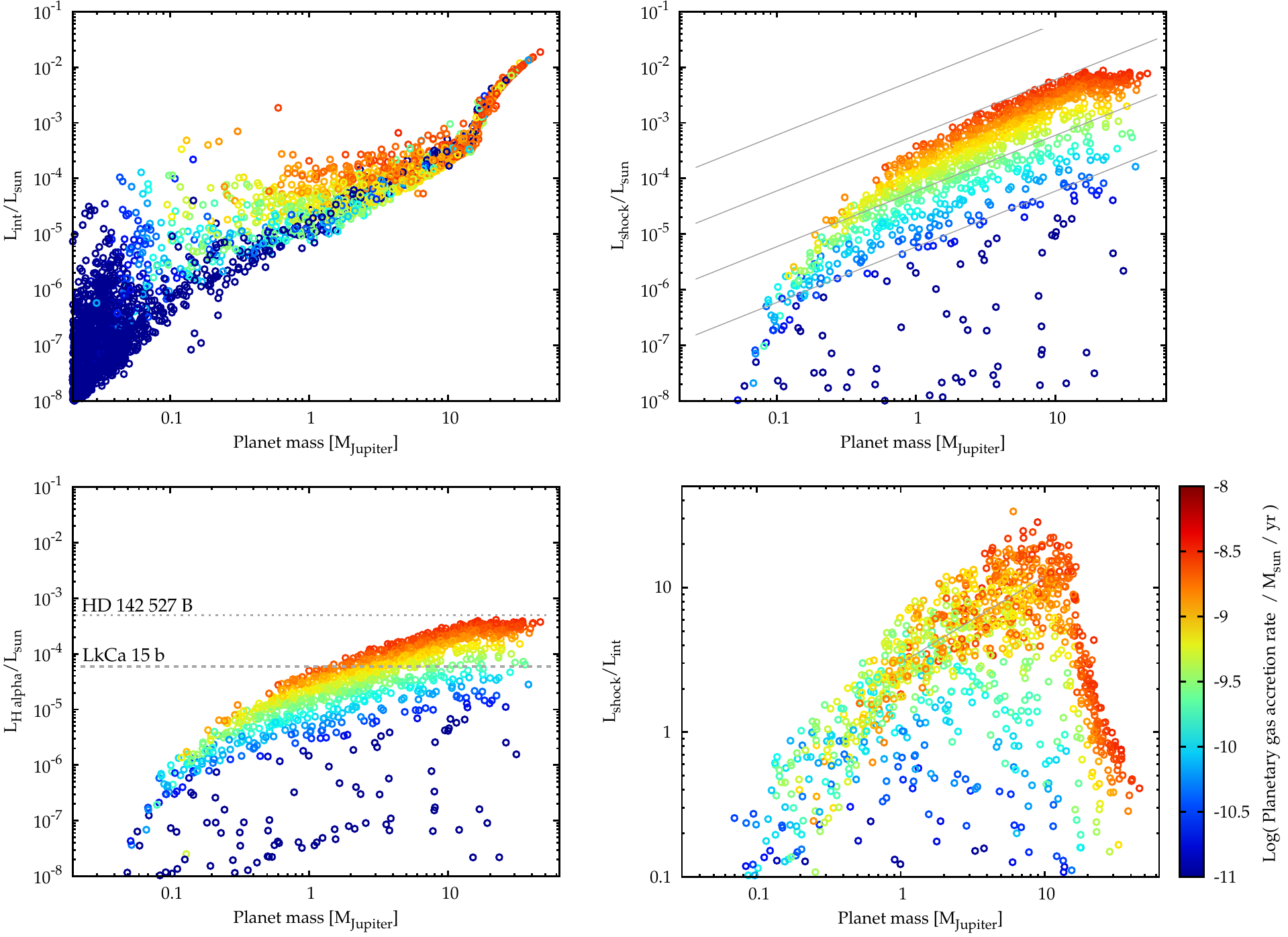}     
                 \caption{The separate contributions to the luminosity of giant planets during the formation phase for the cold nominal accretion case. The internal luminosity $L_{\rm int}$ (top left) and the gas accretion shock luminosity $L_{\rm shock}$ (top right) is shown as a function of planetary mass for the population at 3 Myr. The bottom left panel shows the H$\alpha$ luminosity obtained using the empirical $L_{\rm shock}$ - $L_{\rm H\alpha}$ relation from \citet{rigliaconatta2012} derived for T Tauri stars. We stress that the validity of this relation in the planetary mass range is currently unknown. The observed $L_{\rm H\alpha}$ of HD 142527B \citep{closefollette2014} and LkCa 15b \citep{sallumfollette2015} are also shown for comparison. The bottom right panel shows the ratio of the accretion to the internal luminosity. The colors represent the planets' gas accretion rate. The solid gray lines in the top right panel show $\lshock$ as a function of mass for a fixed radius of 1.5 $\rj$ and accretion rates of $10^{-5}$, $10^{-4}$, $10^{-3}$, and $10^{-2}\mearth$/yr. The gray line in the bottom right panel  shows a $M^{0.5}$ scaling.  }\label{fig:MLLacc3Myr} 
\end{center}
\end{figure*}

The separate detection of H$\alpha$ emission (a typical tracker of accretion) from LkCa 15b \citep{sallumfollette2015} indicates that it is possible to spectroscopically separate $\lint$ from $\lshock$. We therefore plot the separate contributions to the total luminosity in Fig. \ref{fig:MLLacc3Myr}, namely  $\lint$ (originating from the cooling and contraction of matter already in the planet, planetesimal accretion, and potentially deuterium burning) in the top left panel, and $\lshock$ originating from the accretion of the gas in the top right. The cold-nominal population is shown at an age of 3 Myr. In the bottom left we also show an estimate of  $L_{\rm H\alpha}$, while the bottom right panel shows the ratio $\lshock/\lint$. 

The H-$\alpha$ luminosity  is calculated from $\lshock$ via the scaling relation \citep{rigliaconatta2012} for low-mass T Tauri stars (typical $\mstar\sim0.2\msun$)
\beq
\log\left(\frac{L_{\rm H\alpha}}{\lsun}\right)=\frac{1}{a}\left(\log\left(\frac{\lshock}{\lsun}\right)-b\right)
\eeq
where $a=1.49$ and $b=2.99$. We stress that it is currently unknown whether these scaling relations also hold for objects that have masses that are $\sim2$ orders of magnitudes lower \citep[see also][]{zhu2015}.

The colors in the top left panel showing $\lint$ indicate that for accreting giant planets well in the detached phase ($M\gtrsim 1 \mj$) there is a correlation between $\lint$ and the accretion rate $\dot{M}_{\rm XY}$. For planets below the D-burning limit, this can be analyzed by writing \citepalias[\citealt{hartmanncassen1997},][]{mordasinialibert2012b}
\beq
\lint=L-\lshock=(\xi -1) \frac{G M \dot{M}_{\rm XY}}{R}-\frac{\xi G M^{2}}{2 R^{2}}\dot{R}+\frac{GM^{2}}{2 R}\dot{\xi}
\eeq
where 
\beq
\xi=-\frac{2 R E_{\rm tot}}{ G M^{2}}
\eeq
with $E_{\rm tot}=E_{\rm grav}+E_{\rm int}$ the total energy that we determine numerically from the internal structures. We see that for $\xi>1$, the first term is positive and will introduce, if dominant over the two other terms, a $\lint\propto M  \dot{M}_{\rm XY}$. For a $n=1$ polytrope that can been used to model Jupiter nowadays \citep[e.g.,][]{Hubbard1975}, $\xi=1.5$, while in the simulations, during the disk-limited gas accretion phase, we rather see  $\xi\approx1.2$. Thus this term is indeed positive. But if $\lint$ is plotted as a function of $M$ and $\dot{M}_{\rm XY}$, in the simulations, roughly speaking, rather a $\lint\propto M^{0.6} \dot{M}_{\rm XY}^{0.6}$ is found for strongly accreting giant planets, indicating that also the other terms are important.  This scaling is clearly different than the $L\propto M^{2}$ for non-accreting giant planets \citep{burrowsliebert1993} and should be investigated further in dedicated work.  In the panels one furthermore sees the clear imprint of D-burning, strongly increasing $\lint$ for sufficiently massive planets with $M\gtrsim14\mj$. 

The top right panel shows $\lshock$ that can be well approximated with Eq. \ref{eq:lest} for planets between about 1 to 14 $\mj$ assuming a constant radius  of 1.5 $\rj$ (see Fig. \ref{fig:MRcold}). This is illustrated by the gray lines that show Eq. \ref{eq:lest} for a fixed $R=1.5\rj$ and $\dot{M}_{\rm XY}$=$10^{-5}$, $10^{-4}$, $10^{-3}$, and $10^{-2}\mearth$/yr. For lower masses, $\lshock$ is smaller than this expression because the radii are larger. %
The plot also shows that for the most massive planets, $\lshock$ does not further increase. This is a consequence of their increased radii because of  deuterium burning (Fig. \ref{fig:mlform}). 

The bottom right panel shows $\lshock/\lint$. This is an interesting quantity in the context of the detectability of accreting planets. One would maybe expect that from $\lshock\propto M$, and $\lint \propto M^{2}$ for non-accreting planets, the ration should scale as $1/M$. Instead, it increases with mass, in the 1 to 10 $\mj$ domain approximately as $M^{0.5}$. This is the slope of the gray solid line in this panel. This slope can, however, again roughly speaking, be understood with the previously mentioned results because of
\beq
\frac{\lshock}{\lint}\propto\frac{ M \dot{M}_{\rm XY}}{M^{0.6} \dot{M}_{\rm XY}^{0.6}}\propto M^{0.4}.
\eeq
For the second proportionality we have used that in our model, the gas accretion rate in the disk-limited phase is simply a fixed fraction of the local gas accretion rate in the protoplanetary disk, which is independent of planet mass. If the mass of the growing giant planet influences its gas accretion rate in this phase, a different scaling  would result. This is interesting as it may provide observational hints about the process of disk-limited accretion: if the planet remains on a circular orbit in the disk-limited phase, the gas accretion rate is a decreasing function of mass as the gap becomes increasingly wide and deep \citep{lubowseibert1999}. If the planet-disk system becomes instead eccentric because of the eccentric instability \citep{kleydirksen2006}, then the gas accretion rate is high also for massive planets. 

At even higher masses, once deuterium burning sets in, the ratio decreases as $\lint$ increases because of $L_{\rm D}$, whereas $\lshock$ decreases because of the larger radii.

These results indicate that the accretion shock luminosity of  accreting giant planets with masses between about 1 to 14 $\mj$ could be between a factor of a few to more than an order of magnitude higher than their internal luminosity, which would suggest that searching for accretion trackers like H-$\alpha$ or Paschen-$\beta$ line emissions is a promising way to detect forming, detached giant planets. 

However, three caveats must be added: (1) concerning the energetics of the accretion, our model is a simple one-zone model (although it does take the entire radial structure via the numerical calculation of $\xi$ into account) with a radially constant luminosity. While  numerous comparisons with other works that solve the full set of equations  \citep[like][to name just a few]{burrowsmarley1997,bodenheimerhubickyj2000,marleyfortney2007,bodenheimerdangelo2013} showed excellent agreement \citepalias{mordasinialibert2012b}, it cannot be excluded that for some scenarios this approximation leads to different results than a more detailed treatment of the energetics of the accretion process (as in \citealt{berardocumming2016}). This could, in turn, influence the predicted $\lint$ and $\lshock$. (2) Our calculations are for a spherically symmetric accretion flow and neglect the influence of a circumplanetary disk. (3) The shock luminosity of embedded object may not escape unaltered the surrounding protoplanetary disk.

The bottom left panel shows $L_{\rm H\alpha}$, estimated from $\lshock$ with the aforementioned scaling relation. As expected from the relation, it follows a similar morphology as $\lshock$ itself. The comparison with LkCa 15b is found in the next section. The $L_{\rm H\alpha}$ of HD 142527 B \citep{closefollette2014}, which is a low-mass M dwarf \citep{lacourbiller2016}, is also shown for comparison.

\subsubsection{Comparison with observed forming companions:  HD 100546 b and LkCa 15 b}\label{sect:compembedded}
It is interesting to see how the luminosities found theoretically for the accreting or just formed protoplanets compare to the recent observations of the several (candidate) companions that are presumably still forming in their parent disk. In this simple explorative analysis in the context of a statistical study, we will only use the inferred luminosities and ages of the observed objects to search for synthetic counterparts. We thus do not use other available constraints like the metallicity, accretion rate, and mass of the host star (which is for some objects significantly different than the 1 $\msun$ adopted here), the properties of the protoplanetary disk (mass, gaps, SEDs, etc.), and the specific orbital, photometric, and morphological properties of the embedded companion here. Ignoring these constraints means that the results must be understood as only illustrative in nature. At the same time this long list of observational data demonstrates all the available constraints for future dedicated object-by-object comparisons, or in other words the high constraining power of observations of planet formation as it happens for theoretical models. This is especially the case as all these quantities are also available from global planet formation models as presented here, meaning that comprehensive comparisons covering many different aspects are possible.  

\textbf{HD 100546 b}
This object was discovered by \cite{quanzamara2013}, recovered by \citet{curriemuto2014}, and confirmed by \citet{quanzamara2015} {(but see also \citealt{rameaufollette2017})}. The Herbig  Ae/Be host star is a young, actively accreting intermediate-mass B9Vne star ($M=2.4\pm0.1\msun$, $L\approx32\lsun$, $T_{\rm eff}\approx 10500$ K). The star has a complex transition disk, and its age is estimated to be 3-10 Myr \citep{brittaincarr2014}, with a poorly constrained stellar accretion rate of  $5\times 10^{-8} \msun$/yr (corresponding to 0.017 $\mearth$/yr) according to \citet{wrightmaddison2015}, while other estimates give a $\sim$ten times lower value \citep{gradywoodgate2005}. Here we assume a fixed age of 5 Myr, but this could be extended to cover the full allowed age range in an in-depth analysis. 

According to \citet{quanzamara2015}, the co-moving companion detected in L' and M' is embedded in the disk and orbits the star at a deprojected distance of about 53 AU.  The observed emission consists of a point source component surrounded by spatially resolved emission.   By fitting a single-temperature blackbody to the point source, an effective temperature of about $T=932\pm200$ K, a radius of $6.9\pm2.8 \rj$ and a luminosity of $\log(L/\lsun)$ between -3.72 and -3.54 is inferred. This range is indicated in Fig. \ref{fig:mlform} and \ref{fig:mlformhot} at 5 Myr. Here we associate this luminosity with $\lshock+\lint$. The large radius of the emitting zone potentially rather indicates emission from a circumplanetary disk, or an early post-detachment state at a lower mass than shown in Fig. \ref{fig:mlform} and \ref{fig:mlformhot} (see below). For an accreting circumplanetary disk in Keplerian rotation, half of $\lshock$ would  be emitted from the disk \citep[e.g.,][]{frankkingAkkretion}, and the other half from the viscous boundary layer or magnetospheric accretion. This would change the SED \citep{eisner2015,zhu2015}, but the total bolometric luminosity would remain the same.

For cold accretion, synthetic planets within the observed luminosity range have the following properties (ignoring for the moment all other constraints like $T_{\rm eff}$): a mass range between about 0.2 and 9.7 $\mj$ for the planets still actively accreting where typical values are 1-4 $\mj$, and between 9.7 and 13.9 $\mj$ for non-accreting planets (which seems however unlikely in the current embedded case). The planet can thus either have a high mass without (or very little) accretion, or a low (or even very low) mass down to $<1 \mj$ if it is undergoing strong gas accretion. For comparison, at 5 Myr, \citet{quanzamara2015} find a best-fit mass a of non-accreting, hot start planet of about 10 $\mj$.  The gas accretion rates of the synthetic planets with observed luminosities are between $10^{-5}-4\times10^{-4} \mearth$/yr, with the lowest value for the most massive, still accreting planets (for higher $\dot{M}_{\rm XY}$ they would have a too high $L$). For the lowest mass planets, the accretion of planetesimals also contributes significantly, meaning that potentially very luminous impacts may occur frequently on this planet. The radii are 1.9 $\rj$ for the smallest masses (which are still in the process of significant contraction), decreasing to about 1.4 $\rj$ towards higher masses  $\lesssim10\mj$. For even more massive, non-accreting planets, the radius increases again to 1.5-1.6 $\rj$ because of D-burning. The effective temperatures are between 1600-2000 K. Regarding the radius, the values of the synthetic analogs are clearly smaller, and consequently larger for $T_{\rm eff}$ than the ones estimated by \citet{quanzamara2015}, indicating a more extended region of emission. However, it has to be kept in mind that the effective temperature of the HD~100546b is derived from fitting a blackbody spectrum to 4 photometric points, one of which is only an upper limit. The spectral shapes of cool objects may deviate from a blackbody, due to the presence of absorbing molecules like water or methane. This could affect the estimated total luminosity and effective temperature. On the other hand, a significant dust envelope could weaken the effects of molecules because dust opacities are gray absorbers in comparison to the molecules. Clearly, observations at a higher spectral resolution would be very helpful.

The ratio $\lshock/\lint$ of the subset of synthetic planets with the same luminosity as HD~100546b is about 0.1 at the lowest mass, then increases to about 15 at 3 $\mj$, and then decreases again towards more massive but still accreting planets, qualitatively similar to Fig. \ref{fig:MLLacc3Myr}. The accretion luminosity is thus clearly dominant for intermediate mass planets, meaning there is enough luminosity that could be liberated over an extended region around the planet, instead of directly on the planet's surface. This could be a circumplanetary disk or an otherwise heated zone where the hard accretion shock luminosity gets absorbed and then re-emitted at longer wavelength. $L_{\rm D}/\lint$ is about  0.15 for the most massive, but still accreting planets, and completely negligible for the more typical lower masses. For the non-accreting planets, $L_{\rm D}/\lint$ ranges between $10^{-3}$ and 0.8. The planets have entropies between 9 and 11 $k_{\rm b}$/baryon, and core masses between 20 and 400 $\mearth$, with typical values of around 70 $\mearth$. One important constraint can clearly not be met by the synthetic planets: the semimajor axes are just 0.2-16 AU, much less than the 53 AU in the observation. Here, mechanisms currently not included in the formation model must be invoked, which could be scattering \citep{alibertcarron2013} or a faster core growth for example via pebbles \citep{ormelklahr2010,bitschlambrechts2015}.

Due to the core-mass effect, even cold gas accretion leads to quite warm planets in our model, therefore considering hot instead of cold gas accretion (Fig. \ref{fig:mlformhot}) does not strongly change the properties of the synthetic planets with the same luminosities as HD 100546 b. Still, the following numbers change somewhat: the masses are now between 0.2 and 8.5 $\mj$, i.e., the highest mass is smaller because of the still higher entropy. The accreting planets have masses between about 0.2 and 6.5 $\mj$, mainly uniformly distributed between 2 and 5 $\mj$. The radii of the accreting planets are larger than for cold accretion, usually between 1.6 and 2 $\rj$, but still much smaller than derived from the observation.  Because of the lower masses, deuterium burning is of negligible importance, while the entropies are between 10 and 11 $k_{\rm b}$/baryon.

Up to this point we have mainly concentrated on synthetic planets that have the same luminosity as observed, assuming that the observed luminosity originates at least partially from the accretion shock. We have seen that such objects have radii smaller than observed, making it necessary to invoke elements not present in the simple 1D framework, like the presence of a circumplanetary disk or a heated gas blob around the planet \citep{klahrkley2006}, which is certainly possible. 

An alternative hypothesis within the 1D picture that takes, besides the luminosity, also the effective temperature into account is that the object is in an earlier evolutionary state. As mentioned above (Sect. \ref{sect:MLformcold}), protoplanets can also be very bright for a short period during the fast contraction phase just after detachment. Indeed, in the cold-nominal population one finds for example a synthetic planet that has at 4.96 Myr a $\log(L/\lsun)$=-3.54, a radius of about 6.5 $\rj$ and an effective temperature of 915 K, all in agreement with the observed values. This planet has just detached 30'000 year earlier, and is still in the rapid contraction phase which is together with planetesimal impacts the main source of its luminosity. The gas accretion shock luminosity $\propto M/R$ becomes important only later because of the still extended state of the planet and its low mass. The planet has a mass of just 78 $\mearth$ and accretes gas at a rate of about $5\times10^{-4}$ $\mearth$/yr which would be about a third of the lower estimates for the stellar accretion rate. While with 5 AU the semi-major axis of the synthetic planet is much less than the one of the actual object such that this is not a real analog of HD 100546 b, it indicates that such an earlier stage can explain at least some of the observations. The critical point is here that this phase only lasts a few $10^{4}$ years, while the phase were the accretion shock yields the luminosity lasts $10^{5}-10^{6}$ years. We would thus observe HD 100546 b in a special moment, which is not very likely (but neither impossible). Because of its low mass the planet would likely not have yet opened a clear gap in the disk, which is in agreement with the absence of a gap in the polarized light images of \citet{AvenhausQuanz2014}, and it would also not be a significant source of hard radiation like H$\alpha$. Similar objects can also be found in the hot population. These considerations also point towards the fascinating possibility to directly observe the fundamental phases of giant planet formation predicted by the theoretical models (attached, detached, during the rapid contraction, etc.). This will be further addressed in dedicated work.

In summary we see that because of accretion, a wide range of masses is compatible with the observed luminosity. In particular planets with low, potentially sub-Jovian masses could generate the observed luminosity, either due to accretional luminosity, or due to rapid contraction of the gaseous envelope. %

\textbf{LkCa 15 b} 
LkCa 15 is a young (2--5 Myr) solar analog ($L\approx0.74\lsun$, $M=1.01\pm0.03\msun$) in Taurus-Auriga \citep[e.g.,][]{isellachandler2014,pietudutrey2007} that hosts a gas-rich circumstellar disk \citep[][]{andrewswilliams2005,andrewsrosenfeld2011}. Despite a large cavity in the dust extending to about 45 AU, the star is accreting at rate of about $1.3\times10^{-9}$ $\msun$/yr as inferred from the stellar UV excess \citep{hartmanncalvet1998}.

\citet{krausireland2012} have reported the discovery of an unresolved source in K$^{\rm \prime}$ and a surrounding extended emission in L$^{\prime}$  that they interpret as a forming protoplanet surrounded by heated circumplanetary material like a circumplanetary disk. If this candidate protoplanet is coplanar with the disk and on a low-eccentricity orbit, a semimajor axis of 16 AU results. They also estimated a total bolometric luminosity of $\sim2\times10^{-3}\lsun$, and that the unresolved emission could be reproduced naively with the photospheric emission of a non-accreting giant planet of 6 to 10 $\mj$, but stress that the luminosity could rather result from accretion. \citet{sallumfollette2015}  detect two sources, one of which is detected both in Ks and L$^{\prime}$ and in H$\alpha$. They explain the infrared emission by a circumplanetary disk, and the H$\alpha$ by emission from a hot accretion shock with $L_{\rm H\alpha}\approx 6 \times 10^{-5}\lsun$ that they convert via T-Tauri scaling relations \citep{rigliaconatta2012} into an accretional luminosity of $\sim4\times10^{-4}\lsun$.

The bottom left panel  of Fig. \ref{fig:MLLacc3Myr} compares the inferred $L_{\rm H\alpha}$ of Lk Ca 15b with the synthetic population at 3 Myr (a comparison at 2 Myrs yields similar results). It suggests that planets with masses between 1 and several tens of Jupiter masses (i.e., a wide range) can have such accretion luminosities if the luminosity is due to the planetary gas accretion shock and the \citet{rigliaconatta2012} relation extends to these masses, with low masses corresponding to high gas accretion rates and vice versa. However, if it is additionally requested that the planetary gas accretion rate (color coded in Fig. \ref{fig:MLLacc3Myr}) is as least as high as the observed stellar accretion rate as suggested by hydrodynamic simulation \citep{lubowdangelo2006}, then only a narrow mass interval of giant planets with masses between 1 to 2 $\mj$ concurrently fulfills these two constraints. These synthetic planets have a ratio $\lshock/\lint$ of 3-10, radii of about 1.6-2 $\rj$, and escape velocities between 30 to 70 km/s. Their effective temperature to radiate $\lint$ is about 1100 to 1800 K, while the effective temperature to radiate both $\lint+\lshock$ is between 1900 and 2400 K if the accretion shock covers the entire surface of the planet (strictly spherical accretion). 

In the context of magnetospheric accretion onto CTTS one rather observes covering factors $f$ of the accretion columns on the stellar surface that are much less than unity, $f\sim$0.1--30\% \citep[e.g.,][]{calvetgullbring1998,bouvieralencar2007,inglebycalvet2013}. If this picture also applies to planets, for $f=1$\%, these planets would have  shock-heated spots on the photosphere with temperatures of about 6000-8000 K, in good agreement with the estimates of \citet{zhu2015}. We can finally use Eq. 22 of his work (which assumes that the temperatures in a planet's magnetosphere--if there is one--is 8000 K as for CTTS) to very roughly estimate $L_{\rm H\alpha}$. One finds $L_{\rm H\alpha}/\lsun\sim 1-2\times 10^{-5}\times(R_{\rm T}/R_{\jupiter})^{2}$ where $R_{\rm T}$ is the  magnetospheric cavity truncation radius, which depends on the unknown magnetic field strength of the planet. For $R_{\rm T}\sim R_{\jupiter}$ this is at least to order of magnitude consistent with the observationally determined  $L_{\rm H\alpha}\approx 6 \times 10^{-5}\lsun$ that stood at the beginning of this analysis. We conclude that giant planets with rather low masses of 1-2 $\mj$ could be responsible for the observed $L_{\rm H\alpha}$, and that observational determinations of $\lint$, as well as of the magnetic field strength of young planets would be important to further understand the process of planetary gas accretion. We also add the caveat that our synthetic analogs have semimajor axes between 1-5 AU, clearly less than LkCa 15 b, such that they cannot meet also this observational constraint.

\subsubsection{Evolution phase: a spread in the $M-L$ relation at young ages}\label{sect:evophasespread}
\begin{figure*}
\begin{center}
\begin{minipage}{0.5\textwidth}
	      \centering
        \includegraphics[width=0.9\textwidth]{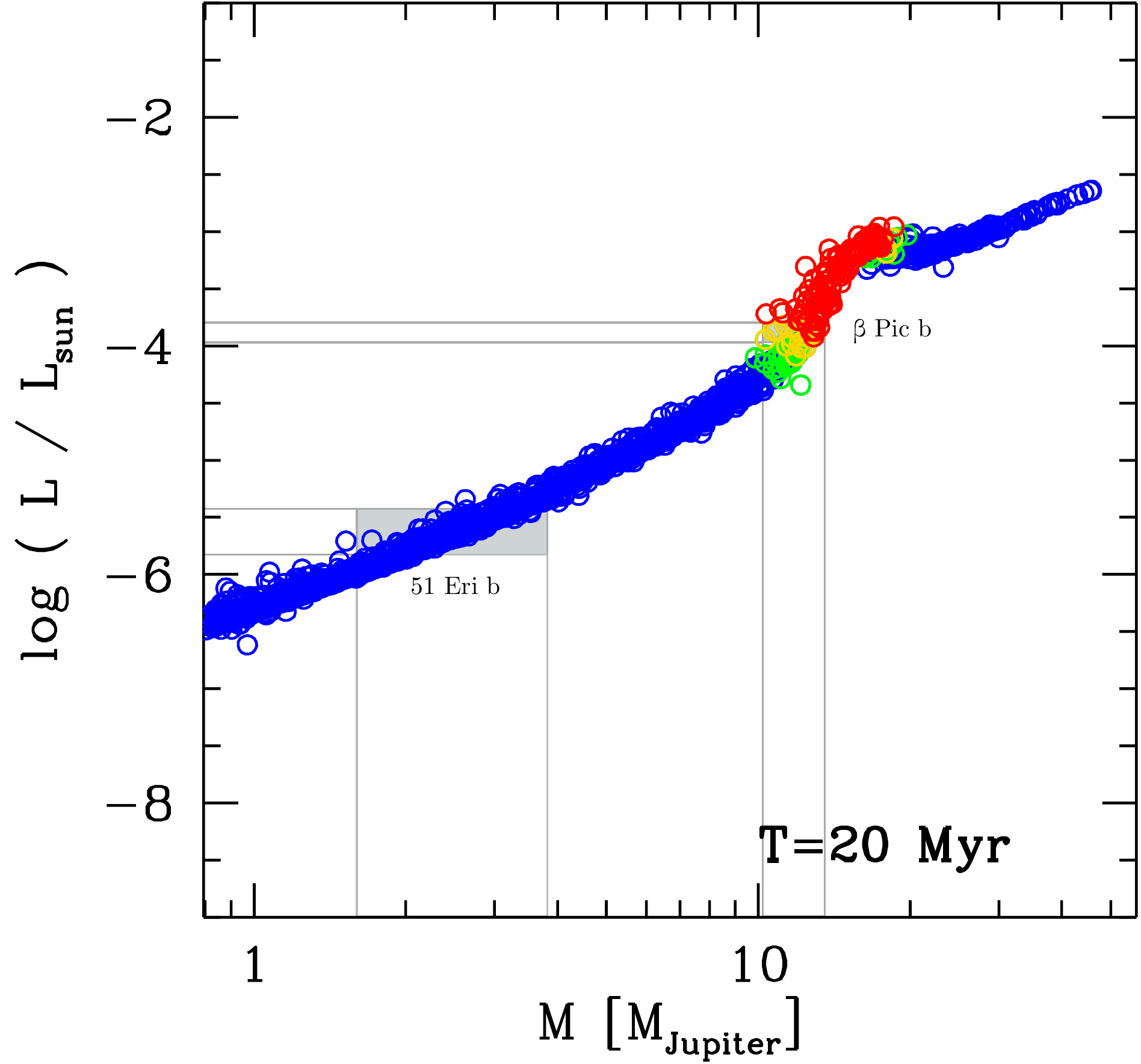}
        \includegraphics[width=0.9\textwidth]{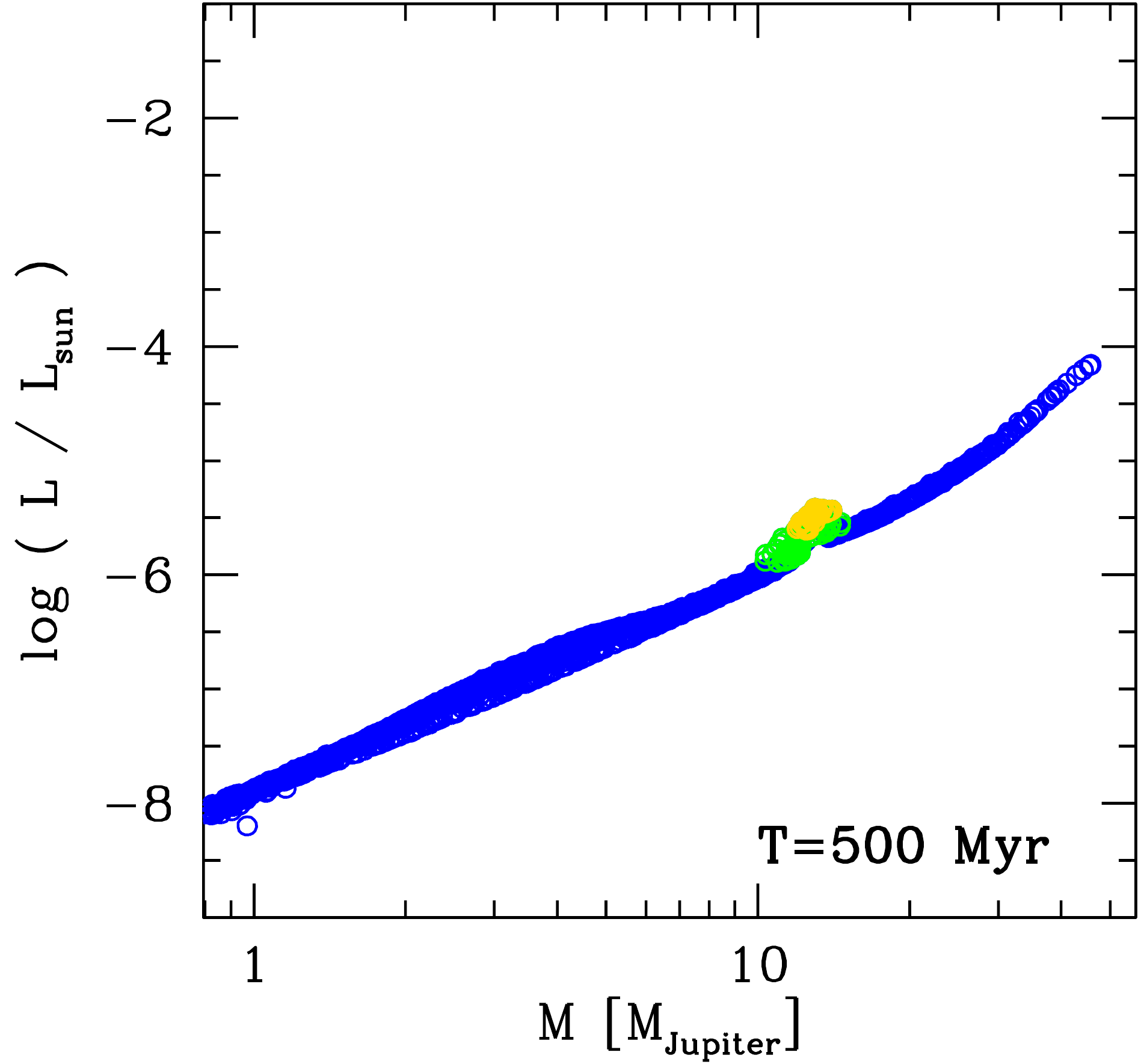}
     \end{minipage}\hfill
     \begin{minipage}{0.5\textwidth}
      \centering
       \includegraphics[width=0.9\textwidth]{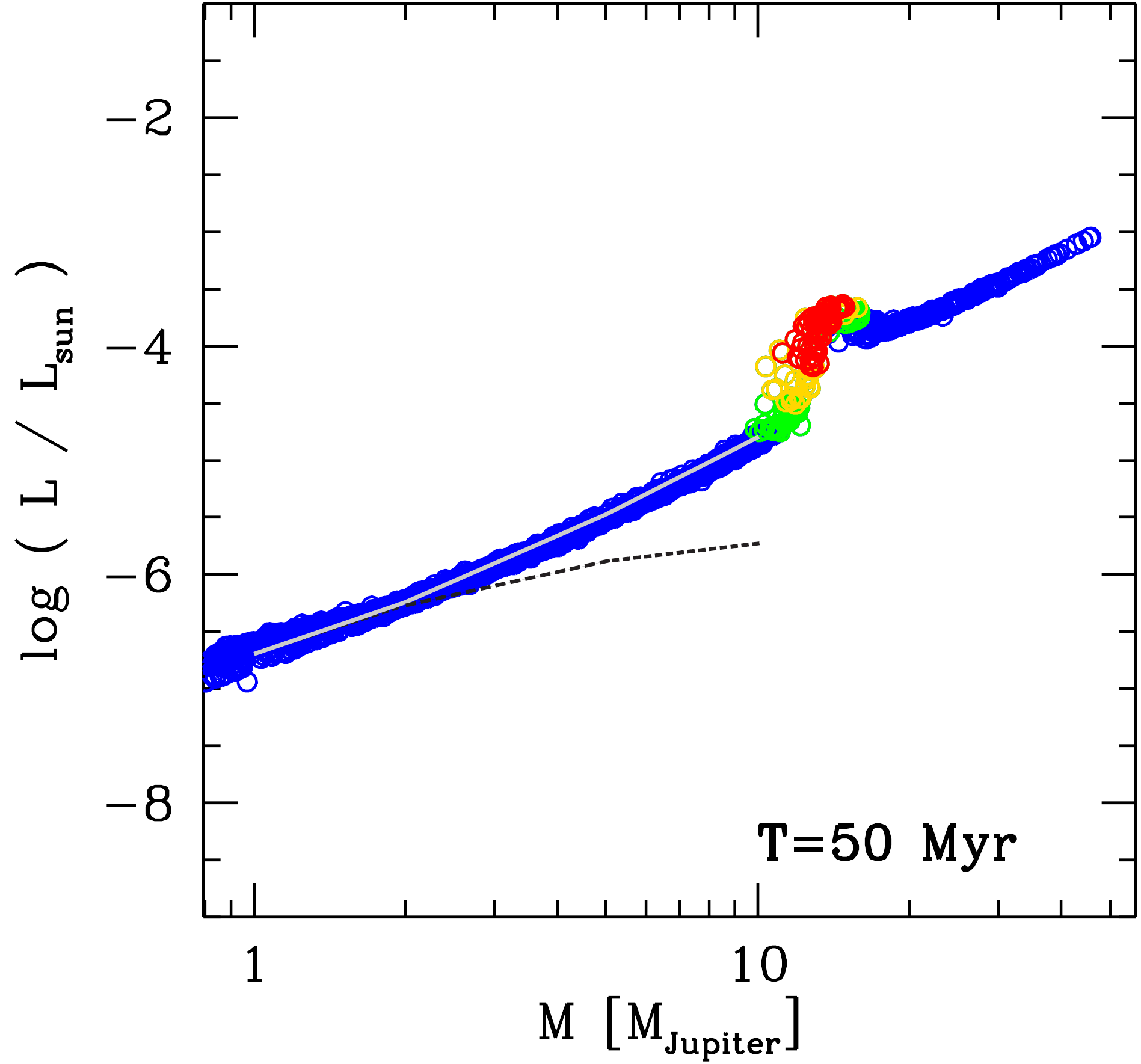}
      \includegraphics[width=0.9\textwidth]{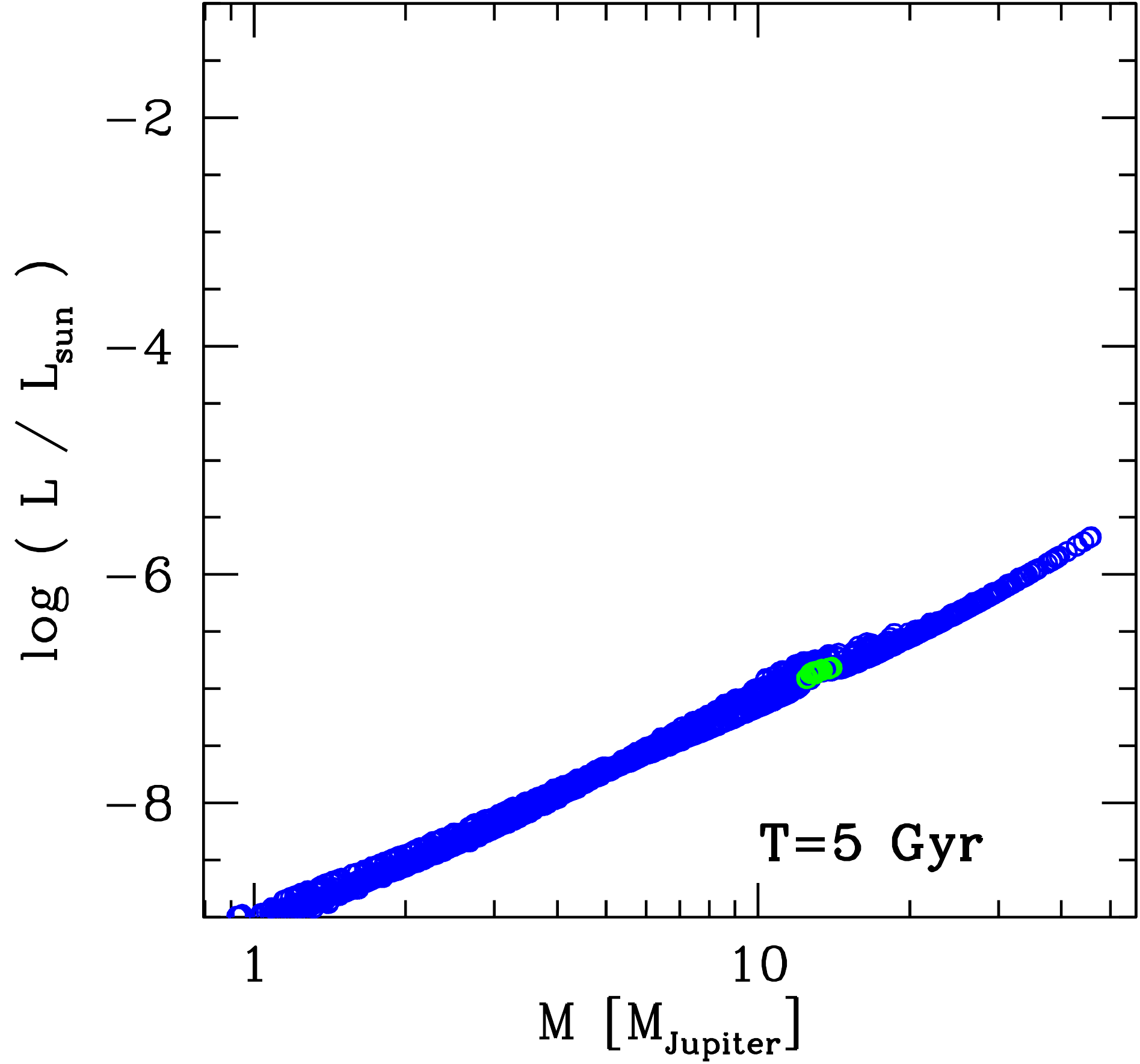}
     \end{minipage}
\caption{The $M$ -  $L$ relationship during the evolutionary phase at constant mass for the cold-nominal population, analogous to Fig. \ref{fig:mlform}. In the top left panel at 20 Myr, the gray lines and shaded regions indicate the observed luminosities and derived masses for 51 Eri b and $\beta$ Pic b.  In the panel at 50 Myr, the gray solid and black dashed lines show the $M-L$ relation of \citet{burrowsmarley1997} and \citet{marleyfortney2007}, respectively.}\label{fig:mlevo}
\end{center}
\end{figure*}

As evident from the last section, it is difficult to constrain the mass from a measured total luminosity  ($\lint+\lshock$) alone during formation because of the possible contribution of accretion. A comprehensive view with separate (spectroscopic ) determination of the different contributions (planetary internal, planetary accretion shock, and circumplanetary disk luminosity) combined with other indicators (like accretion rate or disk structure) should--if possible--be used to get a clearer picture. 

But also during the evolution phase, the diversity of planetary formation histories can still influence a planet's thermodynamical state, meaning that there is no one-to-one mapping from one luminosity to one mass. This has the implication that even if a young planet's luminosity could be measured perfectly without any error bar, one could still only infer its mass from it within a certain interval. This is well known from the cold vs. hot start dichotomy. But even if all planets have a quasi-hot start as it is the case in the cold-nominal population, there is an intrinsic spread in the $M-L$ relation at early ages. We show this in Figure \ref{fig:mlevo} displaying the $M-L$ relation of the cold-nominal population at four moments in time during the evolutionary phase at constant mass. We discuss in the following sections the main reason for the scatter, which is the scatter in the core mass, and exemplify the implications of the non-unique $M-L$ relation by applying it to $\beta$ Pic b and 51 Eri b. The even much bigger spread introduced by core masses that vary much more is discussed in Sect. \ref{sect:coldvshot} which described the $M-L$ relation of the cold-classical and hot population.

At 20 Myr, the spread in the $M-L$ relation leads to a $L$ that can vary by 50\% or more at a given mass. The panel for 50 Myr shows that at this time, the cold-nominal population already follows well the general hot start $M-L$ relation indicated by the gray line, but still it is not unique (see Fig. \ref{fig:mlmcore}).  This panel, together with the one at 0.5 Gyr, also illustrates the effects of deuterium burning: planets undergoing D-burning show up as a clear bump in the $M-L$ relation that in time moves to lower masses and becomes less prominent.  The green symbols at 5 Gyr demonstrate that low-level D-burning (5\% of the total $L$) continues to this age, but only with a very minor imprint on the $M-L$. One can also observe a ``rejuvenating'' effect of D-burning. In the  50 Myr panel, the planets below the D-burning limit and those where the D-burning is already over (blue symbols) both follow the identical $L\propto M^{2}$ scaling. But the planets that previously underwent D-burning have an offset to higher $L$, as if they would be younger objects. 

One furthermore notes the thickening and flattening of the $M-L$ relation at $\log(L/\lsun)\approx-6.5$ that is visible at 50 Myr, 0.5, and 5 Gyrs (where it coincides with the D-burning planets). This is in contrast to the aforementioned spread not an echo of the formation process, but is caused by a change of the internal structure of the planets (a detached radiative zone). We discuss this feature, which is additionally distance (irradiation) dependent, and its impact on the planetary luminosity distribution in Sect. \ref{sect:Ldistro}.

\subsubsection{Impact of $M_{\rm core}$ for the cold-nominal population}\label{sect:impactcoremass}

\begin{figure*}
\begin{center}
\begin{minipage}{0.5\textwidth}
	      \centering
        \includegraphics[width=0.93\textwidth]{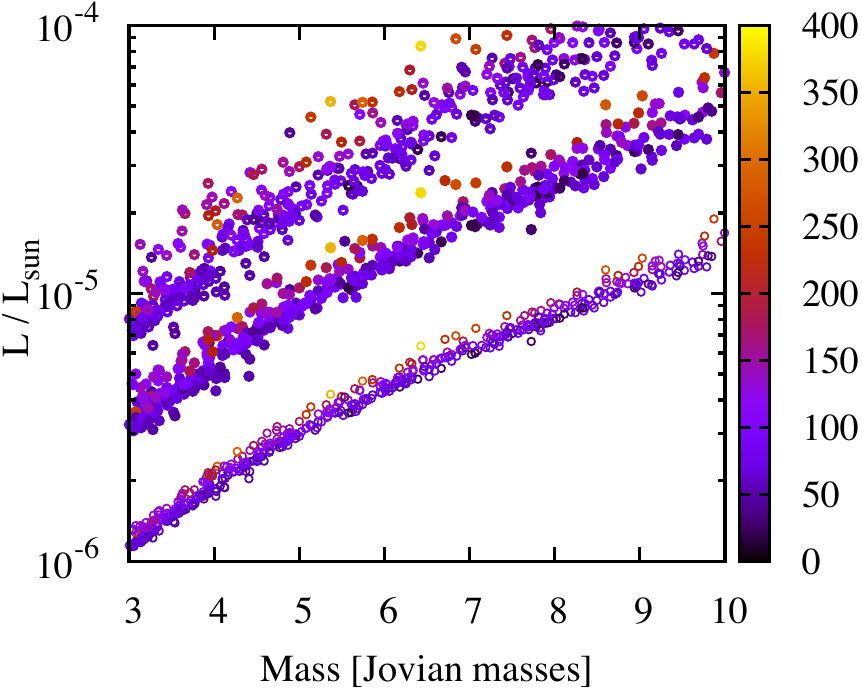}
     \end{minipage}\hfill
     \begin{minipage}{0.5\textwidth}
      \centering
      \includegraphics[width=0.9\textwidth]{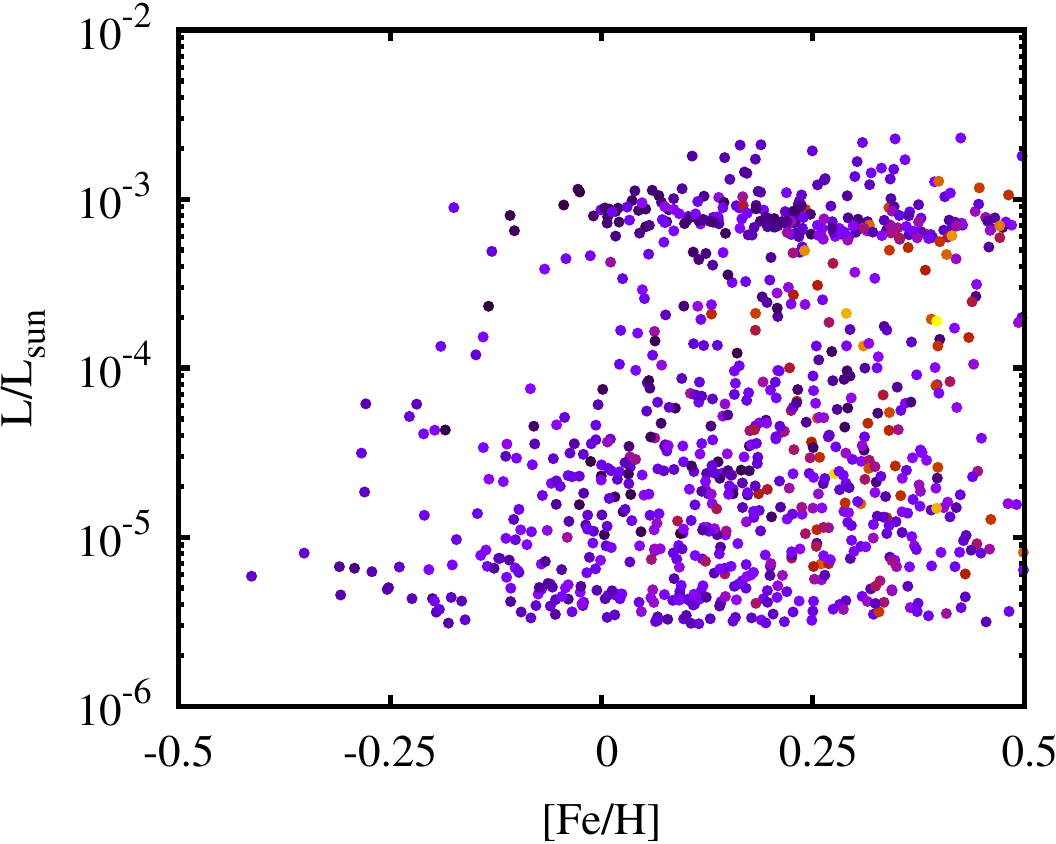}
     \end{minipage}
\caption{Left panel: Internal luminosity as a function of (total) mass (3 to 10 $\mj$) at 10 (half-filled), 20 (filled), and 50 Myr (empty circles) in the cold-nominal population. But the color code now gives the core mass in units of Earth masses. The planets with a more massive core are more luminous for a given total mass and age. Right panel: Internal luminosity as a function of [Fe/H] of the host star (and parent protoplanetary disk) at 20 Myr. The colors give again the core mass. Planets with $M\geq3\mj$ are shown.}\label{fig:mlmcore}
\end{center}
\end{figure*}

The left panel of Fig. \ref{fig:mlmcore} again shows the internal luminosity (excluding $\lshock$ for the handful planets still accreting at 10 Myr) as a function of mass at 10, 20, and 50 Myr for the cold-nominal population. The plot shows three things: 

(1) First, as the plot is zoomed in into a smaller mass and luminosity range than Fig. \ref{fig:mlevo}, it allows to better quantify the intrinsic luminosity spread at a given mass and age. We see that at an age of 20 Myr which is well into the evolutionary phase, there is still a variation in luminosity by a factor of about 1.5 to nearly 2 at a fixed mass. At an age of 50 Myr, the spread is as expected reduced as the hotter planets cool faster leading to convergence, but even then the intrinsic spread is still between 20 to 50\%. This should be critically kept in mind when deriving masses from measured luminosities even at ages of $\sim$50 Myr.

(2) Second, it shows via the color code the planets' (high) core masses. These high core masses are the reason for the high luminosities even for cold gas accretion, via the aforementioned core-mass effect \citep{mordasini2013,bodenheimerdangelo2013}. The core masses in our simulations are, in particular, usually clearly higher than in \citet{marleyfortney2007}. While these authors had core masses of about 18 $\mearth$, all synthetic planets with $M> 0.3 \mj$ in this population have core masses higher than that except for three planets. For 1 $\mj$ planets, e.g., the typical heavy element masses are around 40-60 $\mearth$. Because of the high importance of the core mass for the post-formation luminosity, it is important to see how the heavy element masses in the synthetic planets compare to observational constraints. In the solar system, the highest total heavy element content of Jupiter allowed by internal structure models is about 42 $\mearth$ for adiabatic models and the SCvH EOS \citep{guillotgautier2014}, while for semi-convective models up to 63 $\mearth$  are possible \citep{lecontechabrier2012}. Concerning exoplanets, thanks to recent analyses \citep{millerfortney2011,ThorngrenFortney2016} of the mass-radius relation of warm Jupiters without significant bloating, it is now possible to infer observationally estimates of the heavy element mass contained in planets also outside the solar system. While these analyses are (as those for Jupiter) affected by many uncertainties, e.g., regarding the equation of state and cannot yield (in contrast to the solar system) information about the distribution of the heavy elements within the planet (in the core or mixed throughout the envelope) they still significantly increase the number of planets with heavy element estimates and allow to make statistical inferences. Accepting these caveats, and the fact that these analyses apply to planets inside of 1 AU, while in the context of direct imaging one is mainly interested in planets further out,  then the aforementioned analyses also indicate that rather high heavy element contents seem to be quite common in giant exoplanets. Specifically, for a sample of   {47} planets with masses between about 0.1 and 10 $\mj$, \citet{ThorngrenFortney2016} derive a mean heavy element mass $M_{\rm Z}$ as a function of total mass $M$ of  $M_{\rm Z}/\mearth \approx ({58} \pm {7}) (M/\mj)^{{0.61\pm0.08}}$. A least-squares fit to the synthetic planets in the cold nominal population  with 0.1$\leq M/\mj\le10$ yields, for comparison, $M_{\rm Z}/\mearth \approx 45  (M/\mj)^{0.24}$ if only planets with a semimajor axis of less than 1 AU are included as in the sample of \citet{ThorngrenFortney2016}, and $M_{\rm Z}/\mearth \approx 63  (M/\mj)^{0.26}$ if all orbital distances are included. This shows that the amount of planetesimals accreted in our simulations seems  to be {for approximately Jovian mass planets} in rough agreement with or, for high planetary masses, even lower than the amount indicated by currently available  observational constraints.
This in turn {could indicate} that the high luminosities we
find in the cold-nominal population (almost comparable to those found
for hot accretion) are the outcome for core accretion, and not the very low  luminosities in \citet{marleyfortney2007} or in the low-core mass population where the accretion of planetesimals is artificially shut off. However, there is an important caveat to this{ which is the currently poorly known efficiency of the core-mass effect (see Sect. \ref{sect:limintstruct}).}

(3) Besides the generally high core masses and associated high luminosities, the plot also directly shows the main reason for the intrinsic spread of $L$ at fixed total mass and age. As is visible from colors of the symbols at a given total mass and age, the planets with a higher core mass usually also have a higher luminosity. The correlation between core mass and luminosity  found by \citet{mordasini2013} and \citet{bodenheimerdangelo2013} for idealized conditions (like a fixed semi-major axis, non-evolving disk, externally prescribed gas accretion rate and final planet mass) is thus also  recovered in the population syntheses where these quantities are obtained self-consistently from the disk and migration model, covering also a much larger parameter space. The plot, however, also shows that there are planets that  have a rather low core mass, but still a relatively high luminosity, and vice versa. This illustrates that while the spread in core masses is the main cause for the spread in luminosity, it is not the only one. Also the rate of gas accretion or the moment when most of the accretion happened relative to the disk lifetime influence the luminosity. This means that while there is a general trend of increasing post-formation $L$ with total and core mass, the exact post-formation properties of a planet can only be understood when the complete formation history is considered. In \citet{mordasini2013} it was found that  the post-formation luminosity $\lpf$ increases with core mass at fixed total mass roughly like $\mcore^{2-3}$. It is interesting to explore whether  a similar scaling also exists in the syntheses by studying $\lpf(\mcore)$ in a narrow bin of total mass (say of 1 $\mj$ width). It is found that there is indeed such a scaling, maybe like $\mcore^{1-2}$, but the spread around it is large (see Fig. \ref{fig:MLPostMcoreCD777}).
 
\subsubsection{A correlation of stellar [Fe/H] and planetary $L$?}\label{sect:correlfehL}
In view of these results, it is interesting whether a link can be made between the stellar [Fe/H] which is an observable quantity, and the planetary luminosity which can likely also be observed. The reason is that the observed mass-radius relation of extrasolar hot and warm Jupiters indicates that there is a positive correlation between stellar [Fe/H] and planetary heavy element content \citep{guillotsantos2006,burrowshubeny2007}, even if this correlation appears to be less clear in a bigger more recent data set \citep{ThorngrenFortney2016}. This could lead, via the core-mass effect, to a stellar [Fe/H] - planetary $L$ correlation. Is there such a correlation in the synthetic population?

Therefore, in the right panel of Fig.  \ref{fig:mlmcore} the luminosity at 20 Myr of synthetic planets more massive than 3 $\mj$ is shown as a function of host star's [Fe/H] around which they formed. The stellar [Fe/H] is one of the Monte Carlo initial conditions that is varied in the population syntheses, where it is assumed that the stellar [Fe/H] and disk [M/H] (dust to gas ratio) are proportional to each other \citep[see][]{mordasinialibert2009a}. 

In the plot, several features can be seen. (1) First, there is an approximately horizontal over-density of planets with a luminosity of about $10^{-3}$ $\lsun$. These are planets with masses between about 15 and 25 $\mj$ which are in the process of burning deuterium or which have burnt it recently, and that are now cooling off from it. The tracks of \citet{mollieremordasini2012} show that at this age (both for cold and hot accretion), the luminosity is approximately independent of mass in the $\sim$15-25 $\mj$ range. The reason is the following: within this mass range, more massive objects reach higher luminosities because of burning deuterium (about $10^{-2}$ $\lsun$), but burn most of it earlier (at $\lesssim10^{7}$ years), so that at 20 Myr they have again cooled down by about one order of magnitude in $L$. Lower mass objects in contrast reach less high luminosities because of D-burning (about $10^{-3}$ $\lsun$), but they do it later on, so that at 20 Myr, they are still strongly burning D or have not yet cooled down so much. The combination of the intensity of D-burning and the moment when it happens thus leads to an approximately mass-independent $L$. The luminosity value of the horizontal bar (and the mass of the objects in it) decreases in time which is visible in the distribution of luminosities (Sect. \ref{sect:Ldistro}). %

(2) Second, we see that there is a paucity of very luminous planets (corresponding to masses $\gtrsim10\mj$) at low metallicities. In general, in core accretion, for more common giant planets with lower masses, the final mass of a planet is not correlated with the stellar [Fe/H] \citep{mordasinialibert2012a}. Rather, a higher [Fe/H] leads to a higher fraction of disks that can form a giant, in agreement with the observed metallicity effect \citep{santosisraelian2001,fischervalenti2005}. This effect is also clearly visible in the panel by the high density of points at high [Fe/H]. In other words, [Fe/H] controls whether a giant planet can form, but usually not its mass, which is given by the disk's gas mass \citep{mordasinialibert2012a}. Only for the highest masses, it is different: to grow to a very high mass, a critical core must form before disk evolutions has had time to significantly deplete the disk mass. Such an early start is impossible at low metallicity  where growing a critical core take longer.

(3) Third, the colors of the points show the positive correlation of disk/stellar [Fe/H] and the heavy element content that was mentioned for observations earlier \citep{guillotsantos2006,burrowshubeny2007}.  As discussed already earlier \citep{mordasinialibert2009b}, this is also recovered in the synthetic population. The plot however also shows that the correlation only holds in a statistical sense, as there are also planets at high [Fe/H] without a particularly high heavy element content. 

(4) However, we see that apart of the absence of the highest $L$ at low [Fe/H], there is no correlation of [Fe/H] and $L$. The reason is that the most important quantity determining the luminosity is the total mass, which is, as mentioned, not directly dependent on [Fe/H] for most planets. In this panel, and in contrast to the left panel, the total mass of the planets is not directly visible. But it is more fundamental for $L$ than the effect of $\mcore$ at given total mass. Only if the population is  divided in fine bins in total mass of about 0.5 $\mj$ in width, a positive correlation between $L$ and [Fe/H] becomes visible, but even then, the correlation is weak with a lot of scatter. This is a consequence of the other three  Monte Carlo variables  which are the disk gas mass, disk photoevaporation rate (corresponding to different disk lifetimes), and the starting position of the embryo. They introduce so much variation in the formation tracks that there is only a weak imprint. 

So, in summary we see that the answer to the question posed at the beginning of the section, whether there is a correlation of [Fe/H] and $L$, is no, except for an absence of very luminous (and massive) planets at low [Fe/H]. Regarding them, one should kept in mind that such planets are in the syntheses actually a rare outcome: only about 0.6\% of the embryos reach a mass higher than 13 $\mj$. For comparison \citet{marcybutler2000} have estimated a frequency of companions more massive than 13 $\mj$ within 3 AU of $\lesssim0.5$ \%. In the plots shown here they only appear numerous because we are dealing with a very large synthetic population of about 50,000 planets in total.

\subsubsection{Comparison with $\beta$ Pic b and 51 Eri b}\label{sect:compbetapic51eri}
In the top left panel of  Fig. \ref{fig:mlevo}, the gray lines and shaded regions indicate synthetic planets that have a luminosity compatible with measurements of two important directly imaged extrasolar planets at 20 Myr, namely $\beta$ Pictoris b \citep{lagrangegratadour2009,lagrangebonnefoy2010} and 51 Eridani b \citep{macintoshgraham2015}. Both stars are members of the $\beta$ Pic moving group with an age estimate of $21\pm4$ Myr \citep{binksjeffries2014}, while \citet{macintoshgraham2015} adopt $20\pm6$ Myr for 51 Eri. It is interesting to study the properties of the synthetic analogs of the two planets, but it should be kept in mind that we only selected them based on a compatible luminosity. Additional constraints such as the planetary semimajor axis, the stellar mass (here 1 $\msun$, in reality about 1.75 $\msun$ for both stars) or the stellar [Fe/H] were not considered as constraints. This can, however, be included in specific population syntheses, as demonstrated for $\beta$ Pic (but for older observational constraints) in \citet{bonnefoyboccaletti2013,mordasinimolliere2015}. 

\textbf{$\beta$ Pic b:} From recent spectroscopic and photometric observations, \citet{bonnefoymarleau2014} derive a revised luminosity of $\log(L/\lsun)=-3.90\pm0.07$, a $T_{\rm eff}$ of 1650$\pm$150 K and a $\log g\leq 4.7$~dex. Synthetic planets compatible with the luminosity have the following properties: A mass of 10.4--13.3 $\mj$, which is compatible with, but as expected a somewhat wider range than earlier estimates of about 10.7--12.3 $\mj$ from hot start evolutionary models \citep{bonnefoymarleau2014}. As visible from the figure, even if the luminosity could be determined observationally without any error bar, an uncertainty of about 2 $\mj$ would remain due to the aforementioned intrinsic spread in the $M-L$ relation at this early age. The synthetic planets have a $T_{\rm eff}$ of 1600--1700 K, and a log $g$ of 4.18--4.26 dex which is also compatible with observations. Their radius is 1.28--1.38 $\rj$, in agreement with the observationally inferred values of 1.2--1.7 $\rj$ \citep{bonnefoymarleau2014,bonnefoyboccaletti2013,currieburrows2013}. The planets have a current specific entropy of the interior adiabat of 9.35--9.55 $k_{\rm B}$/baryon\footnote{Again, this corresponds to $S=9.87$--10.07 $k_{\rm B}$/baryon using the published tables of \citet{saumonchabrier1995} tables; see Footnote~\ref{Fussnote:DeltaS}.}, which is a quite narrow range, and a central temperature of 3.7--4.2$\times$10$^{5}$ K. Mixing length theory predicts that there are about 21  layers of eddies transporting the heat in the convective zone of the planet, assuming a mixing length parameter of  $\alpha_{\rm ml}$=1. As can be seen from the colors of the $\beta$ Pic b analogs in Fig. \ref{fig:mlevo}, this planet is at the very interesting transition zone between planets that do not burn deuterium and deuterium-burning planets. At the current epoch, the relative contribution of D-burning to the planet's luminosity is $L_{\rm D}/\lint$ of 0.2 to 0.4 for the lower masses $\leq12.5\mj$, and up to 0.8 if the mass is higher (also depending on the core mass).  Up to now, the planet has however only consumed about 3 to 10 \% of its initial deuterium, assumed  to initially have a standard interstellar medium abundance of D:H of $2\times 10^{-5}$ \citep{spiegelburrows2011}. During the main sequence lifetime of $\beta$ Pic of about 2 Gyr, the planet will typically burn about  10\% of its deuterium if its mass is on the lower side of the allowed range, and up to 90\% if its mass is rather 13 $\mj$. In view of atmospheric spectra, it is interesting to consider the heavy element content in the planets. It is found that the synthetic analogs have typically accreted about 30 to 100 $\mearth$ of solids, but a handful have up to 350 $\mearth$ of metals.  This gives them a wide range of heavy element mass fractions between 0.005 and 0.1, with a typical value of about 0.015. This corresponds to an enrichment relative to solar of about 0.3 to 7, with a typical value of around 1.

\textbf{51 Eridani b:} According to \citet{macintoshgraham2015},  this planet has a luminosity of $\log(L/\lsun)$ of -5.4 to -5.8, which corresponds to synthetic planets with masses between 1.7 and 3.6 $\mj$. This is in agreement with, but covers again (due to the intrinsic scatter), a somewhat wider range than the hot-start masses reported by \citet{macintoshgraham2015}, 1.8-2.9 $\mj$. The plot also shows that even if the luminosity were perfectly known, the intrinsic spread in the $M-L$ relation would lead to a uncertainty of almost 1 $\mj$ in the mass. The synthetic planets have an effective temperature of 570-740 K (630 K at  $\log(L/\lsun)$=-5.6), similar to the values derived from the observations (600-750 K), and most radii are between 1.12 to 1.25 $\rj$, but a handful have lower $R$, down to about $0.85\rj$. This is related to the heavy element content of the planets. Most synthetic planets contain between 30-100 $\mearth$ of metals, but a handful have accreted up to 500 $\mearth$ of solids. Checking the metallicity of the initial conditions that lead to these strongly enriched synthetic planets shows that they only form in clearly supersolar metallicity disks, in contrast to the approximately solar metallicity of 51 Eri A \citep{macintoshgraham2015}. Thus these are not self-consistent analogs, meaning that for an in-depth analysis all observational constrains should be taken into account. The typical bulk heavy element mass fraction then ranges from 0.04 to 0.2, corresponding to about 3 to 16 times solar. If 51 Eri b's atmosphere reflects this bulk composition, it is thus predicted to have supersolar abundances (which new results indicate, see \citealt{samlandmolliere2017}). As discussed in \citet{mordasiniklahr2014,mordasinivanboekel2016}, a higher metal content relative to the star is a typical prediction of the core accretion paradigm for a rather low-mass giant planet like 51 Eri b but not necessarily for a massive planet like $\beta$ Pic b. The specific entropy in the planet's convective zone is about 8.5-8.7 k$_{\rm b}$/baryon. From mixing length theory it is estimated that about 16-18 layers of eddies swirl in the planet's convective zone. Deuterium burning is of course completely negligible, with $\log (L_{\rm D}/ \lint) \approx$ -11. %

\subsubsection{Evolutionary phase: cold vs hot accretion}\label{sect:coldvshot}
\begin{figure*}
\begin{center}
\begin{minipage}{0.5\textwidth}
	      \centering
        \includegraphics[width=0.9\textwidth]{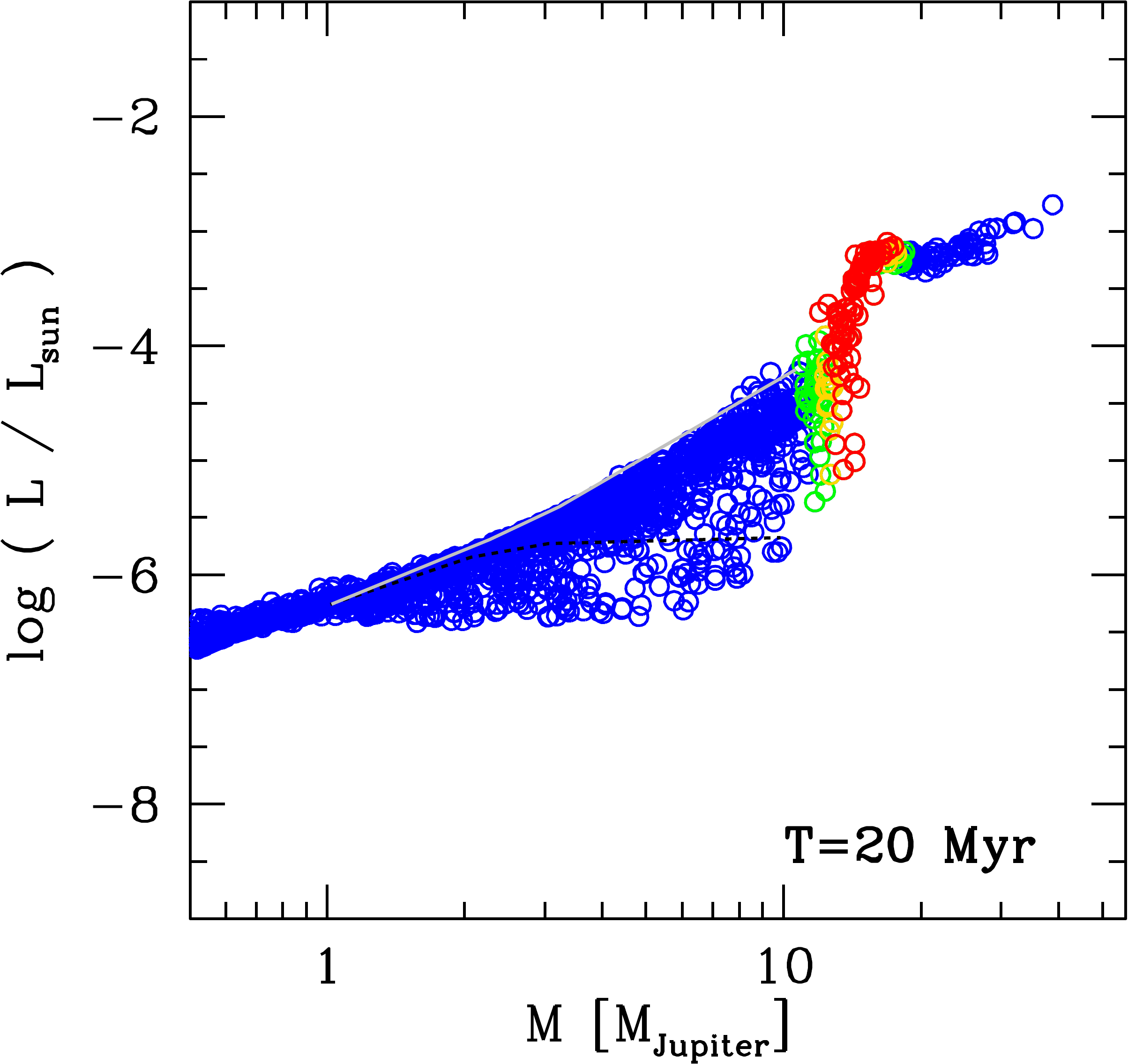}
        \includegraphics[width=0.9\textwidth]{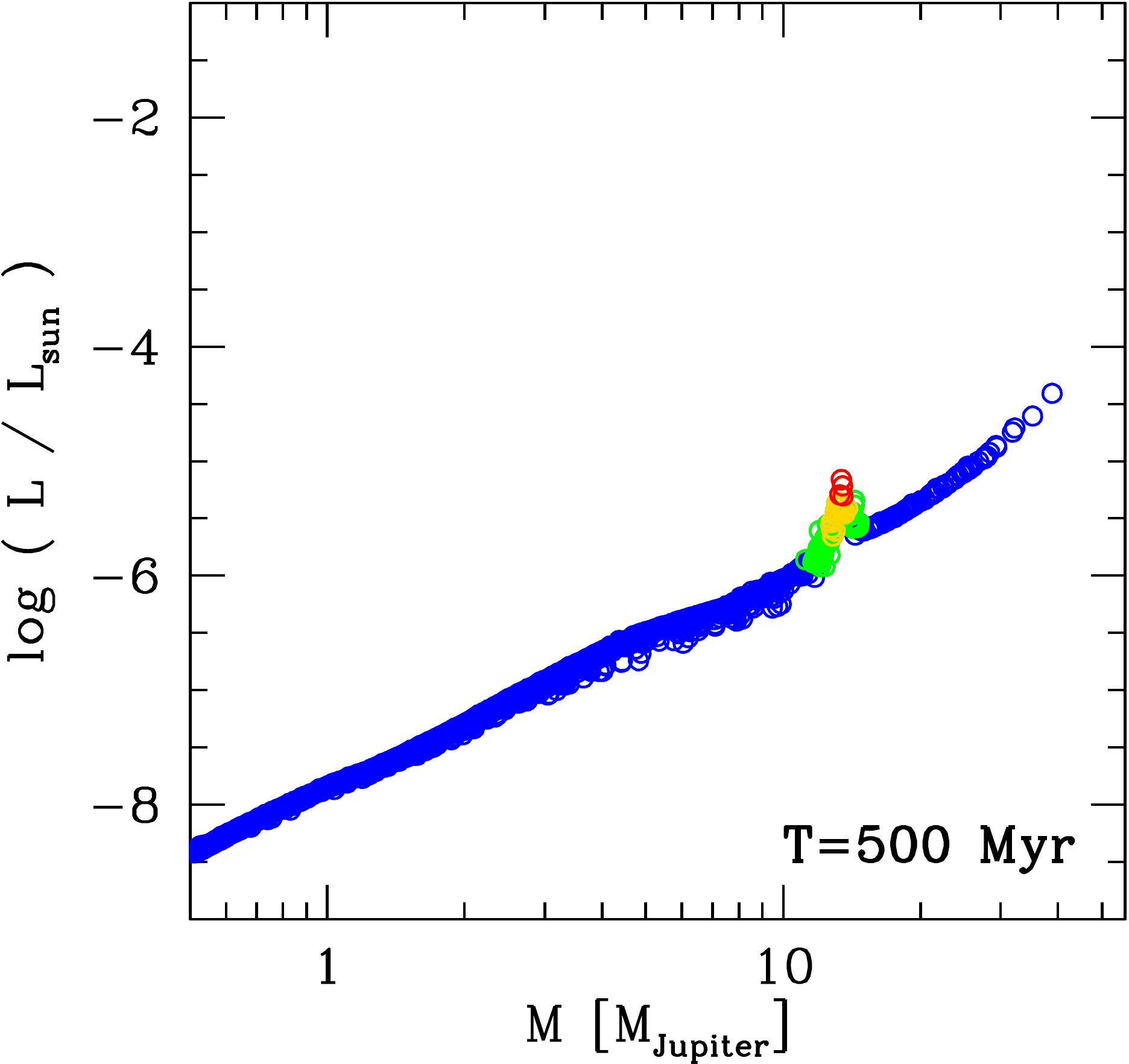}
     \end{minipage}\hfill
     \begin{minipage}{0.5\textwidth}
      \centering
       \includegraphics[width=0.9\textwidth]{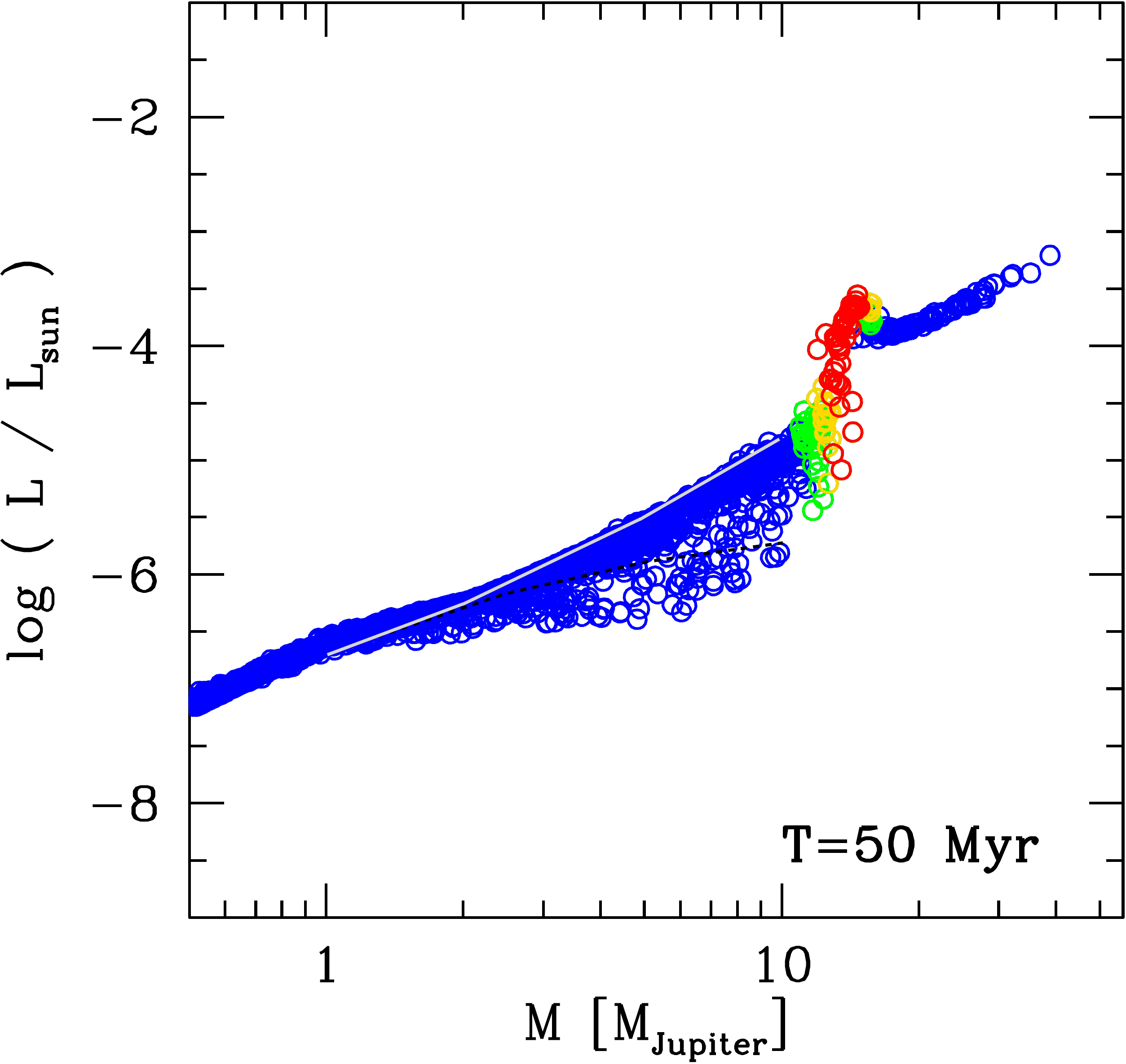}
      \includegraphics[width=0.9\textwidth]{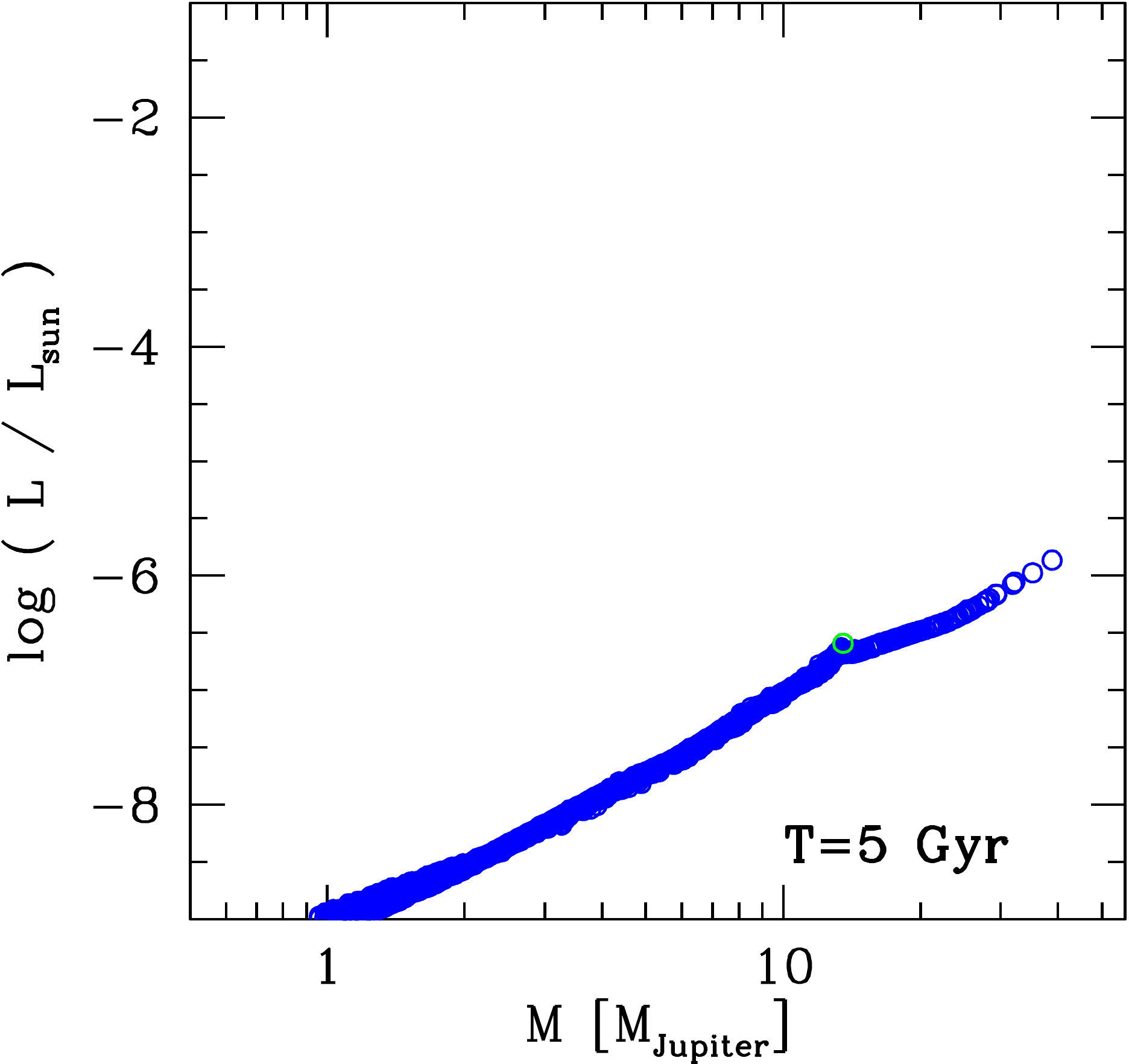}
     \end{minipage}
\caption{The $M$ -  $L$ relationship during the evolutionary phase at constant mass for the cold-classical (low core mass) population, analogous to Fig. \ref{fig:mlevo}. In the panel at 20 and 50 Myr, the gray solid and black dashed lines show the $M-L$ relation of classical hot start \citep{burrowsmarley1997} and cold start \citep{marleyfortney2007} models, respectively. }\label{fig:mlevolimcoldlcm}
\end{center}
\end{figure*}

Figure \ref{fig:mlevolimcoldlcm} shows the mass-luminosity diagram of the cold-classical population. Compared to the cold-nominal and hot population, the cold-classical population contains a characteristic ``bulge'' of planets with masses between about 1 to 7 $\mj$ that have very low luminosities and ``hang'' at early times below the $M-L$ relation of the hot and cold-nominal population. Its temporal evolution can be understood considering \citet{marleyfortney2007}: because of the long Kelvin--Helmholtz timescale of these very cold planets, this ``bulge'' remains almost static in time, until the planets with warm and hot starts have themselves cooled down to these low luminosity values. At this moment the evolution converges and the memory of the initial conditions is lost. Put differently, planets roughly stay at their initial entropy until they have reached an age equal to their initial cooling time, at which point they join the hot-track sequence \citep[see Equation~13 of][]{marleaucumming2014}. The timescale on which this happens increases with planetary mass.

\begin{figure*}
\begin{center}
\begin{minipage}{0.5\textwidth}
	      \centering
        \includegraphics[width=0.9\textwidth]{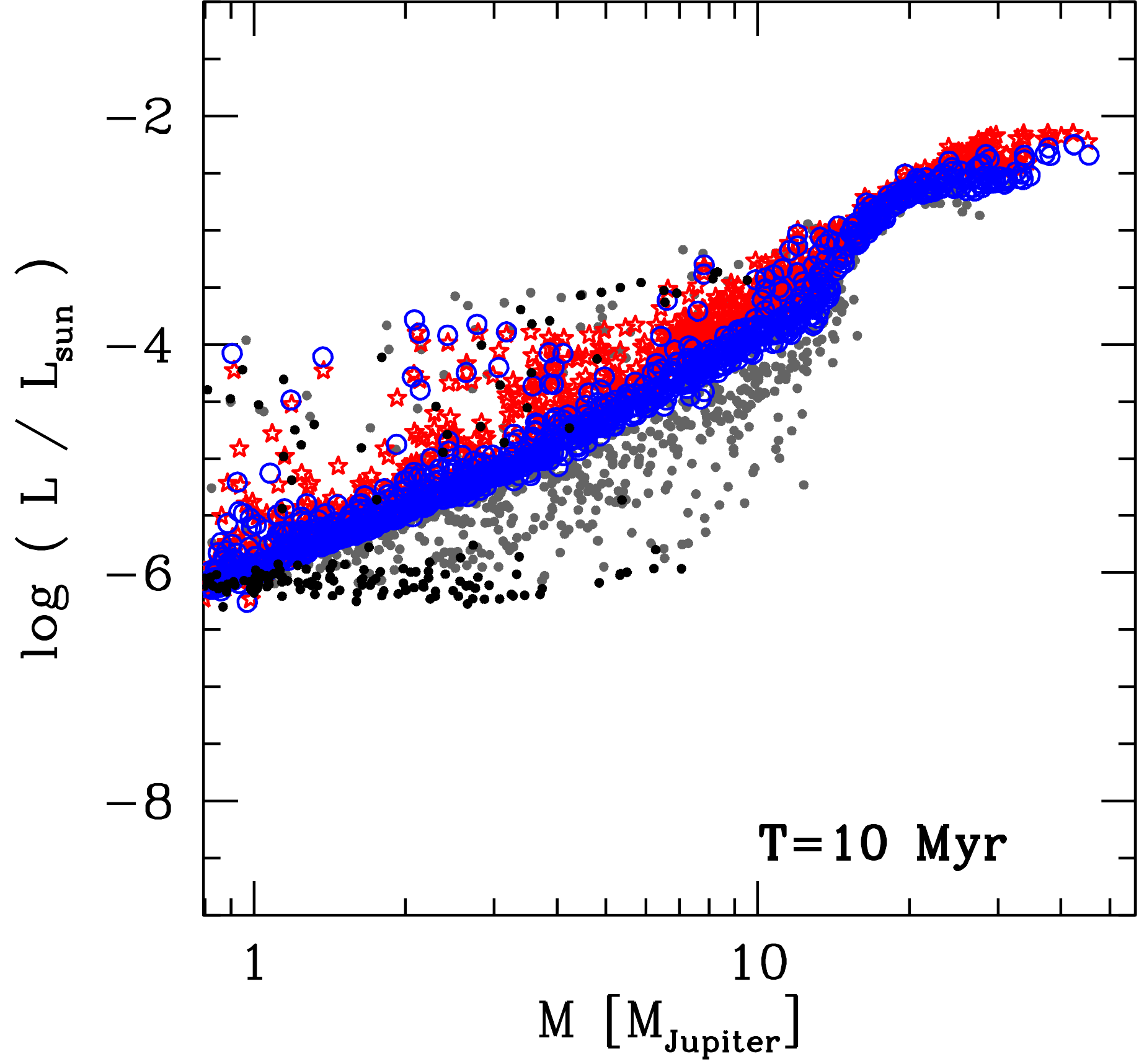}
        \includegraphics[width=0.9\textwidth]{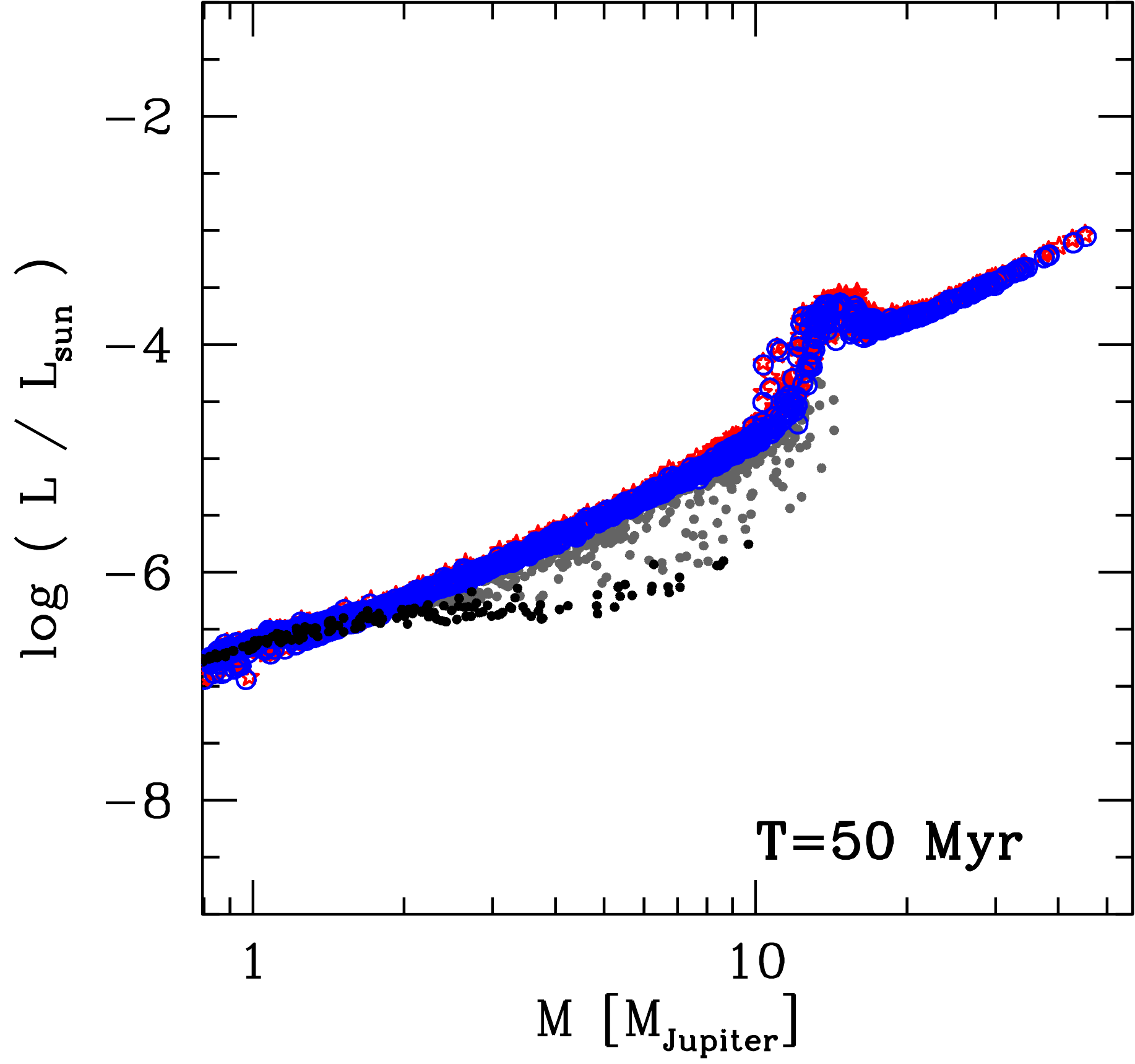}
     \end{minipage}\hfill
     \begin{minipage}{0.5\textwidth}
      \centering
       \includegraphics[width=0.9\textwidth]{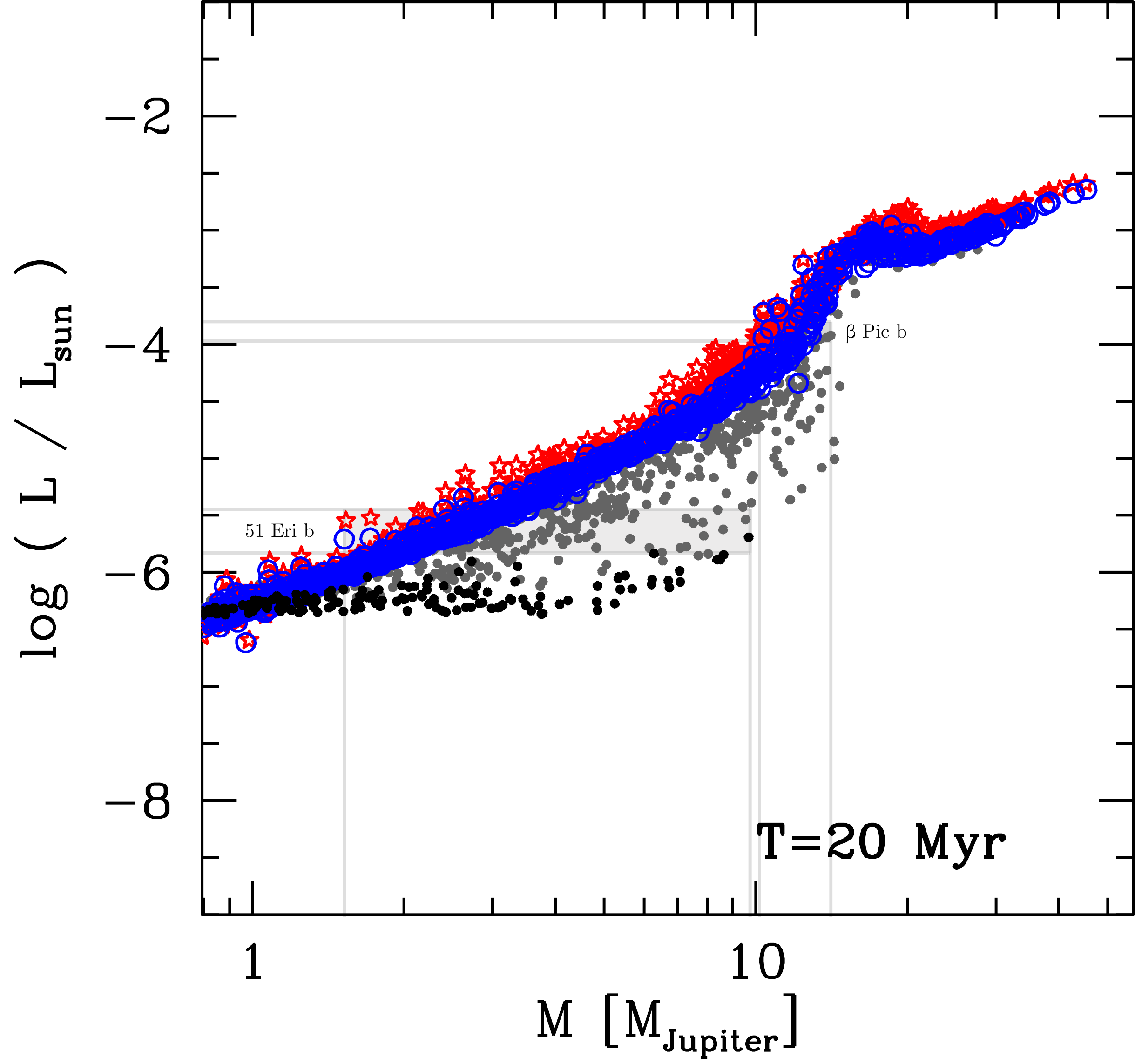}
      \includegraphics[width=0.9\textwidth]{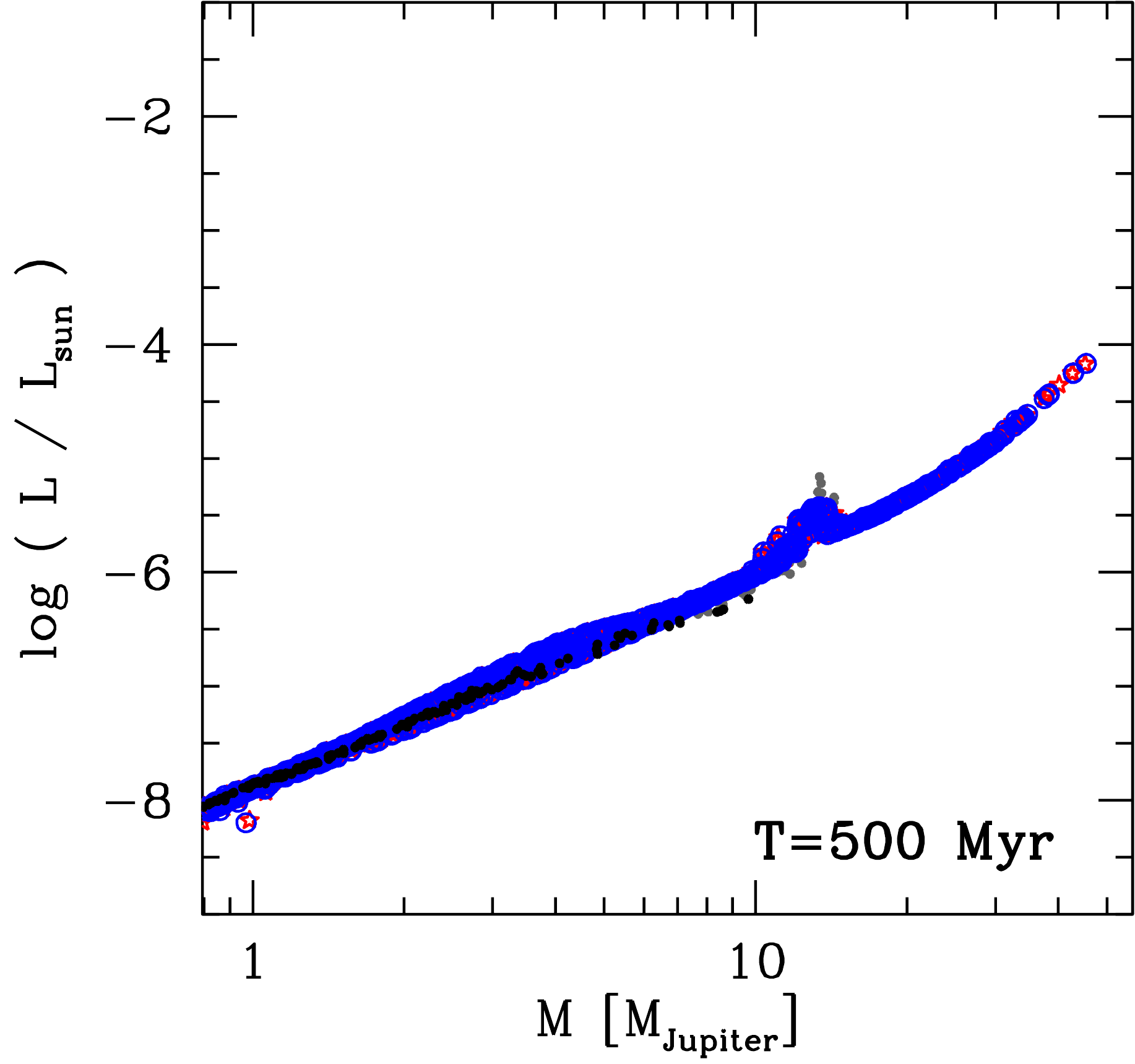}
     \end{minipage}
\caption{Impact of hot and cold accretion, and of the core mass on the temporal evolution of the planetary $M$ - total $L$ diagram. The blue empty circles are the cold-nominal population. The red stars are the hot population with hot gas accretion and otherwise identical assumptions. The black and gray filled circles represent planets in the cold-classical population.   Planets with a core mass of less than 17 $\mearth$ in this population are shown with black dots and those with a core mass above than 17 $\mearth$ in gray. This mimics the classical cold start models of \citet{marleyfortney2007}.  In the panel at 20 Myr the lines shows possible $M-L$ relations for $\beta$ Pictoris b and 51 Eridani b.}\label{fig:mlevocoldhot}
\end{center}
\end{figure*}

Figure \ref{fig:mlevocoldhot} compares the luminosities of the giant planets in the three populations during evolution. In the panel at 10 Myr, it is possible to directly cross-match some accreting planets in the hot and cold-nominal population. This is not possible for the planets in the cold-nominal population as no migration is included in this population, leading to different growth tracks. This cross-matching shows the effect mentioned above that for strongly accreting planets, the cold accretion planet are brighter, at least if  the total luminosity $\lint+\lshock$ is considered. The plot also shows the offset in the timing of deuterium burning.

In the panel at 20 Myr, the luminosities compatible with those of 51 Eri b and $\beta$ Pic b and the corresponding mass intervals are again shown, analogous to Fig. \ref{fig:mlevo} where they were compared with the cold-nominal population only. We see that the situation is quite different for the two planets: 

For 51 Eri b the allowed mass interval is much bigger than when compared  with the cold-nominal population only where a mass between 1.7 and 3.6 $\mj$ was found. Compatible planets from the cold-classical population have in contrast masses between about 1.7 and almost 10 $\mj$. \citet{macintoshgraham2015} estimated for comparison cold-start masses of 2 to 12 $\mj$. The plot furthermore shows that the planets with core masses below 17 $\mearth$ (black dots) cannot (except for two synthetic planets) reproduce the observed luminosity. This means that the lower limit on the heavy-element content is anti-correlated with the (true) total mass.     

The compatible mass interval for $\beta$ Pictoris b is in contrast only marginally increased relative to comparison with the cold-nominal population only, namely by less than 1 $\mj$ to higher masses. This is a consequence of  D-burning that tends to erase the hot vs. cold accretion imprint (Sect. \ref{sect:formationphasehotgasaccretion}).

At 50 Myr, the plot shows that the hot and cold-nominal populations have converged to virtually identical luminosities, while for the cold-classical this takes clearly longer. At 500 Myr, some of the very cold planets just below the deuterium burning limit still have somewhat smaller luminosities. This panels shows that interestingly, some D-burning planets in this population are more luminous than in the hot or cold-nominal population. This can be understood as a mild form of a late D-flash \citep{salpeter1992,marleaucumming2014}.

We finally also comment that for low-mass planets, there is no hot vs. cold accretion difference in the same sense as for giant planets. The more massive a giant planet, the higher the fraction of its final mass it accretes after detachment (at $\sim 100 \mearth$) through an increasingly strong, potentially entropy reducing shock. Detachment occurs when the planetary gas accretion rate as controlled by the planet's Kelvin--Helmholtz contraction rate becomes higher than the rate at which the protoplanetary disk can deliver gas to the planet. As low-mass planets have very long KH-timescales and thus low gas accretion rates \citep[e.g.,][]{ikomanakazawa2000}, they only detach from the nebula when the nebula itself has already almost completely dissipated. This means that only a tiny amount of gas is accreted after detachment through a potentially entropy lowering shock. In this sense, these planets have a hot start.     

However, the post-formation entropies of the low-mass planets ($M\lesssim 1 \mj$) can still vary significantly depending on the formation history because of other mechanisms. As an example, a planet that underwent strong solid accretion not long before the disk disappeared will have a higher entropy at the start of the evolutionary phase than a planet that accreted strongly at the beginning of the disk lifetime, and then already cooled for several million years. The entropy and luminosity of young low-mass planets and their observational consequences will be quantified in future work (Linder et al. in prep.). 

\subsection{Planetary properties at the moment of disk dissipation}\label{sect:spflpf}
In the previous sections, we have analyzed the luminosity of the populations at given uniform ages. An age that is of special interest but different for each planet (or protoplanetary disk) is the moment when the protoplanetary disk disappears. The properties of the planets at this moment are interesting as they set the initial conditions for the evolutionary phase, and are thus important for planet evolution models. Additionally, the properties of the planets at this moment have been previously predicted for specific initial conditions by planet formation models like  \citet{marleyfortney2007,mordasini2013,bodenheimerdangelo2013}. 

\subsubsection{Post-formation entropies $\spf$}\label{sect:postformationentropy}
\begin{figure*}
\begin{minipage}{0.7\textwidth}
	      \centering
       \includegraphics[width=0.9\textwidth]{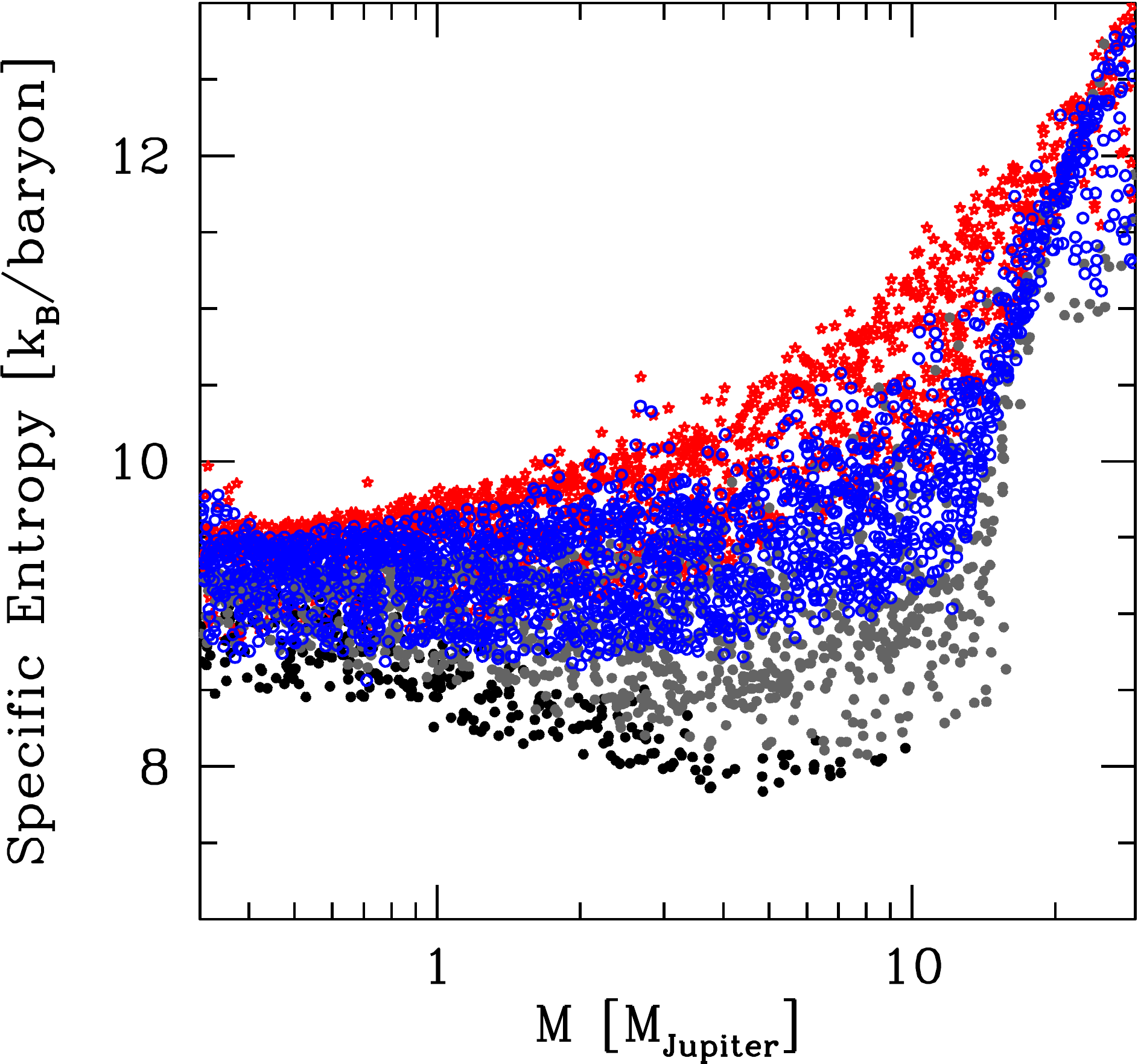}%
      \end{minipage}
     \begin{minipage}{0.3\textwidth}  
     \vfill
     \hfill
     \caption{Entropy ``tuning fork'' diagram, i.e., specific entropy in the convective zone as a function of planet mass at the end of the formation phase when the protoplanetary disk disappears ($t=t_{\rm disk}$).  The blue empty circles  are the cold-nominal population. The red stars are the hot population. The black and gray points show the cold-classical population. For this population,  planets with $M_{\rm core}\leq17~\mearth$ are shown with black points, mimicking  \citet{marleyfortney2007}. Planets with $M_{\rm core}>17~\mearth$ are displayed as gray points.}\label{fig:MSmarleycomp} 
             	\end{minipage}
\end{figure*}
In particular the planetary mass - specific entropy relation at the end of the formation phase is of interest, as the specific entropy of the planet's interior adiabat is the ideal quantity to describe the gravothermal heat content of a planet, in the sense that the mass and the specific entropy of the planet fully describe the interior structure of the planet, at least for a given composition, and under the assumption of a fully convective planet \citep[e.g.,][]{marleaucumming2014}. In the original work of \citet{marleyfortney2007} it was shown that if the specific post-formation entropy $\spf$  is plotted as a function  mass for cold and hot start planets  (their Fig. 2), then the curves take the form of a tuning fork, with an upper prong of high entropies that increase with mass for the hot start planets, and a lower prong of low entropies that decrease with mass for the cold start planets. \citet{spiegelburrows2012} then showed that  intermediate radiative efficiencies of the accretion shock lead to additional prongs of intermediate entropies (warm starts). They also found that lower gas accretion rates lead to lower $\spf$. \citet{mordasini2013} then showed that such intermediate warm states also result from a fully radiatively efficient shock (cold accretion) but a core mass that is higher than in \citet{marleyfortney2007}. This  mechanism was also found by \citet{bodenheimerdangelo2013}.

These results indicate that given the diversity of protoplanetary disks, one expects a significant spread of planetary post-formation entropies (and thus luminosities and radii), and not only two extreme values like in \citet{marleyfortney2007}. To quantify this, we plot the specific entropy of synthetic planets more massive than 0.3 $\mj$ at the moment when their protoplanetary disk disappears in Fig. \ref{fig:MSmarleycomp}.

In the population synthesis calculations (see \citealt{mordasinialibert2009a}), the distribution of external photoevaporation rates is chosen in such a way that the synthetic disks have a lifetime distribution that is approximately in agreement with observations \citep[e.g.,][but see also \citet{pfalznersteinhausen2014}]{haischlada2001}. Consequently, the mean lifetime of all synthetic disks is 3.05 Myr. The mean lifetime of disks which form a giant planet ($M>0.3 \mj$) is, however, longer, 5.1 Myr. This is expected, since giant planet formation by core accretion is facilitated in long-lived disks, as there is more time to build the critical core and to accrete gas \citep{mordasinialibert2012a}. Thus in Fig. \ref{fig:MSmarleycomp} the age of the stars will on average be 5.1 Myr, but it can be as short as 1.6 Myr for the most short-lived disks producing a giant planet, and 16 Myrs for the longest one. 

Considering first the hot (red stars) and the cold-classical mass population (black points), we see the same general picture as found in earlier work. In the hot population, the entropy increases with mass, demonstrating again \citepalias{mordasinialibert2012b} that core accretion with a radiatively inefficient shock leads to high entropies comparable to classical hot start simulations \citepalias[but see][and \citet{berardocumming2016}, for a possible issue when simulating hot accretion with the $\partial L/ \partial r=0$ simplification employed here]{mordasinialibert2012b}. This is traditionally rather associated with gravitational instability \citep{galvagnihayfield2012}, but we show here that this is not necessarily the case. For the cold-classical population, and for core masses less than 17 $\mearth$ which mimics the simulations of \citet{marleyfortney2007}, $\spf$  in contrast decreases with mass. Thus, the fundamental shape of the classical ``tuning fork'' is recovered. However, and this is a new aspect, due to the different formation histories, there is a large spread of about 1  to 1.5 $k_{\rm b}$/baryon in $\spf$ at a given mass in the hot population. In the cold-classical population, the $\spf$ even have a $\sim2$ $k_{\rm b}$/baryon scatter at about 5 $\mj$. One also sees that there is a significant overlap in the covered entropy range between the populations, especially between the hot and the cold-nominal populations, but also between the cold-nominal and the cold-classical populations.

If finally also the cold-nominal population (blue empty circles) is taken into account, the complete covering of a wide entropy range is even more apparent. This population is characterized by a mean entropy that is approximately constant\footnote{In the fit to $\spf$ presented in Appendix \ref{sect:fitS0}, there is actually a slight decrease in $\spf$ for envelope masses between about 0.3 and 2 $\mj$.} at about 9.2 $k_{\rm b}$/baryon for masses $\lesssim5\mj$, followed by an increase of $\spf$ at higher masses.  Also in this population there is an intrinsic scatter in $\spf$ at fixed mass of  about 1 $k_{\rm b}$/baryon at 1 $\mj$ and 1.5 $k_{\rm b}$/baryon at 10 $\mj$. Such a spread in entropy is associated with important luminosity differences.

The mass range in Figure  \ref{fig:MSmarleycomp}  extends to higher masses compared to earlier studies and now includes also D-burning planets. One sees that D-burning leads in the cold population at about 12-13 $\mj$ to a sharp and well-defined upturn in the entropy, as expected \citep{mollieremordasini2012}. Also in the hot population the entropy starts to increase stronger with mass at this point, but the transition is smoother. For the most massive planets ($\gtrsim20\mj$) the entropy is again smaller, as these planets have already burnt most their deuterium when the disk disappears (Sect. \ref{sect:MLformcold}).

It is interesting to compare the hot and cold start entropies in Fig.  \ref{fig:MSmarleycomp} with the ``tuning fork'' in \citet{marleyfortney2007}. One finds that the black points are at even lower entropies, by 0.5--0.7 $k_{\rm B}$/baryon depending on the mass. Also the red points are lower than the upper  line in \citet{marleyfortney2007}, but these entropies were arbitrarily set in  \citet{marleyfortney2007} and not obtained from a formation calculation as here, so that this does not pose an issue. The reason for the difference between the lower prong  in \citet{marleyfortney2007} and the black points is not completely obvious. It is likely a consequence of several different settings in the two models. First, in \citet{marleyfortney2007}, a grain opacity reduced to 2\% of the ISM opacity was used, while here it is reduced to 0.3\%. As demonstrated by \citet{spiegelburrows2012} and \citet{mordasini2013}, a lower opacity leads to lower $\spf$ because the planets can cool more efficiency already during formation. Second, on a related note, the atmospheric boundary conditions used here differ from those used in \citet{marleyfortney2007}. \citet{marleaucumming2014}, who also use an Eddington atmosphere, found that the entropy in their models needed to be increased by 0.14 $k_{\rm B}$/baryon\footnote{Correcting for the physically irrelevant offset; see Footnote~\ref{Fussnote:DeltaS}.} to match a given luminosity in \citet{marleyfortney2007}. Finally, in the synthesis the typical gas accretion rates are (as also seen in the example of Sect. \ref{sect:formevo5MJ}) up to two orders smaller than in \citet{marleyfortney2007} where the $\dot{M}_{\rm XY}$ is (arbitrarily) constant at a high $10^{-2}\mearth$/yr. This is quite different from the behavior here where the disk evolution leads to a gradual reduction of $\dot{M}_{\rm XY}$ down to very low values. As found by \citet{spiegelburrows2012} and \citet{mordasini2013},  lower gas accretion rates during formation  lead to lower $\spf$.

In Appendix \ref{sect:fitS0} we present the post-formation entropy  $\spf$ as a function of H/He envelope mass $\menv$ for the cold and hot population, covering a very wide range in envelope masses. Such a relation is of interest especially as initial conditions for evolutionary models. We therefore provide a least-squares fit to the numerical results in Appendix \ref{sect:fitS0}.

\subsubsection{Post-formation luminosities $\lpf$}
\begin{figure*}
\begin{minipage}{0.7\textwidth}
	      \centering
       \includegraphics[width=0.9\textwidth]{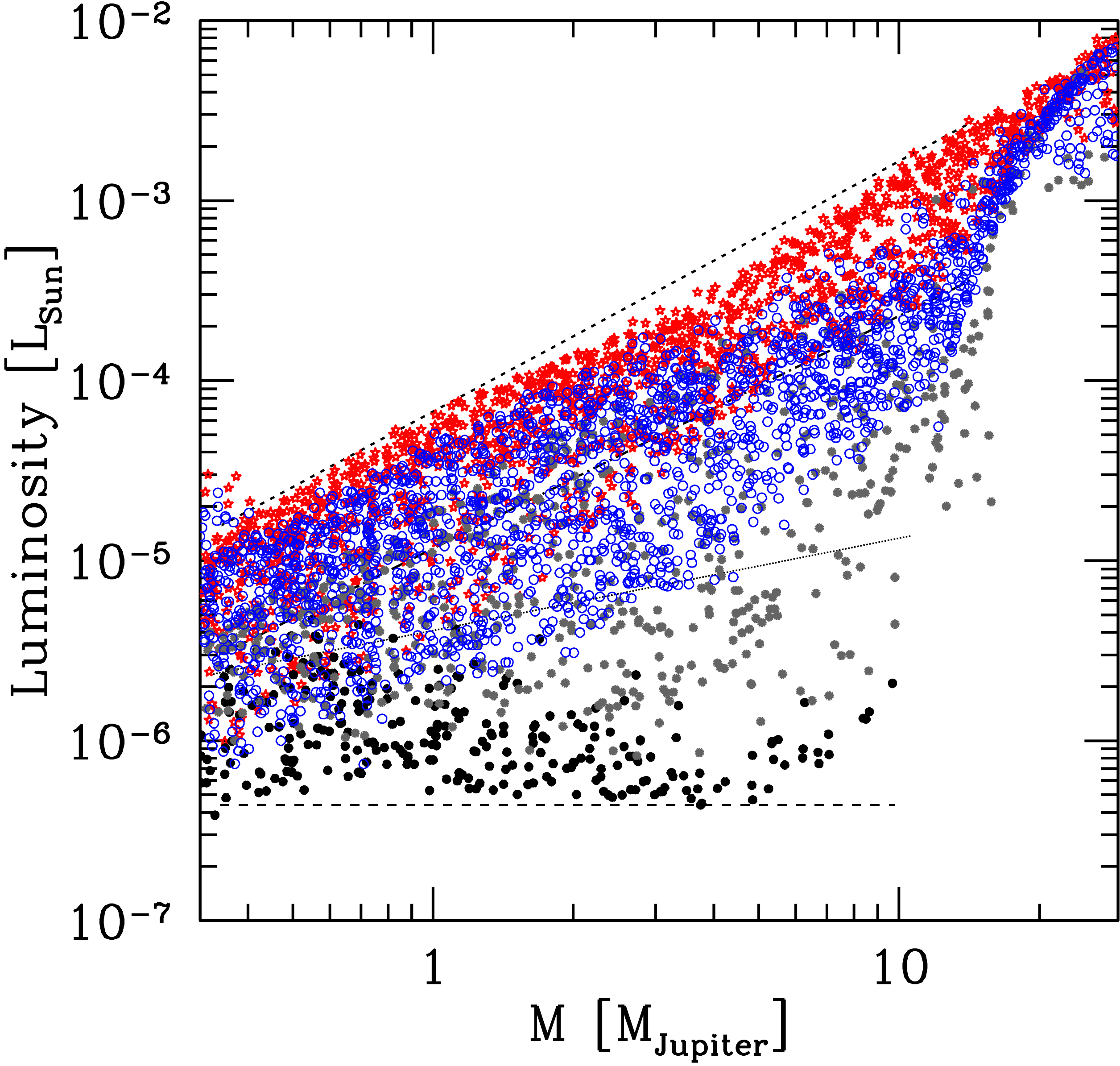}%
      \end{minipage}
     \begin{minipage}{0.3\textwidth}  
     \vfill
     \hfill
     \caption{The mass-luminosity relation of giant planets at the moment when the protoplanetary disk disappears. As in Fig. \ref{fig:MSmarleycomp}, the blue empty circles are the cold-nominal population while the red stars  show the hot population. The black and gray dots show the cold-classical population where black symbols correspond to planets with $M_{\rm core}$$\leq$17 $\mearth$. Note the significant spread in post-formation luminosities of more than two orders of magnitude at a given mass. The four black lines gives  scalings for the hottest, cold-nominal,  cold-classical (warm), and coldest $M-\lpf$ relation discussed in the text.}\label{fig:MLtdisk} 
  
\end{minipage} 
\end{figure*}

A more observational quantification of the thermodynamic state of new-born giant planets is the post-formation luminosity. In analogy to Fig. \ref{fig:MSmarleycomp},  Fig. \ref{fig:MLtdisk} shows the luminosity of giant planets as a function of mass at the moment when the disk disappears. Again, the cold-nominal, hot, and cold-classical populations are shown. The general shape of the $M-\lpf$ is a consequence of the $M-\spf$ relation studied before and the way the luminosity depends on  mass and entropy, $L(M,S)$, as studied in detail in \citet{marleaucumming2014}. In the plot, one can identify four different interesting $M-\lpf$ relations.

(1) First, the hottest planets. The planets with the highest post-formation luminosity are found in the hot population. The upper envelope of points approximately follows the relation $\lpf/\lsun=7\times10^{-5} (M/\mj)^{1.4}$ shown by the short-dashed curve in the figure with a somewhat lower values at intermediate masses.

(2) Second, the  cold-nominal planets. A least-squares fit to the cold-nominal planets (blue circles) gives $\lpf/\lsun=1.2\times10^{-5} (M/\mj)^{1.3}$. This is the dashed-dotted curve in the figure. The (slightly) lower exponent compared to the first case reflects the fact that planets with masses higher than about 3 $\mj$ and up to the D-burning limit have still somewhat lower luminosities because of cold gas accretion which  an increasing difference in luminosities with increasing mass, even though the impact of cold vs. hot gas accretion is strongly reduced because of the core-mass effect.  
 
(3) Third, cold-classical population planets with an intermediate luminosity, shown by the gray symbols. They can be associated with an intermediate warm start, and the dotted line in the plot shows $\lpf/\lsun=4\times10^{-6} (M/\mj)^{0.5}$. In this population there is the largest spread in the post-formation luminosity (a factor 200 at 5-10 $\mj$), with $\lpf$ correlating strongly with the core mass. One furthermore sees that once deuterium burning kicks in a{t} sufficiently high masses, it reduces the spread of the post-formation thermodynamic conditions, with the variation in $\lpf$ for planets more massive than about 15 $\mj$ reduced to about a factor $\leq5$.

(4) Finally, the coldest planets. The coldest planets in the cold-classical population with $M_{\rm core}\leq17\mearth$ have a very low luminosity around $\lpf$=4$\times$10$^{-7}$ $\lsun$ (as shown by the long-dashed line in the figure). The luminosity is approximately independent of mass, with a slight trend of a decreasing luminosity with increasing mass. For comparison, \citet{marleyfortney2007} had also found a nearly mass-independent post-formation luminosity, at a typical value of about 2-3$\times$10$^{-6}$ $\lsun$.

These groups represent physically motivated  initial values for the post-formation luminosities of giant planets obtained from a much higher ensemble of possible formation tracks compared to earlier works. Cooling curves and magnitudes for these different initial conditions employing non-gray atmospheric boundary conditions are in preparation (Marleau et al. in prep.).

\subsubsection{Cold-classical population: impact of $M_{\rm core}$ on $\lpf$}\label{sect:coldclassicalMLMcore}
\begin{figure}%
\begin{center}
	      \centering
            \includegraphics[width=\columnwidth]{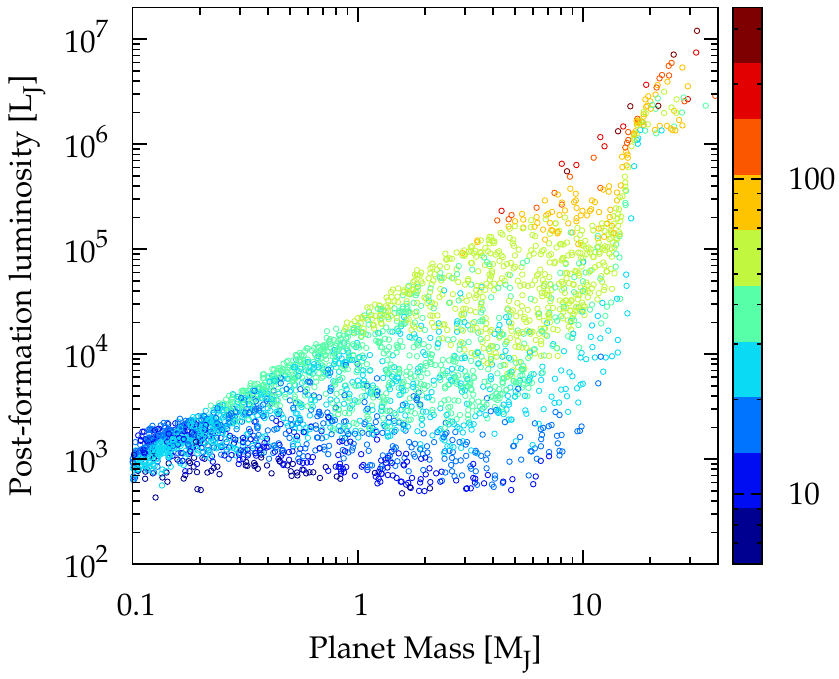}
\caption{Luminosity as a function of mass for the cold-classical population at the moment when the protoplanetary disk disappears. The colors indicate a planet's core mass in units of Earth masses. The increase of $\lpf$ with increasing core mass at a given total mass  is clearly visible.}\label{fig:MLPostMcoreCD777}
\end{center}
\end{figure}

The influence of the core mass for the cold-nominal population was discussed above in Sect. \ref{sect:impactcoremass}. However, if planetesimal accretion is assumed to continue freely after detachment, as is the case in the cold-nominal population, the core masses are effectively high enough to result in relatively high post-formation luminosities, which in part increase with increasing mass, as shown by the blue dots in Fig. \ref{fig:MLtdisk}.  The population where the impact of the core mass is even much larger is the cold-classical population. In this population, giant planets can have a very wide range of core masses from less than 10 $\mearth$ to more than 100 $\mearth$. This leads to a very wide spread of post-formation entropies and luminosity, as shown by the black and gray points in Fig. \ref{fig:MSmarleycomp} and \ref{fig:MLtdisk}.  To directly illustrate the effect of the core mass on $\lpf$ in this population, we again show in Figure \ref{fig:MLPostMcoreCD777} the luminosity at the moment when the protoplanetary disk disappears $\lpf$ as a function of mass, but now color-coding the planets' core mass. The plot shows first the very clear and strong correlation between $\mcore$ and $\lpf$ at a given total mass. It leads to a very large spread in $\lpf$ at one mass, with the spread increasing with total mass for 0.1$\lesssim M/\mj\lesssim7$. At the mass showing the biggest variation (about  7 $\mj$)  $\lpf$ covers almost three orders of magnitude. It is a consequence of the spread in $\mcore$ that varies at this mass by more than a factor 20. 

At even higher masses than about 7 $\mj$, the spread in $\lpf$ becomes smaller again as all these very massive planets also have massive cores, and for even higher masses deuterium burning further reduces the spread in $\lpf$, bringing it down to just about a factor 2 at $M\approx 17 \mj$. For the most massive planets in this population ($M\approx30\mj$) $\lpf$ varies again by about a factor 10 because of different timings of the earlier D-burning. 

The plot furthermore illustrates the correlation between core mass $\mcore$ and total mass $M$ that was extensively studied in \citet{mordasiniklahr2014}. There it was shown that for planets forming in situ and accreting all planetesimals in the feeding zone, one finds analytically that $\mcore\propto M^{1/3}$, and furthermore that $\mcore\propto\Sigma_{P}$ where $\Sigma_{P}$ is the surface density of planetesimals. In the population synthesis, $\Sigma_{P}$ is given as the product of the (initial) disk gas mass and the dust-to-gas ratio, which are both Monte Carlo variables \citep{mordasinialibert2009a}. Their distributions cover about one (for the dust-to-gas ratio) or two orders of magnitude (for the initial disk gas mass), such that the large spread in $\mcore$ and thus $\lpf$ is a direct consequence of the varying initial conditions in protoplanetary disks. We however stress again {(Sect. \ref{sect:limintstruct})} that in reality the core-mass effect might be not as efficient as assumed in the present model, in which accreted planetesimals instantaneously sink to the central core \citep{mordasini2013}. This would then naturally also reduce its impact on $\lpf$. This should be quantified with future work treating self-consistently the thermodynamics and compositional evolution during formation \citep[e.g.,][]{venturinialibert2016}.
 
\subsection{The planetary luminosity distribution}\label{sect:Ldistro}

In the previous sections, we have mostly studied scatter plots involving the luminosity which give an insight into the diversity of it and its dependencies on the mass. However, also the simple distribution of $L$ is of interest, e.g., for statistical comparisons of the theoretically predicted luminosity distribution with observational data as soon as there will be a higher number of directly imaged planets, or to estimate the necessary sensitivity to enter a certain discovery parameter space. Compared to the mass-luminosity relation, the advantage here is that the mass is not necessary for a comparison. The interest in the luminosity distribution is analogous to the situation of radial velocity surveys in the past  which allowed to compare the observed planetary mass distribution \citep[][]{howardmarcy2010,mayormarmier2011} with theoretical predictions from population syntheses  \citep{mordasinialibert2009b,benzida2013}. 

\subsubsection{Luminosity distribution as a function of time for the cold-nominal population}
\label{sect:L(t)fuerkalt}
\begin{figure*}
\begin{center}
\begin{minipage}{0.49\textwidth}
\includegraphics[width=0.98\textwidth]{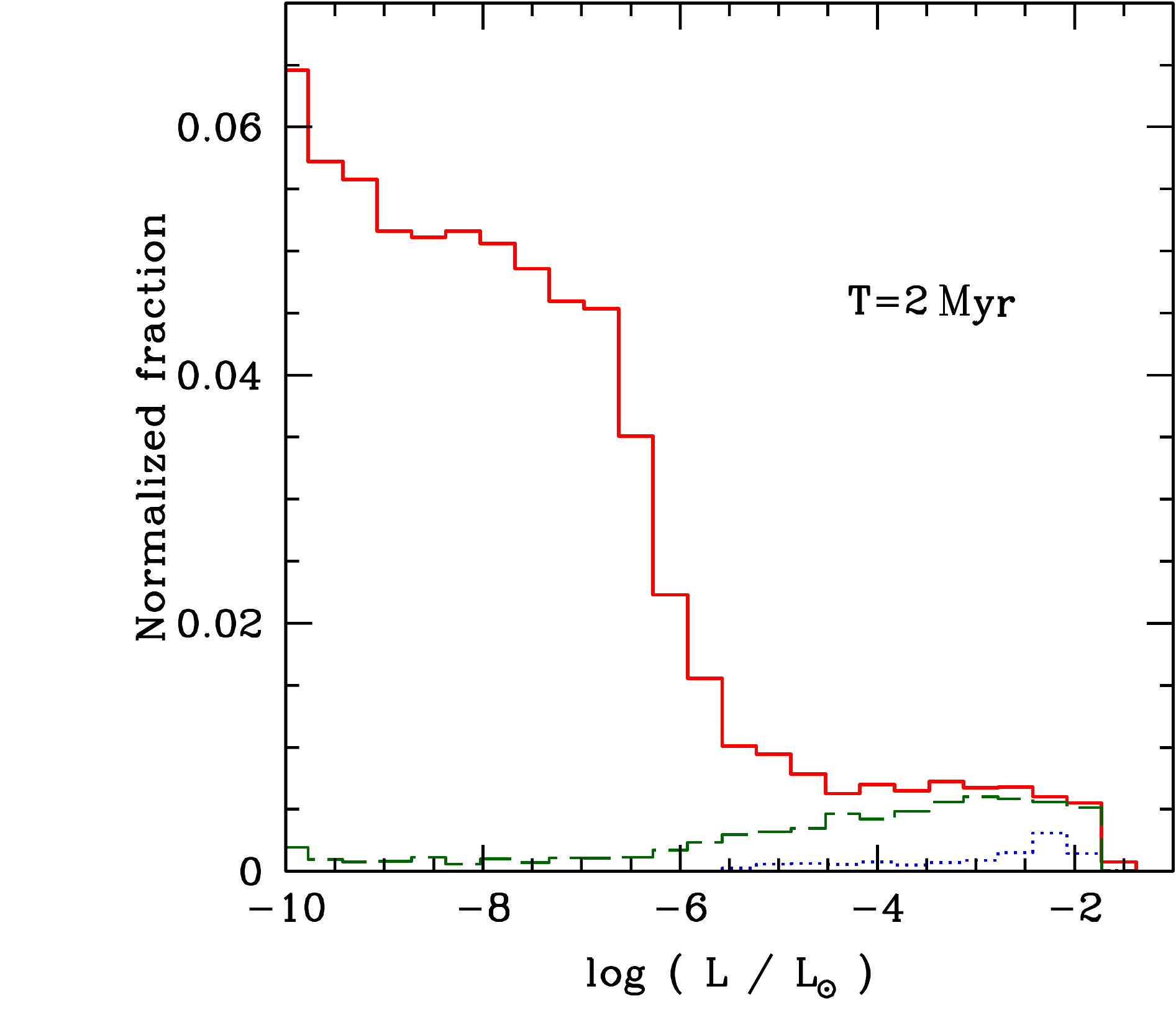}
\end{minipage}%
\begin{minipage}{0.49\textwidth}
\includegraphics[width=0.98\textwidth]{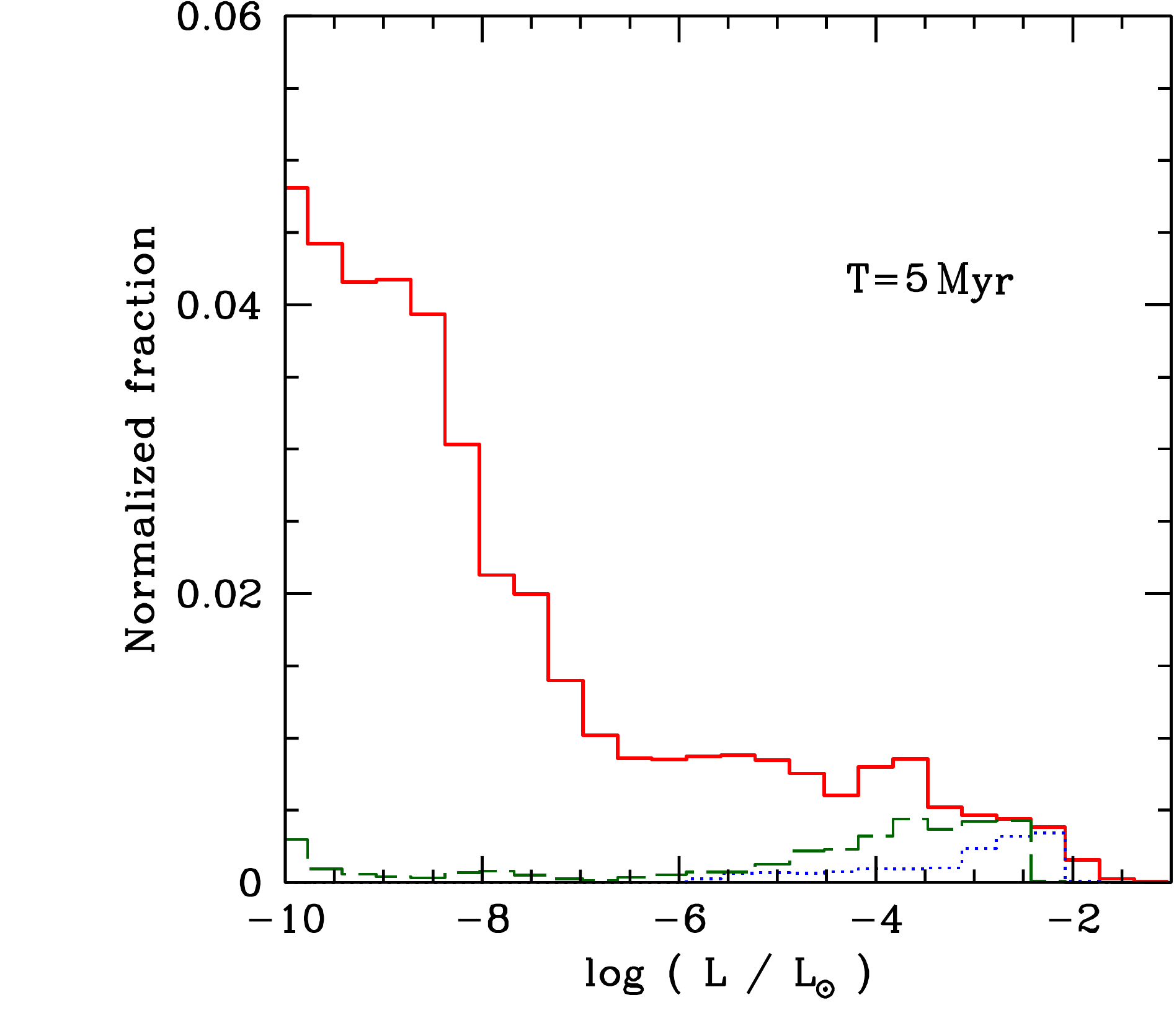}
\end{minipage}%

\begin{minipage}{0.49\textwidth}
\includegraphics[width=0.98\textwidth]{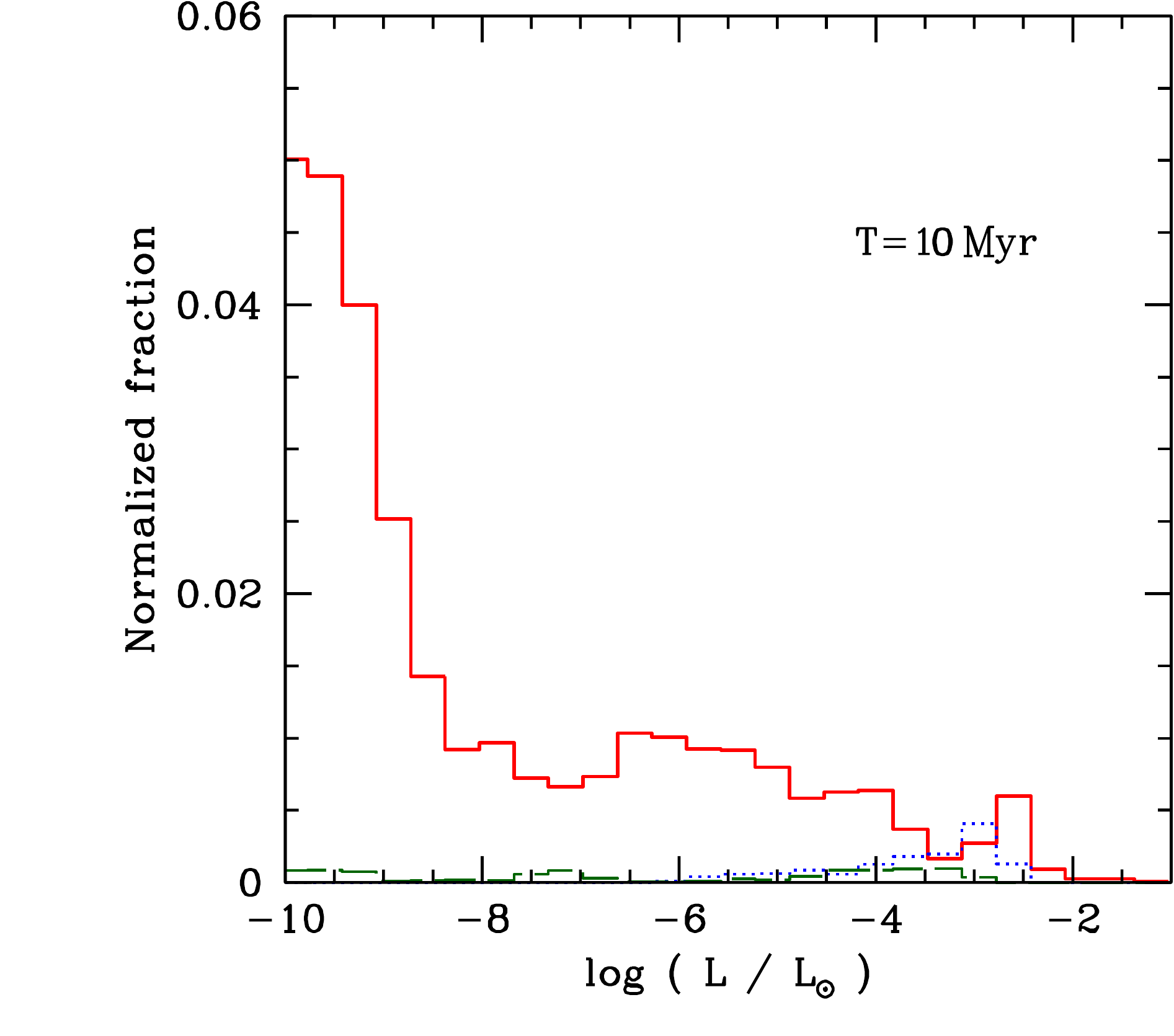}
\end{minipage}%
\begin{minipage}{0.49\textwidth}
\includegraphics[width=0.98\textwidth]{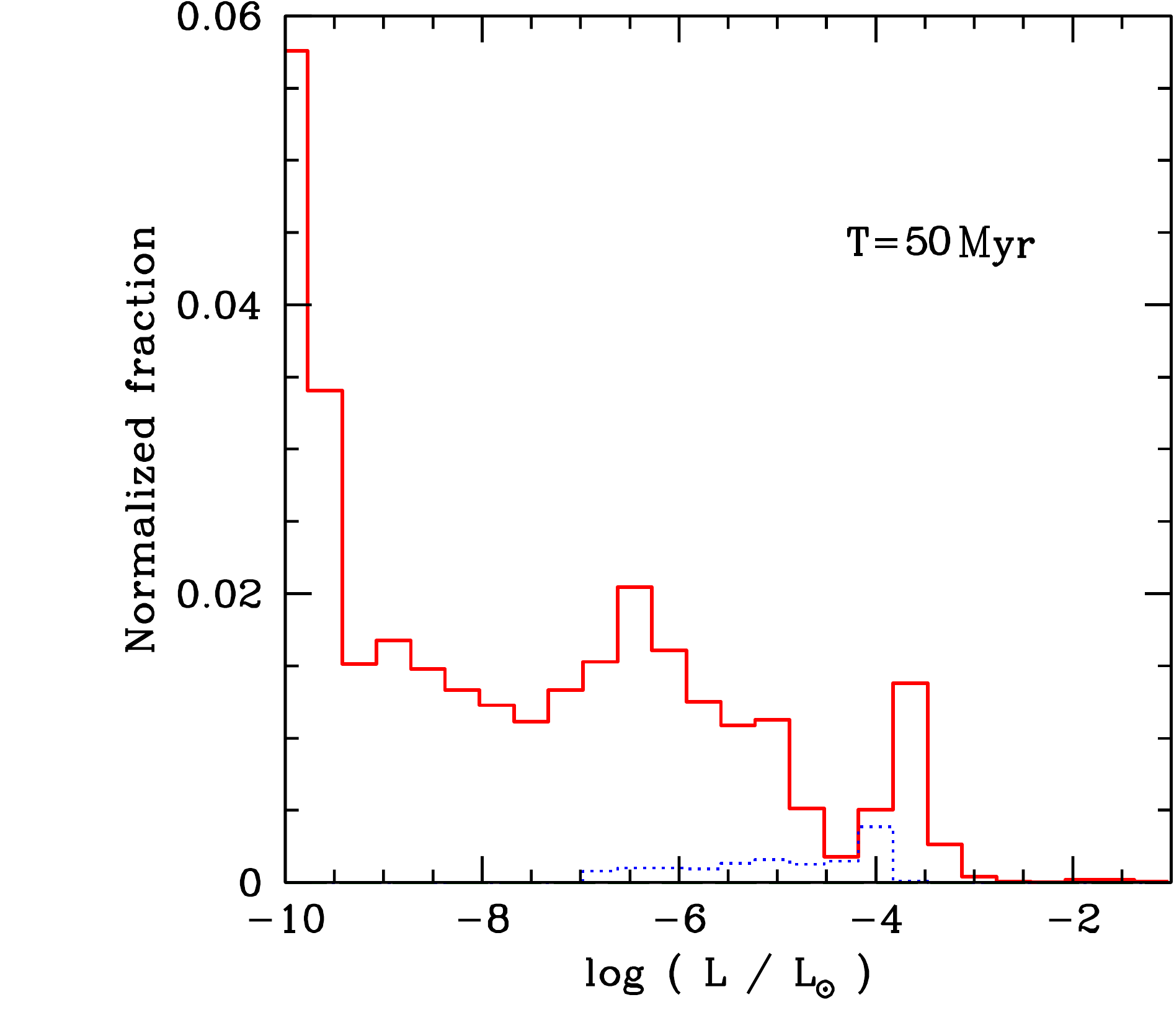}
\end{minipage}%

\begin{minipage}{0.49\textwidth}
\includegraphics[width=0.98\textwidth]{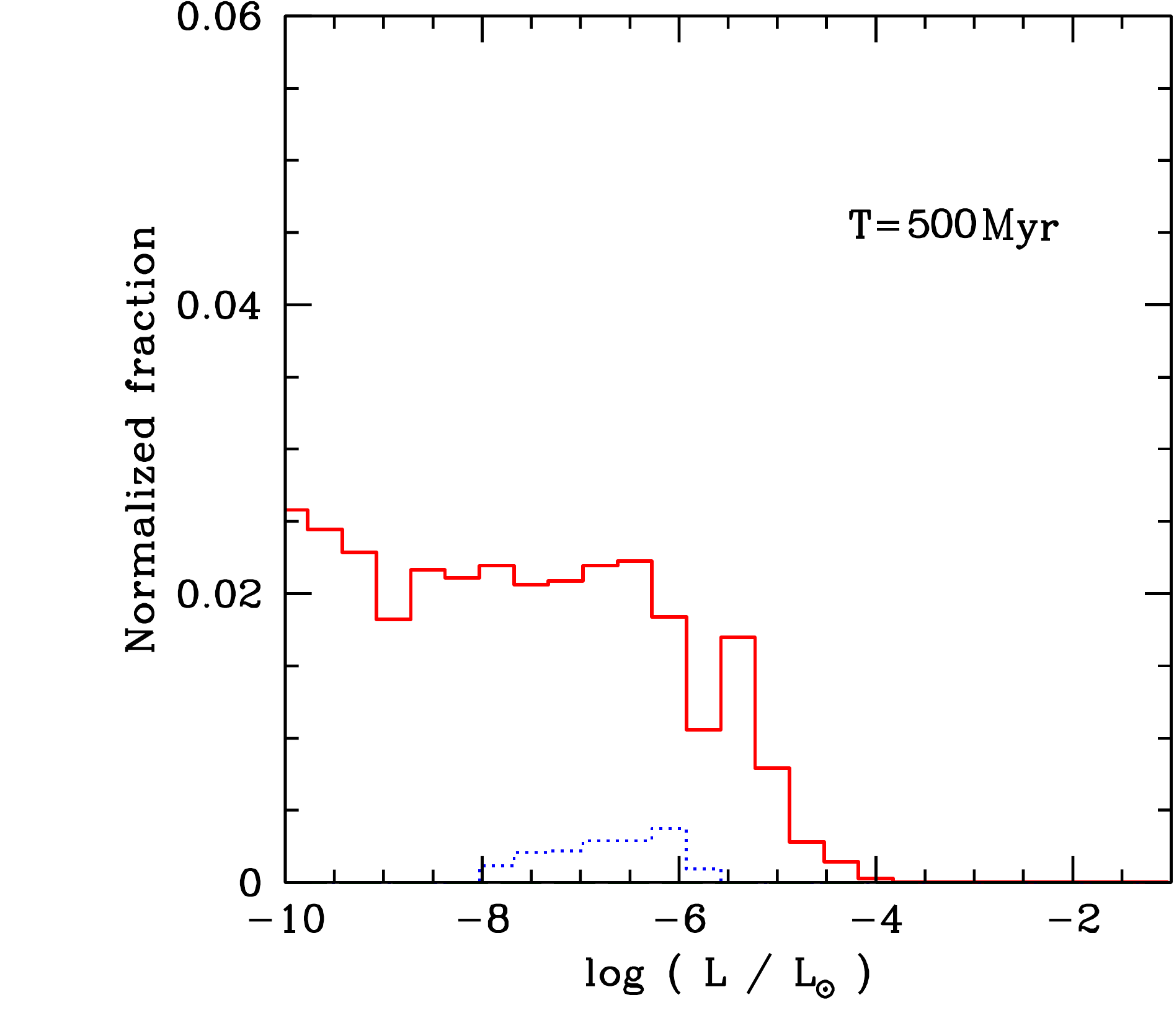}
\end{minipage}
\begin{minipage}{0.49\textwidth}
\includegraphics[width=0.98\textwidth]{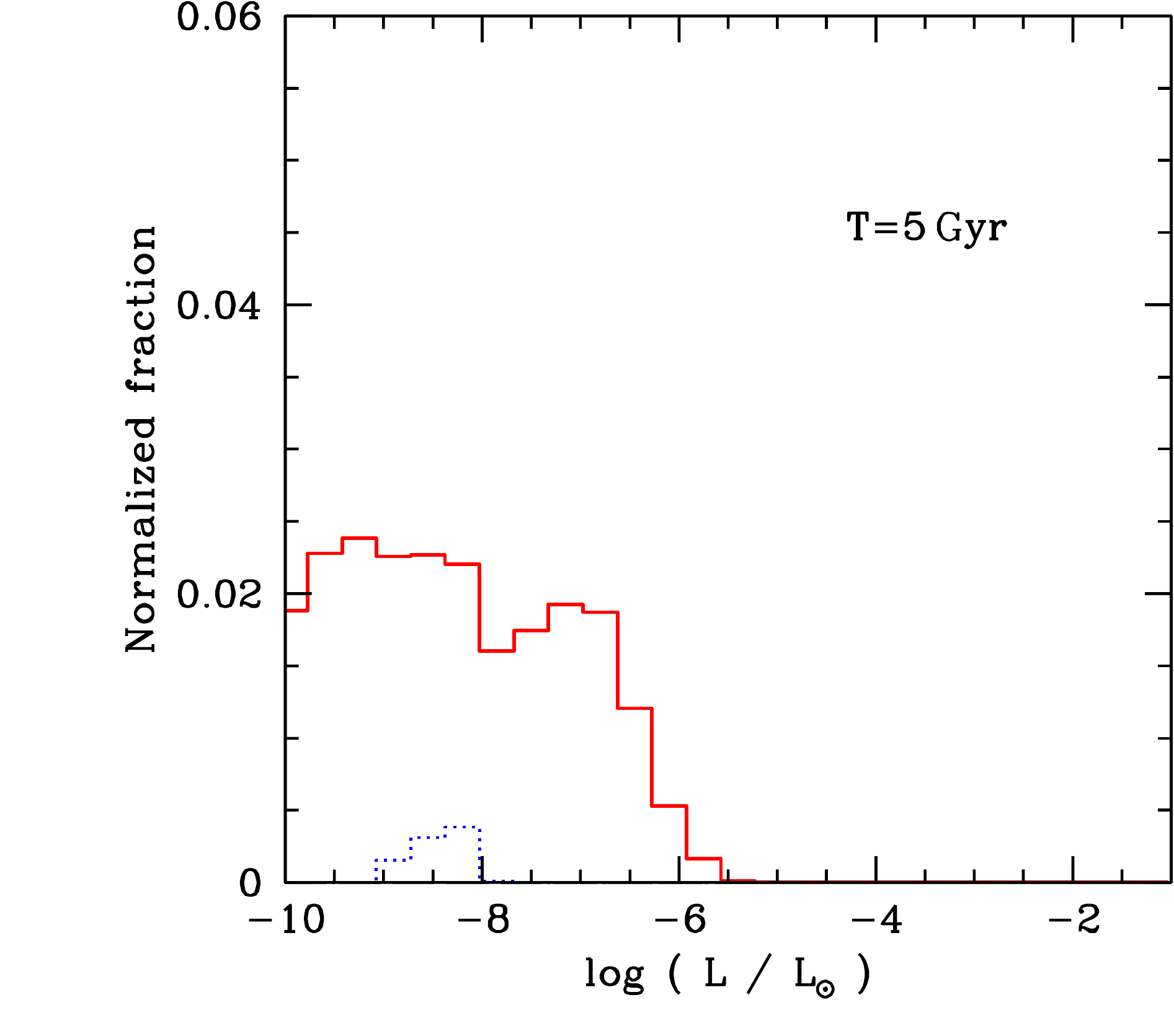}
\end{minipage}%
\caption{Distribution of  planetary luminosities in the cold-nominal population at six moments in time as indicated in the panels. Planets with $a>0.11$ AU are included.  The red solid line shows the total luminosity that contains the internal  $\lint$ and (during formation) shock luminosity $\lshock$. The green long-dashed line is the shock luminosity separately, while the blue dotted line is the deuterium burning luminosity $L_{\rm D}$ alone. It is only shown if $L_{\rm D}/L>0.01$. }\label{fig:histosr}  
\end{center}
\end{figure*}

In Fig. \ref{fig:histosr} we show the distribution of the luminosities in the  cold-nominal population at six moments in time. The histogram extends down to a luminosity of about 0.1 present time Jovian luminosities. As expected, the general trend is that the distribution shifts to lower $L$ as time goes on because of cooling. At a fixed moment in time, the $L$ distribution mainly reflects the mass distribution, and how the luminosity depends on it. The  distribution of $\log(M)$ in the synthetic populations is characterized by a relatively flat distribution in the giant planet regime (see Fig. \ref{fig:histoMcomp} and \citealt[][]{mordasinialibert2009b}), corresponding to a distribution scaling approximately with $M^{-1}$ in linear units, in good agreement with observations of extrasolar giant planets inside of 6~AU \citep[e.g.,][]{marcybutler2005}. Below 10-40 $\mearth$, there is a sharp upturn of the mass function, reflecting the transition from giant planets that underwent gas runaway accretion to the much more numerous lower mass planets that did not do so. Such a transition is also seen in the observational data \citep{howardmarcy2010,mayormarmier2011}. However, there are also a number of features that are specific to the luminosity distribution, especially at young ages. We now discuss each age in turn.

\textbf{2 Myr} In the middle of the formation phase, one sees an approximately flat part (in log) of high luminosities, and strong upturn starting at $\log(L/\lsun)\approx-6$. This upturn divides lower mass subcritical planets (masses of less than a few 10 $\mearth$) that are in the attached phase from forming giant planets that are detached. The luminosity of  these lower mass planets is mainly powered by the accretion of planetesimals $L_{\rm pla}$, whereas the luminosity of the forming giant planets originates predominately from the gas accretion, as visible from the green dashed line showing $\lshock$. To order of magnitude, one can estimate the luminosity resulting from the accretion of a core of mass $M_{\rm core}$, radius $R_{\rm core}$ with a rate $\dot{M}_{\rm core}$ as 
\beq\label{Gl:Lpmal}
L_{\rm pla} \approx\frac{G M_{\rm core} \dot{M}_{\rm core}}{R_{\rm core}}\sim 10^{-6}~\lsun\, \left(\frac{M_{\rm core}}{10 \, \mearth}\right)^{1.73}\left(\frac{t_{\rm form}}{1\, \mathrm{Myr}}\right)^{-1} %
\eeq
where we have assumed rocky cores following the mass-radius relation of \citet{valenciaoconnell2006}  and a constant accretion timescale of $t_{\rm form}\equiv M_{\rm core}/\dot{M}_{\rm core}$ for the right hand side. In reality, the accretion rate depends on the surface density of planetesimals, the distance from the star etc., and varies in time (see Sect. \ref{sect:formevo5MJ}). The equation nevertheless illustrates that for the accretion of solid cores of 1--10 $\mearth$ on a timescale of a few Myr, one expects luminosities between $\log(L/\lsun)\approx-8$ and $-6$. At even lower luminosities, there is also an important contribution by low-mass planets ($M\lesssim5\mearth$) that are not significantly accreting planetesimals. Their luminosity is mainly given by the cooling and contraction of the low-mass gaseous envelope.

Both high precision RV surveys \citep[e.g.,][]{mayormarmier2011} and the Kepler satellite \citep[e.g.,][]{fressintorres2013a} have shown that Neptunian and super-Earth planets are very abundant around solar like stars, a result recovered in the syntheses. These planets have luminosities $\lesssim 10^{-6}\lsun$ during formation. If it would be possible to detect and disentangle them form other disk features despite the fact that these planets are still embedded in the disk (attached phase) and cold, one would expect to find them in most protoplanetary disks. Their indirect impact on the disk, e.g., heating it up locally, which increases the vertical scale height \citep[][]{klahrkley2006}, and which  changes the disk's  local molecular abundances (e.g., in HCN) could possibly be detectable with ALMA \citep{cleevesbergin2015}.

\textbf{5 Myr} At this age, the general shape is similar to the one at 2 Myr,  except for a general reduction of the luminosities that are related to accretion. The peak of the luminosity generated from gas accretion has fallen by about 1 order of magnitude to now around   $\log(\lshock/\lsun)\approx-3$, while the upturn related to the subcritical planets  now rather starts at  $\log(L/\lsun)\lesssim-7$.

\textbf{10, 50 Myr} In these panels, almost all planets have entered the evolutionary phase, meaning that the underlying mass distribution is constant. At 10 Myr the green line indicates that in some long-lived disk, giant planet formation is still ongoing, but this affects only a few planets. Thus the luminosity distribution mainly reflects the mass distribution via the mass-luminosity relation for objects not undergoing accretion. We thus see an upturn at about $\log(L/\lsun)\approx-8.5$ and -9.5 at 10 and 50 Myr, respectively, corresponding to luminosities of planets with masses of a few 10 $\mearth$ shortly after formation.

At  higher luminosities (e.g., $\log(L/\lsun)\approx-8$ to -3.5 at 10 Myr), the luminosity distribution is to first order flat\footnote{More precisely, it is slightly decreasing, reflecting a similar slight decrease in the giant planet mass function (see Fig. \ref{fig:histoMcomp}).}, modulated by a local maximum at about  $\log(L/\lsun)\approx-6.5$. This range contains the ``normal'' lower mass giant planets that do not burn deuterium, and is therefore of key importance from an observational point of view. The flat distribution and the local maximum can be understood in the following way: 

(a) Regarding the flat distribution. The planetary mass function in the giant planet regime from about 0.3 to 10 $\mj$  scales as mentioned approximately as $1/M$ followed by a faster decrease at even higher masses, both in the observations \citep{marcybutler2005,cummingbutler2008} and the synthetic population studied here (Fig. \ref{fig:histoMcomp}, see also \citealt{mordasinialibert2009b}). If the luminosity is a power law in $M$, like in particular $L\propto M^{2}$  \citep{burrowsliebert1993} during evolution, then the probability distribution of $L$ will also be proportional to $1/L$, or uniform in $\log(L)$ from the fundamental transformation law of probabilities. This explains why in the giant planet regime, the luminosity distributions are to first order flat over almost four orders magnitude. As the $L\propto M^{2}$ relation is independent of time, this flat distribution can be seen in several panels of Fig. \ref{fig:histosr} and Fig. \ref{fig:histolevocoldhot} below. This is an important results result of this study, and has important implications for example for the expected yield of direct imaging searches, especially once they start to probe closer-in giant planets where radial velocity surveys have found the aforementioned mass distribution.

\begin{figure}
	\centering
       \includegraphics[width=0.45\textwidth]{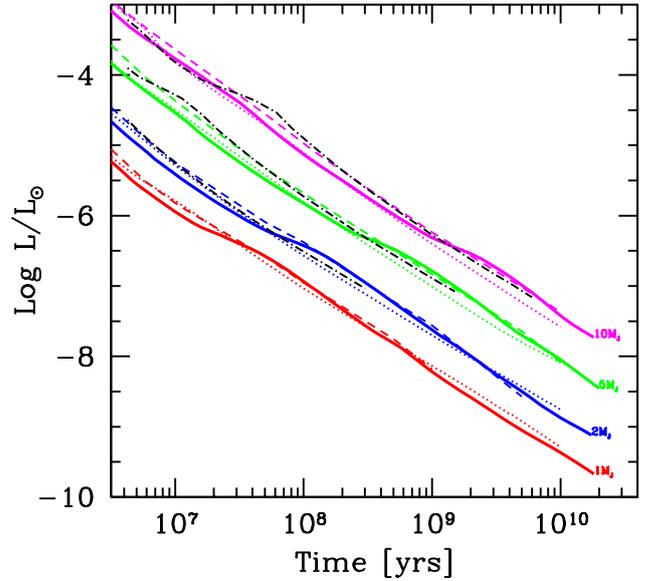}
    \caption{Luminosity as a function of time for giant planets at 5.2 AU during the evolutionary phase. The {thick} solid lines show the model used for this work. Note the bump in the luminosity at $\log(L/\lsun)\approx-6.5$. The dotted lines show for comparison the results of  \citet{burrowsmarley1997}, while the dashed lines show \citet{baraffechabrier2003}. {The black dash-dotted lines finally are the more modern  hybrid models of \citet{saumonmarley2008}.}  }\label{fig:tlevo4masses} 
\end{figure}

(b) Regarding the maximum at about $\log(L/\lsun)\approx-6.5$. This feature is seen at several ages in both Fig. \ref{fig:histosr}  and \ref{fig:histolevocoldhot} and is not related to the underlying mass distribution. It rather related to a feature in the luminosity evolution as a function of time of giant planets which is in turn controlled by the microphysics (opacity) as we will see now. It can be understood from considering Fig. \ref{fig:tlevo4masses}, which shows the luminosity as a function of time for giant planets of 1 to 10 $\mj$ during the evolutionary phase. The plot, which is a zoomed in version of Fig. 10 of \citetalias{mordasinialibert2012b}, shows the cooling curves of the model used in this paper as solid lines. In the figure one sees that at a luminosity of about $\log(L/\lsun)\approx-6.5$, there is a phase where the luminosity decreases less rapidly as a function of time than otherwise. This feature is very similar to the one  in Fig. 3 and 8 of \cite{marleaucumming2014} who use very similar atmospheric boundary conditions  as in this work (gray atmosphere with \citet{freedmanmarley2008} opacities). Since the value of $L$ where this flattening of the $L(t)$ curves occurs is independent of mass (but it occurs at different moments in time for different masses), it has the population-wide consequence that at a give time (like 50 Myr), there is a flattening of the mass-luminosity relation (a lower $dL/dM$). This is, e.g., visible in Fig. \ref{fig:mlevo} where there is a flattening of the $M-L$ relation at $\log(L/\lsun)\approx-6.5$ at 50 Myr, 0.5, and 5 Gyr. This flattening means that compared to otherwise, a larger mass interval maps approximately into the same luminosity interval,  which means for the luminosity histogram that a larger number of planets falls into the same luminosity bin of  $\log(L/\lsun)\approx-6.5$ (given the underlying flat mass distribution). From this consideration we can also see that at 50 and 500 Myr, the maximum in $N(L)$ is formed by planets of masses around 1 and 5 $\mj$, respectively. 

The underlying physical reason for the change in slope of $L(t)$ is a change in the atmospheric structure and thus the efficiency of cooling, caused by the existence of a detached radiative zone \citep{fortneyikoma2011}. A related, but much deeper detached radiative zone was originally proposed to exist in Jupiter to this day \citep{guillotchabrier1994,guillotchabrier1994b}. The  later addition of  the opacity provided by alkali metals, that are important opacity sources where  other elements (hydrogen, helium, water, methane and ammonia) have opacity windows, has since much reduced the possible $p-T$ domain where radiative transport is possible \citep{guillotstevenson2004}. Figure \ref{fig:tlevo4masses} indeed shows that for a Jupiter-mass planet, the detached radiative zone already disappears at around 20 Myr, when the opacity provided by the alkali metals is included as it is the case in the \citet{freedmanmarley2008} opacities we use. More generally, it is found that when the effective temperature of a planet has fallen to about 400 K (and the temperature at the radiative-convective boundary is $\sim$1500 K), the detached  radiative zone in the planet disappears, and the radiative convective boundary jumps from about 10 bar to 0.3 bar, much closer to the photosphere. This is visible in the temporal evolution of the $p-T$ diagram of Jupiter shown in Fig. 5 of  \citetalias{mordasinialibert2012b} or for other masses in Fig. 1 of \citet{marleaucumming2014}. As discussed by \citet{marleaucumming2014}, the disappearance of the detached convective zone is due to the change of the behavior of  the Rosseland opacity that changes from decreasing along the $p-T$ profile to increasing with it once the effective temperature has fallen below about 400 K. This is in turn due to the appearance of CH$_{4}$ which is crucial for the opacity in this regime. Interestingly, we thus see an imprint of the microphysics into the planetary luminosity distribution. 

\begin{figure*}
\begin{center}
\begin{minipage}{0.5\textwidth}
	      \centering
        \includegraphics[width=0.9\textwidth]{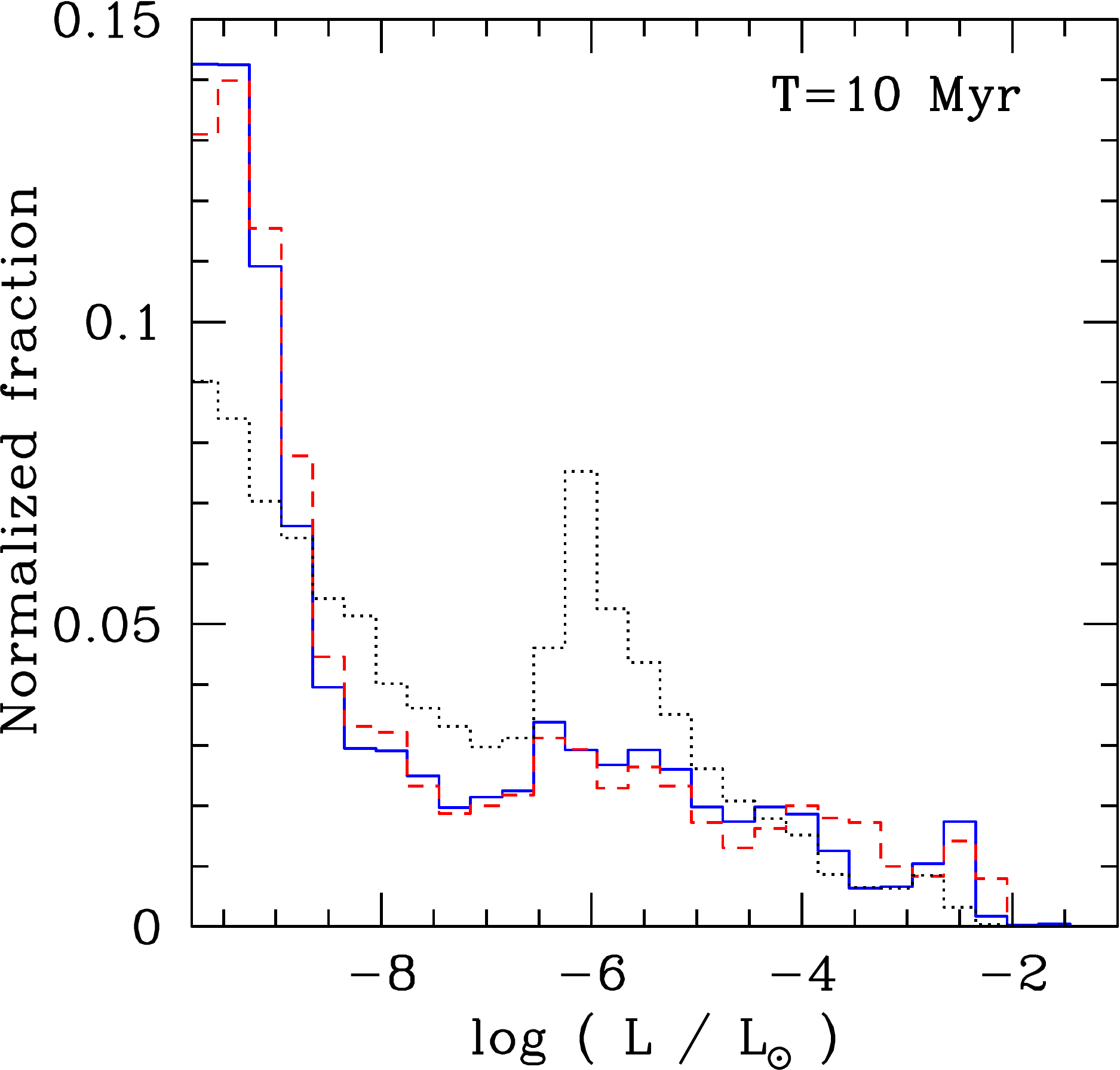}
        \includegraphics[width=0.9\textwidth]{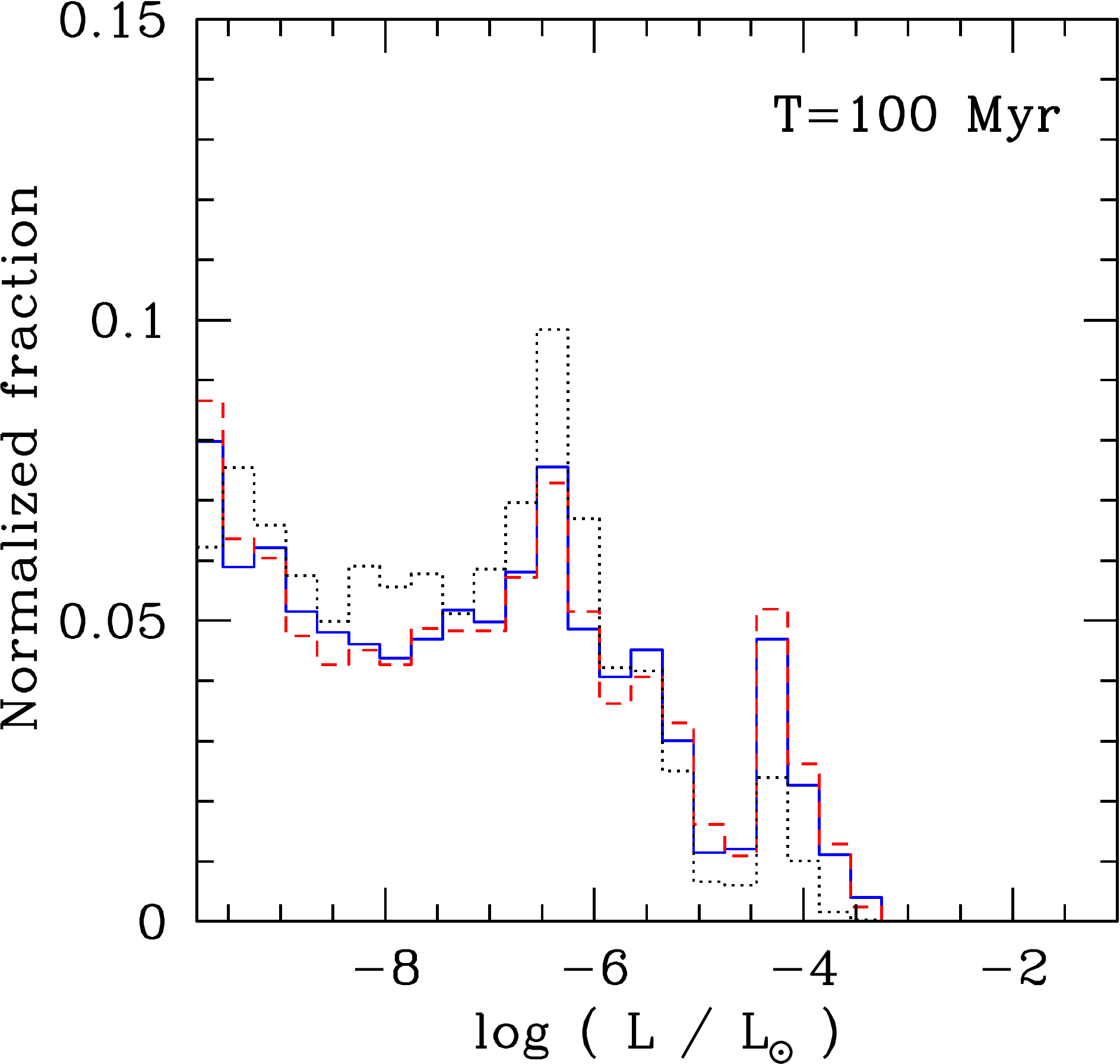}
     \end{minipage}\hfill
     \begin{minipage}{0.5\textwidth}
      \centering
       \includegraphics[width=0.9\textwidth]{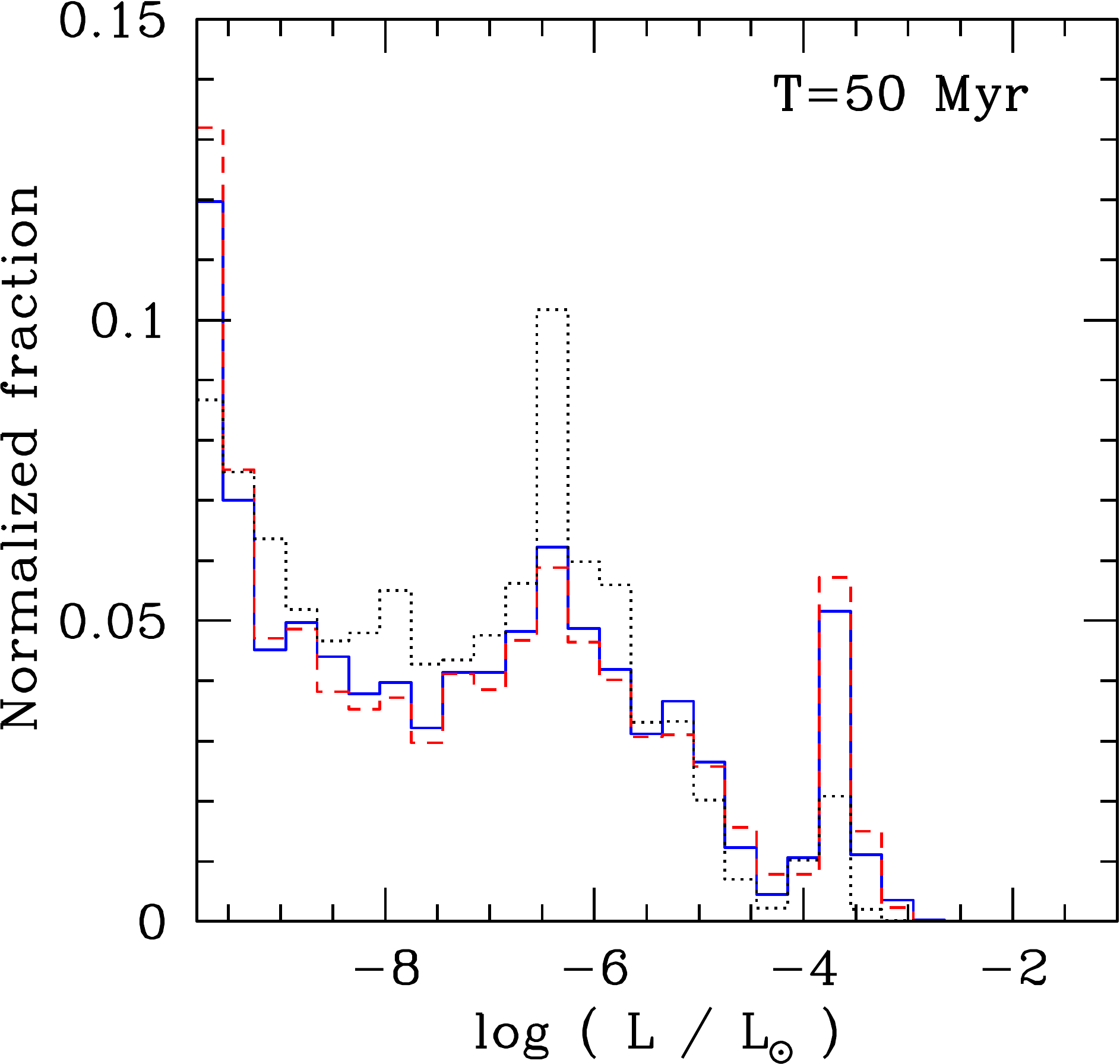}
      \includegraphics[width=0.9\textwidth]{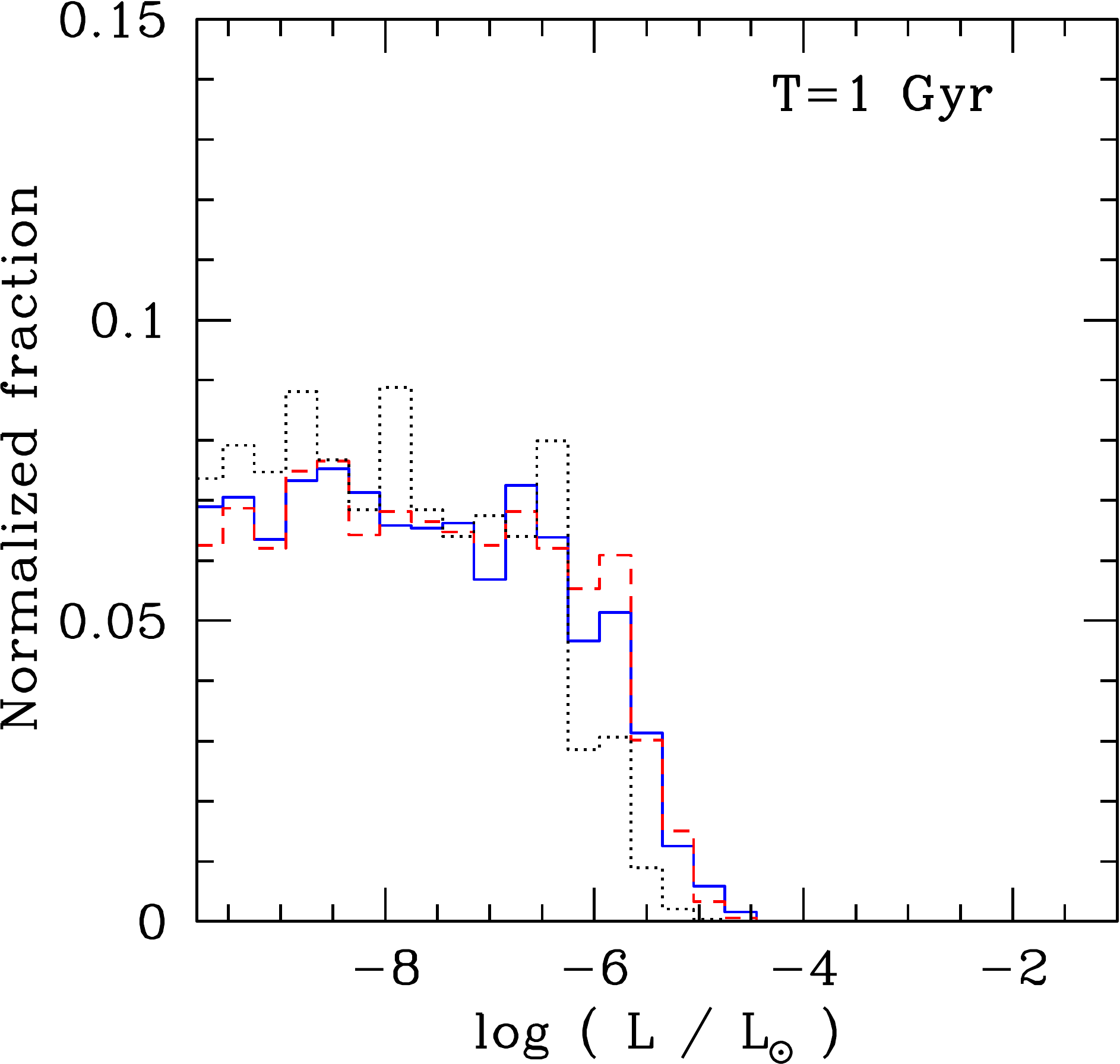}
     \end{minipage}
\caption{Distribution of  planetary luminosities during the evolutionary phase at 10, 50, 100 Myr and 1 Gyr.  The cold-nominal (blue solid), the hot (red dashed), and the cold-classical population (black dotted line) are shown. All these populations have similar mass distributions (except for an certain absence of the most massive planets in the cold-classical population) such that the different luminosity distributions are a consequence of the different entropies, and not masses, of the planets in the three populations. The peak at around $\log(L/\lsun)=-6$ in the cold-nominal population clearly distinguishes this population from the other two.}\label{fig:histolevocoldhot}
\end{center}
\end{figure*}

However, Figure \ref{fig:tlevo4masses} also shows that for the atmospheric models of \citet{burrowsmarley1997} and \citet{baraffechabrier2003}, there is no comparable bump in the $L(t)$ at around $\log(L/\lsun)\approx-6.5$. These models use both non-gray atmospheres and different underlying line lists for the opacities. {The absence of the bump is also seen in the more modern hybrid models of \citet{saumonmarley2008}. These models include the transition from cloudy L dwarfs to cloudless T dwarfs at a $T_{\rm eff}$ of 1400-1200 K\footnote{{This transition causes the bump in the black luminosity tracks at around $\log(L/\lsun)$=-4.5 to -4. Interestingly, it leads in the brown dwarf population syntheses of \citet{saumonmarley2008} to a pile-up of objects in the transition, in analogy to the pile-up found in the planetary population synthesis here.}}}. If the evolution of the population would be calculated with these boundary conditions, there would not be a maximum in the luminosity distribution at $\log(L/\lsun)\approx-6.5$. The recent models of \citep{fortneyikoma2011} on the other hand contain a small bump in $L(t)$ at a comparable, but somewhat lower luminosity of $\log(L/\lsun\approx-7$. Interestingly enough, these non-gray models also find a detached radiative zone that disappears at $T_{\rm eff}\lesssim 400$ K, or about 30 Myr for Jupiter. This feature thus depends on the specific abundances and opacity tables. We can thus speculate that the diversity of atmospheric compositions, which is identical (solar composition) in all synthetic planets in an artificial way, could  lead to a smearing out, or even disappearance of the peak. We therefore stress that the key result for the luminosity distribution of giant planets is that it is, except for the ``D-peak'' discussed below, roughly flat in $\log(L)$ (and not the potential maximum at $\log(L/\lsun)\approx-6.5$).

It should be noted that the maximum at $\log(L/\lsun)\approx-6.5$ would similarly not be present if one were to consider only closer-in planets in the synthetic population where irradiation prevents the temperature in the atmosphere to fall below about 400 K, i.e., where the equilibrium temperature that is given by the stellar irradiation is larger than this value. The temporal evolution of the atmospheric structure proceeds differently in this case, in particular  no detached radiative zone forms (see Fig. 8 in \citealt{mordasinivanboekel2016}). Rather, there is just one exterior radiative zone that deepens in time. Therefore, there is also no change of the slope of $L(t)$ as seen for the less irradiated planets when the detached radiative zone disappears. This in turn leads to very smooth $L(t)$ curves similar to the ones seen in the \citet{burrowsmarley1997} and \citet{baraffechabrier2003} models. This is exemplified by the luminosity as a function of time after formation of the planet shown in  Fig. \ref{fig:tML5MJ}, which has a final semimajor axis of 0.24 AU, and thus an equilibrium temperature of about 570 K. This distance dependence explains why the $M-L$ relation (e.g., in Fig. \ref{fig:mlevo}) not only flattens at $\log(L/\lsun)\approx-6.5$, but also widens.

Coming back to the luminosity distribution shown in Fig. \ref{fig:histosr}, at luminosities larger than $\log(L/\lsun)\approx-5$ at 50 Myr, there is a decrease in the luminosity distribution with increasing $L$, reflecting the further decrease of the number of planets with increasing mass. 
Finally at even higher luminosities ($\log(L/\lsun)\gtrsim-3$, we see another feature that is not visible in the mass distribution. As discussed in Sect. \ref{sect:correlfehL}, for massive companions, D-burning causes the luminosity to be similar for a relatively wide mass range, as more massive planets burn D earlier and at a higher $L$, while less massive ones burn it later and at a lower luminosity, leading to overlapping $L(t)$ tracks \citep[see the tracks in ][]{mollieremordasini2012}. This manifests in the $L$ distribution with a ``D-peak'', i.e., a narrow local maximum in the luminosity. At 10 Myr, the peak is at  $\log(L/\lsun)\approx-2.5$ formed by planets with masses of about 18 to 35 $\mj$ (Fig. \ref{fig:mlform}), while at 50 Myr, it is at  $\log(L/\lsun)\approx-3.8$, containing planets with masses between about 12 and 25 $\mj$ (Fig. \ref{fig:mlevo}). There is a paucity of luminosities just below these values, as the usual planetary cooling is delayed by the D burning.   

\textbf{0.5, 5 Gyr} In this panels the luminosity of the non-giant planets has fallen below the lowest plotted value, so that the increase of the luminosity distribution at the transition from solid to gas-dominated planets is not visible anymore. The flat part in the distribution formed by giant planets extends from about $\log(L/\lsun)\approx-10$ to -6 at 0.5 Gyr. The ``D-peak'' is still well visible at 0.5 Gyr at $\log(L/\lsun)\approx-5.5$ consisting of planets with masses between about 11 and 19 $\mj$, whereas at 5 Gyr, the contribution of D-burning has decreased so much that the ``D-peak'', which would at this moment be at $\log(L/\lsun)\approx-7$ (see Fig. \ref{fig:mlevo}), is no more clearly visible. The luminosity distribution therefore simply reflects the giant planet mass distribution and the upper end of the planetary mass function.

\subsubsection{Luminosity distribution: comparison of the cold-nominal, hot, and cold-classical populations}\label{sect:lumidistcompcoldhot}

\begin{figure}
	      \centering
       \includegraphics[width=0.45\textwidth]{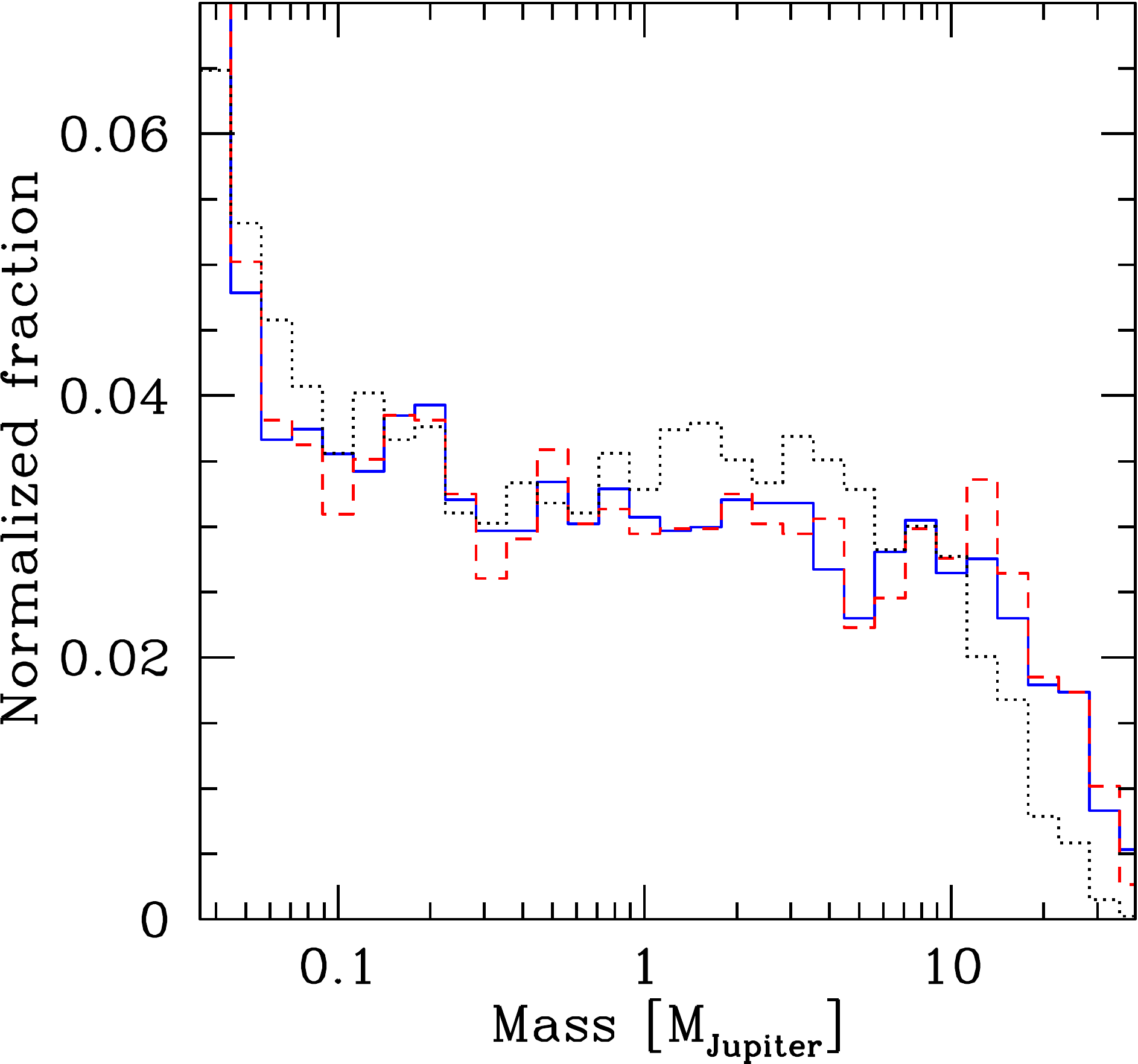}
    \caption{Comparison of the final mass distribution of giant planets for the cold-nominal  (blue solid), hot (red dashed) and cold-classical population (black dotted line).}\label{fig:histoMcomp} 
\end{figure}

After the detailed study of the distribution of $L$ in the cold-nominal population as a function of time, we show in Fig. \ref{fig:histolevocoldhot} a comparison of the luminosity distributions of the cold-nominal, hot, and cold-classical populations at four moments in time during the evolutionary phase. It is important to note that the cold-nominal and hot populations have nearly identical mass distributions (as mentioned above there is no impact of hot/cold accretion for the mass accretion rate included in the formation model). Also the cold-classical population has a very similar mass function in the giant planet range, except for a somewhat lower number of very massive planets with $M\gtrsim10\mj$. This is shown in Fig. \ref{fig:histoMcomp} which compares the mass distributions in the giant planet domain after accretion has essentially stopped. These similar mass functions mean that the differences in the luminosity distributions between the three populations are only due to difference in the thermodynamic state of the planets. 

As expected from Fig. \ref{fig:mlevocoldhot}, there is only a small difference between the luminosity distributions of the cold-nominal and hot population, namely a somewhat higher number of planets with high $\log(L/\lsun)\approx-4$ in the hot population at 10 Myr. Otherwise, one recognizes the same features as seen in the distribution of the cold-nominal population (Fig. \ref{fig:histosr}). The ``D-peak'' is, e.g., particularly well visible in both cases. The cold-classical population has, in contrast, a very different luminosity distribution: its distribution contains a strong local maximum  at about $\log(L/\lsun)\approx-6$ at 10 Myr as all giant planets with masses between about 1 and 10 $\mj$ and core masses of less than 10-20 $\mearth$ in this population approximately have a mass-independent luminosity after formation (\citealt{marleyfortney2007}, Fig. \ref{fig:mlevocoldhot}). Luminosities exceeding this value correspond to planets with higher core masses (Fig. \ref{fig:MLPostMcoreCD777}).  This strong local maximum remains visible in the distribution up to an age of at least 100 Myr, additionally amplified by the bump in $L(t)$ at  $\log(L/\lsun)\approx-6.5$ discussed in the previous section.  This very different luminosity distribution with a peak can clearly serve as a statistical diagnostic of the thermodynamics of the formation process once a higher number of directly imaged exoplanets is known. 

\subsection{Mass - radius relation with deuterium burning for cold and hot accretion}\label{sect:MRrelation}
\begin{figure*}
\begin{center}
\begin{minipage}{0.5\textwidth}
	      \centering
        \includegraphics[width=0.9\textwidth]{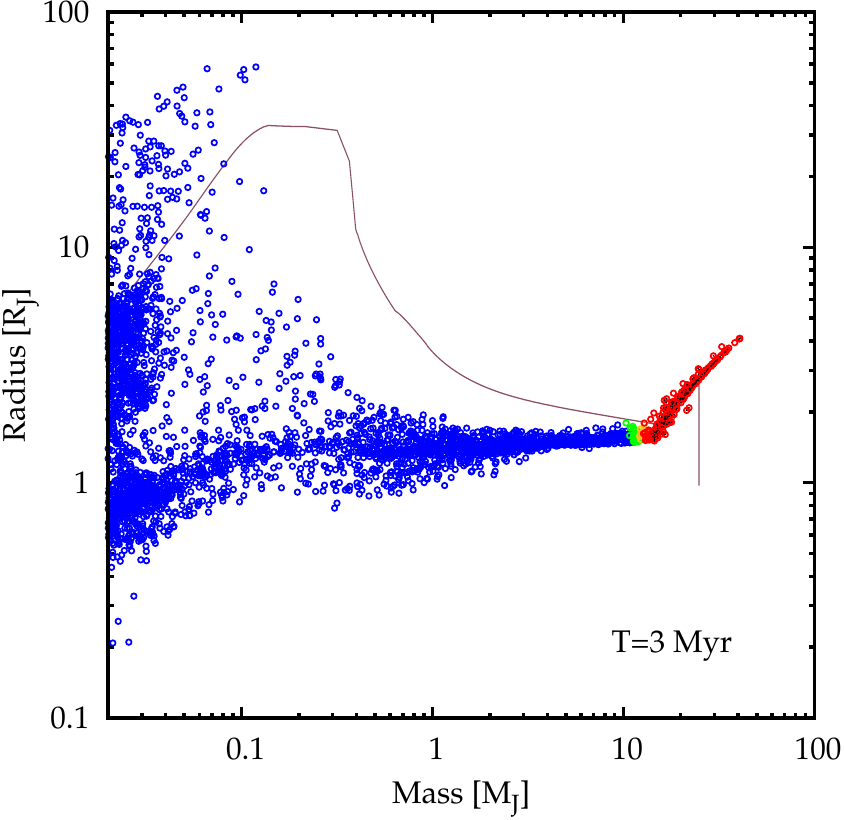}
        \includegraphics[width=0.9\textwidth]{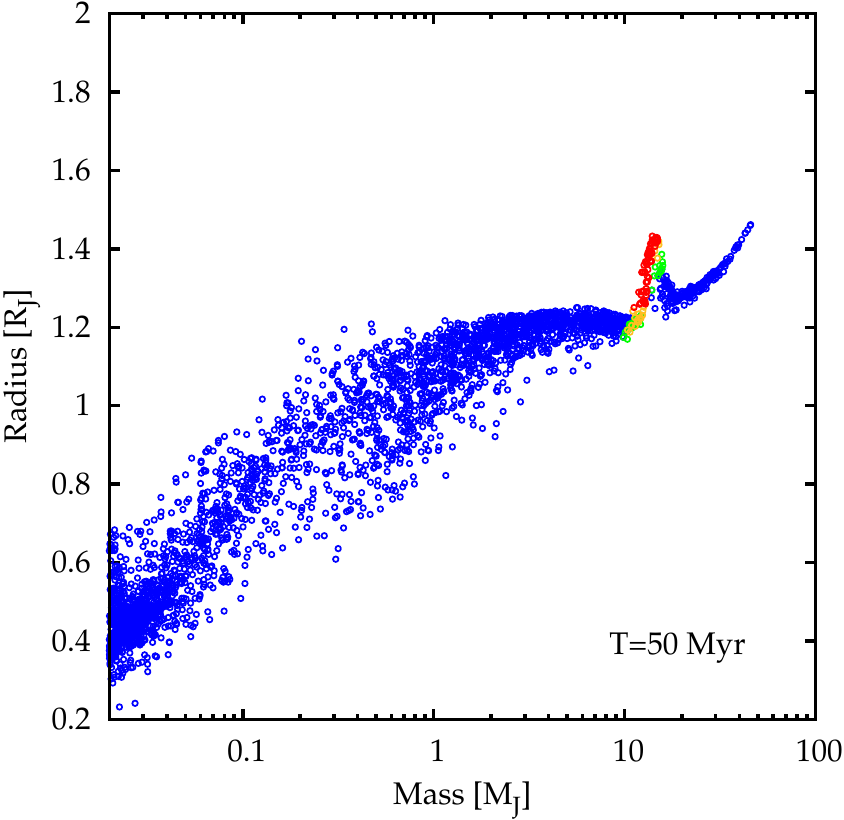}
     \end{minipage}\hfill
     \begin{minipage}{0.5\textwidth}
      \centering
       \includegraphics[width=0.9\textwidth]{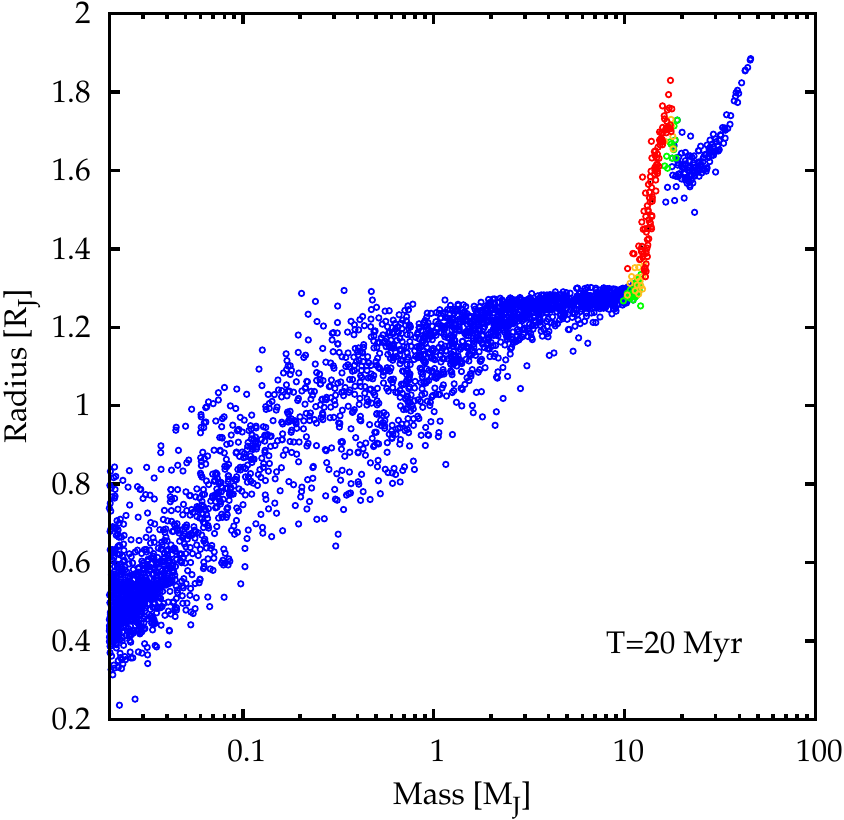}
      \includegraphics[width=0.9\textwidth]{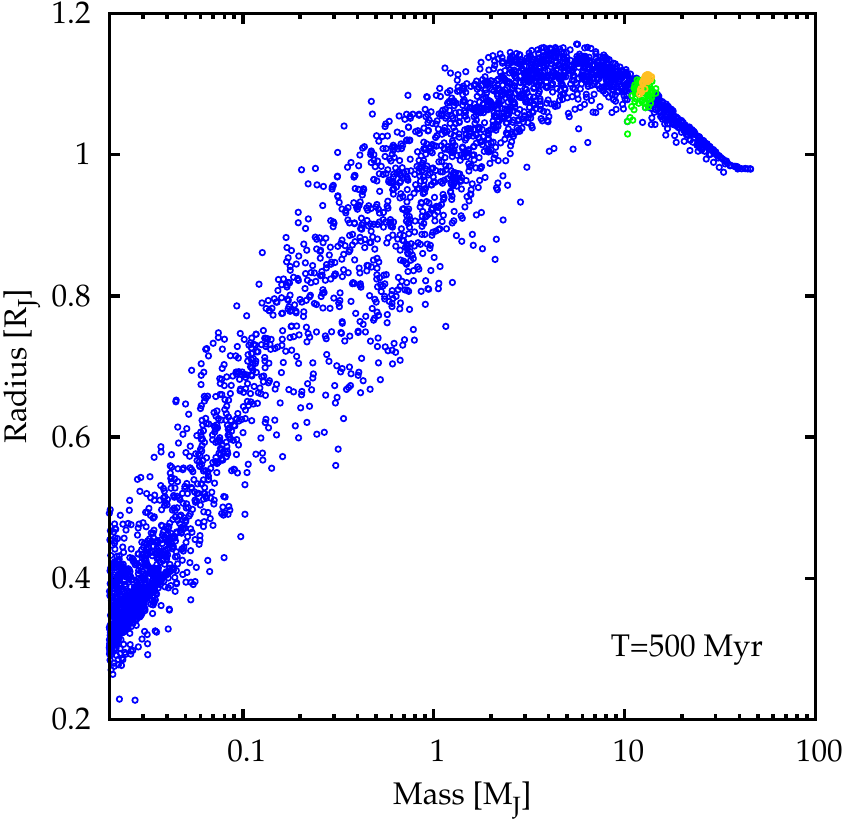}
     \end{minipage}
\caption{Impact of D-burning on the mass-radius relation of the cold-nominal population at 3 Myr (formation phase, top left) and during the evolutionary phase, namely at 20 (top right), 50 (bottom left), and 500 Myr (bottom right).  Planets with a semimajor axis between 0.2 and 5 AU are shown. As in Figure \ref{fig:mlform}, green, yellow, and red points correspond to planets with $L_{\rm D}/L_{\rm int}$ of at least 0.05, 0.1, and 0.5. Small black dots additionally show planets with $L_{\rm D}/L_{\rm int}>1$, i.e., where the D-burning leads to an expansion of the planet's radius.  The thin brown line in the top left panel shows the example of a M-R track of a planet eventually undergoing deuterium burning (see discussion in Sect.~\ref{sect:MR Cold gas accretion}). }\label{fig:MRcold}
\end{center}
\end{figure*}

The planetary mass-radius relation for cold accretion was extensively studied in \citetalias{mordasinialibert2012c}, but these calculation neglected deuterium burning. After a (giant) planet has finished its rapid contraction occurring after it detaches from the disk, its radius only changes by a factor of a few (about 3-5) during the remainder of its lifetime, while the luminosity still changes by many orders of magnitude. This weak dependency of the radius is a consequence of the partially degenerate interiors of the giant planets that are characterized by an EOS that is only weakly temperature dependent. The $M-R$ relation nevertheless reflects the thermodynamic state of a planet and depends on the cold/hot accretion mode and on the occurrence of D-burning.

\subsubsection{Cold gas accretion (cold-nominal population)}
 \label{sect:MR Cold gas accretion}

In Figure \ref{fig:MRcold} we therefore show the impact of deuterium burning on the temporal evolution of the $M$-$R$ relation of the cold-nominal population. The plot shows the mass-radius relation during formation at 3 Myr and at three times during the evolutionary phase, at 20, 50, and 500 Myr. Green, yellow, and red points correspond to planets with $L_{\rm D}/L_{\rm int}$ of at least 0.05, 0.1, and 0.5, as in Figure \ref{fig:mlform}. Small black dots additionally show planets with $L_{\rm D}/L_{\rm int}>1$.

The plot illustrates the population-wide impact of a main finding of \citet{mollieremordasini2012}, namely that for cold-accretion, D-burning does not merely delay the contraction of the radius as it is the case in classical hot start simulations \citep[e.g.,][]{burrowsmarley1997}, but leads to a re-inflation of the radius.  This is illustrated by the example of one $M-R$ track of a deuterium burning planet shown by the brown line in the top left panel. After the rapid decrease of the radius occurring after the detachment from the protoplanetary starting here at about 0.3 $\mj$, the planet\footnote{This planet has already reached a mass of about 24 $\mj$ at 3 Myr, and detached already at about 0.5 Myr. At that time, the gas accretion rate in the protoplanetary disks was higher, explaining why it detached at a higher mass than indicated by the points in the plot showing planets undergoing detachment after 3 Myr only, at about 0.1 $\mj$.}  then grows from 1 to 17 $\mj$ while the radius decreases from about 3.5 to 1.7 $\rj$, a typical sign of cold gas accretion. At this moment, the deuterium burning becomes so strong that it starts to drive an expansion of the outer radius. In this phase, the  luminosity radiated at the surface is lower than the internally produced deuterium luminosity. This excess is used to increase the planets radius and thus gravothermal energy, which is then again radiated at later times. The planet's radius reaches a maximum of almost 3 $\rj$, after which is evolves vertically down in the $M-R$ plane.  We thus see that for cold accretion, D burning has a clear and quite severe impact on the radii during formation, with an increasing radius with mass at 3 Myr for masses higher than about 15 $\mj$. 

The top left panel of Figure \ref{fig:MRcold} furthermore shows that for cold accretion, the typical radii of fully collapsed giant planets not undergoing D-burning (masses between about 1 and 13 $\mj$) in the flat part of the  $M-R$ relation cover at 3 Myr a quite narrow range of about 1.4 to 1.6 $\rj$. At 1 Myr, the situation is qualitatively similar, but the typical radius is now rather 1.8 $\rj$, and the spread around this value is larger. At 8 Myr, the typical radius is 1.3 - 1.4 $\rj$. These nearly mass-independent radii are interesting because the first facilitate estimating the accretion shock luminosity (Sect. \ref{sect:MLformcold}) and second because this is different from hot accretion, as we will see below. This could be interesting from an observational point of view if future spectroscopic observations of accreting planets can determine the surface gravity observationally.

In the panel at 20 Myr, gas accretion has stopped. The specific shape of the $M-R$ reflects now the following effects \citep{mollieremordasini2012,bodenheimerdangelo2013}: Because of their higher central temperature, the most massive planets with $M\gtrsim20\mj$ have already fully burned their deuterium at this time, and are back at cooling and contracting, while lower mass planets only now undergo significant D-burning, with the highest D-burning luminosities occurring at a mass of about 17 $\mj$. The radius distribution now bends upwards at mass of about 12.5 $\mj$, less than the often used 13 $\mj$ D-burning boundary, as the presence of massive cores shifts the limit downwards. This means that a ``wave'' of planets undergoing D-burning and expanding moves to lower masses in time. Lower mass planets burn the deuterium later, on a longer timescale, and with a lower $L_{\rm D}$. Once the deuterium in the planet is consumed the planet re-contracts. %

This leads to two local maxima of radius values in the population at 20 and 50 Myr, at 14.5 $\mj$ and for the most massive planets at about 45 $\mj$. Finally, at 500 Myr, only two subtle imprints of deuterium burning remain in the $M-R$ relation: first, there is a barely noticeable bump of about 0.02 $\rj$ at a mass of about 13 $\mj$ formed by planets that still undergo mild burning. Second, the most massive planets ($M\gtrsim 35 \mj$) still bear the imprint of their strong deuterium burning during formation. This causes a flat $M-R$ relation diverging from the usual pattern of a decreasing radius with mass for $M\gtrsim5\mj$. At even later times, also these most massive planets  follow this trend that stems from the higher compressibility of more degenerate objects \citep{chabrierbaraffe2009}. 

\subsubsection{Hot gas accretion}

\begin{figure}%
\begin{center}
	      \centering
            \includegraphics[width=\columnwidth]{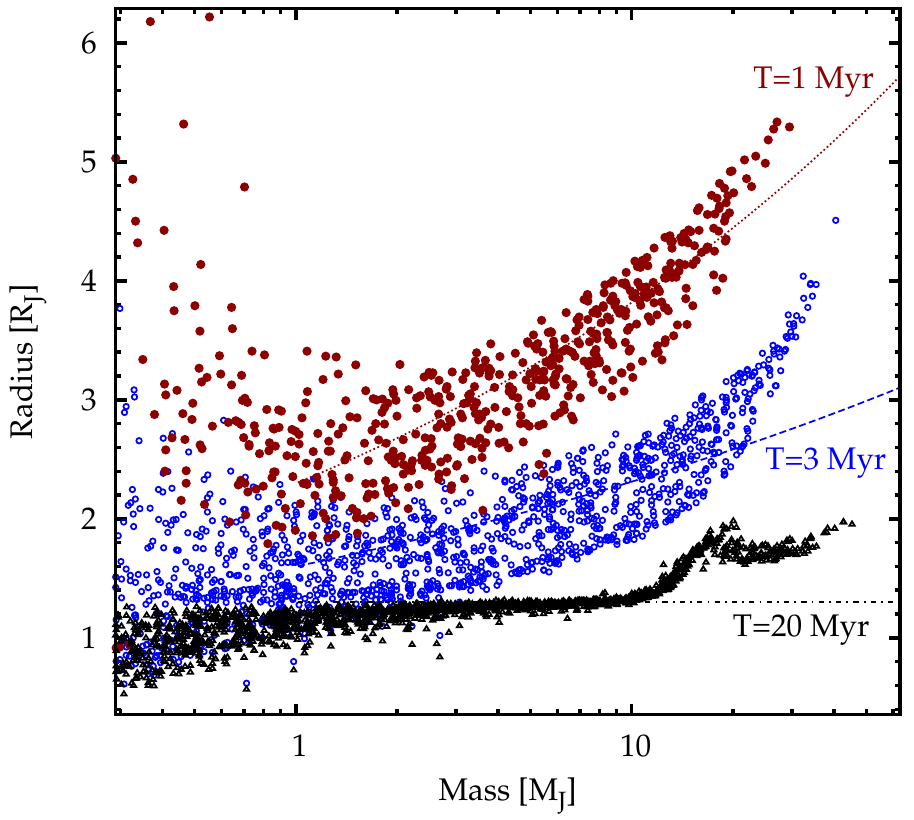}     
\caption{Mass-radius relationship for hot accretion during the formation phase at 1 and 3 Myr and during the early evolution phase at 20 Myr. The lines show empirical relations discussed in the text. }\label{fig:MRhot}
\end{center}
\end{figure}

Figure \ref{fig:MRhot} shows the mass-radius relation for hot gas accretion during the formation phase at 1 and 3 Myr when most planets are  accreting, and during the early evolutionary phase at constant mass at 20 Myr.  Due to the deposition of the shock luminosity into the planet, the radius is, for planets that have completely undergone the fast contraction after the detachment, an increasing function of the mass, in contrast to the cold accretion case. Such an increase of $R$ with $M$ is also seen in classical hot start models which do not calculate the formation of the planet but assume a very hot state of the planet as an arbitrary initial condition. 

In the plot we have also included three empirical mean $M-R$ relations. For planets with $1<M/\mj<10$, the mean radius is approximately  $R\approx 2.3 \rj \times (M/\mj)^{0.22} $ at 1 Myr which means that these planets have quite large radii. Even larger radii are seen for planets in the fast contraction phase ($M<1\mj$), but this phase is short, a few 10$^{4}$ years.  At 3 Myr, we find  a mean radius of  about $R\approx 1.6 \rj (M/\mj)^{0.16}$. Interestingly, this is quite similar to the radii in the hot start simulations with arbitrary initial conditions of \citet{marleyfortney2007} at 1 Myr (their Fig. 4), but for already fully formed planets.  At 20 Myr, the radii are about 1.3 $\rj$, already weakly dependent on mass as it is the case for mature planets \citepalias{mordasinialibert2012c}, with a difference in $R$ between 1 and 10 $\mj$ of only about 0.2 $\rj$. This is again similar to the classical hot start models. The flat $M-R$ in this phase is a consequence of a competition between the higher entropy in the more massive planets increasing $R$ and their higher compressibility due to a weaker ionic contribution in the EOS, decreasing $R$. During formation, there is a spread of about 1 $\rj$ around the mean value. 

During formation, no very clear imprint of D-burning is seen in contrast to the cold accretion case. This is not surprising, because for hot accretion, D burning only delays the contraction, but does not lead to a re-increase of the planetary radius after the collapse \citep{mollieremordasini2012}. Therefore, one notes that for planets undergoing D-burning, there is an absence of small planetary radii, but in general, the $M-R$ is relatively smooth for $M\gtrsim 1 \mj$, at least during the formation phase. During evolution, at 20 Myr, the $M-R$ is, in contrast, similar to the cold case with the characteristic upturn of the radii at about 10-12 $\mj$. Compared to the cold case, the radii of these planets are slightly bigger, by about 0.1 to 0.2 $\rj$. We thus see that for planets that are sufficiently massive, D-burning tends to partially erase the effect between cold and hot accretion \citep{mollieremordasini2012} as in both cases the planets approach the gravothermal state obtained after the complete deuterium reservoir has been burned. The way this state is approached, and at which moment it happens, is however affected by the hot/cold accretion mode.

\section{Summary and conclusions}\label{sect:conclusions}

In this paper, we studied for the first time
the statistics of planetary luminosities during formation
and during evolution in the framework of the canonical core accretion scenario for giant planet formation
\citep{pollackhubickyj1996,bodenheimergrossman1980} using the tool of population synthesis.
Since the effects of the accretion shock are thought to be important,
but are currently not definitely predicted by theory (see \citealp{szulagimordasini2016,marleauklahr2016})
nor constrained by observations,
we considered both a cold-nominal and a hot population,
in which the kinetic energy of the incoming gas is respectively
fully absorbed by the planet or fully radiated away.
We also calculated a cold-classical case where,
as in the pioneering work of \citet{marleyfortney2007},
it was additionally assumed that
planetesimal accretion stops
artificially once a giant planet enters the disk-limited gas
accretion (detached) phase
and that the planets form in situ.

We discussed three fundamental statistical properties of the formed planets:
the planetary mass--luminosity relation  during both formation and evolution (Sect.~\ref{sect:MLrelation}),
the mass--entropy diagram at the moment when the protoplanetary disk disappears (Sect.~\ref{sect:spflpf}),
and finally the luminosity distribution as a function of time (Sect.~\ref{sect:Ldistro}).
We furthermore revisited the mass--radius relation that was extensively discussed in \citetalias{mordasinialibert2012c} in Sect.~\ref{sect:MRrelation}, now including the effect of deuterium burning
as implemented in \citet{mollieremordasini2012}. Our main findings are as follows: 
\begin{enumerate}

  \item
The planetary mass--luminosity relation during the formation phase is shown
for the cold-nominal, hot, and cold-classical populations
in Figs.~\ref{fig:mlform}, \ref{fig:mlformhot}, and~\ref{fig:mlformlimited}, respectively. Important features include:

\begin{enumerate}
  \item
At a stellar age of 1~Myr, the accretion luminosity of giant cold-nominal planets almost always dominates over the internal luminosity. If the planetary shock luminosity $L_{\rm shock}$ can be measured observationally from accretion trackers,  and the gas accretion rate $\dot{M}_{\rm XY}$ can be constrained from disk and stellar observations like the stellar accretion rate, the planet's mass can be estimated by (Eq.~\ref{eq:lest})
\beq
M_{\rm est}=\frac{R L_{\rm shock}}{G\dot{M}_{\rm XY}}
\eeq
with a radius $R$$\approx$1.8$\rj$ (at 1 Myr; 1.5 $~\rj$ at 3 Myr; and 1.3 $~\rj$ at 8 Myr) approximately independently of mass.
For known $\dot{M}_{\rm XY}$, this is accurate to about 20\,\%\ for $M=3$--17~$\mj$
(Sect.~\ref{sect:MLformcold}).

  \item
At $\approx$3--5~Myr, the $L$--$M$ diagram is characterized
by two groups: an upper accreting sequence
of protoplanets (with $L\propto M$),
whose parent disk has not yet dissipated,
and an evolving sequence (with $L\propto M^2$),
which is already on standard cooling tracks.
Over time, all luminosities drop due to decreasing accretion rates
and normal cooling, respectively,
and planets move onto the evolving sequence (Sect.~\ref{sect:MLformcold}).

  \item
Comparing the highest total luminosities at a given mass (strongest accretors),
planets in the cold-nominal population have  during formation a higher total luminosity than the brightest planets in  the hot population
of the same mass, which only have the interior luminosity. This situation is inverted  relative to the subsequent evolutionary phase at constant mass.

This also means that the effective
temperature of accreting planets is higher for cold accretion than
for hot accretion at a given planetary mas.
This inversion comes from energy conservation
(Sect.~\ref{sect:formationphasehotgasaccretion}).

  \item
In the cold-classical population, we recover for a significant group of planets the \citet{marleyfortney2007} result of a low luminosity (cold start) independent of planet mass (Sect.~\ref{sect:formationphasecoldclassicalgasaccretion}).
However, most planets exhibit a warm start because of the
core-mass effect, from the disk mass and metallicity distribution.
The cold-classical population is the population
with the greatest variation in the (total) luminosity at a given mass.
A large spread could also result if the core-mass effect is not as efficient as assumed here (leading to lower $\lint$)
but if the efficiency of the accretion shock in radiating away the accretional energy is less than 100\%.

  \item
For giant planets in their main accretion phase, the shock luminosity $\lshock$ is higher than the internal luminosity $\lint$ by a factor 2 to more than one order of magnitude. The ratio $
\lshock/\lint$ increases with mass up to the deuterium burning limit (Sect.~\ref{sect:lshocklint}). %
\end{enumerate}

  \item
We performed simple comparisons to observations of embedded (and potentially accreting) planets in the formation phase and older purely cooling planets in the evolutionary phase and found the following:
\begin{enumerate}

  \item
Comparing the cold-nominal population to HD~100546~b,
the observed luminosity could be matched by {a mass of} $0.2$--$9.7~\mj$, depending on the (unknown) relative importance of the accretional and internal luminosity.
For the lowest mass planets,
the accretion of planetesimals contributes significantly to the luminosity,
meaning that potentially very luminous impacts may occur frequently on
this planet.
Results are similar for hot starts due to the core-mass effect
(Sect.~\ref{sect:compembedded}).

\item
The same exercise for LkCa~15~b yielded the result that
masses of 1--2~$\mj$ are consistent with the inferred H\,$\alpha$ luminosity
\textit{and} the (estimated) minimum gas accretion rate (Sect.~\ref{sect:compembedded}).

\item
We found that it is difficult to constrain the
mass from a measured total luminosity ($L_{\rm int} + L_{\rm shock}$) alone during
formation because of the possible contribution of accretion.
A comprehensive view with spectroscopic determination
of the different contributions (planetary internal, planetary
accretion shock, and circumplanetary disk luminosity) combined
with other indicators (accretion rate, disk structure, etc.)
should be used to get a clearer picture (Sect.~\ref{sect:compembedded}).

\item
Looking at $\beta$ Pic b, we find it is with 10--13~$\mj$
at the interesting transition near the deuterium-burning limit.
During the main-sequence lifetime of $\beta$ Pic of about
2~Gyr, $\beta$ Pic b will typically burn about 10\,\% of its deuterium
reservoir if its mass is on the lower side of the allowed range and up to
90\,\% if its mass is rather $13~\mj$.
The synthetic analogs have a wide range of heavy element
mass fractions of 0.005--0.1, with a typical value of
about 0.015. This corresponds to an enrichment relative to solar
of about 0.3 to 7, with a typical value of around 1
(Sect.~\ref{sect:compbetapic51eri}).

\item
For 51 Eri b, the luminosity measurement
implies $M=1.7$--3.6~$\mj$,
a somewhat wider range than the hot-start masses
reported by \citet{macintoshgraham2015}
(Sect.~\ref{sect:compbetapic51eri}). This wider range is a consequence of the intrinsic scatter in the $M-L$ relation at young ages. It stems from different formation histories. 
Planets in the cold-classical population have in contrast
masses between about 1.7 and almost 10~$\mj$
(Sect.~\ref{sect:coldvshot}).

\end{enumerate}

  \item  
One of the key findings of this work is the importance of the
core-mass effect \citep{mordasini2013,bodenheimerdangelo2013}:
Due to it %
and to the large mean amount of heavy elements expected in
giant planets \citep{ThorngrenFortney2016},
the luminosities in the
cold-nominal population are almost comparable to those found for
hot accretion.
Hot or at least warm planets could thus indeed be the expected outcome for core
accretion {(but see also Sect. \ref{sect:limintstruct} and Appendix \ref{app:coremasseffect})}, and not the very low luminosities in \citet{marleyfortney2007}
where the accretion of planetesimals is artificially shut off (Sect. \ref{sect:evophasespread}).

Nevertheless, even at an age of 20 or 50~Myr, well into the evolutionary phase,
there is still an intrinsic scatter in luminosity in the cold-nominal population by a factor of respectively $\approx1.5$--2 or 20--50\,\%\ at a given total mass, mainly due to the different core masses {(Fig. \ref{fig:mlmcore})}. {The actual ``spread'' in the evolution could in reality be even wider than found in our model. Relevant factors could be different opacities, envelope metal enrichments, non-solar stellar masses, and other effects that are not included in this current generation of models (such as sub-unity shock efficiencies, complex infall geometries, or planetary magnetic fields)}.
This should be critically kept in mind when deriving masses from measured luminosities. It means that even if the luminosity could be determined observationally with a vanishing error bar, there is no one-to-one translation into a single planet mass (Sect. \ref{sect:impactcoremass}). %

  \item
In %
Fig.~\ref{fig:MSmarleycomp},
we presented for all three populations
a key  %
diagnostic outcome of a formation model,
the entropy ``tuning fork'' diagram of exoplanets.
This shows the specific entropy in the convective
zone as a function of planet mass at
the end of the formation phase when
the protoplanetary disk disappears.
Fig.~\ref{fig:MLtdisk} shows the corresponding luminosities.

A new aspect is that due to the different formation histories,
there is a large spread of about 1 to 1.5 $\kB$/baryon
at a given mass in the hot and cold-nominal population. In
the cold-classical population, the post-formation entropies
even have a $\approx2~\kB$/baryon
scatter at about $5~\mj$ {which results primarily from the spread in core masses as demonstrated by Fig. \ref{fig:MLPostMcoreCD777}.}
This means that the entire phase space
from cold through warm to hot initial states is populated even if
only the limiting cases of completely hot and cold gas accretion
are considered.
{The appendices \ref{sect:fitS0} and }\ref{app:fit10MyrLSD} contain fit{s} to the post-formation properties of the synthetic planets which are of interest as initial condition for evolution models.

 \item Next, we examined an important statistical result
of this work, the planetary luminosity function
(i.e., the $M$--$L$ distribution marginalized over mass)
during formation and evolution.
The main results are (Figs. \ref{fig:histosr} and \ref{fig:histolevocoldhot}):

\begin{enumerate}
  \item
At a given time, the $L$ distribution mainly reflects the mass
distribution. The distribution of $\log(M)$ in the synthetic populations
is relatively flat in the giant planet regime
(see Fig.~\ref{fig:histoMcomp} and \citealt{mordasinialibert2009b}),
corresponding to a distribution
scaling approximately with $M^{-1}$ in linear units,
in good agreement with observations.

  \item
Up to ca.~5~Myr, %
one sees an approximately flat part (in $\log L$) of high luminosities,
and strong upturn starting at $\log(L/\lsun) \approx -6$.
This upturn divides lower mass subcritical planets
that are in the attached phase from forming giant planets that are detached.
The luminosity of these lower mass planets ($M\lesssim10$--$40~M_\oplus$)
is mainly powered by the accretion of planetesimals.
(Sect.~\ref{sect:L(t)fuerkalt}).

  \item
In the pure cooling phase (after $\sim$10 Myr),
the luminosity distribution for non-deuterium-burning giant planets
is, to first order, uniform in the logarithm of the luminosity over four orders of magnitude.
This flatness comes from the $1/M$ mass function from about 0.3 to 10~$\mj$
combined with the fixed-time $L\propto M^2$ scaling \citep{burrowsliebert1993}.
This
is an important results result of this study and has important implications
for example for the expected yield of direct imaging
searches, especially once they start to probe closer-in giant planets,
which are more likely to be formed by core accretion
(Sect.~\ref{sect:L(t)fuerkalt}).

  \item
{At even higher masses, t}here is  a clearly visible, $\approx0.5$~dex-wide peak in the 
luminosity distribution that is separated from the remaining distribution. The peak is located at $\log(L/\lsun) \approx -4$ and~$-5.5$ at 50 and 500~Myr, respectively
(Sect.~\ref{sect:L(t)fuerkalt}). It is due to deuterium burning, which slows the cooling or even reverses it and thus leads to a flattening of $L(t)$.

\end{enumerate}

These points pertained to the cold-nominal and hot population,
with no significant difference between the two (Sect. \ref{sect:lumidistcompcoldhot}).

The cold-classical population has a very different luminosity distribution,
with a strong local maximum near
$\log(L/\lsun) \approx -6$ at 10~Myr {instead of a flat distribution}, as all giant planets with masses between
about 1 and 10~$\mj$ and core masses of less than 10--$20~\mearth$
in this population approximately have a mass-independent luminosity
after formation, as seen in \citet{marleyfortney2007}.
This strong local maximum remains visible
in the distribution up to an age of at least 100 Myr.
This very different luminosity distribution
with a peak can clearly serve as a statistical diagnostic of the
thermodynamics of the formation process once a higher number
of directly imaged exoplanets is known
(Fig. \ref{fig:histolevocoldhot}).

\item
Finally, we studied the impact of deuterium burning on the mass--radius relationship (Sect. \ref{sect:MRrelation}).
We saw, as in \citet{mollieremordasini2012} and \citet{bodenheimerdangelo2013}, that for cold accretion, deuterium burning does not merely delay the contraction of the radius
as for classical hot starts but also brings about a re-inflation of the planet.
This leads to an increasing radius with mass at 3~Myr
for masses higher than about $15~\mj$.
The $M$--$R$ relation was additionally seen to (i)~be nearly flat from 1 to~10~$\mj$
up to $\approx50$~Myr,
and (ii)~show a local peak (higher radii)
near 15~$\mj$.

In the hot population, the radii increase  in contrast during formation with mass and can reach 4 to 5 $\rj$. D burning (which sets in later) here only delays the contraction,
but does not lead to a re-increase of the planetary radius;
therefore, no very clear imprint of D-burning is seen in the $M-R$ relation during formation
in contrast to the cold-accretion case.

\end{enumerate}

One important point to take away from this work is that
core accretion cannot be excluded as the formation mechanism
based on an observed high luminosity alone, as seen in Sects.~\ref{sect:MLformcold} and~\ref{sect:impactcoremass}.
However, one should bear in mind that the self-amplifying "core-mass effect"
which is responsible for this as described in \citet{mordasini2013} (see also \citealt{bodenheimerdangelo2013}) requires
that the planetesimals be accreted rapidly during the late attached and early detached phase and that the solids sink quickly deep into
the potential well \citep{mordasini2013}. This has not yet been studied with giant planet formation models tracking both the thermodynamical and compositional evolution of the interior {(Sect.~\ref{sect:limintstruct}). A preliminary analysis (Appendix \ref{app:coremasseffect}) indicates that relative to sinking, the heating by impacting planetesimals could be reduced by factors 2-3 for homogeneous mixing into the envelope, and 3-8 for no sinking. }


%
Furthermore, our results indicate that during formation a rough estimation of the planetary mass may be possible if the planetary gas accretion rate and its accretion shock luminosity can be determined, at least for cold gas accretion where the radius is nearly independent of mass. If it is unknown whether the planet still accretes gas, the total luminosity (accretional and internal) spread at a given mass may be as large as two orders of magnitude, therefore inhibiting the mass estimation in this way. Due to the core-mass effect even planets which underwent cold gas accretion can have large post-formation entropies and luminosities. This means that alternative formation scenarios such as gravitational instabilities do not need to be invoked from a (high) luminosity point of view alone. Once the number of self-luminous exoplanets with known ages and luminosities increases thanks to future survey with small inner working angles, the observed luminosity distribution may be compared with our theoretical predictions. This comparison will eventually allow to develop a better understanding of the thermodynamics of giant planet formation, and the  planet formation process overall. 

Looking ahead, it will be interesting to repeat the statistical analysis as the theoretical description of the underlying physical processes in the planet formation and evolution models are improved.
Of primary importance will be the coupling of predictive models  \citep[e.g.,][]{marleauklahr2016}
for the accretion shock's efficiency in radiating away $\lshock$ to structure calculations. This will eliminate the current need to arbitrarily assume a fully cold or hot gas accretion. Another point is to include the radial non-constancy of the luminosity in accreting objects
\citep{stahler1988,berardocumming2016}, or the replacement of the gray atmospheric boundary conditions with more realistic non-gray models (Marleau et al. in prep.). This will lead to a direct prediction of the magnitudes instead of the total luminosity only. 
Also, considering multiple embryos per disk
to take their dynamical interactions into account \citep{alibertcarron2013}
will enable us to also look at the luminosity--separation diagram,
of crucial importance for direct detections.
Finally, it will be interesting to perform dedicated simulations
for specific recently discovered systems, tuning in particular the stellar mass
and metallicity, as was done for $\beta$~Pic~b \citep{bonnefoyboccaletti2013}.

We finish commenting on two recently observed extrasolar protoplanet candidates. They are {potentially} among the first examples where we may directly observe planet formation as it happens. Comparing HD 100546 b and LkCA 15 b, it seems possible that the two protoplanets represent two different evolutionary stages of giant planet formation. HD 100546 b could still be in the earlier, more extended, and cooler phase during the transition from the attached to the detached phase at a lower mass without significant hard accretion-driven radiation, and without a (deep) gap in the protoplanetary disk. LkCa 15 b would in contrast correspond to the later, fully detached phase characterized by a higher mass, higher temperatures, gap formation, and hard accretion shock radiation. To firmly establish this, more observations seem necessary, also to disentangle the impact of the circumplanetary disk. But directly observing the different theoretically predicted main phases of (giant) planet formation would represent an important step forward to a better understanding of this process. We note that the model predicts (see Sect. \ref{sect:Ldistro}) that protoplanetary disk should contain a much higher number of still fully embedded, even lower mass protoplanets that are in the even earlier attached phase. Past radial velocity and transit surveys have shown that such low-mass planets are frequent. They are characterized during formation by low temperatures (a few 100 K) and low luminosities $\lesssim10^{-6}\lsun$, making them difficult to detect and disentangle from disk features. As shown by \citet{vanboekelhenning2016}, comparisons of the synthetic populations presented here with the expected detection limits of the Mid-Infrared ELT Imager and Spectrograph METIS \citep{brandlfeldt2014} indicate that at least the more massive planets in this stage could potentially nevertheless be detected with this instrument. This opens up the fascinating possibility to directly observe the ``when'' and ``where'' of this crucial phase of planet formation, which would yield observational constraints of uttermost importance for planet formation theory.

\acknowledgements{3M thank Y.~Alibert, W.~Benz, B.~Biller, M.~Bonavita, M.~Bonnefoy, E.~B\"uenzli, K.-M.~Dittkrist, {R.~Helled}, H.~Klahr, A.-M.~Lagrange, M.~Meyer,  S.~Quanz , and R.~van Boekel for useful discussions. {We also thank the referee J. Fortney for a helpful report}. C.M. and G.-D.M acknowledge the support from the Swiss National Science Foundation under grant BSSGI0$\_$155816 ``PlanetsInTime''. Parts of this work have been carried out within the frame of the National Center for Competence in Research PlanetS supported by the SNSF. }

\appendix

\section{Empirical fit to the post-formation entropy $\spf$}\label{sect:fitS0}

\begin{figure*}
\begin{center}      
	      \begin{minipage}{0.45\textwidth}
	      \centering
        \includegraphics[width=0.99\textwidth]{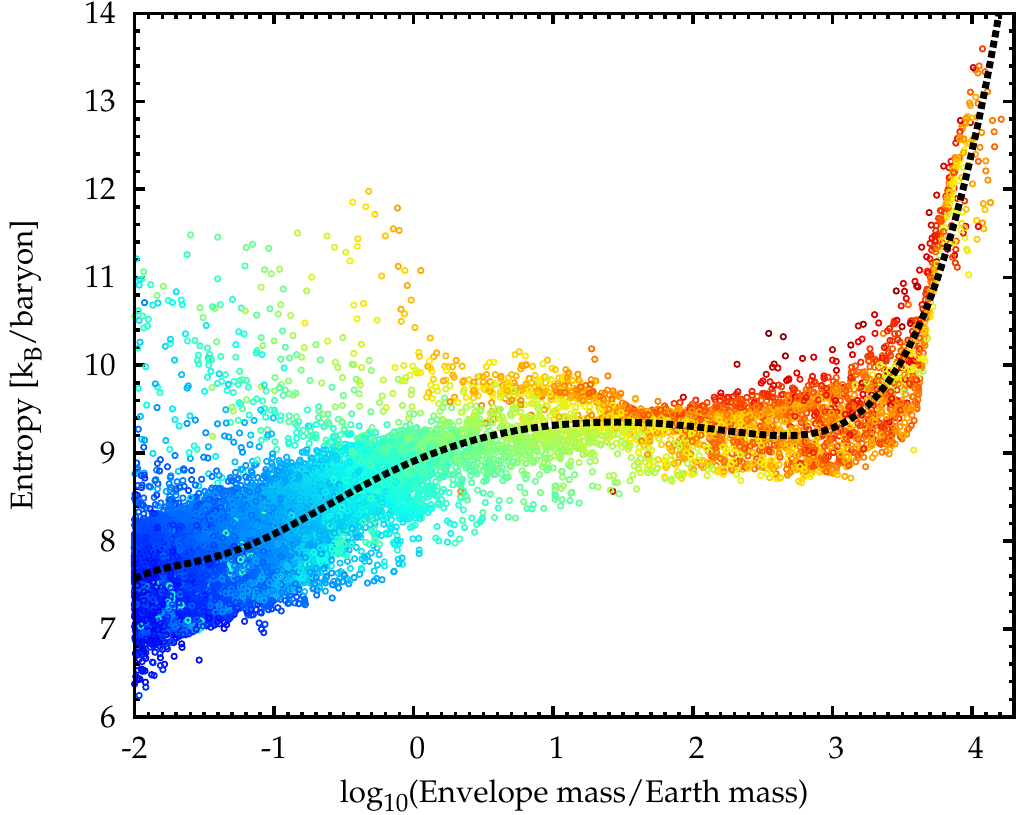}
     \end{minipage}\hfill
     \begin{minipage}{0.55\textwidth}
      \centering
        \includegraphics[width=0.92\textwidth]{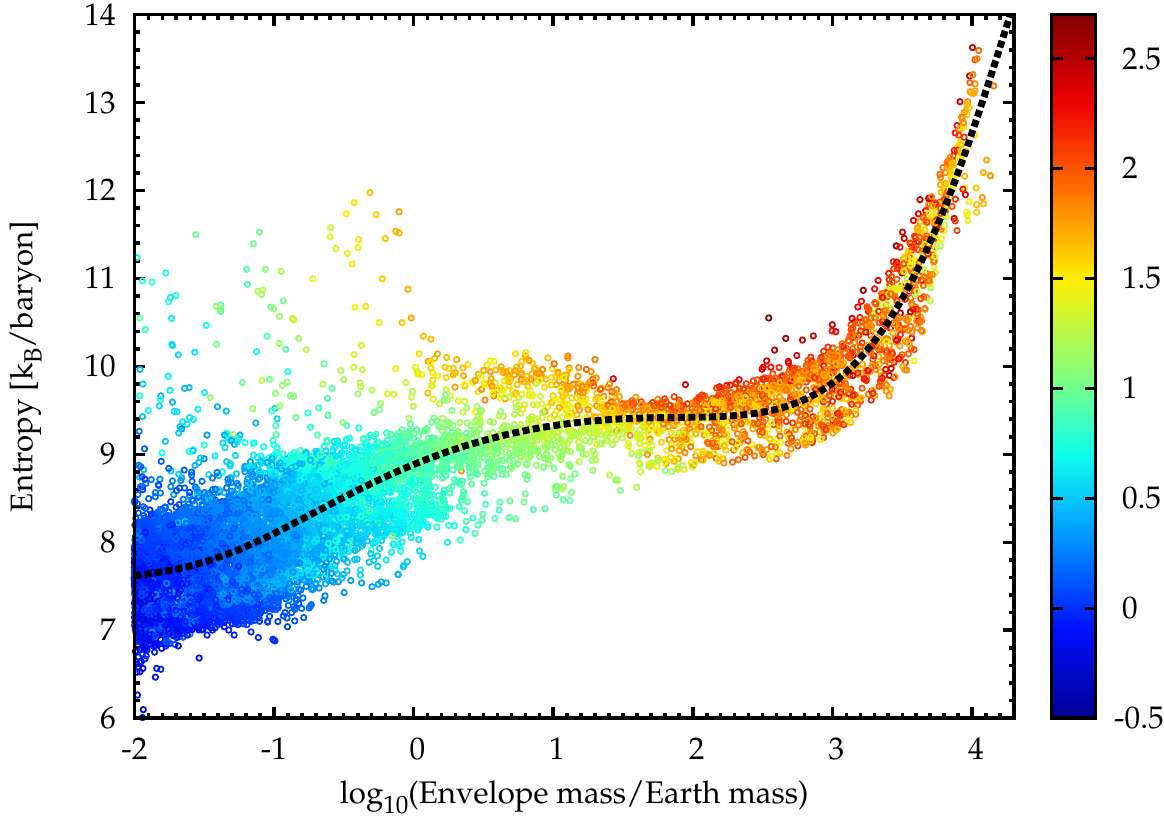}
     \end{minipage}
\caption{Specific entropy $\spf$ at the bottom of the gaseous envelope as a function of envelope mass at the end of the disk lifetime (i.e., formation phase) for the cold-nominal (left) and hot population (right panel).  The colors represent the core mass of the planets in units of $\log_{10}(\mcore/\mearth)$. The black lines show the least-squares fit of Eq. \ref{eq:S0fit}. }\label{fig:mstdisk}
\end{center}
\end{figure*}

As an initial condition for evolutionary models and cooling tracks we provide a least-squares fit to the relation of the planetary envelope mass $\menv$ to the post-formation specific entropy in the inner convective zone at the moment when the disk disappears, $\spf$.  For giant planets, the envelope mass is to first order equal to the total mass, or can be estimated by using the observationally inferred relation of core and total mass of \citet{ThorngrenFortney2016}. Theoretically predicted $M_{\rm core}-\menv$ relations for low-mass planets can for example be found in \citet{mordasiniklahr2014} or \citet{leechiang2015}. Considering $\spf$ as a function of $\menv$ is found to reduce the scatter compared to considering it directly as a function of the total mass.

The $\menv-\spf$ relation is shown in Fig. \ref{fig:mstdisk}.  For the fit, the entropy at the end of the formation phase $\spf$ in units of $k_{\rm B}$/baryon can written as a polynomial of degree 9 in terms of $\chi=\log_{10}({\menv/\mearth})$ as
\beq\label{eq:S0fit}
\spf(\chi)=\sum_{i=0}^{9} a_{i} \chi^{i}.
\eeq
The coefficients $a_{i}$ for the  cold-nominal and hot populations are given in Table \ref{tab:S0fit}. It is important to note that these fits can only be used in the domain $-2\leq\chi\leq4.25$ and diverge outside. The rms of residuals around the fit is in both cases 0.46 $k_{\rm B}$/baryon, reflecting a total spread of typically about 1 $k_{\rm B}$/baryon around the mean value for a given mass due to the different formation histories. Some low-mass planets ($\menv\leq 1 \mearth$) have clearly larger entropies than predicted by the fit. These are planets at large distances which were in the process of significant planetesimal accretion at the moment when the disk disappears. They have not yet accreted all planetesimals in the feeding zone as it is otherwise usually the case for planets at smaller separations. The associated high $L_{\rm pla}$ leads to the high $\spf$.  

The two fits overlap as expected for low-mass planets with $\menv\lesssim 30 \mearth$. For such low-mass planets no distinction of cold or hot starts exists because they do not accrete gas through a potentially entropy reducing shock, as discussed in Sect. \ref{sect:coldvshot}. In the giant planet mass regime, the fit for the cold population predicts a slight decrease of $\spf$, before it increases again due to deuterium burning. In the hot population, no such decrease occurs (see Fig. \ref{fig:mstdisk}).  It should be noted that the $\spf$ for low envelope masses is likely affected by the lack of including  in our model the core's thermal contribution to energy budget.

\begin{table}
\caption{Coefficients for the polynomial fit to the post-formation entropy in Eq. \ref{eq:S0fit}. }\label{tab:S0fit}
\begin{center}
\begin{tabular}{l c c}
\hline \hline
Coefficient & Cold & Hot  \\ \hline
$a_{0}$ & 8.9077 & 8.88378\\
$a_{1}$ & 0.67848 & 0.65215\\
$a_{2}$ & -0.27768 & -0.20626\\
$a_{3}$ & -0.04525 & -0.02744 \\
$a_{4}$& 0.07491 & 0.03693\\
$a_{5}$& -0.01991 & -0.01401 \\
$a_{6}$ & -0.00712 & -0.00084 \\
$a_{7}$ & 0.003967 & 0.002074 \\
$a_{8}$ & -0.000404 & -0.000297 \\ \hline
\end{tabular}
\end{center}
\end{table}%

\section{Empirical fits to $L$, $s$, and [D/H] at 10 Myrs}\label{app:fit10MyrLSD}

For some low-mass planets with a envelope mass of less than $\sim$1 $\mearth$, the post-formation entropy is as mentioned still heavily increased as the planet was undergoing an intense accretion of planetesimals near the end of the disk lifetime. These planets shows up in Fig. \ref{fig:mstdisk} with a $\spf\gtrsim 10$ $k_{\rm B}$/baryon. We therefore also give empirical fits to the planetary $\menv-L$, $\menv-s$, and $\menv-[D/H]$ relations at 10 Myr when accretion has stopped for almost all planets (only a small fraction of synthetic disks has a lifetime exceeding 10 Myr, namely 390 out of 50968), such that no exceptionally high entropies exist any more. A time of 10 Myr has often been used as a starting time for evaporation in evolutionary calculations \citep[e.g.,][]{lopezfortney2013,ChenRogers2016b}.

\begin{figure*}
\begin{center}      
	      \begin{minipage}{0.32\textwidth}
	      \centering
        \includegraphics[width=0.98\textwidth]{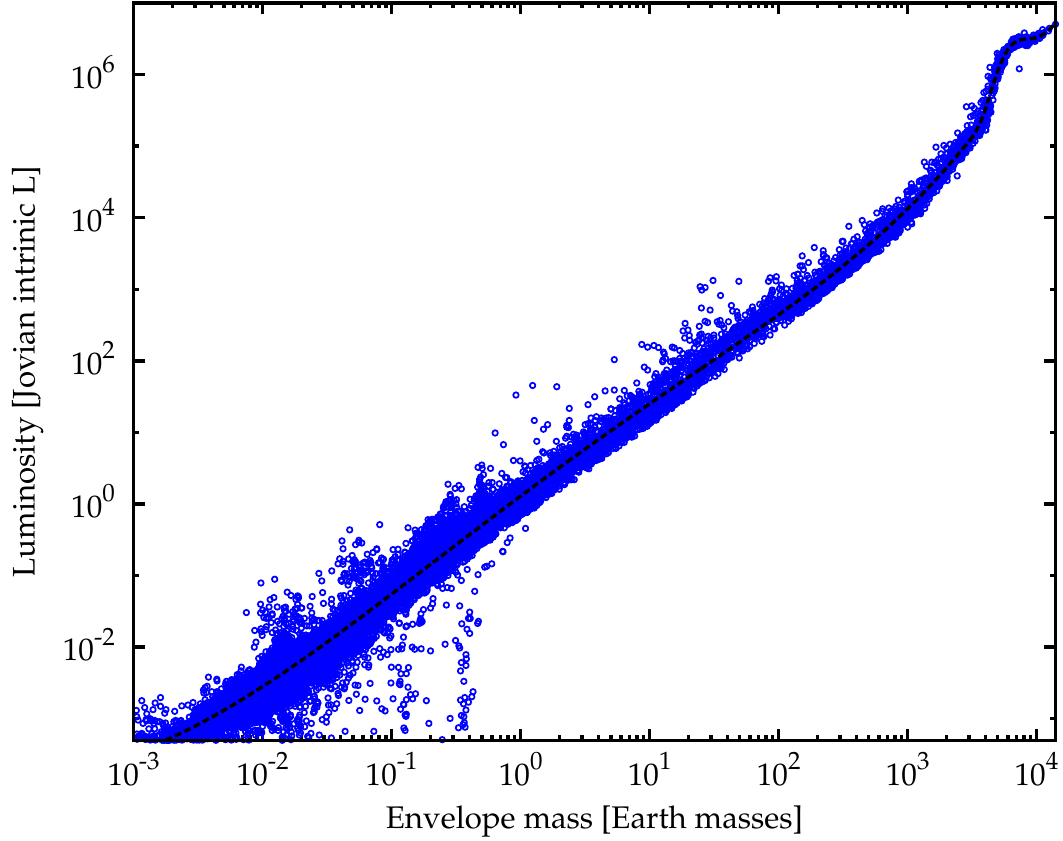}
     \end{minipage}\hfill
     \begin{minipage}{0.32\textwidth}
      \centering
        \includegraphics[width=0.98\textwidth]{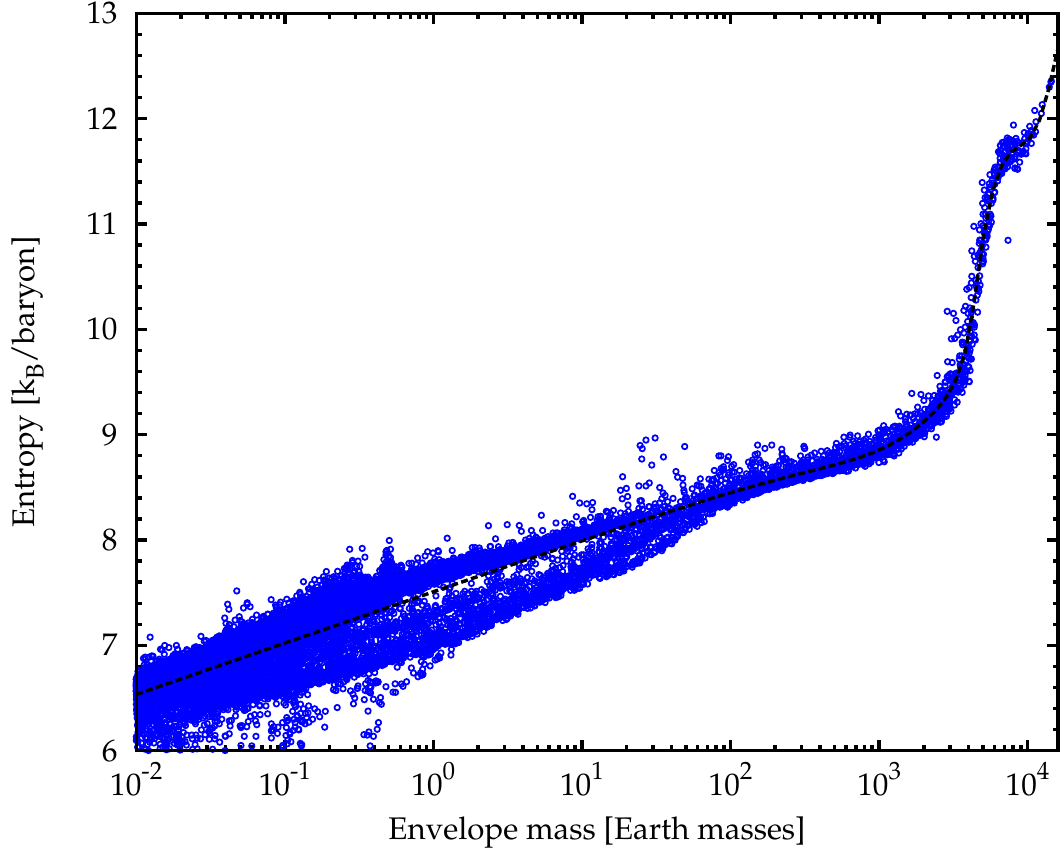}
     \end{minipage}\hfill
  \begin{minipage}{0.33\textwidth}
      \centering
        \includegraphics[width=0.99\textwidth]{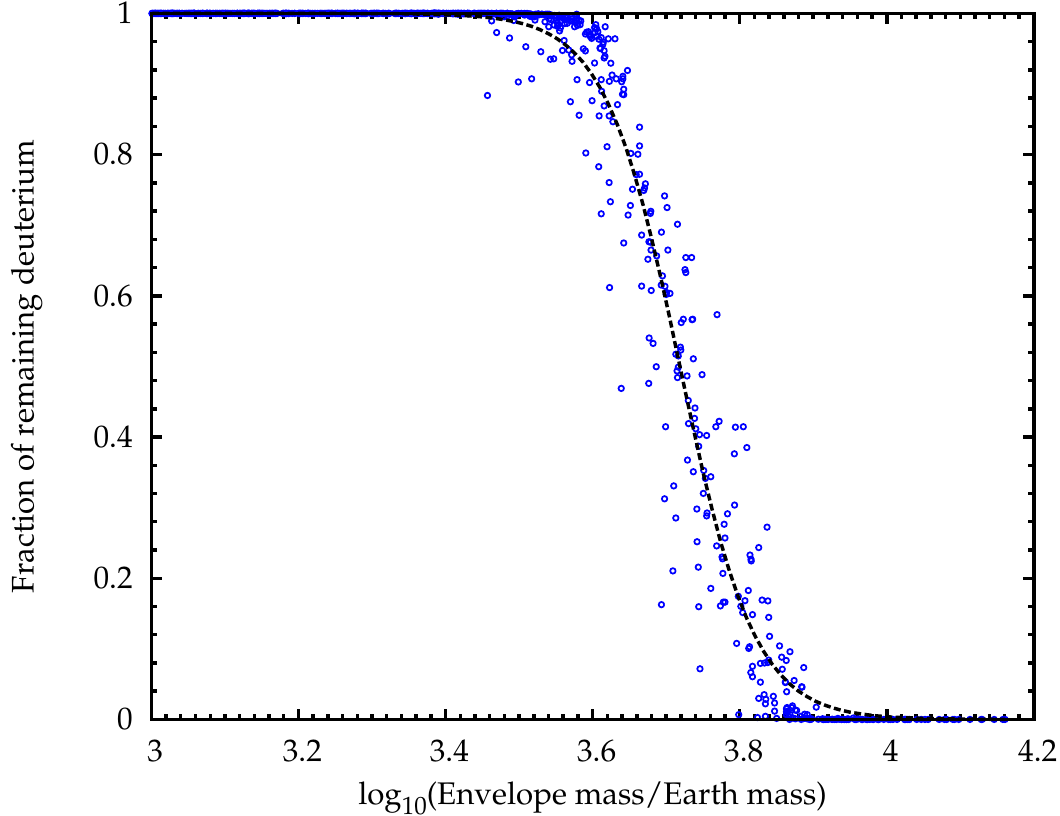}
     \end{minipage}     
\caption{Total luminosity (left), specific entropy at the bottom of the gaseous envelope (middle panel), and fraction of remaining deuterium (right panel) as a function of envelope mass for the cold-nominal population at 10 Myr. Only planets that do not accrete any more are included. The black lines show the least-squares fits.}\label{fig:fitMLMSMD}
\end{center}
\end{figure*}

Figure \ref{fig:fitMLMSMD} shows the total luminosity $L_{10}$ (which in the evolutionary phase is equal to the internal luminosity), the specific entropy $s_{10}$ at the bottom of the gaseous envelope, and for massive planets, the fraction of remaining deuterium $f_{\rm RD10}$ for the cold-nominal population at 10 Myr. The black lines are non-linear least-squares fits. The three quantities are plotted as a function of envelope mass rather than total mass, because this is found to reduce the scatter around the fit compared to using the total mass. 

The logarithm of the luminosity can be fitted as a function of $\chi=\log_{10}({\menv/\mearth})$ with a polynomial of degree 5,  
\beq\label{eq:L10fit}
\log_{10}(L/\lj)=\sum_{i=0}^{5} l_{i}\, \chi^{i}.
\eeq
Two sets of coefficients are needed depending on the envelope mass. For lower envelope masses in the range  $-3.0\leq\chi\leq3.55$, the coefficients in the second column of Table \ref{tab:L10fit} are used, while for $3.55\leq\chi\leq4$.1 those in the third column apply.

\begin{table}
\caption{Coefficients for the polynomial fit for the luminosity at 10 Myr in Eq. \ref{eq:L10fit}. }\label{tab:L10fit}
\begin{center}
\begin{tabular}{l c c}
\hline \hline
Coefficient & -3.0$\leq\chi\leq$3.55 &   3.55$\leq\chi\leq$4.1\\ \hline
$l_{0}$ & 0.12671 & -641.717\\
$l_{1}$ & 1.32098 & 478.447\\
$l_{2}$ & -0.06015 & -117.778\\
$l_{3}$ & -0.00258 & 9.67127 \\
$l_{4}$& 0.01155 & 0.0\\
$l_{5}$& -0.00106 & 0.0 \\ \hline
\end{tabular}
\end{center}
\end{table}%

The entropy at 10 Myr $s_{\rm 10}$ as a function of envelope mass has a similar shape as the luminosity, but the upturn at higher masses is more prominent. Below this upturn at about $\menv\approx1000\mearth$, the entropy roughly lies on a straight line, indicating a logarithmic dependency. For these planets, a simple fit is given as $s_{10}\approx7.5+0.21 \ln( \menv/{\mearth})$ k$_{\rm B}$/baryon.  %

The full fit to $s_{10}$ in units of  k$_{\rm B}$/baryon as function of  $\chi=\log_{10}(\menv/\mearth$) for -3$\leq\chi\leq$4.1 can be written as 
\beq\label{eq:S10fit}
s_{10}=\sum_{i=0}^{4} s_{i}\, \chi^{i}.
\eeq
where for $\chi \leq 3.55$ the coefficients in the second columns of Tab. \ref{tab:S10fit} must be used, and for higher masses the ones in the third column.

\begin{table}
\caption{Coefficients for the fit for the specific entropy at 10 Myr in Eq. \ref{eq:S10fit}.}\label{tab:S10fit}
\begin{center}
\begin{tabular}{l c c}
\hline \hline
Coeff. & -3$\leq\chi\leq$3.55 &   3.55$\leq\xi\leq$4.1\\ \hline
$s_0$ & 7.542 & 20115.1 \\
$s_1$ & 0.45712 & -21339.9 \\
$s_2$ &  -0.03777 &   8470.78 \\
$s_3$ &  0.00235 &  -1490.64 \\
$s_4$& 0.00378 & 98.1429\\
\end{tabular}
\end{center}
\end{table}%

In the panel showing the entropy, one notes that for a  given envelope mass between about 0.5 and 20 $\mearth$, there is an upper and a lower group of points, and a depletion of points between them. It is found that this is related to the semimajor axis of the planet, with the upper group of points being formed by close-in planets ($a\leq0.1-1$ AU) while the lower entropy group belongs to planets at larger semimajor axes. The fact that the fit runs through the depleted region reflects that such a dependency cannot be captured by the simple one parameter fit presented here.

Finally, we also give a fit of the remaining fraction of deuterium $f_{\rm RD10}$ as a function of $\chi=\log_{10}({\menv/\mearth})$, since for deuterium burning planets, also this quantity must be specified for a complete description of the initial conditions for the evolutionary phase. The initial deuterium number fraction [D/H] is $2\times10^{-5}$ \citep{mollieremordasini2012}. As one would expect, the right panel of Figure \ref{fig:fitMLMSMD} shows that below a certain $\chi$, no deuterium has been burned, while above a certain limit, no deuterium remains. This suggest a functional form of the fit of  
\beq
f_{\rm RD10}=\frac{1}{2}\left(1-\tanh\left[\frac{\chi}{a}-b\right]\right).
\eeq
The fit parameters $a$ and $b$ are again determined by the least-squares method which yields $a=0.101561$ and $b=36.616373$. 

\section{Considerations on the efficiency of the core-mass effect}\label{app:coremasseffect}
In this section we present an exploratory study of the efficiency of the core-mass effect in the case that planetesimals do not sink to the core. It is clear the future work will have to investigate this in a self-consistent fashion \citep[see][for works that do this for the composition of protoplanets without addressing the luminosity and the solid accretion during gas runaway, though.]{venturinialibert2016,lozovskyhelled2017}.

First, we assess whether much less potential energy is liberated as $L_{\rm pla,mix}$ by planetesimals that  mix throughout the envelope instead of sinking, i.e., whether we are greatly overestimating the heating by impacting planetesimals $L_{\rm pla}$ that in the sinking approximation we use in the syntheses is given approximately by 
\beq\label{eq:Lplasink}
L_{\rm pla,sink}\approx\frac{G M_{\rm core} \dot{M}_{\rm Z}}{R_{\rm core}}.
\eeq
In the equation $G$ is the gravitational constant, $M_{\rm core}$, $R_{\rm core}$ the core mass and radius, and $\dot{M}_{\rm Z}$ the accretion rate of solids.
This expression is strictly speaking only valid if the envelope mass is negligible. Otherwise, the numerically obtained value of the difference of the gravitational potential between the surface of the core and infinity must be used to obtain the luminosity. This is done in the numerical calculations, but for the phase we are mostly interested in (total mass $M\lesssim100\mearth$) where core and envelope mass are comparable, Eq. \ref{eq:Lplasink} only differs by less than 30\% from the numerically found value for the case we studied. Therefore we use it here because of its simple analytical form. 

If in reality the planetesimals do not reach the core but get dissolved into the envelope less potential energy gets liberated. No direct core hits are expected for, e.g., $\lesssim100$-km-sized rocky planetesimals plunging radially into planetary envelopes of $\gtrsim 3~\mearth$ \citep{mordasinialibert2006,mordasinimolliere2015}.

We finally also consider  that solids do not sink at all, but stay at the place where the planetesimals were destroyed in the envelope. This should give a lower limit on the impact heating, $L_{\rm pla, no sink}$.

\subsection{The case of a n=1 polytrope}
Before we study more realistic (numerical) models, we calculate the difference in energy deposition in a n=1 polytrope when matter is added at the outer surface and at the center. While an n=1 polytrope can clearly not capture all aspects of Jupiter like the presence of a core, it nevertheless represent a useful analytical representation of Jupiter's basic structure nowadays \citep{hubbard1974,depaterlissauer2010}. The density as a function of radius $r$ in a n=1 polytrope is given as
\beq
\rho(r)=\rho_{\rm c}\frac{\sin(\pi r/R)}{\pi r / R}
\eeq
where $\rho_{\rm c}$ is the central density and $R$  the total radius. From this, the gravitational potential $\Phi$ is found by integrating the Poisson equation in spherical symmetry 
\beq
\frac{1}{r}\frac{d^{2}}{dr^{2}}(r \Phi)= 4 \pi G \rho.
\eeq
Setting the zero point of $\Phi$ at infinity, and requesting that  $\Phi$ and its spatial derivative are continuous at $R$, we find a potential 
\beq
  \Phi(r) = \left.
  \begin{cases}
    -\frac{G M}{R} \left(1+\frac{R}{\pi r}\sin(\pi r/R)\right) & \text{for } r \leq R  \\
    -\frac{G M}{R}  & \text{otherwise }\\
  \end{cases}
  \right.
\eeq
So if mass is added at the surface of the planet at a rate $\dot{M}$, an accretion luminosity of $G M  \dot{M}/R$ is generated, as expected. If the mass is instead added at the center, we have to consider $\lim_{r \to 0} \Phi(r) $. As $\lim_{x \to 0} \sin(x)/x $=1, one finds $\lim_{r \to 0} \Phi(r)=-2 GM/R $. This means that if material is sinking to the center, twice as much luminosity is generated. This result that the two expressions only differ by a factor of a few is repeated also in the numerical results below. It reflects that for the acceleration of the sinking material, the enclosed (and thus accelerating) mass diminishes, but also $1/r^{2}$ becomes larger.

\subsection{Uniform mixing: $L_{\rm pla, mix}$}
For the full mixing case, we calculate the change in potential energy of the planet due to increasing its metallicity uniformly in every layer, yielding $L_{\rm pla, mix}$. We address the assumption of uniformness afterwards by considering several timescales describing the system.

Let $\rxy(r)$ and $\rz(r)$ be the time-dependent local densities of hydrogen and helium
and of metals, respectively, with total density $\rho=\rxy+\rz$,
and let the total mass of hydrogen and helium and of metals in the envelope only 
be $\mxy$ and $\mzenv$, respectively. 
The local and global metallicities are defined as
\begin{align}
 \Z(r) & \equiv \frac{\rz(r)}{\rxy(r)+\rz(r)},\\
 Z &\equiv \frac{\mzenv}{\mxy+\mzenv}.
\end{align}
For concurrent accretion of gas and solids, the changes in the metallicity are therefore given by
\begin{align}
\frac{\partial\Z(r)}{\partial t} &= \frac{\rdotz(r)}{\rho(r)} \left[1-\Z(r)\frac{\rdotxy(r)+\rdotz(r)}{\rdotz(r)} \right],\label{Gl:dZlokdt allgemein}\\
 \frac{dZ}{dt} &= \frac{\mdotz}{\menv}\left[1-Z\frac{\mdotxy+\mdotz}{\mdotz}\right],\label{Gl:dZdt allgemein}
\end{align}
where the total envelope mass is $\menv=\mxy+\mzenv$.
It should be noted that since Eq.~(\ref{Gl:dZlokdt allgemein}) is written at fixed radius (and not in the Lagrange frame at fixed mass coordinate),
homologous contraction without global accretion ($\mdotxy=\mdotz=0$) can be sufficient
to have $\frac{\partial\Z(r)}{\partial t}\neq0$.
When $\mdotxy=0$,
Equation~(\ref{Gl:dZdt allgemein}) reduces to
\begin{equation}
 \label{Gl:dZdt b}
 \frac{dZ}{dt} = \frac{\mdotz}{\menv}\left[1-Z\right].
\end{equation}
Similarly, ignoring contraction and if the hydrogen--helium component of the mixture is not redistributed,
the local change in metallicity reduces to
\begin{equation}
 \frac{\partial\Z(r)}{\partial t} = \frac{\rdotz}{\rho}\left[1-\Z(r)\right]. \label{Gl:dZlokdt b}
\end{equation}

In the case that the metals are mixing uniformly throughout the planet, it must hold that
$\Z(r) = Z$ at all radii.

Thus in particular
\begin{equation}
 \label{Gl:dZlokdt dZdt}
 \frac{\partial\Z(r)}{\partial t} = \frac{dZ}{dt},
\end{equation}
implying, when $\mdotxy=0$, with Eqs.~(\ref{Gl:dZdt b}) and~(\ref{Gl:dZlokdt b}) that 
\begin{equation}
 \label{Gl:drhoZdt}
 \rdotz(r) = \rho(r) \frac{\mdotz}{\menv}.
\end{equation}
This result will be needed shortly.

We now compute the luminosity generated by adding $\dot{M}_{Z}$ homogeneously into the envelope. To simplify, let us assume that the potential $\Phi(r)$ inside the planet is fixed and that the core does not contract.
For total mass $M=\menv+\mcore$ and radius $R=\rcore+R_{\rm env}$, the luminosity is given by
\begin{align}
 L_{\rm pla,mix} &= -\frac{d\Epot}{dt} = -\frac{d}{dt}\left( \int_0^M \Phi \,dm\right)\\
           &= -\frac{d}{dt}\left( \int_0^{\rcore} \Phi \, \frac{dm}{dr}\,dr + \int_{\rcore}^R \Phi \, \frac{dm}{dr}\,dr\right)\\
           &= 0 -\int_{\rcore}^R \Phi\, \frac{\partial}{\partial t}\left(4\pi r^2\rho\right) \,dr\\
           &= -\int_{\rcore}^R \Phi\, 4\pi r^2\frac{\partial}{\partial t}\left(\rxy+\rz\right)  \,dr\\
           &= -\int_{\rcore}^R \Phi\, 4\pi r^2\left(0+\rho(r) \frac{\mdotz}{\menv}\right) \,dr \;\;\mbox{(by Eq.~\ref{Gl:drhoZdt})}\\
           &= -\frac{\mdotz}{\menv}\int_{\rcore}^R \Phi \, \frac{dm}{dr} \,dr
\end{align}
These quantities are obtained numerically in the formation model at each timestep, and will be used below to compare $L_{\rm pla,sink}$, $L_{\rm pla,mix}$, and $L_{\rm pla,nosink}$.

\subsection{No sinking: $L_{\rm pla, no sink}$}
Besides the case of homogeneous mixing, we can also consider the limiting case that the solids do not sink at all from the point where they are deposited into the envelope by the impact. From the calculation of the trajectories of impacting planetesimals \citep{mordasinialibert2006} we know the (enclosed) mass and radius in the envelope of maximum energy deposition, $M_{\rm max E}$ and $R_{\rm max E}$. For the large impactors we consider, this is also the place of maximum mass deposition \citep[e.g.,][]{mordasini2014}. When the planet is still attached or early after detachment, $M_{\rm max E}$ is similar to the total mass, but $R_{\rm max E}$ is much smaller than the total radius, as the planet has a very voluminous thin outer atmosphere through which the planetesimal penetrate. In late stages when the planet has fully contracted, both the mass and radius of maximum energy deposition are similar to the total mass and radius.

\begin{figure}
\centering
\includegraphics[width=0.49\textwidth]{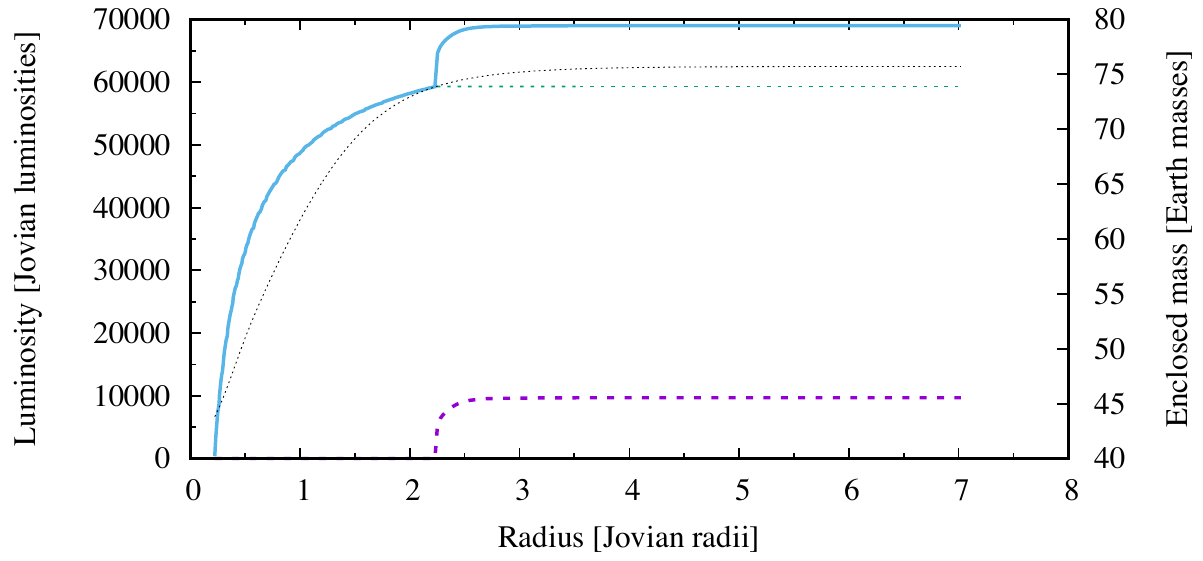}
    \caption{Luminosity caused by planetesimal accretion as a function of radius in the planet discussed in Sect. \ref{sect:formevo5MJ} at 1.95 Myr, i.e., shortly after the planet has detached from the disk. The luminosity directly from the impact (violet dashed), from the settling to the core (green short dashed), and the sum of the two is shown (blue solid). These quantities belong to the left y-axis. The thin dotted black line is the enclosed mass in the protoplanet's envelope (right y-axis).  }\label{fig:RLumistruct} 
\end{figure}

Figure \ref{fig:RLumistruct} shows the luminosity caused by radial planetesimal impacts at 1.95 Myr shortly after detachment for the planet discussed in Sect. \ref{sect:formevo5MJ}. The properties of the planet at this moment are shown in Table \ref{tab:prop1p95Myr}.  

\begin{table}
\caption{Planet properties at 1.95 Myr.}\label{tab:prop1p95Myr}
\begin{center}
\begin{tabular}{l c}
\hline \hline
Quantity & Value  \\ \hline
Total mass $M$ & 76.7 $\mearth$\\
Core mass  $M_{\rm core}$&  43.8 $\mearth$\\
Total radius $R$ & 6.98 $\rj$ \\
Core radius $R_{\rm core}$ & 0.22 $\rj$ \\
Radius of rad.-conv. boundary  & 5.35 $\rj$ \\
Mass in the radiative zone & 0.016 $\mearth$ \\
Radius of max. energy dep. $R_{\rm maxE}$ & 2.23 $\rj$ \\
Mass of max. energy dep. $M_{\rm maxE}$ & 74.7  $\mearth$ 
\end{tabular}
\end{center}
\end{table}%

The plot shows the two contribution to $L_{\rm pla}$ separately. The first results directly from the impact of the planetesimal itself. The second contribution is due to the subsequent settling of the debris to the core after the planetesimal was destroyed. In the sinking approximation, the sum of the two is assumed to heat the planet. The shape of the direct contribution shows that the energy and mass is deposited essentially in one narrow location at 2.2 $\rj$ because of the terminal explosion that is typical for large aerodynamically disrupted bodies as it was also the case for Shoemaker-Levy 9 \citep{maclowzahnle1994}. This location corresponds to the locus of maximum energy deposition $M_{\rm max E}$ and $R_{\rm max E}$ and also the location where the planetesimal's mass is deposited. The plot and Table \ref{tab:prop1p95Myr} show that in terms of radius, the planetesimals penetrate deep into the planet, in particular into the convective zone. But in terms of enclosed mass, they are destroyed relatively close to the surface ``below'' about 2 $\mearth$ of envelope. We also see that the settling contribution is about 6 times as large as the direct contribution (cf. Fig. \ref{fig:LumiComp3}).

With $M_{\rm max E}$ and $R_{\rm max E}$ we obtain the planetesimal luminosity in the no sinking case as
\beq
L_{\rm pla, no sink}\approx\frac{G M_{\rm max E} \dot{M}_{\rm Z}}{R_{\rm max E}} 
\eeq
In Fig. \ref{fig:RLumistruct} this corresponds to the violet curve, i.e., the direct contribution. The reduction of the luminosity in the no sinking case relative to the sinking case is 
\beq\label{eq:lplanosinglplasink}
\frac{L_{\rm pla, no sink}}{L_{\rm pla, sink}}\approx\frac{M_{\rm max E}}{M_{\rm c}}\frac{R_{\rm c}}{R_{\rm max E}}
\eeq
When the planet has not yet accreted a significant envelope and planetesimals penetrate to the core, this expression becomes unity, as can be seen in Fig. \ref{fig:LumiComp3} below.  At crossover, $M_{\rm max E}/{M_{\rm c}}\sim2$, and the syntheses show that at that point ${R_{\rm core}}/{R_{\rm max E}}$ lies between 0.08 and 0.24, with a typical value of about 0.15. This corresponds to a planetesimal heating that is reduced relative to the sinking case by a factor $\sim$3 with a spread between about 2 to 6. When $M_{\rm env}\gg M_{\rm core}$, $M_{\rm max E}\approx M$ and $R_{\rm max E}\approx R$. Using the result from the syntheses that for typical giant planet during formation with masses between 100 and 1000 $\mearth$, ${R_{\rm core}}/{R_{\rm max E}}\sim0.2$, we see that $L_{\rm pla, no sink}/L_{\rm pla, sink}$ could become larger than unity. This is however an artefact of approximating the potential at the surface of the core as $G M_{\rm core}/R_{\rm core}$ which breaks down exactly if $M_{\rm env}\gg M_{\rm core}$. In reality, the ratio is given by the ratio of the actual gravitational potentials, $\Phi_{\rm max E}/\Phi(R_{\rm core}) \approx \Phi(R)/\Phi(R_{\rm core})$. This expression is always smaller than unity, as otherwise, parts of the planet would be accelerated outward, which is of course not the case.

\subsection{Comparison of sinking, uniform mixing, and no sinking}
\begin{figure*}
\begin{center}
\begin{minipage}{0.5\textwidth}
	      \centering
        \includegraphics[width=0.9\textwidth]{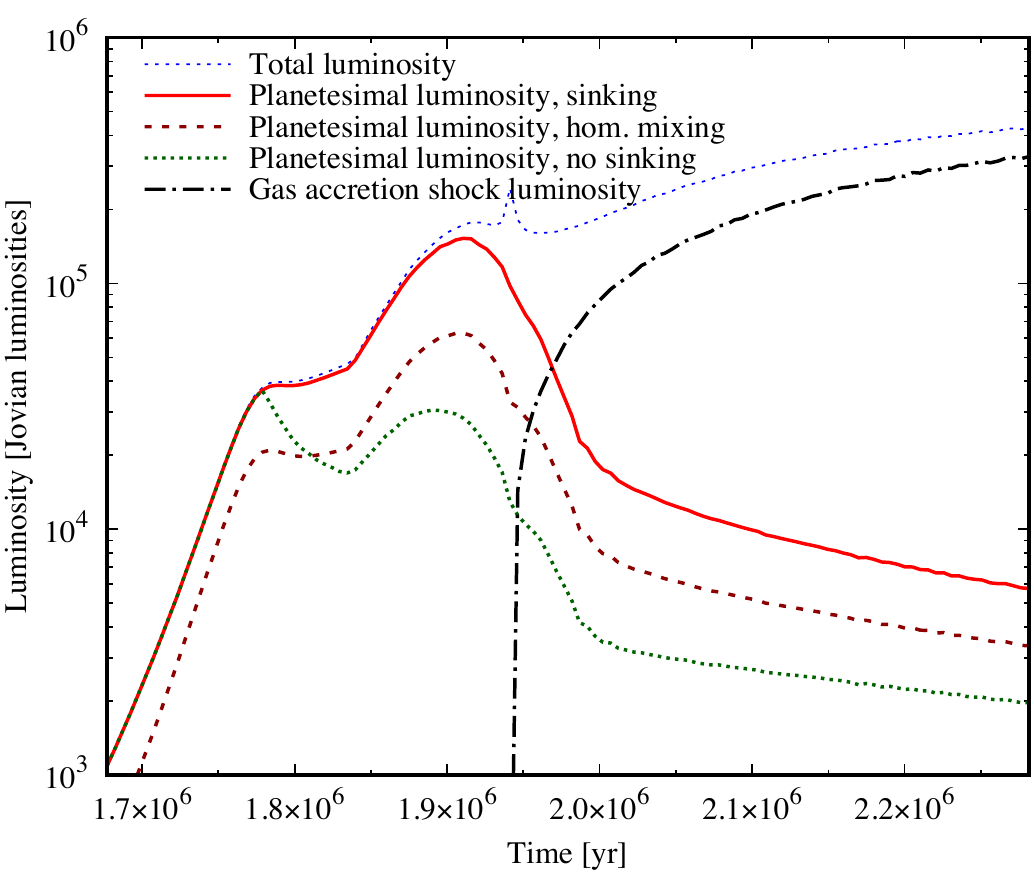}
     \end{minipage}\hfill
     \begin{minipage}{0.5\textwidth}
      \centering
       \includegraphics[width=0.9\textwidth]{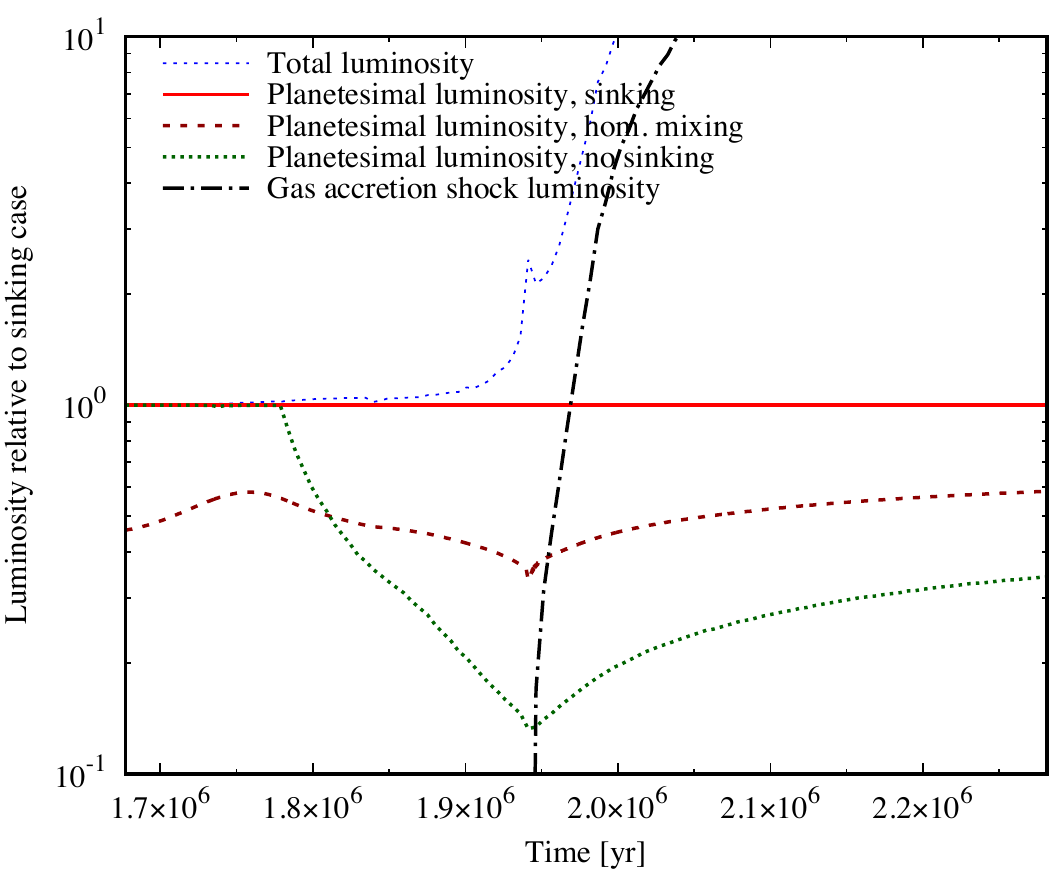}
     \end{minipage}
\caption{Luminosity as a function of time during the early formation phase of the 5 $\mj$ planet discussed in Sect. \ref{sect:formevo5MJ}.  The left panel shows the total luminosity $L$, the gas accretion shock luminosity $L_{\rm shock}$, and the planetesimal accretion luminosity in the sinking approximation $L_{\rm pla,sink}$. These luminosities are the ones used in the simulations. The plot also shows the planetesimal accretion luminosity for the cases that the solids would homogeneously mix into the envelope, and if they would stay where they were deposited into the envelope by the impacting planetesimals (no sinking). The right panel shows the same quantities, but normalized to $L_{\rm pla,sink}$. Relative to sinking,  the heating is reduced by a factor 2 to 3 for  mixing, and by a factor 3 to 7 for no sinking.}\label{fig:LumiComp3}
\end{center}
\end{figure*}

Figure \ref{fig:LumiComp3} show $L$, $L_{\rm shock}$, $L_{\rm pla,sink}$, $L_{\rm pla,mix}$ and $L_{\rm pla,nosink}$ during the early formation phase of the planet discussed in Sect. \ref{sect:formevo5MJ}. This includes the phase when the planet goes into gas runaway and then disk-limited gas accretion. The planet detaches from the disk at about 1.94 Myr. This is the time when the core-mass effect could be acting. During the time interval shown, the planet's core grows from about 2 to 48 $\mearth$, almost the final core mass, while the total mass increases from 2 to 335 $\mearth$.

In the left panel we see that after the initial build-up of the core, $L_{\rm pla,sink}$, $L_{\rm pla,mix}$ and $L_{\rm pla,nosink}$ follow a similar temporal pattern. We also see that before significant gas accretion starts at around 1.9 Myr, $L_{\rm pla,sink}$ provides the dominant luminosity source of the planet. The assumption of sinking has therefore important consequences for the efficiency of gas accretion and the moment when runaway gas accretion occurs, as already demonstrated by \citet{pollackhubickyj1996}. After detachment, the gas accretion shock luminosity $L_{\rm shock}$ grows quickly.
 
The plot shows that for homogeneous mixing and no sinking, the heating of the planet by  impacting planetesimals is reduced relative to sinking, as expected. The reduction is stronger in the no sinking case relative to mixing, which is  expected as well, as in the mixing case, some material still sinks deep into the potential well. In the right panel, we have divided the different luminosities  by $L_{\rm pla,sink}$ so that we can see how strongly the energy input is reduced. One sees that relative to sinking,  the heating is reduced by a factor 2 to 3 for homogeneous mixing, and by a factor 3 to 7 for no sinking.  We thus find that also a homogeneous mixing of the solids into the envelope instead of full sinking provides a significant energy source, at least for this case. 

\begin{figure}
\centering
\includegraphics[width=0.5\textwidth]{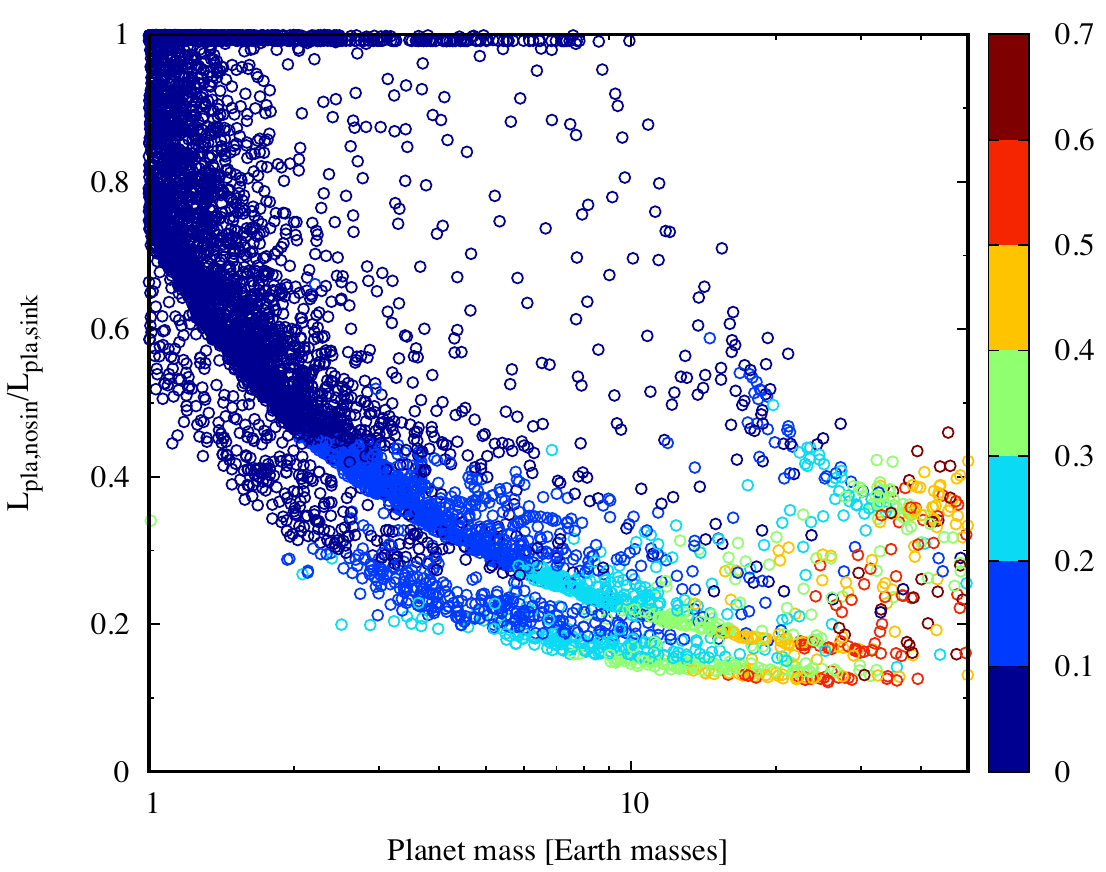}
    \caption{Reduction of the planetesimal impact heating in case of no sinking relative to the sinking case (Eq. \ref{eq:lplanosinglplasink}). The dots show $L_{\rm pla, no sink}/L_{\rm pla, sink}$ as a function of mass for synthetic planets with $M_{\rm env}\leq 2 M_{\rm core}$ in the CD777 population at 2 Myr. The color code shows the envelope mass fraction $M_{\rm env}/M$. Near crossover, the heating efficiency is reduced by typically a factor $\approx$3 with a spread of 2.5-8. }\label{fig:LnosinkLsinkCD7772e6} 
\end{figure}

To give an impression how general these results are, we show in Figure \ref{fig:LnosinkLsinkCD7772e6} the ratio ${L_{\rm pla, no sink}}/{L_{\rm pla, sink}}$ (Eq. \ref{eq:lplanosinglplasink}) as a function of planet mass for the CD777 synthetic population at 2 Myr. Only planets where $M_{\rm env}\leq 2 M_{\rm core}$ are included such that the approximation of the core potential as $G M_{\rm core}/R_{\rm core}$ still approximately holds. The color code gives the envelope mass fraction, $M_{\rm env}/M$. We see that in the no sinking case, the heating by planetesimals is about 2.5 to 8 times less strong than for sinking when the planets are near runaway, i.e., when $M_{\rm env}/M\gtrsim 0.5$. This is comparable to the result of Fig. \ref{fig:LumiComp3}. 

An interesting point to address in future work is to check whether such a non-central energy input could shut down convection. For this it is however necessary to consider that impacts are not a spherically symmetric phenomenon, as indicated by the timescale estimates below. 

\def\RmaxE{R_{\rm max E}}

\begin{figure*}
\centering
\includegraphics[width=0.45\textwidth]{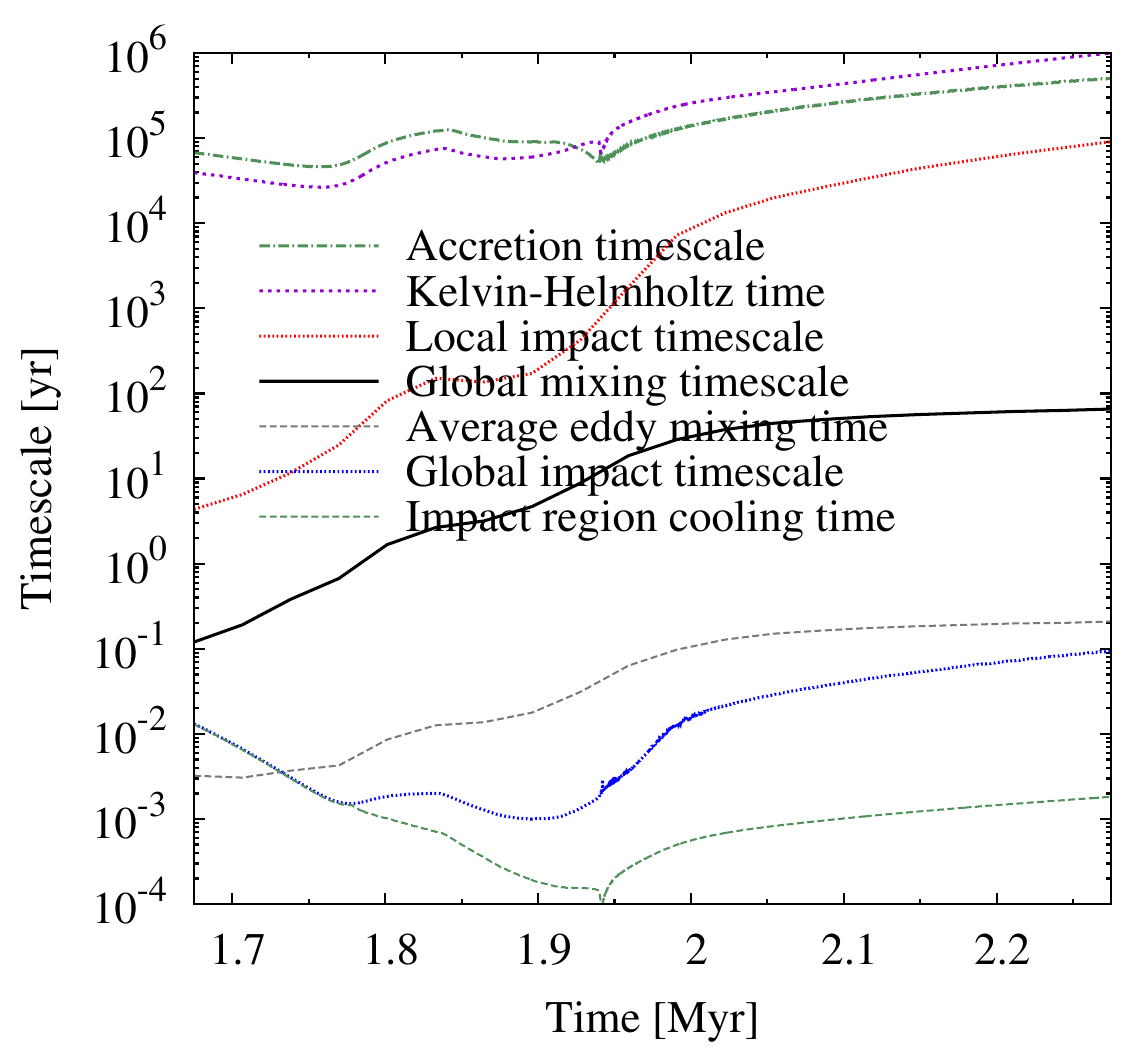}
\includegraphics[width=0.45\textwidth]{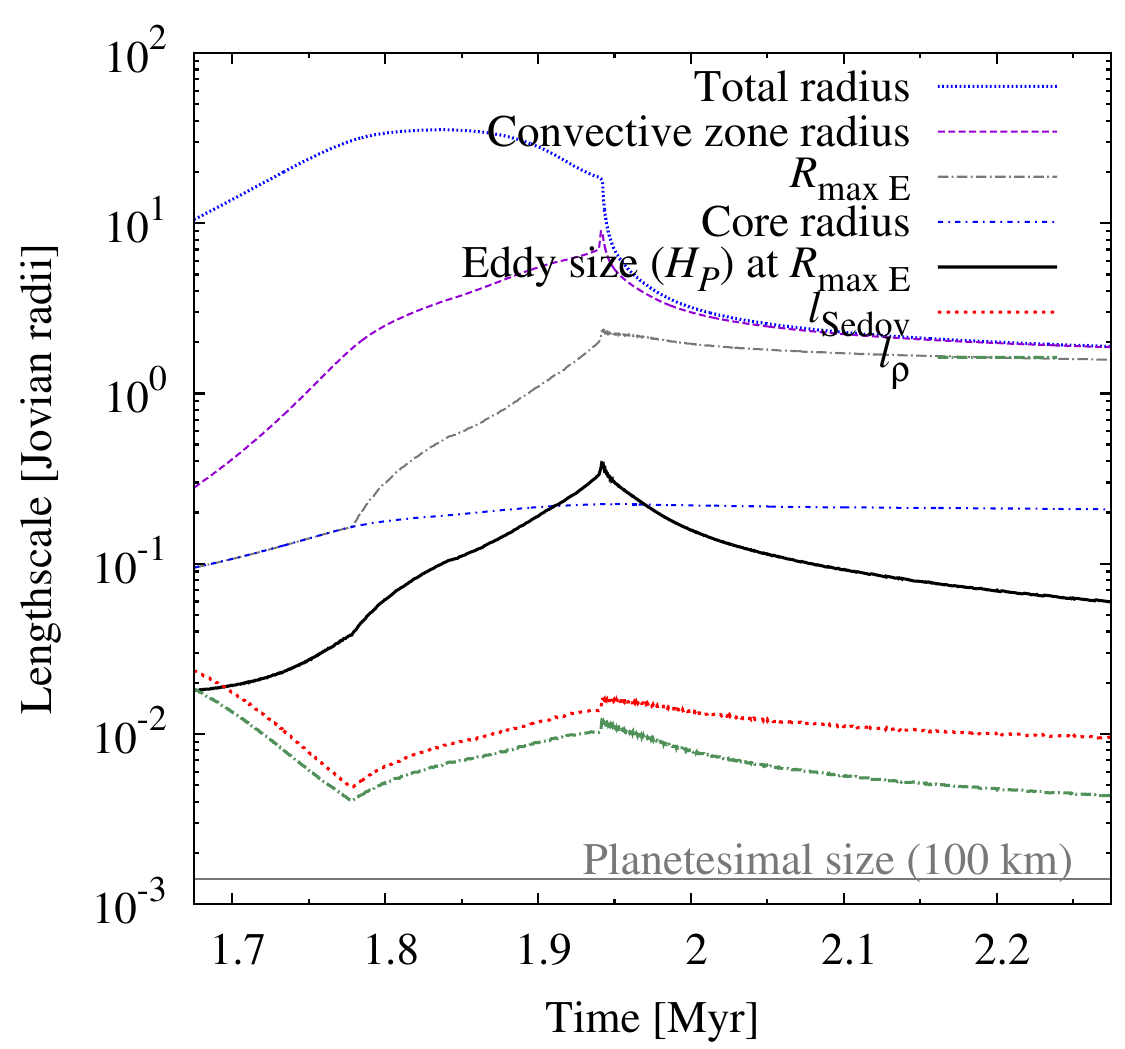}
 \caption{
 \textit{Left panel:} Timescales for the 5-$\mj$ example of Sect.~\ref{sect:formevo5MJ}
 studied in this Appendix.
 The timescales are, from top to bottom in the legend, the
   accretion timescale $\tau_{\rm acc} = M/(\mdotxy+\mdotz)$;
   Kelvin--Helmholtz time $\tauKH = E_{\rm tot}/L_{\rm int}$;
   local impact timescale from Eq.~\ref{Gl:tauEinschlaglokal},
     using the geometric mean of $\ell_\rho$ and $\ell_{\rm Sedov}$;
   global and eddy mixing times given by Eq.~(\ref{Gl:tauMisch});
   global impact timescale given by Eqs.~(\ref{Gl:tauEinschlag});
   and cooling time of the region affected by one impact, $\tau_{\rm cool}=E_{\rm impact}/L_{\rm int}$.
   \textit{Right panel:}
   Relevant lengthscales: From top to bottom,
     total radius $R$;
     (outer) radiative--convective boundary;
     radius of maximal energy deposition $\RmaxE$;
     core radius $\rcore$;
     size of a convective eddy at $\RmaxE$, which is equal to the pressure scale height since $\alpha_{\rm MLT}=1$;
     size of the Sedov blast wave estimate $\ell_{\rm Sedov}$;
     size of the density-contrast estimate $\ell_\rho$;
     and the size (diameter) of the planetesimals.
   }\label{Abb:Zeitskalen und Skalenlaengen}
\end{figure*}

\subsection{Mixing timescale of metals in the envelope}

We have showed in the previous section that the uniform mixing of the metals provided by
thermally and mechanically destroyed
planetesimals
represents a source of luminosity comparable to the case when the planetesimals hit the core.
The question is now whether this mixing can happen fast enough for it to be relevant for the cooling,
i.e., whether the mixing timescale for the whole envelope $\tauMisch$
is much smaller than the Kelvin--Helmholtz timescale $\tauKH$.
It should be noted that contrary to what one might intuitively imagine,
the mixing timescale throughout the envelope 
does not need to be shorter than the time between impacts at a given location
(as it nevertheless is, as Fig.~\ref{Abb:Zeitskalen und Skalenlaengen}a will show);
in the case it is not, a ``pile-up'' of metals in one layer still only needs to mix with the rest of the envelope
faster than the envelope cools in order to contribute significantly to the luminosity.

The time needed for the composition to become uniform throughout the planet can be calculated
from the the mixing time within one convective eddy, equal to its lifetime,
and the picture of diffusion across the eddies.
Thus, as implemented in the Bern planet formation code by \citet{mollieremordasini2012},
\begin{align}
\label{Gl:tauMisch}
 \tauMisch &= {N_{\rm eddy,~r}}^2 \tauMischlok 
           = {N_{\rm eddy,~r}}^2 \left\langle\frac{\alpha_{\rm MLT}\HP}{v_{\rm conv}}\right\rangle\\
           &= {N_{\rm eddy,~r}}^2 \frac{1}{\menv}\int_{\rm conv}\frac{\alpha_{\rm MLT}\HP}{v_{\rm conv}}\, dm,
\end{align}
which implicitly defines the average mixing time in one eddy $\tauMischlok$;
the integral is computed over the convective region and weighs by mass,
$v_{\rm conv}$ is the convective velocity in every layer,
the mixing-length parameter is $\alpha_{\rm MLT}=1$,
and the total number of eddies in the radial direction is given by
\begin{equation}
 N_{\rm eddy,~r}=\int_{\rm conv}\frac{dr}{\alpha_{\rm MLT}\HP}.
\end{equation}
Equation~(\ref{Gl:tauMisch}) represents an approximate upper bound since in general
$\RmaxE$ could be closer to the middle of the planet, reducing the number of eddies through which
the metals must diffuse.

Figure~\ref{Abb:Zeitskalen und Skalenlaengen}a shows the relevant timescales as a function of time
for the 5-$\mj$ example studied in the previous sections.
The global mixing timescale varies from $\tauMisch\sim1$~yr at 1.8~Myr,
when the planetesimals do not reach the core anymore,
to $\tauMisch\sim100$~yr at 2.3~yr, when the planet has grown to 1~$\mj$.
Over this period, there are always between 15 and 20 convective eddies.
The reciprocal of the mass-weigthed inverse convective velocity is 0.003--0.01~km\,s$^{-1}$
decreasing with time\footnote{This perhaps surprisingly low value comes
  from the low convective velocities near the core which dominate the inverse average,
  compounded by the fact that most of the mass sits near the core.
  The mass-weigthed average velocity, however, is closer to 2~km\,s$^{-1}$.},
the mass-weighted pressure scale height is of order $0.2~\rj$,
and the mass-weigthed average mixing time for a single eddy is $\tauMischlok\sim{\rm few}$~days at 1.8~Myr,
going up to $\tauMischlok\approx2.5$~months at 2.3~Myr and 6~months at the end of formation.
Thus the global mixing time according to Eq.~(\ref{Gl:tauMisch}) is between $\tauMisch\sim1$~yr at 1.8~Myr
and 100~yr at 2.3~Myr.
However, the Kelvin--Helmholtz timescale\footnote{For very extended objects,
  the approximation $\tauKH\approx GM^2/(RL)$ can be off by orders of magnitude.}
$\tauKH=E_{\rm tot}/L_{\rm int}$, with $E_{\rm tot}$ and $L_{\rm int}$ the total energy and luminosity,
goes over this period from around $\tauKH\sim0.1$ to 1~Myr,
for a ratio $\tauKH/\tauMisch\sim10^5$--$10^4$.
Thus, it is a robust result that  
the mixing should happen much faster than the planet cools.

We note that we ignore the possibility that a sufficiently strong metallicity gradient could
inhibit convection on the largest scales \citep{lecontechabrier2012,lecontechabrier2013,vazanhelled2016},
preventing a homogenization of the composition. See also \citet{lozovskyhelled2017,berardocumming17}.

\subsection{Impact timescale of planetesimals}
We now estimate whether the impacts could shut off or at least impair convection by generating
a spherically symmetric hot layer of material at $\RmaxE$, which is close to the outer radius
but within the convective zone (see Fig.~\ref{Abb:Zeitskalen und Skalenlaengen}a shows).
Heating by impacts will hinder convection if there is not enough time between impacts
for the effect of a planetesimal to become washed out before the next one arrives.
This means that the convection would ``see'' a spherically-symmetric blanket of planetesimals
falling down.
We thus need to determine the global impact timescale, the size of the region affected by an impact,
and from this then the local impact timescale.

The global average time between impacts is $\tauEin = {\mpla}/{\mdotz}$
for a monodispersive planetesimal mass distribution of $\mpla$.
Inside of the iceline, where our example planet remains during its formation,
the planetesimals are rocky, with a material density $\rhopla=3.2$~g\,cm$^{-3}$.
With a diameter $\dpla=100$~km, the mass of each planetesimal is $\mpla=1.7\times10^{21}$~g,
implying
\begin{equation}
 \tauEin = 1.0~{\rm day}\,\left(\frac{\dpla}{100~{\rm km}}\right)^3\left(\frac{\mdotz}{10^{-4}~\mearth\,{\rm yr}^{-1}}\right)^{-1}\frac{\rhopla}{3.2~{\rm g\,cm}^{-3}}.\label{Gl:tauEinschlag}
\end{equation}
The results of Eq.~(\ref{Gl:tauEinschlag}) are shown in Fig.~\ref{Abb:Zeitskalen und Skalenlaengen}b.
We note in passing that these impacts are thus quite frequent and might be observable by dedicated monitoring campaigns.
The caveats are however that at this stage the protoplanet might be hidden
by the circumstellar (and possibility also the circumplanetary) disc(s)
but mostly that the phase in which these impacts are expected is very short compared to the typical age
of young stars.

One can estimate the size of the region affected by a single impact in at least two ways.
One measure $\ell_\rho$ is given by 
the volume at which the mean density of the dissolved metals
drops to the background density at the impact location $\rho_{\rm amb}$:
\begin{equation}
 \rhopla (\dpla/2)^3 = \rho_{\rm amb} {\ell_{\rho}}^3.
\end{equation}
Another estimate of the size of the explosion region is given
by the Sedov blast wave solution for a point explosion, 
\begin{equation}
 \ell_{\rm Sedov} = \left(\frac{3E_{\rm impact}}{4\pi e_{\rm int,~amb}}\right)^{1/3}
       = \left(\frac{3(\gamma_{\rm eff,~amb}-1)E_{\rm impact}}{4\pi P_{\rm amb}}\right)^{1/3},\label{Gl:Sedov ell}
\end{equation}
where $E_{\rm impact}$ is the energy deposited by the impact of one planetesimal
(which is very localized; see Fig.~A.1b of \citealp{mordasini2014}, even though for icy planetesimals)
and
$e_{\rm int,~amb}$, $\gamma_{\rm eff,~amb}$, and $P_{\rm amb}$
are respectively
the ambient internal energy density,
effective adiabatic exponent $\gamma_{\rm eff}=P/(\rho e_{\rm int})+1\approx1.1$--1.4 at the relevant conditions,
and pressure
at the blast location.
The explosion or disruption energy is
\begin{equation}
  E_{\rm impact} \approx\frac{1}{2}\mpla {v_{\rm esc}(\RmaxE)}^2 = \frac{GM_{\rm max E}\mpla}{\RmaxE},
\end{equation}
with $v_{\rm esc}(\RmaxE)$ the escape velocity at $\RmaxE$.
For completeness, the density at $\RmaxE$ varies from $\rho_{\rm amb}=10^{-3}$ at the onset of detachment
to $10^{-2}$~g\,cm$^{-3}$ at 2.3~Myr,
while $P$ and $T$ range typically from 200 to 10$^4$~bar and 5000 to $10^4$~K,
respectively.
The energy deposition is typically around $5\times10^{33}$--$10^{34}$~erg per planetesimal
and is at any time a factor $10^9$--$10^{10}$ smaller than the planet's binding energy.

As Fig.~\ref{Abb:Zeitskalen und Skalenlaengen}b shows,
the density-based and the energy-based lengthscales ($\ell_{\rm Sedov}$ and $\ell_{\rho}$) agree quite well
and yield a typical $\ell\approx300$--1000~km.

The eddy size at $\RmaxE$ is also shown and is roughly one order of magnitude larger
than $\ell$.
Also, the mass contained in the background gas in the impact volume $m_{\rm impact}=4\pi\ell^3/3\times\rho_{\rm amb}$
is a factor of two to ten larger than the planetesimal mass,
with $m_{\rm impact}/\mpla = \gamma_{\rm amb}(\gamma_{\rm eff,~amb}-1)\mathcal{M}^2$ where $\gamma$ is the heat capacity ratio
and $\mathcal{M}$ the Mach number of the impact\footnote{
   The ratio of $\mpla$ to the mass in a convective eddy is orders of magnitude smaller,
   but it is a priori not obvious what the implications for the Ledoux stability of the eddy are.}.
Therefore, by all counts the effects of the impact are very localized in the sense that $\ell\ll\HP\ll\RmaxE$.
This also justifies a posteriori our use of the point explosion expression in Eq.~(\ref{Gl:Sedov ell})
instead of the expression for a line charge.

We can now estimate the time between impacts into the same region of size $\sim\ell^3$ at a height $\RmaxE$.
Since there are $N_\ell=4\pi\RmaxE^2/(\pi \ell^2)\sim10^5$--$10^6$ (from 1.8 to 2.3~Myr)
such regions in a shell at height $\RmaxE$,
the timescale is approximately
\begin{equation}
\label{Gl:tauEinschlaglokal}
 \tauEinlok = \tauEin \left(\frac{4\RmaxE}{\ell}\right)^2.
\end{equation}
This is also shown in Fig.~\ref{Abb:Zeitskalen und Skalenlaengen}a and is roughly $\tauEinlok\sim1000$~yr at crossover
to 0.1~Myr at 2.3~Myr (within a factor of ten since $\ell_\rho$ and $\ell_{\rm Sedov}$ differ by at most a factor of four,
at 2.3~Myr).
The local impact timescale is thus $10^3$--$10^6$ times longer than the eddy mixing time.
Also, the local cooling time $\tau_{\rm cool} = E_{\rm impact}/L_{\rm int} = 4\pi\ell^3/3\times e_{\rm int}$
is 1--10~hours.
Therefore, the impacts are not spherically symmetric as far as the convection is concerned,
in the sense that each impact can be locally ``forgotten'' before the next one takes place.
As consequence, one would not expect the energy deposition of planetesimals to be a thermal impediment to convection.

\subsection{Summary}
In summary we find with this preliminary analysis that relative to the sinking case, the planetesimal impact heating during runaway could be reduced by a factor 2 to 3 for homogeneous mixing, and a factor 3 to 8 for no sinking. How this translates into post-formation luminosities needs to be assessed with future work. It seems that the reduction by less than an order of magnitude would still lead to luminosities  higher than in the classical models of \citet{marleyfortney2007} where no planetesimal impact heating at all occurs during this phase. A more definitive answer will however require a fully self-consistent treatment of the problem which is not trivial (non-spherical symmetry of the impacts and thus mass deposition, modification of the EOS,  local variation of the opacity, transport processes under the influence of compositional gradients, etc). However, such studies are important given the significant implications for the formation and evolution itself \citep[e.g.,][]{venturinialibert2016,vazanhelled2016,lozovskyhelled2017} as well as for the detectability via direct imaging.

\bibliographystyle{aa} %
\bibliography{biball2015.bib}

\end{document}